%% file: stokes_imaging_and_calibration.tex
\documentclass[a4paper,fleqn,usenatbib]{mnras}
\usepackage{newtxtext,newtxmath}
\usepackage[T1]{fontenc}
\usepackage{ae,aecompl}

\usepackage{graphicx}	% Including figure files
\usepackage{amsmath}	% Advanced maths commands
\usepackage{amssymb}	% Extra maths symbols
\usepackage{cancel}
\usepackage{amsfonts}
\usepackage{algorithm}
\usepackage{algpseudocode}
\usepackage{graphbox}
\usepackage{float}

\makeatletter
\def\BState{\State\hskip-\ALG@thistlm}
\makeatother

\usepackage{listings}
\usepackage{enumitem}
\usepackage{calc}

\usepackage{bm}
\usepackage{times}
\usepackage{cases}
\usepackage{textgreek}
\usepackage{caption}
\usepackage{tikz}
\usetikzlibrary{calc, shapes.geometric, arrows,patterns, matrix,fit,positioning, decorations.pathreplacing}
\usepackage{amsmath}
\tikzset{
    ncbar angle/.initial=90,
    ncbar/.style={
        to path=(\tikztostart)
        -- ($(\tikztostart)!#1!\pgfkeysvalueof{/tikz/ncbar angle}:(\tikztotarget)$)
        -- ($(\tikztotarget)!($(\tikztostart)!#1!\pgfkeysvalueof{/tikz/ncbar angle}:(\tikztotarget)$)!\pgfkeysvalueof{/tikz/ncbar angle}:(\tikztostart)$)
        -- (\tikztotarget)
    },
    ncbar/.default=0.5cm,
}
\tikzset{square left brace/.style={ncbar=0.2cm}}
\tikzset{square right brace/.style={ncbar=-0.2cm}}

\tikzset{round left paren/.style={ncbar=0.5cm,out=120,in=-120}}
\tikzset{round right paren/.style={ncbar=0.5cm,out=60,in=-60}}

\tikzstyle{arrow} = [thick,->]

\tikzstyle{cbox_blue} = [rectangle, minimum width=1.2cm, minimum height=0.05cm, fill=blue]
\tikzstyle{cbox_or} = [rectangle, minimum width=1.2cm, minimum height=0.05cm, fill=orange]
\tikzstyle{cbox_mag} = [rectangle, minimum width=1.2cm, minimum height=0.05cm, fill=magenta]
\tikzstyle{cbox_br} = [rectangle, minimum width=1.2cm, minimum height=0.05cm, fill=brown]

\tikzstyle{cbox_blue_light} = [rectangle, minimum width=1.2cm, minimum height=0.1cm, pattern=grid, pattern color=blue]

\tikzstyle{cbox_or_light} = [rectangle, minimum width=1.2cm, minimum height=0.1cm, pattern=grid, pattern color=orange]
\tikzstyle{cbox_mag_light} = [rectangle, minimum width=1.2cm, minimum height=0.1cm, pattern=grid, pattern color=magenta]
\tikzstyle{cbox_br_light} = [rectangle, minimum width=1.2cm, minimum height=0.1cm, pattern=grid, pattern color=brown]

\tikzset{
    >=stealth',
    punkt/.style={
           rectangle,
           rounded corners,
           draw=black, very thick,
           text width=6.5em,
           minimum height=2em,
           text centered},
    pil/.style={
           ->,
           thick,
           shorten <=2pt,
           shorten >=2pt,}
}

\graphicspath{{final_images/}}

\usepackage[english]{babel}
\usepackage{dashrule}
\usepackage{multirow}
\usepackage{tabularx}

\input alphabet
\input functions
\DeclareMathSymbol{\phi}{\mathalpha}{operators}{8}

\title[Joint DDE calibration and Stokes imaging for RI]
{Polca SARA - Full polarization, direction-dependent calibration and sparse imaging for radio interferometry}

\author[J. Birdi et al.]{
Jasleen Birdi,$^{1}$\thanks{E-mail: jb36@hw.ac.uk}
Audrey Repetti,$^{1,2}$
and Yves Wiaux$^{1}$
\\
$^{1}$ Institute of Sensors, Signals and Systems, Heriot-Watt University, Edinburgh EH14 4AS, UK \\
$^{2}$ Department of Actuarial Mathematics and Statistics, Heriot-Watt University, Edinburgh EH14 4AS, UK}

\date{Accepted XXX. Received YYY; in original form ZZZ}

\pubyear{2019}

\begin{document}
\label{firstpage}
\pagerange{\pageref{firstpage}--\pageref{lastpage}}
\maketitle

% Abstract of the paper
\begin{abstract}
New generation of radio interferometers are envisaged to produce high quality, high dynamic range Stokes images of the observed sky from the corresponding under-sampled Fourier domain measurements. In practice, these measurements are contaminated by the instrumental and atmospheric effects that are well represented by Jones matrices, and are most often varying with observation direction and time. These effects, usually unknown, act as a limiting factor in achieving the required imaging performance and thus, their calibration is crucial. 
To address this issue, we develop a global algorithm, named Polca SARA, aiming to perform full polarization, direction-dependent calibration and sparse imaging by employing a non-convex optimization technique.
In contrast with the existing approaches, the proposed method offers global convergence guarantees and flexibility to incorporate sophisticated priors to regularize the imaging as well as the calibration problem. Thus, we adapt a polarimetric imaging specific method, enforcing the physical polarization constraint along with a sparsity prior for the sought images. We perform extensive simulation studies of the proposed algorithm.
While indicating the superior performance of polarization constraint based imaging, the obtained results also highlight the importance of calibrating for direction-dependent effects as well as for off-diagonal terms (denoting polarization leakage) in the associated Jones matrices, without inclusion of which the imaging quality deteriorates. 
\end{abstract}

\begin{keywords}
techniques: interferometric -- techniques: polarimetric -- techniques: image processing -- techniques: calibration
\end{keywords}

%%%%%%%%%%%%%%%%%%%%%%%%%%%%%%%%%%%%%%%%%%%%%%%%%%

%%%%%%%%%%%%%%%%% BODY OF PAPER %%%%%%%%%%%%%%%%%%

\input{intro.tex}

%-------------------------------------------

\vspace{-0.2cm}
\input{obs_model.tex}
%-------------------------------------------

\input{nutshell.tex}

%-------------------------------------------

\vspace{-0.5cm}
\input{imaging.tex}
%-------------------------------------------

\vspace{-0.3cm}
\input{calibration.tex}
%-------------------------------------------

\vspace{-0.4cm}
\input{proposed_app.tex}

%-------------------------------------------

\input{sim_results.tex}
%-------------------------------------------

\vspace{-0.4cm}
\input{conclusion.tex}

%-------------------------------------------

\vspace{-0.2cm}
\section*{Acknowledgements}
We would like to thank R. Perley from NRAO for providing the two sets of images used for simulation studies. This work was supported by the UK Engineering and Physical Sciences Research Council (EPSRC, grant EP/M011089/1).

%%%%%%%%%%%%%%%%%%%%%%%%%%%%%%%%%%%%%%%%%%%%%%%%%%
%%%%%%%%%%%%%%%%%%%% REFERENCES %%%%%%%%%%%%%%%%%%

\bibliographystyle{mnras}
\bibliography{library_article} 

%%%%%%%%%%%%%%%%%%%%%%%%%%%%%%%%%%%%%%%%%%%%%%%%%%

%%%%%%%%%%%%%%%%%%%%%%%%%%%%%%%%%%%%%%%%%%%%%%%%%%
%%%%%%%%%%%%%%%%% APPENDICES %%%%%%%%%%%%%%%%%%%%%

\appendix

\section{Table of notations}
A list of useful notations and algorithmic parameters to be configured are provided in Table~\ref{tab:notations}.

\vspace{-0.5cm}
\begin{table}
\begin{center}
\begin{tabular}{l  l}
\hline 
${\Sbs}_0$ & Stokes matrix containing the high amplitude, \\
& thresholded part of $\overline{\Sbs}$ obtained with 1GC \\
${\Ebs}$ & Stokes matrix containing the unknown part of $\overline{\Sbs}$ \\
$\overline{\Dbs}_{t,\alpha}$ & DDEs for antenna $\alpha$ and time instant $t$ \\%, of size $2 \times 2N$ \\
$\overline{\Ubs}_{\alpha,1} = \overline{\Ubs}_{\alpha,2} = \overline{\Ubs}_\alpha$ & Compact-support Fourier kernels corresponding \\
& to $\overline{\Dbs}_{t,\alpha}$ \\
$\overline{\Ubs}_1 \, (\text{and}\, \overline{\Ubs}_2) $ & Concatenation of matrices $(\overline{\Ubs}_{\alpha,1})_{1 \leq \alpha \leq n_a}$ \\
& (and $(\overline{\Ubs}_{\alpha,2})_{1 \leq \alpha \leq n_a})$)\\
$K \, \text{and}\, P$ & Support sizes in spatial and temporal  \\
& Fourier domains of DDEs, respectively \\
$L^{(i)} \, \text{and}\,\, J^{(i)}$ & Number of inner-loop FB iterations for DDEs and \\
& images updates, respectively \\
$\bm{\eta} = [\eta_1, \eta_2, \eta_3]$ & Regularization parameters for the images\\
${\gamma}$ & Regularization parameter for the DDEs  \\
$\varepsilon_\Us \, \text{and}\, \varepsilon_\Es $ & Stopping criterion for the DDEs and images \\
& update, respectively \\
$\varepsilon_0$ & Global stopping criterion for Algorithm~\ref{algo_polcal} \vspace{0.2cm} \\ 
\hline
\end{tabular}
\caption{\label{tab:notations}
Important variable notations and configurable algorithmic parameters for Polca SARA method (Algorithm~\ref{algo_polcal}).}
\end{center}
\end{table}

\bsp	% typesetting comment
\label{lastpage}
\end{document}

%% file: functions.tex
\newcommand{\minimize}[2]{\ensuremath{\underset{\substack{{#1}}}
{\mathrm{minimize}}\;\;#2 }}
\newcommand{\prox}{\ensuremath{\operatorname{prox}}}

\renewcommand{\leq}{\ensuremath{\leqslant}}
\renewcommand{\geq}{\ensuremath{\geqslant}}
\renewcommand{\le}{\ensuremath{\leqslant}}
\renewcommand{\ge}{\ensuremath{\geqslant}}

%% file: intro.tex
\section{Introduction} \label{sec:intro}

In radio interferometry (RI), the acquired measurements are related to the Fourier coefficients of the sky brightness distribution under observation \citep{Thompson2001}. These sampled frequency points in the Fourier domain are dictated by the separations between the antenna pairs constituting the interferometer. A finite number of antennas within an interferometer leads to under-sampling of the Fourier domain associated with the sought images. Given such measurements, the task to recover the images of interest can be performed by using a suitable imaging technique in RI. In fact, very often, the target radio sources generate polarized radiations \citep{Ginzburg1965}. Study of these polarized emissions is crucial for various scientific goals (e.g. to determine the magnetic field distributions around the source and along the signal's propagation path \citep{Pacholczyk1970}), offering a way to probe the details in addition to those provided by the total intensity alone. For this purpose, Stokes parameters- $I,Q,U$ and $V$ provide a representation of the sky intensity distribution, where the first parameter denotes the total intensity whereas the others characterize the polarization state of the radio emissions. In particular, while the linear polarization is described by the Stokes parameters $Q$ and $U$, the circular polarization is denoted by Stokes $V$. Conventionally speaking, although the main focus has been to develop algorithms for total intensity imaging, for the purpose of imaging these parameters, referred to as polarimetric imaging, the same techniques as developed for Stokes $I$ imaging have been applied for imaging each of the Stokes parameters.

In practice, rather than providing an accurate representation of the true sky distribution, the RI measurements are often corrupted both by the atmospheric effects encountered by the radio emissions en route to the receiving antennas and by the antenna-based errors. These measurement-contaminating effects are represented in the form of Jones matrices \citep{Smirnov2011}. Only in the cases when these effects are absent or known accurately, image recovery amounts to use of any RI imaging algorithm. In practical cases, where these effects are unknown, these need to be estimated along with imaging to achieve good quality reconstruction. Calibration is this process of estimation of the Jones matrices. Moreover, these time-variable calibration terms in general also exhibit direction dependency, thereby encompassing both direction-independent and -dependent effects (DIEs and DDEs, respectively). For instance, while the antenna-based gains denote the DIEs, the antenna primary beam pattern, atmospheric phase delays etc. exhibit spatial variation within the field of view and thus constitute DDEs. Historically, only the DIE calibration has been performed \citep{Thompson2001}. However, the calibration of DDEs has become a crucial issue especially for the new generation radio telescopes including the LOw Frequency ARray (LOFAR) ({http://www.lofar.org}) and the upcoming Square Kilometre Array (SKA) ({https://www.skatelescope.org}), aiming to produce sky images at unprecedented resolution with high sensitivity. Without efficient DDE calibration, these telescopes cannot be exploited to their fullest potential. The incorporation of the DDEs in the calibration process is also essential to produce high quality images without limiting their dynamic ranges \citep{Bhatnagar2013}. Thus, the resultant problem of image recovery from the acquired corrupted data consists in estimating both the sought images and the DDE calibration terms. We now review the imaging and calibration techniques in RI, which can then be combined to solve the overall problem.

In the context of RI imaging techniques, the standard $\textsc{clean}$ algorithm implements a non-linear, greedy approach to perform deconvolution in an iterative manner \citep{Hogbom1974}. Working pixelwise, \textsc{clean} implicitly assumes sparsity of the sought image, removing at each iteration, a fraction of the maximum intensity pixel convolved with the dirty beam from the computed residual image. Over the past years, many variants of this celebrated algorithm have been developed, for instance, its multi-scale version \citep{Cornwell2008}. In addition to these \textsc{clean} based approaches, recent developments in the field of compressive sensing (CS) applied for astronomical imaging have given birth to many imaging algorithms, particularly for radio interferometry \citep{Wiaux2009, Carrillo2012, Garsden2015, Onose2016}. Leveraging optimization framework, these methods aim to solve the underlying image recovery problem by enforcing sparsity of the image of interest in some suitable domain. As a matter of fact, these techniques have shown the potential to surpass the image reconstruction quality obtained by \textsc{clean} based approaches \citep{Carrillo2014, Onose2016, Pratley2016, Onose2017, Dabbech2018}. While the aforementioned methods have been developed mainly for Stokes $I$ imaging, as previously mentioned these can be extended to polarimetric imaging by following the same imaging approach for all the Stokes parameters. In the context of sparsity regularized approaches, one such extension for polarimetric imaging has been presented in \cite{Akiyama2017a}, promoting sparsity of each of the Stokes parameters using $\ell_1$ norm combined with total variation (TV) regularization \citep{Rudin1992}. A point worth noting here is that the above mentioned approaches, whether \textsc{clean} and its variants or the sparsity regularized method, solve for the Stokes parameters totally independently. However, these images are physically linked via polarization constraint, that is the polarized intensity cannot be greater than the total intensity. Within the CS framework, this constraint has been exploited by a recently proposed approach, namely Polarized SARA, estimating jointly the Stokes parameters \citep{Birdi2018a, Birdi2018b}. This approach has been shown to provide better reconstruction quality in comparison with the case when the constraint is not accounted for.  

Concerning the calibration problem, it can be formulated as a non-linear least squares problem and to solve it, most of the existing solvers employ gradient-based techniques like the Levenberg-Marquardt (LM) algorithm \citep{More1978}. But it suffers from its high computational cost. To overcome this issue, an efficient and fast solver named StEFCal has been proposed in \cite{Salvini2014}, which relies on an alternating direction implicit method. However, it solves only for the DIEs. In the context of DDE calibration, several methods have been proposed in the last years. 
In particular, the Source Peeling and Atmospheric Modeling (SPAM) method proposed in \cite{Intema2009} models the ionosphere as a phase screen, whose parameters are estimated by fitting the calibrated phases obtained from peeling \citep{Noordam2004} with the modelled phases, using the LM algorithm. Subsequently, this model can be used for the prediction of correction phases in any arbitrary desired direction within the field of view. To apply these corrections, SPAM relies on a facet-based approach, applying the DDE calibration solution for the facet centre, defined by a bright source or approximate centre of a cluster of closely located bright sources, to the whole facet. This scheme works under the assumption that the DDEs show smooth variation over the field of view, thereby having (approximately) constant value across a given facet. This approach has analogies with the facet calibration which consists in partitioning the sky into a number of facets and solving for the DDE solution of the facet centre using a self-calibration (SelfCal) loop, which in turn is applied to the whole facet \citep{van2016}. Despite some similarities, these two approaches are different in the way they obtain the calibration solutions, i.e. unlike SPAM, facet calibration does not make use of global phase screen model. In contrast to these approaches solving for the calibration solutions source by source (or in terms of facets), SAGE calibration algorithm solves for the calibration solutions for all sources at once, employing an extension of the Expectation Maximization algorithm \citep{Yatawatta2009,Kazemi2011}. It therefore provides increased convergence speed in comparison with the traditional LM solvers. It is to be noted that within a given iteration, SAGE algorithm requires to define partitioning over the unknown Jones parameter space depending on the source characteristics. Another approach for DDE calibration arises from the convolutional nature of the DDEs. More precisely, the multiplication of the DDEs in the image domain can be equivalently seen as their convolution in the Fourier domain. The A-Projection algorithm makes use of the latter characteristic and offers a way to correct for those \textit{known} DDEs whose associated Jones matrix are (approximately) unitary, by embedding them into the forward (degridding) and backward (gridding) operators \citep{Bhatnagar2013}. While this approach was only developed for the correction of \textit{known} DDEs, it has been incorporated in the Pointing SelfCal algorithm to solve for antenna pointing errors \citep{Bhatnagar2017}.

In order to estimate the sky model while having imperfect knowledge of calibration terms, any of the above mentioned calibration algorithms is generally combined with \textsc{clean} (or its variants) for image recovery step. As such, the global algorithm fails to have any convergence guarantees. The first step in the direction of addressing this issue has been made by the joint calibration and imaging algorithm proposed in \cite{Repetti2017}. This algorithm leverages non-convex optimization techniques to present a global algorithm which alternates between the estimation of the image of interest and the DDEs, while benefitting from the convergence guarantees. In addition to this, unlike the approaches requiring sky partitioning, which may not always be best achieved in an automatic manner, this technique works on the whole image all at once with minimal user intervention. However, this approach has been developed only for Stokes $I$ imaging and calibration, without dealing with the full polarization model. As a matter of fact, in the case of full Stokes imaging and calibration, even the global algorithm consisting of combination of \textsc{clean} based imaging and any of the previously mentioned calibration techniques, do not adopt any polarimetric imaging specific approach. These factors raise the scope for the development of a globally convergent algorithm which not only incorporates full polarization model, but also uses advanced approaches specific for full Stokes imaging to produce good quality images.

The work presented here lies within this scope and we propose Polca SARA - a full polarization, calibration and sparse imaging algorithm with proven convergence guarantees. In particular, we build our method on the approach developed in \cite{Repetti2017}, and later on extended in \cite{Repetti2017b} and \cite{Thouvenin2018}, assuming spatial and temporal smoothness of the DDEs. On the one hand, we generalize this approach to the full polarization model, developing an algorithm alternating between the estimation of the DDEs and the Stokes images. On the other hand, thanks to the underlying non-convex optimization technique, the proposed approach can deal with sophisticated priors suited to the images under consideration as well as to the DDEs. Leveraging this flexibility, we adapt Polarized SARA method, specifically designed for Stokes imaging, to be used for imaging step in the considered case.

The remainder of the article is organised as follows. {Section~\ref{sec:obs_model} describes the full polarization model encountered in RI. The proposed approach to solve the underlying joint calibration and imaging problem is presented briefly in Section~\ref{sec:overview}. A more formal and mathematical presentation of the proposed approach is provided in later sections. Particularly, the imaging problem to be solved is posed in Section~\ref{sec:im_prob}, followed by the introduction of the calibration problem in Section~\ref{sec:cal_prob}.} A blend of these problems to give the resultant joint calibration and imaging problem and the proposed algorithm to solve it are detailed in Section~\ref{sec:cal_im}. To assess the performance of the proposed algorithm, we perform various tests which are presented in Section~\ref{sec:results}. We give the concluding remarks and discuss about possible directions of future work in Section~\ref{sec:conc}.

%% file: obs_model.tex
\section{Full polarization observation model}
\label{sec:obs_model}

The radio interferometric measurements, termed as visibilities, are acquired by antenna pairs. Within $n_a$ number of total antennas, consider an antenna pair $(\alpha, \beta) \in \{1, \ldots, n_a\}^2$, at time instant $t \in \{1,\ldots,T\}$, associated with the baseline components $(u_{t, \alpha, \beta}, v_{t, \alpha, \beta}, w_{t, \alpha, \beta})$ expressed in observation wavelength units. Here, $\bm{u}_{t, \alpha, \beta} = (u_{t, \alpha, \beta}, v_{t, \alpha, \beta})$ represents the components in the plane normal to the target source direction, while $w_{t, \alpha, \beta}$ is the component in the source direction. Furthermore, let $\overline{\bm{S}} = \bigl[\begin{smallmatrix} I & Q\\ U & V\end{smallmatrix} \bigr]$ encompasses the Stokes parameters of the target sky area, described in terms of the direction cosines $(l,m,n)$ with $\bm{l} = (l,m)$ lying in the tangential plane to the celestial sphere and $n(\bm{l}) = \sqrt{1 - l^2 - m^2}$ lying in the line of sight. Then, the measurements $\Ybs_{t,\alpha,\beta, \nu} \in \eC^{2 \times 2}$ made by the antenna pair $(\alpha, \beta)$ at time instant $t$ and observation frequency $\nu$ is given by the radio interferometric measurement equation (RIME) \citep{Hamaker1996, Smirnov2011} as follows
\begin{equation} \label{eq:cont_prob}
 \displaystyle \Ybs_{t, \alpha,\beta, \nu} = \int \overline{\bm{D}}_{t,\alpha, \nu}(\bm{l}) \, \mathcal{L}\big(\overline{\bm{S}}_{t, \nu}(\bm{l}) \big) \, \overline{\bm{D}}_{t,\beta, \nu}^\dagger (\bm{l}) \es^{-2 \is \pi \bm{u}_{t, \alpha, \beta}  \cdot \bm{l}} d^2 \bm{l},
\end{equation} 
where $(\cdot)^\dagger$ denotes the Hermitian conjugate of its argument and $\mathcal{L}\big(\overline{\bm{S}}_{t, \nu}(\bm{l}) \big) = \overline{\bm{B}}_{t, \nu} (\bm{l})$ is the brightness matrix formed by the application of the linear operator $\mathcal{L}$ on the components of the Stokes matrix $\overline{\bm{S}}$. For instance, considering linear feeds, the brightness matrix is given by 
$\overline{\bm{B}}_{t, \nu} (\bm{l}) = \biggl[ \begin{smallmatrix}
I + Q & U + \text{i} V \\
 U - \text{i} V & I - Q
\end{smallmatrix} \biggr]_{t, \nu}  (\bm{l})$, whereas for circular feeds, it reads as $\overline{\bm{B}}_{t, \nu} (\bm{l}) = \biggl[ \begin{smallmatrix}
I + V & U + \text{i} Q \\
 U - \text{i} Q & I - V
\end{smallmatrix} \biggr]_{t, \nu}  (\bm{l})$.
Furthermore, in equation~\eqref{eq:cont_prob}, for every antenna $\alpha \in  \{1,\ldots,n_a\}$, the direction-independent and -dependent effects (DIEs and DDEs, respectively) at time $t$ and frequency $\nu$ are encoded in the Jones matrix $ \overline{\bm{D}}_{t,\alpha, \nu}$. It is worth mentioning that equation~\eqref{eq:cont_prob} presents a general case where the Stokes images and the DIEs/DDEs can vary with respect to the observation frequency and time. However, in the current work, we consider the Stokes images without having any time or frequency dependence. In addition, for the DIEs/DDEs, we deal with their temporal dependency considered at a single observation frequency. Hence, hereafter, we take $\overline{\bm{S}}_{t, \nu}(\bm{l}) = \overline{\bm{S}}(\bm{l})$ and drop the frequency index from the other variables.

\begin{figure}
    \begin{center}
    
\begin{tikzpicture}
\draw [black, line width=0.2mm] (0,1) to [square left brace ] (0,2.4);
    
    \begin{tabular}{c c}
      \node (n1) at (0.4,1.6) {\small $(\overline{\Dbs}_{t,\alpha})_{11}$}; & 
      \node (2) at (1.4,1.6) {\small $(\overline{\Dbs}_{t,\alpha})_{12}$}; \\
      \node (11) at (0.5,1.8) [cbox_blue] {}; & 
      \node (12) [cbox_or, right of=11, xshift=0.4cm] {}; \\
      \node at (0.4,1.65) {\small $(\overline{\Dbs}_{t,\alpha})_{21}$}; & 
      \node at (1.4,1.65) {\small $(\overline{\Dbs}_{t,\alpha})_{22}$}; \\
      \node (21) at (0.5,1.16) [cbox_br] {}; & 
      \node (22) [cbox_mag, right of=21, xshift=0.4cm] {}; \\
    \end{tabular}
    \draw [black, line width=0.2mm] (2.5,1) to [square right brace] (2.5,2.4);
\vspace{2cm}
\centering \node at (0.5,0.7) {$\overline{\Dbs}_{t,\alpha}$};

\draw [decorate,decoration={brace,amplitude=2mm},yshift=10pt,thick,red] (0,2) -- (1.1,2) node [black,midway,xshift=-0.8cm, yshift=0.35cm] {\footnotesize
$N$};

\node at (2.5,1.8) {$\overset{\textbf{FT}}{\bm{\longrightarrow}}$};

\draw [black, line width=0.2mm] (4,1) to [square left brace ] (4,2.4);
    
     \begin{tabular}{c c}
      \node at (3.5,1.6) {\small $(\widehat{\overline{\Dbs}}_{t,\alpha})_{11}$}; & 
      \node at (4.5,1.6) {\small $(\widehat{\overline{\Dbs}}_{t,\alpha})_{12}$}; \\
      \node (11f) at (4.5,1.8) [cbox_blue] {}; & 
      \node (12f) [cbox_or, right of=11f, xshift=0.4cm] {}; \\
      \node at (3.5,1.65) {\small $(\widehat{\overline{\Dbs}}_{t,\alpha})_{21}$}; & 
      \node at (4.5,1.65) {\small $(\widehat{\overline{\Dbs}}_{t,\alpha})_{22}$}; \\
      \node (21f) at (4.5,1.16) [cbox_br] {}; & 
      \node (22f) [cbox_mag, right of=21f, xshift=0.4cm] {}; \\
     \end{tabular}
     \draw [black, line width=0.2mm] (6.5,1) to [square right brace] (6.5,2.4);
 \vspace{2cm}
 \centering \node at (3.6,0.7) {$\widehat{\overline{\Dbs}}_{t,\alpha}$};
\end{tikzpicture}
 \end{center}
 \vspace{-0.5cm}
    \caption{Schematic representation of a $2 \times 2 N$ matrix, $\overline{\bm{\mathsf{D}}}_{t,\alpha}$ (left) that can be seen as a $2 \times 2$ block matrix, with each block $(\overline{\bm{\mathsf{D}}}_{t,\alpha})_{i i^\prime} , \{i, i^\prime\} \in \{1,2\}^2$ containing a vectorized image of dimension $N$. The Fourier transformation (FT) of this matrix computes 2D Fourier transform of each of the block images. For each block in matrix $\overline{\bm{\mathsf{D}}}_{t,\alpha}$ on left, its corresponding FT is shown by same colored vectors in $\widehat{\overline{\bm{\mathsf{D}}}}_{t,\alpha}$ on right.}
    \label{fig:FT_Jones}
    \label{fig:block_matrix}
\end{figure}
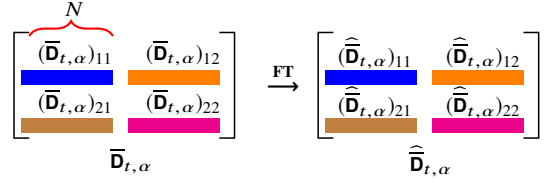

To recover the Stokes parameters from the given measurements, we formulate the inverse problem~\eqref{eq:cont_prob} in the discrete domain. It amounts to sampling the continuous variables such that the field of view is discretized into a grid, with the vectorized form of this grid being represented by the index $n \in \{-N/2, \ldots, N/2-1\}$. Within this representation, we thus have the Stokes matrix $\overline{\Sbs} \in \eR^{2 \times 2 N}$, and for each antenna $\alpha$ and time instant $t$, the DDEs $\overline{\Dbs}_{t, \alpha} \in \eC^{2 \times 2 N}$. As illustrated in Fig.~\ref{fig:block_matrix}, these matrices can also be seen as $2 \times 2$ block matrices\footnote{For any such matrix $\overline{\Sbs}$, $\overline{\Sbs}(n)$ denotes the $2 \times 2$ matrix consisting of the $n^{\text{th}}$ elements of each of the blocks in the parent block matrix $\overline{\Sbs}$. Furthermore, for $(i,i^\prime) \in \{1,2\}^2$, $[\overline{\Sbs}(n)]_{i i^\prime}$ refers to the $n^{\text{th}}$ element of the row vector block contained in the $i^{\text{th}}$ row and ${i^\prime}^{\text{th}}$ column of the argument block matrix.}, whose each block is a row vector of dimension $N$.  
In particular, for the Stokes matrix $\overline{\Sbs} = \bigl[ \begin{smallmatrix} \overline{\bm{s}}_1 & \overline{\bm{s}}_2 \\ \overline{\bm{s}}_3 & \overline{\bm{s}}_4 \end{smallmatrix}\bigr]$, the elements $\overline{\bm{s}}_1, \overline{\bm{s}}_2, \overline{\bm{s}}_3$ and $\overline{\bm{s}}_4$, each of size $N$, are respectively the discretizations of the Stokes parameters $I, Q, U$ and $V$. On the other hand, for the DDEs, for each index $n$, we have the so called $2 \times 2$ Jones matrix $\overline{\Dbs}_{t,\alpha}(n) \in \eC^{2 \times 2}$. The diagonal elements of this matrix correspond to the antenna's complex voltage pattern, whereas the off-diagonal elements account for the polarization leakage terms arising from instrumental leakage, transmission effects, etc \citep{Bhatnagar2013}. Moreover, if for every $(i ,i^\prime) \in \{1,2\}^2$, $[\overline{\Dbs}_{t,\alpha}]_{i i^\prime} = \delta_{t,\alpha} \bm{1}_N$ with $\delta_{t,\alpha} \in \eC$ and $\bm{1}_N$ being $N$ dimensional unitary row vector, then $\overline{\Dbs}_{t,\alpha}$ reduces to a DIE.

Following the introduced notations, each component of the visibility matrix $\Ybs_{t, \alpha,\beta}$, indexed by $[\cdot]_{i i^\prime}$ with $(i,i^\prime) \in \{1,2\}^2$, at discrete spatial frequency $k_{t,\alpha, \beta}$ can be represented as
%%-----------------------------------
\begin{multline} \label{eq:obs_model}
$$[\Ys_{t,\alpha,\beta}]_{i i^\prime} = 
\displaystyle \sum_{j, j^\prime= 1}^{2} \sum_{n=-N/2}^{N/2-1} \big[\overline{\Dbs}_{t,\alpha}(n)\big]_{ij} \big[\mathcal{L}\big(\overline{\Sbs}(n) \big) \big]_{j j^\prime} \big[\overline{\Dbs}^*_{t,\beta}(n) \big]_{j^\prime i^\prime} 
\\ \es^{-2\is \pi (k_{t,\alpha,\beta}) \frac{n}{N}} + [{\Omega}_{t,\alpha, \beta}]_{i i^\prime},$$
\end{multline}
where $(\cdot)^*$ stands for complex conjugation of its argument and the visibilities are corrupted by an additive Gaussian noise ${\bm{\Omega}}_{t,\alpha, \beta} \in \eC^{2 \times 2}$. For $n \in \{-N/2, \ldots, N/2 - 1\}$, as for the continuous version~\eqref{eq:cont_prob}, $\mathcal{L}\big(\overline{\Sbs}(n) \big) = \overline{\Bbs} (n) \in \eC^{2 \times 2}$ is the brightness matrix. 

%% file: nutshell.tex
\section{A brief overview of the proposed method}
\label{sec:overview}

 While a more formal description of the proposed approach is presented in the later sections, this section is designed to give an intuitive explanation of the approach. Particularly, the description in this section is targeted to facilitate readers to understand the results and conclusion sections without delving into the mathematical details provided in Sections~\ref{sec:im_prob}-~\ref{sec:cal_im}. 
 
 Given the measurements of the form~\eqref{eq:obs_model}, the aim is to recover the images of interest. The problem of image recovery becomes more challenging when the DDEs $\overline{\Dbs}$ are unknown and need to be estimated along with the images. The current work is devoted to this more challenging yet practical case, for which we propose a joint calibration and imaging algorithm to solve the underlying problem. The resultant approach is dubbed as Polca SARA.  
 
 In order to calibrate for the DDEs, we consider their smooth variation both across the field of view and in time, implying that they are spatially and temporally band-limited \citep{Repetti2017, Thouvenin2018}. Exploiting this characteristic, we instead represent the DDEs $\overline{\Dbs}$ by their compact-support kernels $\overline{\Ubs}$ in the spatial and temporal Fourier domain. Then, only these non-zero Fourier coefficients need to be estimated. It reduces significantly the dimension of the underlying problem. Furthermore, we note that while equation~\eqref{eq:obs_model} is linear with respect to $\overline{\Sbs}$ (when the DDEs are kept fixed), it is non-linear with respect to the DDEs. To counteract this issue, we draw on the notion of bi-linearity proposed in \cite{Salvini2014} and \cite{Repetti2017} in the context of RI calibration. It consists in introducing $\Ubs_1 = \Ubs_2 = \Ubs$ such that problem~\eqref{eq:obs_model} becomes bi-linear with respect to the DDEs. The resultant problem can then be written as
 \begin{equation}
     \Ybs = \phi (\Sbs, \Ubs_1, \Ubs_2) + \bm{\Omega},
 \end{equation}
 where $\phi$ is the tri-linear measurement operator, mapping the images of interest coupled with the DDE Fourier kernels $\Ubs_1$ and $\Ubs_2$ to the acquired visibilities $\Ybs$. 
 On the imaging front, we further consider that calibration transfer has been performed and thus, the Jones matrices in equation~\eqref{eq:obs_model} can be taken as identity as a first instance, without any directional dependency. Doing so, we first solve the problem to obtain an estimation $\Sbs^\prime$ of the unknown images. Nevertheless, the used measurement model being devoid of DDEs, the estimated images are likely to contain artefacts. We thus instead keep a thresholded version $\Sbs_0$ of $\Sbs^\prime$, considering that $\Sbs = \Sbs_0 + \Ebs$, where $\Ebs$ is unknown and needs to be estimated. 
 
 In view of the discussion above, we propose to estimate the variables of interest by solving the following problem:
 \begin{equation} \label{eq:brief_min}
     \underset{\Ebs, \Ubs_1, \Ubs_2}{\operatorname{minimize}} \,\,
     h(\Ebs, \Ubs_1, \Ubs_2) +   r(\Ebs) +  p(\Ubs_1, \Ubs_2),
 \end{equation}
 where the data-fidelity term $h$ is given by a least-squares criterion (equations~\eqref{eq:data_fid_im}-\eqref{eq:data_fid_cal2}) and the regularization terms are defined both for the images ($r$ in  equation~\eqref{eq:reg_im_final}) and the DDEs ($p$ in  equation~\eqref{eq:reg_cal}).

 \begin{figure}
\centering 
\includegraphics[width=8cm]{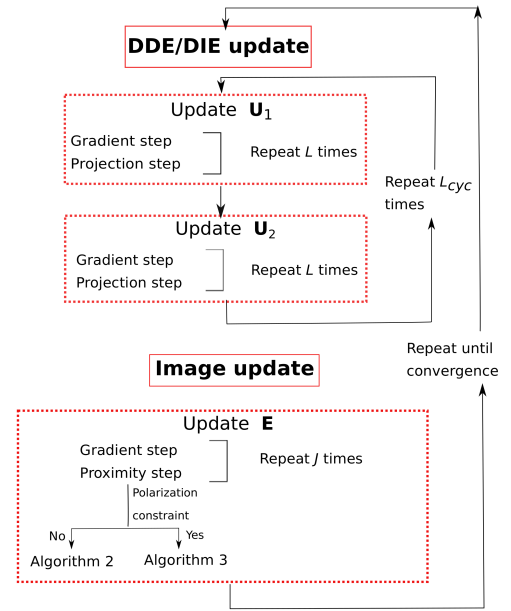}
\vspace{-3.4cm}
\caption{{Block diagram illustrating the proposed Polca SARA method that alternates between the DDEs/DIEs update and the Stokes images update until convergence. For DDEs/DIEs update, $L \in \eN$ forward-backward iterations are performed for the estimation of each of the matrices $\Ubs_1$ and $\Ubs_2$. Similarly, for the image update, $J \in \eN$ forward-backward steps are executed, wherein depending on the chosen regularization, the proximity step is either evaluated using Algorithm~\ref{algo:dualfb} or Algorithm~\ref{algo:pd_fb}}.
} 
\label{fig:block_diag}
\end{figure}

 For the images of interest, we exploit the compressive sensing framework \citep{Candes2006} and consider $\ell_1$ regularization term to promote sparsity of the sought images in a suitable dictionary. More precisely, inspired by the good performance of the sparsity averaging reweighted analysis (SARA) prior for Stokes $I$ imaging in \cite{Carrillo2013, Carrillo2014} and for full Stokes imaging in \cite{Birdi2018b}, we employ this prior consisting in promoting average sparsity over a concatenation of Dirac basis and first eight Daubechies wavelet bases. It is coupled with a reweighting scheme, solving sequentially the $\ell_1$ regularized problems \citep{Candes2008}. Although the current work does not deal with the reweighting procedure, it can be easily incorporated. In practice, other sophisticated priors injecting additional information about the images of interest can also be taken into account in the regularization term, eg. polarization constraint in \cite{Birdi2018b}. In the current work, we consider both the cases without and with the enforcement of the polarization constraint. For the DDEs, the regularization term $p$ is defined to impose values of the Jones matrices' elements to lie within specified bounds as well as to control the distance between $\Ubs_1$ and $\Ubs_2$ which ideally should be equal. Additional details for the image (resp. DDEs) estimation are provided in Section~\ref{sec:im_prob} (resp. Section~\ref{sec:cal_prob}).
 
 The resultant problem~\eqref{eq:brief_min} is solved using a block-coordinate forward-backward approach \citep{Chouzenoux2016}. As the name suggests, this iterative procedure relies on a forward-backward (FB) scheme to update each of the blocks/variables of interest $(\Ubs_1, \Ubs_2, \Ebs)$. The FB scheme consists in treating the smooth term with a gradient step and the non-smooth term via a proximity step. The global method is detailed in Section~\ref{sec:cal_im}. For a better understanding, the overall algorithmic structure is depicted in Fig.~\ref{fig:block_diag}. It has similarities with the traditional self-calibration method in the sense that it alternates between the imaging and calibration steps. Yet the two approaches are different in terms of their underlying principle and implementation. First, the imaging and calibration steps in our approach use the same optimization toolbox and hence, the global algorithm benefits from convergence guarantees. Second, to ensure this global convergence, each of these updates is obtained by performing only a finite number of iterations and not until convergence. {Moreover, these updates must follow a cyclic rule, that is each variable must be updated at least once every given number of iterations. To this end, $L_{cyc}$ global iterations are performed to update the DDEs $(\Ubs_1,\Ubs_2)$, alternating between $L$ FB iterations on $\Ubs_1$ and $L$ FB iterations on $\Ubs_2$. After $L_{cyc}$ global iterations, the Stokes images $\Ebs$ are updated by executing $J$ FB iterations.}

 The results obtained by implementing Polca SARA are presented in Section~\ref{sec:results}. Particularly, we compare between the following cases: on the calibration front- considering normalized DIEs (i.e. identity Jones matrices), DIE calibration, Jones matrices DDE calibration without considering off-diagonal terms, and full Jones matrices DDE calibration; and on the imaging front- in the absence and presence of the polarization constraint. While the calibration cases are chosen to judge the importance of calibrating for the off-diagonal terms as well as for the DDEs, the two imaging cases are to test the efficiency of the polarization constraint in the context of joint calibration and imaging. Furthermore, concerning these imaging cases, when the polarization constraint is not imposed, the proximity step for the update of $\Ebs$ is evaluated using Algorithm~\ref{algo:dualfb}, whereas in the case of enforcing this constraint, Algorithm~\ref{algo:pd_fb} is employed to compute the corresponding proximity step.

 A detailed mathematical formulation of the proposed approach is presented in the next sections, starting with the description of the imaging problem followed by the calibration problem and finally combining the two. On the other hand, with the intuitive description provided in this section at disposal, the reader might jump directly to the simulations and results section (Section~\ref{sec:results}) averting the mathematical details.

%% file: imaging.tex
\section{Imaging problem}
\label{sec:im_prob}

The case of having either pre-calibrated data or knowledge of DIEs/DDEs beforehand amounts to solving the problem only for image recovery. The RI literature is brimming with such methods aiming to estimate the images of interest from the underlying imaging problem. Nevertheless, most of these methods were initially proposed mainly for Stokes $I$ imaging, including $\textsc{clean}$ based methods \citep{Hogbom1974, Bhatnagar2004, Cornwell2008} and CS techniques \citep{Wiaux2009, Carrillo2014, Garsden2015, Onose2016, Onose2017}. In order to apply these methods for full polarization imaging, these can be extended to the estimation of each of the Stokes parameters totally independently using the same techniques as developed for Stokes $I$ imaging. Such an extension of CS based Stokes $I$ imaging to full polarization imaging has been presented in \cite{Akiyama2017a}, solving independently for the sought images. A more evolved approach aiming to jointly estimate the Stokes parameters has been very recently proposed in \cite{Birdi2018a, Birdi2018b}. In this section, we aim to summarize this work which was proposed in the context of DIEs, and generalize it to the DDEs case, which will be useful for the understanding of the joint imaging and calibration method developed in Section~\ref{sec:cal_im}.

In this context, a reformulation of equation~\eqref{eq:obs_model} is more commonly used where the measurements made by an antenna pair $(\alpha, \beta)$ at time instant $t$ are represented by a vector $\yb_{t,\alpha,\beta} \in \eC^{4}$, whose each element $q \in \{1,\ldots,4\}$ is given by
%%-----------------------------------
\begin{multline} \label{eq:obs_model_im}
    [\yb_{t,\alpha,\beta}]_q = \displaystyle \sum_{z= 1}^{4} \sum_{n=-N/2}^{N/2-1} [\overline{\Mbs}_{t,\alpha,\beta}(n)]_{qz} \big[\mathcal{\widetilde{L}}\big(\overline{\Sbs}(n) \big) \big]_{z} \\
    \es^{-2\is \pi (k_{t,\alpha,\beta}) \frac{n}{N}} + [{\omega}_{t,\alpha, \beta}]_q.
\end{multline}
with $\bm{\omega}_{t,\alpha, \beta} \in \eC^{4}$ being a realization of additive Gaussian noise, and $\mathcal{\widetilde{L}}\big(\overline{\Sbs}(n) \big)$ operates on the Stokes matrix to produce the brightness matrix $\overline{\Bbs}(n)$, followed by its vectorization. 
Furthermore, $\overline{\Mbs}_{t,\alpha,\beta} = (\overline{\Dbs}_{t,\alpha} \otimes \overline{\Dbs}_{t,\beta}^\dagger) \in \eC^{4 \times 4N}$ is the Mueller matrix formed by the outer product of the Jones matrices for antennas $\alpha$ and $\beta$ at time instant $t$, and in turn, this matrix can be seen as a $4 \times 4$ block matrix with each block given by $N$ dimensional row vector. Let us note that, for the Fourier transformed Mueller matrix, denoted by $\widehat{\overline{\Mbs}}_{t,\alpha,\beta}$, each block is obtained by the complex convolution of the Fourier transforms of the corresponding Jones matrices \citep{Bhatnagar2013}.

Using equation~\eqref{eq:obs_model_im}, the overall measurement model can be written as
\begin{equation} \label{eq:im_model}
    \bm{y} = \Phi(\overline{\Sbs}) + \bm{\omega},
\end{equation}
where $\bm{y} \in \eC^{4 M}$ is the resultant measurement vector corrupted by the noise vector $\bm{\omega} \in \eC^{4 M}$. The measurement operator $\Phi$ mapping the images of interest to the acquired visibilities is modelled as
\begin{equation} \label{eq:meas_op}
    \Phi (\overline{\Sbs}) = \Gbs \Fbs \widetilde{\Zbs} \mathcal{\widetilde{L}} (\overline{\Sbs}),
\end{equation}
that can be explained as follows. The operation $ {\mathcal{\widetilde{L}}}\big(\overline{\Sbs}\big)$ generates the brightness vector $\bm{b} \in \eC^{4N}$ that needs to be Fourier transformed at the sampled spatial frequencies. To evaluate these Fourier transforms in a computationally efficient manner, we make use of the non-uniform fast Fourier Transform that relies on interpolation of the Fourier coefficients from discrete to continuous domain \citep{Fessler2003}. In this context, the zero-padding matrix $\widetilde{\Zbs} \in \eC^{\gamma 4 N \times 4 N}$ oversamples the images in $\bm{b}$, by factor $\gamma$ in each dimension and also accounts for the scale factors to pre-compensate for the interpolation convolution kernels. A fast Fourier transform operator $\Fbs \in \eC^{\gamma 4N \times \gamma 4N}$ is then applied to compute the 2D Fourier transform of each of these oversampled images. To interpolate these discrete Fourier coefficients to the continuous frequency points, the compact support convolution kernels (the so called degridding kernels) are embedded in the matrix $\Gbs \in \eC^{4M \times \gamma 4 N}$. More precisely, the matrix $\Gbs$ is modelled to take into account the degridding kernels as well as the components of the Fourier transformed Mueller matrix. In particular, each of the four rows of $\Gbs$ associated with a frequency $k_{t,\alpha,\beta}$ contains the convolution of the corresponding elements of $\widehat{\overline{\Mbs}}_{t,\alpha,\beta}$ centred on $k_{t,\alpha,\beta}$ with the degridding kernel. Then, the application of this matrix on the Fourier transformed over-sampled images produce the measurements. When working in the absence of the DDEs, a direct correspondence of the measurement operator in equation~\eqref{eq:meas_op} can be noticed with the measurement operator used for Stokes $I$ imaging \citep{Onose2017, Pratley2017} and more recently, for full Stokes imaging \citep{Birdi2018b}.

It is to be mentioned here that most of the imaging techniques in RI literature rely on taking the Jones matrices and hence Mueller matrix as identity in equation~\eqref{eq:obs_model_im}. In other words, they work under the assumption of absence of any calibration errors or availability of pre-calibrated data. However, in presence of calibration errors, the imaging techniques should account for these terms within their models. Working in this direction, we can generalize the model proposed in \cite{Birdi2018b} and employ it to solve inverse problem~\eqref{eq:im_model}. The corresponding imaging strategy leverages the CS theory.  
In particular, CS based methods offer a way from optimization point of view to solve the underlying inverse problem. The main idea is to obtain the estimation of the sought images by solving a minimization problem of the form \citep{Birdi2018a,Birdi2018b}
\begin{equation} \label{eq:const_min}
    \minimize{\Sbs}
 r(\Sbs) \,\, \text{subject to} \,\,  \|\Phi(\Sbs) - \bm{y} \|_2 \leq \epsilon,
\end{equation}
where $\epsilon$ is related to the norm of the additive noise,  $\|\cdot\|_2$ denotes the $\ell_2$ norm of its argument vector. The term $r(\Sbs)$ is the regularization term, injecting prior information in the reconstruction process, whereas the data fidelity is ensured by the constraint $\|\Phi(\Sbs) - \bm{y} \|_2 \leq \epsilon$, implying that the residual is situated within an $\ell_2$ ball whose radius is determined by $\epsilon$. 
In the current work, due to technical assumptions related to the proposed joint imaging and calibration algorithm \citep{Chouzenoux2016}, we propose to solve an unconstrained version of problem~\eqref{eq:const_min}, given by 
\begin{equation} \label{eq:unconst_min_im}
    \minimize{\Sbs}
\bar{h} \big( \Sbs \big)
+  r(\Sbs),
\end{equation}
where $\bar{h}(\Sbs)$ is the data fidelity term.
Using this formulation and assuming that the additive noise is i.i.d. Gaussian, $\bar{h}(\Sbs)$ is given by a least squares criterion 
\begin{equation} \label{eq:data_fid_im_orig}
     (\forall \Sbs \in \eR^{2 \times 2 N}) \,\, \bar{h}(\Sbs) = \frac{1}{2}\|\Phi(\Sbs) - \bm{y} \|_2^2.
\end{equation}
This formulation has been introduced in \cite{Repetti2017} in the particular case of joint Stokes $I$ imaging and calibration. In order to incorporate various polarimetric imaging specific prior informations in the regularization function, \cite{Birdi2018b} proposed to use a hybrid regularization term of the form
\begin{equation} \label{eq:r_im}
    r(\Sbs) = g(\Sbs) + r^\prime(\Sbs),
\end{equation}
where the function $g\colon \eR^{2 \times 2N} \rightarrow ]-\infty, +\infty]$ imposes sparsity of the underlying Stokes images and the function $r^\prime\colon \eR^{2 \times 2N} \rightarrow ]-\infty, +\infty]$ constrains the domains of the argument images as per some physical constraints. In particular, the first term is inspired by the CS framework which proposes to exploit the sparsity of the target images in some dictionary $\Psib \in \eC^{N \times J}$ \citep{Candes2006, Donoho2006}. The choice of this dictionary is dependent on the images under scrutiny. To give an intuitive idea, images which are already sparse, for instance point sources images, are well represented by dirac basis taking $\Psib$ to be identity. Otherwise, sparsity can be imposed in some other domain, such as TV regularization for piece-wise constant images \citep{Rudin1992, Wiaux2010, Akiyama2017a}, wavelet basis for smooth images \citep{Mallat2009}, etc. In particular for astronomical images, a collection of wavelet bases is shown to be a good candidate for sparsifying dictionary $\Psib$ \citep{Carrillo2012, Carrillo2014}, both for Stokes $I$ imaging \citep{Onose2016,Onose2017} and specifically for polarization imaging in \cite{Birdi2018a,Birdi2018b}.
Formally, for any matrix $\widetilde{\Sbs} \in \eR^{N \times 4}$, the sparse representation of the sought images in a chosen dictionary $\Psib$ is given by $\Psib^\dagger \widetilde{\Sbs}$. The regularization function is then defined to impose sparsity of this term. This can be achieved by using the $\ell_0$ pseudo-norm, counting the number of non-zero components of its argument \citep{Donoho1995}. A more common approach is to use $\ell_1$ norm which overcomes the problem of non-convexity of the $\ell_0$ norm \citep{Chen2001}. An even better approximation of the $\ell_0$ norm is provided by the (re)weighted $\ell_1$ norm, which tends to diminish the magnitude dependency of the $\ell_1$ norm \citep{Candes2008, Carrillo2012}. The (re)weighted $\ell_1$ norm is given by
\begin{align} \label{eq:weight_l1}
    g(\Sbs)  = \| \Psib^\dagger \mathcal{R} \big(\Sbs \big) \|_{\Wbs,1}
     = \displaystyle \sum_{i=1}^4 \sum_{j=1}^J \Ws_{j,i} \big|[\Psib^\dagger \mathcal{R} \big(\Sbs \big) ]_{j,i} \big| \, ,
\end{align}
where the subscripts $j$ and $i$ in the notation $[\cdot]_{j,i}$ stand respectively for the row and column indices of the argument matrix, and $\mathcal{R}\colon \eR^{2 \times 2 N} \rightarrow \eR^{N \times 4}$ is the operator consisting in placing the four Stokes images contained in the matrix $\Sbs$ in four columns, whereas its adjoint $\mathcal{R}^\dagger:\eR^{N \times 4} \rightarrow \eR^{2 \times 2 N}$ do the contrary, i.e. storing the four images in the rows of a $2 \times 2$ block matrix. In equation~\eqref{eq:weight_l1}, $\Wbs \in \eR_+^{J \times 4}$ is the weighting matrix. In the case when this matrix is chosen to be identity, the $\ell_1$ regularization term is obtained. 
 Another useful piece of information which can be used to regularize the problem is the polarization constraint to be satisfied by the recovered Stokes images \citep{Birdi2018b}. This constraint comes from a physical point of view, that the polarized intensity should be smaller or at most equal to the total intensity and mathematically, can be described by the following set:
\begin{multline}
    \mathbb{P} = \big\{ \widetilde{\Sbs} = \mathcal{R}\big(\Sbs \big) \in \eR^{N \times 4} \, \big| \, (\forall n \in \{1,\ldots,N\}) \\ - \widetilde{\Ss}_{n,1} + \|\widetilde{\Sbs}_{n, 2:4} \|_2 \leq 0 \big\}.
\end{multline}
As can be noticed, this constraint also imposes implicitly the positivity of the total intensity image (Stokes $I$).
The enforcement of this constraint amounts to incorporation of the indicator function of the set $\mathbb{P}$ in the regularization function, i.e.
\begin{equation} \label{eq:r_pol}
    r^\prime(\Sbs) = \iota_{\mathbb{P}} (\Sbs).
\end{equation}
The indicator function of any such set $\mathbb{P}$ is defined as 
\begin{equation}
    \iota_{\mathbb{P}}(\Sbs) = \begin{cases}
    0, \quad  \quad  \text{if} \, \Sbs \in \mathbb{P} \,, \\
    +\infty, \quad \text{otherwise}.
    \end{cases}
\end{equation} 

%% file: calibration.tex
\section{Calibration problem}
\label{sec:cal_prob}

In practice, the DDEs are often unknown and need to be estimated. In this section, we formulate the calibration problem to be solved.
We assume that the DDEs exhibit a smooth variation both across the field of view and in time. It implies that DDEs are band-limited spatially as well as temporally. This is enforced by considering compact-support kernels of the DDEs in both spatial \citep{Repetti2017b, Repetti2017} and temporal Fourier domains \citep{Thouvenin2018}. More specifically, for each antenna $\alpha \in \{1,\ldots,n_a\}$, the DDEs $(\overline{\Dbs}_{t,\alpha})_{1 \le t \le T}$ are represented by the Fourier kernels $\overline{\Ubs}_{\alpha} \in \eC^{2 \times 2 K \times P}$, where $K$ and $P$ are the support sizes in spatial and temporal Fourier domain, respectively, with $K \ll N, P \ll T$. 
Then, the task is to estimate only non-zero Fourier coefficients of the DDEs, thereby reducing the dimension of the underlying problem significantly. 
Analogous to the calibration inverse problem in \cite{Repetti2017}, problem~\eqref{eq:obs_model} can be reformulated as
\begin{equation} \label{eq:cal_mod}
\Ybs_{t,\alpha,\beta} = \mathcal{D}_{t}(\overline{\Ubs}_\alpha)\, \mathcal{X}_{t, \alpha, \beta} \big( \Fbs \, \widetilde{\Zbs} \, \mathcal{\widetilde{L}} (\overline{\Sbs}) \big)\, \mathcal{D}_{t}^\prime(\overline{\Ubs}_\beta) + \bm{\Omega}_{t,\alpha,\beta},
\end{equation}
{where $\mathcal{D}_{t} \colon \eC^{2 \times 2 K \times P} \rightarrow \eC^{2 \times 2N}$ is the operator acting on $\overline{\Ubs}_\alpha$ to give a sparse matrix $\mathcal{D}_{t}(\overline{\Ubs}_\alpha)$ containing the compact support kernels in $\widehat{\overline{\Dbs}}_{t, \alpha}$, flipped and centred in zero spatial frequency. Similarly, the operator $\mathcal{D}_{t}^\prime \colon \eC^{2 \times 2  K \times P} \rightarrow \eC^{2N \times 2}$ is defined such that $\mathcal{D}_{t}^\prime(\overline{\Ubs}_\beta)$ is a sparse matrix consisting of the compact support kernels in $\widehat{\overline{\Dbs}}^\dagger_{t, \beta}$ centred in zero spatial frequency. Finally,
$\mathcal{X}_{t, \alpha, \beta} \big( \Fbs \, \widetilde{\Zbs} \, \mathcal{\widetilde{L}} (\overline{\Sbs}) \big) \in \eC^{2N \times 2N}$ is a $2 \times 2$ block matrix, with each block of size $N \times N$. Each row/ column of such a block consists of a shifted version of the Fourier transform of the corresponding image in reshaped brightness vector $\widetilde{\mathcal{L}}(\overline{\Sbs})$, mimicking the convolution operation. Moreover, to account for the continuous sampled frequencies, these Fourier transforms are convolved with the degridding kernels centred at the associated frequency $k_{t,\alpha,\beta}$. }

It can be observed that problem~\eqref{eq:cal_mod} is non-linear with respect to the compact-support kernels $(\overline{\Ubs}_\alpha)_{1 \leq \alpha \leq n_a}$. Following the approach proposed in \cite{Salvini2014} and \cite{Repetti2017}, we linearize it by introducing the matrices $\overline{\Ubs}_{\alpha,1}$ and $\overline{\Ubs}_{\alpha,2} $ such that $\overline{\Ubs}_{\alpha,1} = \overline{\Ubs}_{\alpha,2} = \overline{\Ubs}_{\alpha}$ for every $\alpha \in \{1, \ldots, n_a\}$. Using this strategy, the problem becomes bi-linear and then the objective is to estimate both the matrices $\overline{\Ubs}_1 = (\overline{\Ubs}_{\alpha,1})_{1 \leq \alpha \leq n_a} \in \eC^{2 \times 2 K \times P \times n_a}$ and $\overline{\Ubs}_2 = (\overline{\Ubs}_{\alpha,2})_{1 \leq \alpha \leq n_a} \in \eC^{2 \times 2 K \times P \times n_a}$, where $\overline{\Ubs}_1$ (resp. $\overline{\Ubs}_2$) concatenates the non-zero Fourier coefficients  $\overline{\Ubs}_{\alpha,1}$ (resp. $\overline{\Ubs}_{\alpha,2}$) for all antennas. The estimation of these matrices is achieved by solving the following minimization problem for the DDE calibration:
\begin{equation} \label{eq:unconst_min_cal}
    \minimize{\Ubs_1, \Ubs_2}
\widetilde{h} \big( \Ubs_1, \Ubs_2 \big)
+ p(\Ubs_1, \Ubs_2),
\end{equation}
where $\widetilde{h}$ is the least-squares data fidelity term and $p$ is the regularization function for $\Ubs_1$ and $\Ubs_2$. 
The data fidelity term for DDE calibration reads as
\begin{align} 
    \widetilde{h} \big( \Ubs_1, \Ubs_2 \big) & = \ \sum_{\underset{1 \leq \alpha \leq {n_a}}{}} \frac{1}{2}\| \mathcal{G}_{\alpha,1} (\Ubs_{\alpha,1}) - \Ybs_{\alpha} \|_F^2, \label{eq:cal_U1}\\
    & = \sum_{\underset{1 \leq \alpha \leq {n_a}}{}} \frac{1}{2}\| \mathcal{G}_{\alpha,2} (\Ubs_{\alpha,2}) - \Ybs_{\alpha} \|_F^2, \label{eq:cal_U2}
\end{align}
where $\|\cdot\|_F$ stands for the Frobenious norm of the matrix argument, and considering $t$ and $\beta$ taking all the values respectively in the ranges $\{1,\ldots,T\}$ and $\{1, \ldots, n_a\}$, the following hold: $\Ybs_{\alpha} = (\Ybs_{t,\alpha,\beta})_{t,\beta \neq \alpha}$ and the operator $\mathcal{G}_{\alpha,1} (\Ubs_{\alpha,1})$ in equation~\eqref{eq:cal_U1} generates a concatenation of the terms of the form $\mathcal{D}_{t}({\Ubs}_{\alpha,1})\, \mathcal{X}_{t, \alpha, \beta} \big( \Fbs \, \widetilde{\Zbs} \, \mathcal{\widetilde{L}} ({\Sbs}) \big)\, \mathcal{D}_{t}^\prime({\Ubs}_{\beta,2})$. Similarly, $\mathcal{G}_{\alpha,2} (\Ubs_{\alpha,2})$ in equation~\eqref{eq:cal_U2} consists in the operation $\mathcal{D}_{t}({\Ubs}_{\beta,1})\, \mathcal{X}_{t, \beta, \alpha} \big( \Fbs \, \widetilde{\Zbs} \, \mathcal{\widetilde{L}} ({\Sbs}) \big)\, \mathcal{D}_{t}^\prime({\Ubs}_{\alpha,2})$ to produce the corresponding measurements.

On the other hand, the regularization term $p$ in problem~\eqref{eq:unconst_min_cal} is given by
\begin{equation} \label{eq:reg_cal}
p\big( \Ubs_1, \Ubs_2 \big)
=  \gamma \| \Ubs_{1} - \Ubs_{2} \|_F^2 
%+ \mu \Big( \| \Ubs_{1} - \Thetab \|_2^2
%+  \| \Ubs_{2} - \Thetab \|_2^2 \Big)
+ \iota_{\mathbb{D}} \big( \Ubs_1 \big) + \iota_{\mathbb{D}}\big( \Ubs_2 \big),
\end{equation}
where $\gamma >0$ is the regularization parameter and the first term in equation~\eqref{eq:reg_cal} controls the distance between the matrices $\Ubs_1$ and $\Ubs_2$, thereby imposing the constraint that these two matrices should be equal. The set $\mathbb{D}$ is defined to constrain the values of the Fourier coefficients of the DDEs to lie within the specified bounds. In particular, $\mathbb{D}$ is defined such that for each $t \in \{1,\ldots,T\}$, the Fourier kernels stored in diagonal terms of $\widehat{{\Dbs}}_{t, \alpha}$ have the central coefficients \big($[\widehat{{\Dbs}}_{t, \alpha}(0)]_{11} \, \text{and} \, [\widehat{{\Dbs}}_{t, \alpha}(0)]_{22}$\big) belonging to an $\ell_\infty$ complex ball centred in 1 with radius $\theta_1 > 0$, whereas the rest of the coefficients belong to an $\ell_\infty$ complex ball centred in 0 with radius $\theta_1$. On the other hand, for the Fourier kernels in the off-diagonal terms, the central and the other coefficients are assumed to be contained in $\ell_\infty$ balls centred in 0 with radius $\theta_2$ and $\theta_3$, respectively. 
Intuitively, this can be understood as follows. Since we work in the scenario after the calibration transfer is performed, the zero spatial frequency coefficient of the DDEs (i.e. the DIEs) encoded in the diagonal terms of the Jones matrices are normalized to 1, thus lying in a complex neighbourhood of $1 + \text{i}0$. Moreover, while the central coefficient in the spatial Fourier domain represents the mean gain, the higher order spatial frequencies characterize the gain variations across the field of view with respect to this mean gain. Therefore, these coefficients have smaller values in comparison with the central coefficient. Lastly, concerning the off-diagonal terms which encompass the polarization leakage, their values are usually much smaller than the diagonal terms.

%% file: proposed_app.tex
\section{Proposed calibration $\&$ imaging approach}
\label{sec:cal_im}

The current work deals with a practical case when neither the Stokes images nor the DDEs associated with the antennas are known. In the particular case when the sky is considered to be unpolarized, only Stokes $I$ imaging needs to be performed and the $2 \times 2$ Jones matrices are often replaced by a scalar value. In this scenario, \cite{Repetti2017} recently proposed a method to perform joint calibration and imaging. Motivated by the good performance obtained by this flexible method with proven convergence guarantees, we extend this approach for full polarization model and propose a joint calibration and imaging algorithm to solve the problem under scrutiny. This leads to considering a global minimization problem aiming to solve for the Stokes parameters $\Sbs$ and the calibration matrices $\Ubs_1$ and $\Ubs_2$. This non-convex problem can be cast by combining the minimization problems~\eqref{eq:unconst_min_im} and calibration~\eqref{eq:unconst_min_cal} which were proposed in the earlier sections solely for Stokes imaging and DDE calibration, respectively.

Given the non-convexity of the underlying minimization problem, choice of initialization is crucial. In this context, we exploit the fact that the calibration transfer has been performed, to obtain an initial estimate of the Stokes images, as suggested in \cite{Repetti2017b} (although for Stokes $I$ imaging only). Such first imaging step consists in considering the Jones matrices as identity and solving the associated minimization problem~\eqref{eq:unconst_min_im}. 

Let $\Sbs^{\prime}$ be the Stokes parameters estimated by solving problem~\eqref{eq:unconst_min_im}. Since these are obtained ignoring the DDEs, in general, these may contain artefacts. Therefore, we instead use a thresholded version of $\Sbs^{\prime}$, denoted by $\Sbs_0$, which contains only the high amplitude coefficients of $\Sbs^{\prime}$. With this first approximation of the images at hand, the original unknown image can be seen as a sum of $\Sbs_0$ and $\Ebs \in \eR^{2 \times 2N}$, where the latter is unknown and need to be estimated.

Finally, we propose to define the estimates $(\Ebs, \Ubs_1, \Ubs_2)$ as solutions to the following global, non-convex minimization problem:
\begin{equation} \label{eq:global_min}
\minimize{\Ebs, \, \Ubs_{1}, \Ubs_{2}}
h \big( \Ebs, \Ubs_{1}, \Ubs_{2} \big)
+  r(\Ebs)
+  p\big( \Ubs_{1}, \Ubs_{2} \big),
\end{equation}
where $h$ is the least squares data fidelity term associated with the data model,
$r$ and $p$ are the regularization terms for the image and the DDEs, respectively. {In particular, with $\Sbs = \Sbs_0 + \Ebs$, the data fidelity term is given by the least squares criterion as proposed in equations~\eqref{eq:data_fid_im_orig}, ~\eqref{eq:cal_U1} and~\eqref{eq:cal_U2}, i.e.
\begin{align} 
 h \big( \Ebs, \Ubs_{1}, \Ubs_{2} \big)  
 & = \|\Phi(\Sbs_0 + \Ebs) - \bm{y} \|_2^2 \label{eq:data_fid_im}\\
% \| \mathcal{G}(\Ubs_1, \Ubs_2) \mathcal{F}( \Sbs_0 + \Ebs) - \Ybs \|_2^2, \\
&  = \sum_{1 \leq \alpha \leq {n_a}} \frac{1}{2}\| \mathcal{G}_{\alpha,1} (\Ubs_{\alpha,1}) - \Ybs_{\alpha} \|_F^2, \label{eq:data_fid_cal1} \\
& = \sum_{1 \leq \alpha \leq {n_a}} \frac{1}{2}\| \mathcal{G}_{\alpha,2} (\Ubs_{\alpha,2}) - \Ybs_{\alpha} \|_F^2, \label{eq:data_fid_cal2}
\end{align}
where operator $\Phi$ in equation~\eqref{eq:data_fid_im} is formed using fixed values of $(\Ubs_1, \Ubs_2)$. Similarly, $\mathcal{G}_{\alpha,1} (\Ubs_{\alpha,1})$ in equation~\eqref{eq:data_fid_cal1} (and $\mathcal{G}_{\alpha,2} (\Ubs_{\alpha,2})$ in equation~\eqref{eq:data_fid_cal2}, resp.) is determined by fixed $(\Ubs_{\beta,2}, \Ebs)$ ($(\Ubs_{\beta,1}, \Ebs)$, resp.) with $\beta \in \{1,\ldots,n_a\}$ and $\beta \neq \alpha$. While estimating the Stokes parameters, equation~\eqref{eq:data_fid_im} is employed as the data fidelity term. In particular, keeping $(\Ubs_1, \Ubs_2)$ fixed, the convexity of this term with respect to $\Ebs$ can be noticed. In the same manner, equation~\eqref{eq:data_fid_cal1} (\eqref{eq:data_fid_cal2}, resp.) is chosen while updating $\Ubs_1$ ($\Ubs_2$, resp.) which is convex with respect to $\Ubs_1$ ($\Ubs_2$, resp.) keeping the other two variables fixed. It is then straightforward to see that the non-convex function $h$ is in fact convex for each of the variables while fixing the others. }

Concerning the regularization terms, the function $p\big( \Ubs_{1}, \Ubs_{2} \big)$ is given from equation~\eqref{eq:reg_cal}, whereas the function $r(\Ebs)$ for the images is associated with the priors introduced in equations~\eqref{eq:r_im},~\eqref{eq:weight_l1} and~\eqref{eq:r_pol}. In particular, one can choose whether to take the polarization constraint into account or not. In the former case, the regularization term boils down to
\begin{equation} \label{eq:reg_im_final}
    r(\Ebs) = g(\Ebs) +  \iota_{\mathbb{P}}(\Sbs_0 + \Ebs) +  \iota_{\mathbb{K}} ( \Ebs),
\end{equation}
{where $g(\Ebs) = \sum_{i=1}^4 \eta_i \| \big(\Psib^\dagger \mathcal{R}( \Sbs_0 + \Ebs) \big)_{\colon,i}\|_{1} $ denotes the SARA prior considering $\Psib$ to be a concatenation of Dirac basis and first eight Daubechies wavelets \citep{Carrillo2014, Birdi2018b}. Here, we have taken the weighting matrix to be equal to identity, i.e. without adapting the reweighting scheme and, for every
$i \in \{1, 2, 3, 4\}$, $\eta_i > 0$ is the regularization parameter.} The set $\mathbb{K}$ is chosen to take into account the errors that might appear on the estimated non-zero coefficients of $\Sbs_0$. Formally, it is defined as
\begin{equation}
    \mathbb{K} = \big\{ \Ebs \in \eR^{2 \times 2 N} | (\forall n \in \eS_0) \, \, \Es (n) \in [-\vartheta \, \Ss_{0} (n) , \vartheta \,  \Ss_{0} (n)] \, \big\},
\end{equation}
where $\eS_0$ is the support of $\Sbs_0$. The parameter $\vartheta \in [0,1]$ is chosen according to the error percentage considered on $\Sbs_0$.

On the other hand, in the absence of the polarization constraint, the positivity of Stokes $I$ image needs to be imposed explicitly, that can be accounted for by replacing the set $\mathbb{K}$ in equation~\eqref{eq:reg_im_final} by the set $\mathbb{K}^\prime$, defined as
\begin{multline}
   \mathbb{K}^\prime = \big\{ \Ebs \in \eR^{2 \times 2 N} | (\forall n \in \eS_0) \, \, \Es(n) \in [-\vartheta \, \Ss_{0} (n) , \vartheta \,  \Ss_{0} (n)] \, , \\
     (\forall n \not \in \eS_0) \, \, [\Es (n)]_{1,1} \geq 0 \big\}.
\end{multline}
In such a case, 
the function $r$ reads as
\begin{equation} \label{eq:reg_im_final_wo_const}
    r(\Ebs) = g(\Ebs) +  \iota_{\mathbb{K}^\prime} ( \Ebs).
\end{equation}

\subsection{Proposed algorithm}
In order to solve problem~\eqref{eq:global_min}, we observe that it has a block-variable structure with $\Ubs_1, \Ubs_2$ and $\Ebs$ being the three blocks constituting the problem. On top of it, although the global problem is non-convex, it is convex with respect to each of these blocks. Leveraging this block-variable structure, we propose to use an iterative algorithm based on a block-coordinate forward-backward approach \citep{Chouzenoux2016} to solve problem~\eqref{eq:global_min}. It consists in alternating between the estimation of the DDEs and the Stokes images. In turn, for each of these estimations, FB iterations are employed, i.e. the smooth terms are dealt by their gradient, whereas the non-smooth terms are managed by their proximity operators. 
Formally, given a proper, lower-semicontinuous function $f \colon \eR^{2 \times 2 N} \rightarrow ]-\infty, +\infty]$, its proximity operator \citep{Moreau1965} at $\Sbs \in \eR^{2 \times 2 N}$ is defined as
\begin{equation}
    \operatorname{prox}_f(\Sbs) = \underset{\Vbs \in \eR^{2 \times 2 N}}{\operatorname{argmin}} \, f(\Vbs) + \frac{1}{2} \|\Vbs - \Sbs \|_F^2.
\end{equation}
In the particular case of $f$ being an indicator function of a set $\eE$, the proximity operator reduces to commonly known projection operator $\mathcal{P}_{\eE}$, defined as
\begin{equation}
    \mathcal{P}_{\eE}(\Sbs) = \underset{\Vbs \in \eE}{\operatorname{argmin}} \, \|\Vbs -\Sbs\|_F^2.
\end{equation}
In light of the discussion above, we present the proposed algorithm as Algorithm~\ref{algo_polcal}. It consists of a global loop and inner iteration loops. At each iteration of the global loop, indexed by $i \in \eN$ (step~\ref{algo1:step:global_loop}), we choose either to update the DDEs or the image following an essentially cyclic rule, that is each of the variables must be updated at least once within a given finite number of iterations. {For every $i^{\text{th}}$ iteration, this is taken care by the choice of number of inner loop iterations $L^{(i)} \in \eN$ and $J^{(i)} \in \eN$ to update the DDEs and the image, respectively.} To be more precise, in the former case, each of the calibration matrices $\Ubs_1^{(i)}$ (step~\ref{algo1:step:update_u1}) and $\Ubs_2^{(i)}$ (step~\ref{algo1:step:update_u2}) are updated by performing $ L^{(i)}$ number of FB iterations in the inner loop, using the images estimated at the previous iterate. When the images are chosen to be updated in the global loop, the updated DDEs from the previous iterate are used to estimate the image in step~\ref{algo1:step:update_e}, executing $J^{(i)}$ FB iterations. The overall algorithm can then be understood by splitting it into two parts: Calibration and Imaging. These two parts are explained in what follows.

%%%%%%%%%%%%%%%%%%%%%%%%%%%%%%%%%%%%%%%%%%%%%%%%%
%
%%%%%%%%%%%%%%%%%%%%%%%%%%%%%%%%%%%%%%%%%%%%%%%%%
\begin{algorithm}
\caption{Joint calibration and imaging algorithm}\label{algo_polcal}
\begin{algorithmic}[1]
\State \textbf{Initialization}: $\Ebs^{(0)} \in \eR^{2 \times 2 N}$, $(\Ubs_1^{(0)}, \Ubs_2^{(0)}) \in \big(\eC^{2 \times 2 K \times P \times n_a}\big)^2$.  \quad  \quad Let, for every $i \in \eN$, {$(L^{(i)}, J^{(i)}) \in \eN^2$}

\vspace{0.2cm}
\For{$i = 0, 1, \ldots$	} \label{algo1:step:global_loop}
\Statex {\text{Choose to update either the DDEs or the images.}}	

\vspace{0.2cm}

\Statex {\textbf{If the DDEs are updated:}} 

\vspace{0.1cm}

\State \big( $\Ubs_{1}^{(i,0)},  \Ubs_{2}^{(i,0)} \big) = \big( \Ubs_{1}^{(i)},  \Ubs_{2}^{(i)}$ \big) \label{algo1:step:update_u_start}

\vspace{0.1cm} 

\For{$\ell = 0, \ldots, L^{(i)}-1$} \label{algo1:step:update_u1_fb_start}

\vspace{0.1cm}
\State	\label{algo1:step:update_u1}
\hspace*{-0.2cm}$\displaystyle \begin{array}{ll}
\Ubs_1^{(i,\ell+1)}
= {\Pc_{\mathbb{D}}} \bigg( \Ubs_1^{(i,\ell)} - & \hspace{-0.2cm}  {\Gamma_1^{(i)}} \cdot \nabla_{\Ubs_1} h \big( \Ebs^{(i)}, \Ubs_1^{(i,\ell)}, \Ubs_2^{(i)} \big) 	\\
& \displaystyle -  \Gamma_1^{(i)} \cdot \gamma  \big( \Ubs_1^{(i,\ell)} -  \Ubs_2^{(i)} \big)  \bigg) 
\end{array}$

\EndFor \label{algo1:step:update_u1_fb_end}
\vspace{0.1cm}

\State $\Ubs_1^{(i+1)} = \Ubs_1^{(i,L^{(i)})}$

\vspace{0.1cm}

\For{$\ell = 0, \ldots, L^{(i)}-1$}	\label{algo1:step:update_u2_fb_start}

\vspace{0.1cm}
\State	\label{algo1:step:update_u2}
\hspace*{-0.2cm}$\displaystyle \begin{array}{ll}
\Ubs_2^{(i,\ell+1)} 
= {\Pc_{\mathbb{D}}} \bigg( 
\Ubs_2^{(i,\ell)} 
- & \hspace{-0.25cm}  \Gamma_2^{(i)} \cdot 
 \nabla_{\Ubs_2} h \big( \Ebs^{(i)}, \Ubs_1^{(i+1)}, \Ubs_2^{(i,\ell)} \big)  \\
& \displaystyle -  \Gamma_2^{(i)} \cdot \gamma  \big( \Ubs_2^{(i,\ell)} - \Ubs_1^{(i+1)} \big)  \bigg)
\end{array}$

\EndFor \label{algo1:step:update_u2_fb_end}

\vspace{0.1cm}

\State $\Ubs_2^{(i+1)} = \Ubs_2^{(i,L^{(i)})}$	
\vspace{0.1cm}

\State \label{algo1:step:update_u_end}
$\Ebs^{(i+1)} = \Ebs^{(i)}$

% ---------------------------------------------------------------------------------------
\vspace{0.1cm}

\Statex {\textbf{If the Stokes images are updated:}}  

\vspace{0.1cm}

\State \label{algo1:step:update_e_start}
$\Ebs^{(i,0)} = \Ebs^{(i)}$

\vspace{-0.15cm}

\hspace{-0.15cm}\For{$j = 0, \ldots, J^{(i)}-1$}

\vspace{0.1cm}

\State \label{algo1:step:update_e}
$
\hspace{-0.22cm} \Ebs^{(i,j+1)}
\hspace{-0.12cm} = { \prox_{\sigma^{(i)} r} } \hspace{-0.1cm}\left( \Ebs^{(i,j)}
\hspace{-0.1cm} - {\sigma^{(i)}} {\nabla_{\Ebs} h \big( \Ebs^{(i,j)}, \Ubs_1^{(i+1)}, \Ubs_2^{(i+1)} \big)} \right)$

\hspace{-0.15cm}\EndFor
\vspace{0.1cm}

\State $\Ebs^{(i+1)} = \Ebs^{(i, J^{(i)})}$ 

\vspace{0.1cm}

\State \label{algo1:step:update_e_end}
${\big( \Ubs_1^{(i+1)}, \Ubs_2^{(i+1)} \big) = \big( \Ubs_1^{(i)}, \Ubs_2^{(i)} \big).}$

\vspace{0.1cm}

\EndFor

% ---------------------------------------------------------------------------------------

\end{algorithmic}
\end{algorithm}
%%%%%%%%%%%%%%%%%%%%%%%%%%%%%%%%%%%%%%%%%%%%%%%%%
%
%%%%%%%%%%%%%%%%%%%%%%%%%%%%%%%%%%%%%%%%%%%%%%%%%
% 
\paragraph*{Calibration:}
It comprises of the estimation of the matrices $\Ubs_1$ and $\Ubs_2$. 
In this case, while the data fidelity term $h$ is differentiable, the regularization term $p$ consist of both smooth and non-smooth terms. Thus, in each $\ell^{\text{th}}$ FB iteration to estimate either of these matrices, the gradient step for the differentiable terms is coupled with the proximity operator of the non-smooth term, as shown in steps~\ref{algo1:step:update_u1} and~\ref{algo1:step:update_u2}. In particular, for the gradient step, for every $q \in \{1,2\}$, the step size $\Gamma_q^{(i)}\in  \eR^{2 \times 2K \times P \times n_a}$ is chosen as 
\begin{equation}\label{eq:step_size_U}
    \Gamma_q^{(i)} = \big(\zeta_{q,\alpha}^{(i)} \, \bm{1}_{2 \times 2K \times P} \big)_{1 \leq \alpha \leq n_a}, 
\end{equation}
where $\bm{1}_{R}$ is a matrix of ones of dimension $R$, and $\zeta_{q,\alpha}^{(i)}$ is given by
$
 0 < \zeta_{q,\alpha}^{(i)} < 1/(\gamma + \upsilon_{q, \alpha}^{(i)}),
$
with $\upsilon_{q, \alpha}^{(i)}$ denoting the Lipschitz constant of the partial derivative of $h$ with respect to $\Ubs_{q,\alpha}^{(i)}$. 

Furthermore, since in this case the non-smooth term is the indicator function of the set $\mathbb{D}$, as explained previously, this reduces to performing projection on this set, $\mathcal{P}_{\mathbb{D}}$ which basically ensures that the values of the estimated DDEs Fourier coefficients lie within the bounds specified earlier.

\paragraph*{Imaging:}
This step updates the Stokes images while using the DDEs estimates from the previous iterate. As shown in step~\ref{algo1:step:update_e}, it involves computing the gradient of the data fidelity term, followed by the proximity operator of the regularization function $r$. In this case, the step size $\sigma^{(i)}$ for the gradient step is chosen such that it satisfies
\begin{equation} \label{eq:step_size_im}
0 < \sigma^{(i)} < 1/\|\Phi\|_{2},
\end{equation}
{where $\|\Phi\|_{2}$ computes the spectral norm of $\Phi$, and $\Phi$ is generated using the updated values $(\Ubs_1^{(i+1)}, \Ubs_2^{(i+1)})$. }

Regarding the proximity step, the regularization term $r(\Ebs)$ is a hybrid term incorporating a mixture of prior information. Particularly, based on the choice of inclusion or exclusion of the polarization constraint, the computation of the proximity operator differs. For comparison purposes, here we consider both the cases. 

\paragraph*{Regularization without polarization constraint:} When we work in the absence of the polarization constraint, the regularization term $r(\Ebs)$ is given by equation~\eqref{eq:reg_im_final_wo_const}. Then, the proximity operator of $r$ evaluated at any point $\Rbs \in \eR^{2 \times 2 N}$ amounts to
\begin{equation} \label{eq:prox_r_wo_const}
    {\operatorname{prox}_{\sigma^{(i)}r}(\Rbs) = \underset{\widetilde{\Ebs}}{\operatorname{argmin}} \, \, \sigma^{(i)}} g(\widetilde{\Ebs}) +  \iota_{\mathbb{K}^\prime} (\widetilde{\Ebs}) + \frac{1}{2} \|\widetilde{\Ebs} - \Rbs \|_F^2,
\end{equation}
which does not have an explicit formulation and thus requires sub-iterations for its computation. One such possibility is to employ dual forward-backward algorithm \citep{Combettes2011, Combettes2011b}, as presented in Algorithm~\ref{algo:dualfb}.

In particular, step~\ref{algoDFB:proj} of Algorithm~\ref{algo:dualfb} performs the projection onto the set $\mathbb{K}^\prime$. It is followed by the computation of the proximity operator of function $g$ in step~\ref{algoDFB:prox}, which in the current case of $g$ being the $\ell_1$ norm, corresponds to the \textit{soft-thresholding operator} \citep{Chaux2007}. This is evaluated as
\begin{equation} \label{eq:soft_thresh1}
\operatorname{prox}_{\varepsilon^{(i)} g}\big(\mu^{-1} \Hbs^{(k)}\big)
= \widetilde{\Rbs},
\end{equation}
where $\varepsilon^{(i)} = \mu^{-1} \sigma^{(i)} $ and the soft-thresholding operation is a component-wise operation such that $\widetilde{\Rbs} \in \eR^{J \times 4}$ is defined for every $q \in \{1,2,3,4\}$ and $j \in \{1,\ldots,J\}$ as
\begin{equation} \label{eq:soft_thresh2}
 \widetilde{\Rbs}_{j,q} = \begin{cases}
              - {\mu^{-1} {\Hs}}_{j,q}^{(k)}  + {\varepsilon^{(i)} \eta_q}  \, , & \text{if} \,\, \mu^{-1} {\Hs}_{j,q}^{(k)} < -{\varepsilon^{(i)} \eta_q}\, , \\
              0, & \text{if} \,\, - {\varepsilon^{(i)} \eta_q} \leq \mu^{-1} {\Hs}_{j,q}^{(k)} \leq {\varepsilon^{(i)} \eta_q}\, , \\
              {\mu^{-1} {\Hs}}_{j,q}^{(k)}  -  {\varepsilon^{(i)} \eta_q}\, , & \text{otherwise}.
              \end{cases} 
\end{equation} 
In essence, this operation sets the elements smaller than a threshold to zero, whereas shrinks all the other elements by this value. Therefore, it promotes sparsity of its argument variable.  

%%%%%%%%%%%%%%%%%%%%%%%%%%%%%%%%%%%%%%%%%%%%%%%%%%%%%%%%%%%%%%
%---------Dual FB---------------------
%%%%%%%%%%%%%%%%%%%%%%%%%%%%%%%%%%%%%%%%%%%%%%%%%%%%%%%%%%%%%%
\begin{algorithm}
\caption{Dual Forward-Backward algorithm to compute \eqref{eq:prox_r_wo_const}}
\label{algo:dualfb}
\begin{algorithmic}[1]

\State \textbf{Initialization:} Let $\widetilde{\Pbs}^{\text{(0)}}\in \eR^{J \times 4}$, $\bar{\epsilon} \in\, ]0, \text{min}\{1, 1/\| \Psib^\dagger\|^2\}[$, and ${\mu \in [\bar{\epsilon} , 2/\|\Psib^\dagger\|^2 - \bar{\epsilon}]}$

\vspace{0.1cm}

\State 
\textbf{for} $k = 0, 1, \ldots$

\vspace{0.1cm}

\State \label{algoDFB:proj}
\quad $\Vbs^{(k)} = \mathcal{P}_{\mathbb{K}^{\prime}} \bigg(\Rbs - \mathcal{R}^\dagger \big(\Psib  \widetilde{\Pbs}^{(k)} \big)\bigg)$

\vspace{0.05cm}

\State \label{algoDFB:change_dom}
\quad $\Hbs^{(k)} = \widetilde{\Pbs}^{(k)} + \mu \Psib^\dagger \mathcal{R} \big(\Vbs^{(k)} \big) $

\vspace{0.1cm}

\State \label{algoDFB:prox}
\quad $\widetilde{\Pbs}^{(k+1)} = \Hbs^{(k)} - \mu \, \operatorname{prox}_{\mu^{-1} \sigma^{(i)} g}\big(\mu^{-1} \Hbs^{(k)}\big)$

\vspace{0.1cm}

\State 
\textbf{end for}

\vspace{0.1cm}

\State 
\textbf{Return:} $ \widetilde{\Ebs} =  \lim_k \Vbs^{(k)}$

\end{algorithmic}
\end{algorithm}
%%%%%%%%%%%%%%%%%%%%%%%%%%%%%%%%%%%%%%%%%%%%%%%%%
%
%%%%%%%%%%%%%%%%%%%%%%%%%%%%%%%%%%%%%%%%%%%%%%%%%

\paragraph*{Regularization with polarization constraint:}
In the case when the polarization constraint is to be enforced, the regularization term is given from equation~\eqref{eq:reg_im_final} and the associated proximity operator at a point $\Rbs$ is given by
\begin{equation} \label{eq:prox_r_with_const}
    \operatorname{prox}_{{\sigma^{(i)}} r}(\Rbs) = \underset{\widetilde{\Ebs}}{\operatorname{argmin}} \, \, {\sigma^{(i)}} g(\widetilde{\Ebs}) +  \iota_{\mathbb{K}} (\widetilde{\Ebs}) + \iota_{\mathbb{P}}(\widetilde{\Ebs} + \Sbs_0) + \frac{1}{2} \|\widetilde{\Ebs} - \Rbs \|_F^2.
\end{equation}
The evaluation of this operator requires projection onto the set $\mathbb{P}$, that does not have a closed form solution. To circumvent this difficulty, we have recently proposed a method, named Polarized SARA  \citep{Birdi2018a,Birdi2018b}, that enforces this constraint leveraging the epigraphical projection techniques \citep{Chierchia2015}. In this context, with the introduction of an auxiliary variable $\Zbs \in \eR^{N \times 2}$, the polarization constraint set is splitted into simpler, easily manageable constraint sets, thereby performing the projection onto these sets. Formally, it corresponds to the following reformulation of problem~\eqref{eq:prox_r_with_const}:
\begin{subequations}
\begin{equation} \label{eq:prox_r_epi}
    \operatorname{prox}_{{\sigma^{(i)}} r}(\Rbs) = 
    \underset{\widetilde{\Ebs}, \Zbs}{\operatorname{argmin}} \, \, {\sigma^{(i)}} g(\widetilde{\Ebs}) +  \iota_{\mathbb{K}} (\widetilde{\Ebs}) + \frac{1}{2} \|\widetilde{\Ebs} - \Rbs \|_F^2
\end{equation}
subject to ($\forall  n \in \{ 1, \ldots, N \}$) 
\begin{numcases}{}
h_1({\mathcal{R} (\widetilde{\Ebs} + \Sbs_0)}_{n,1}) = - ({\widetilde{\Es} + \Ss_0)}_{n,1}
	\leq {{\mathsf{Z}}_{n,1}}, \label{eq:h1_cons} \\
h_2({\mathcal{R} (\widetilde{\Ebs} + \Sbs_0)}_{n,2:4}) = \| {\mathcal{R} (\widetilde{\Ebs} + \Sbs_0)}_{n,2:4} \|_2
\leq {{\mathsf{Z}}_{n,2}},\label{eq:h2_cons} \\
{{\mathsf{Z}}}_{n,1} + {{\mathsf{Z}}}_{n,2} \leq 0.	\label{eq:zeta_cons}
\end{numcases}
\end{subequations}
% 
% 

%%%%%%%%%%%%%%%%%%%%%%%%%%%%%%%%%%%%%%%%%%%%%%%%%
%
%%%%%%%%%%%%%%%%%%%%%%%%%%%%%%%%%%%%%%%%%%%%%%%%%
\begin{algorithm}
\caption{Primal-dual algorithm to solve problem~\eqref{eq:prox_r_epi_final}}\label{algo:pd_fb}
\begin{algorithmic}[1]
% -------------------------------------------------------------

\vspace*{0.1cm}
\State \textbf{Initialization:}  $\widetilde{\Ebs}^{(0)} \in \mathbb{R}^{2 \times 2 N}, \bm{\mathsf{Z}}^{(0)} \in \mathbb{R}^{N \times 2}, \bm{\mathsf{A}}^{(0)} \in \mathbb{R}^{J \times 4},$ $\bm{\mathsf{C}}^{(0)} \in \mathbb{R}^{N \times 4}, \, \bm{\mathsf{B}}^{(0)} \in \mathbb{R}^{N \times 2}, (\rho_1, \rho_2, \tau) \in \eR_+^3$ such that $\rho_1 {\|\Psib^\dagger \|_{2}^2} + \rho_2 < \tau^{-1}$

\vspace*{0.1cm}

\State \textbf{repeat for} $k = 0, 1, \ldots$
\vspace*{0.1cm}
\Statex \quad \, \fbox{\textbf{Primal updates}}

\vspace{0.05cm}

\State  \label{alg:primal_update}
\quad \, $\widetilde{\Ebs}^{(k+1)} = {\mathcal{P}}_{\mathbb{K}} \bigg(\widetilde{\Ebs}^{(k)} - \tau \, \big(\widetilde{\Ebs}^{(k)} - \Rbs \big) - \tau \mathcal{R}^\dagger \big( \Psib \bm{\mathsf{A}}^{(k)} + \bm{\mathsf{C}}^{(k)} \big) \bigg)$

\vspace*{0.05cm}

\State \label{alg:primal_update_zeta}
\quad \, $\bm{\mathsf{Z}}^{(k+1)} = \bm{\mathcal{P}}_{\mathbb{V}} \bigg(\bm{\mathsf{Z}}^{(k)} - \tau \bm{\mathsf{B}}^{(k)} \bigg)$

\vspace*{0.05cm}
\Statex \quad \, \fbox{\textbf{Dual updates}}

\vspace*{0.05cm}

\Statex \, \quad \, \underline{Promoting sparsity:}

\vspace*{0.05cm}
\State \label{alg:a_update}
\quad \, $\widetilde{\bm{\mathsf{A}}}^{(k)} = \bm{\mathsf{A}}^{(k)} + \rho_1 \mathsf{\bm{\Psi}}^\dagger \mathcal{R} \bigg( 2 \, \widetilde{\Ebs}^{(k+1)} + \Sbs_0 - \widetilde{\Ebs}^{(k)} \bigg)$

\vspace*{0.05cm}

\State \label{alg:prox_l1}
\quad \, $\bm{\mathsf{A}}^{(k+1)} = \widetilde{\bm{\mathsf{A}}}^{(k)} - \rho_1 \operatorname{prox}_{\rho_1^{-1} {\sigma^{(i)}} g} \bigg(  \rho_1^{-1} \widetilde{\bm{\mathsf{A}}}^{(k)} \bigg)$

\vspace*{0.05cm}

\Statex \, \quad \, \underline{Performing epigraphical projection:}

\vspace{0.1cm}
\State \label{alg:c_update}
\quad \, $ \widetilde{\bm{\mathsf{C}}}^{(k)} = \bm{\mathsf{C}}^{(k)} + \rho_2 \, \mathcal{R} \bigg( 2 \, \widetilde{\Ebs}^{(k+1)} + \Sbs_0 - \widetilde{\Ebs}^{(k)} \bigg)$

\vspace*{0.05cm}

\State \label{alg:c_zeta1_update}
\quad \, $\widetilde{\bm{\mathsf{B}}}^{(k)} = \bm{\mathsf{B}}^{(k)} + \rho_2 \bigg( 2 \, \bm{\mathsf{Z}}^{(k+1)} - \bm{\mathsf{Z}}^{(k)} \bigg)$

\vspace*{0.1cm}

\State \label{alg:proj_epi_h1}
\quad \, $\left[\begin{matrix}
\bm{\mathsf{C}}_{:,1}^{(k+1)} \\
\vspace*{-0.2cm} \\
 \bm{\mathsf{B}}_{:,1}^{(k+1)}
\end{matrix}\right]
= 
\left[\begin{matrix}
\widetilde{\bm{\mathsf{C}}}_{:,1}^{(k)} \\
\vspace*{-0.2cm} \\
\widetilde{\bm{\mathsf{B}}}_{:,1}^{(k)}
\end{matrix}\right]
- \rho_2 \, \bm{\mathcal{P}}_{\mathbb{E}_1}
\left(\frac{1}{\rho_2} \left[\begin{matrix}
\widetilde{\bm{\mathsf{C}}}_{:,1}^{(k)} \\
\vspace*{-0.2cm} \\
\widetilde{\bm{\mathsf{B}}}_{:,1}^{(k)}
\end{matrix}\right]
\right)$

\vspace*{0.1cm}

\State \label{alg:proj_epi_h2}
\quad \, $\left[\begin{matrix}
\bm{\mathsf{C}}_{:,2:4}^{(k+1)} \\
\vspace*{-0.2cm} \\
\bm{\mathsf{B}}_{:,2}^{(k+1)}
\end{matrix}\right]
= 
\left[\begin{matrix}
\widetilde{\bm{\mathsf{C}}}_{:,2:4}^{(k)} \\
\vspace*{-0.2cm} \\
\widetilde{\bm{\mathsf{B}}}_{:,2}^{(k)}
\end{matrix}\right]
- \rho_2 \, \bm{\mathcal{P}}_{\mathbb{E}_2}
\left(\frac{1}{\rho_2} \left[\begin{matrix}
 \widetilde{\bm{\mathsf{C}}}_{:,2:4}^{(k)} \\
  \vspace*{-0.2cm} \\
 \widetilde{\bm{\mathsf{B}}}_{:,2}^{(k)} 
\end{matrix}\right]
\right)
$

\vspace*{0.1cm}

%\State
\State \textbf{until convergence} \label{alg:d2_update_end_TV}

\end{algorithmic}
\end{algorithm}
%%%%%%%%%%%%%%%%%%%%%%%%%%%%%%%%%%%%%%%%%%%%%%%%%
%
%%%%%%%%%%%%%%%%%%%%%%%%%%%%%%%%%%%%%%%%%%%%%%%%% 
% 
In order to impose the constraints~\eqref{eq:h1_cons}-~\eqref{eq:zeta_cons}, we make use of the indicator functions of the corresponding sets. In particular, the constraints~\eqref{eq:h1_cons} and~\eqref{eq:h2_cons} are enforced using indicator functions of the epigraphs\footnote{The epigraph of a proper, lower semi-continuous function $\psi \colon \mathbb{R}^N \to ]-\infty, +\infty]$ corresponds to $\operatorname{epi} \psi = \left\{ (\bm{z}, \kappa) \in \mathbb{R}^N \times \mathbb{R} \, \middle\vert \, \psi(\bm{z}) \leq \kappa \right\}.$} $\mathbb{E}_1, \mathbb{E}_2$ respectively of the functions $h_1$ and $h_2$. On the other hand, the set 
$\mathbb{V} = \left\{ \bm{\mathsf{Z}} \in \mathbb{R}^{N \times 2} \middle\vert \, \big(\forall  n \in \{ 1, \ldots, N \} \big)  \; {{\mathsf{Z}}}_{n,1} + {{\mathsf{Z}}}_{n,2} \leq 0  \right\}$ is used to impose the constraint~\eqref{eq:zeta_cons}. For further details, we refer the reader to \cite{Birdi2018b}. 

Doing so, the proximity operator~\eqref{eq:prox_r_epi} results into
\begin{align} \label{eq:prox_r_epi_final}
    \operatorname{prox}_{{\sigma^{(i)}} r}(\Rbs)  = 
    \underset{\widetilde{\Ebs}, \Zbs}{\operatorname{argmin}} & \, \, {\sigma^{(i)}} g(\widetilde{\Ebs}) +  \iota_{\mathbb{K}} (\widetilde{\Ebs}) + \frac{1}{2} \|\widetilde{\Ebs} - \Rbs \|_F^2 + \iota_{\mathbb{V}} (\bm{\mathsf{Z}}) \nonumber \\
    & + \iota_{\mathbb{E}_1} \left(\mathcal{R} (\widetilde{\Ebs} + \Sbs_0)_{\colon,1}, \bm{\mathsf{Z}}_{:,1} \right)  \nonumber \\
    & +  \iota_{\mathbb{E}_2} \left( \mathcal{R}(\widetilde{\Ebs} + \Sbs_0)_{\colon, 2:4} ,  \bm{\mathsf{Z}}_{:,2} \right).  
\end{align}
For the computation of this proximity operator, we employ Polarized SARA method which is based on primal-dual forward-backward algorithm \citep{Condat2013,Vu2013,Pesquet2015}. Particularly, leveraging the flexibility and parallelizability offered by the primal-dual algorithms, we can easily adapt Polarized SARA method to solve our underlying problem~\eqref{eq:prox_r_epi_final}. For the sake of completeness, we also present this adapted version in Algorithm~\ref{algo:pd_fb}. It comprises of solving for the variables of interest, the primal variables, along with the associated auxiliary variables, the dual variables. In this context, the update of the primal variables is similar to the forward-backward strategy, wherein the gradient of the differentiable terms is followed by their proximity step. For the current case, it implies projection onto the set $\mathbb{K}$ (step~\ref{alg:primal_update}) and $\eV$ (step~\ref{alg:primal_update_zeta}) for the update of variables $\widetilde{\Ebs}$ and $\Zbs$, respectively. It is to be noted that these updates incorporate an additive term based on the corresponding dual variables. More precisely, the update of $\widetilde{\Ebs}$ involves the variables $\Abs \in \eR^{J \times 4}$ and $\Cbs \in \eR^{N \times 4}$ related to the sparsity prior and the epigraphical constraints, respectively. Similarly, step~\ref{alg:primal_update_zeta} comprises of the variable $\Bbs \in \eR^{N \times 2}$ associated with the epigraphical constraints. These dual variables are in turn updated by the computation of their associated proximity operators, that is the soft-thresholding operator for $\Abs$ (step~\ref{alg:prox_l1}) using the definition in equations~\eqref{eq:soft_thresh1} and~\eqref{eq:soft_thresh2}, and the projections onto the sets $\eE_1$ and $\eE_2$ for the variables $\Cbs$ and $\Bbs$. A detailed description of these updates is provided in \cite{Birdi2018b}.

\subsection{Convergence properties}
The convergence properties of the proposed joint calibration and imaging algorithm for full polarization model (Algorithm~\ref{algo_polcal}) can be deduced from \cite{Chouzenoux2016}. Particularly, if a finite number of iterations are performed to update each of the variables, that is for every $i^{\text{th}}$ iteration, choosing $L^{(i)}$ and $J^{(i)}$ to be finite in Algorithm~\ref{algo_polcal}, and if the
\begin{itemize}[labelindent=0pt, labelwidth=\widthof{(\ref{algo:prox_step})}, itemindent=1em, leftmargin=!]
    \item blocks $(\Ebs^{(i)}, \Ubs_1^{(i)}, \Ubs_2^{(i)})$ in Algorithm~\ref{algo_polcal} are estimated at least once in every finite number of given iterations, 
    
    \item step sizes $\Gamma_1^{(i)}$ and  $\Gamma_2^{(i)}$ for the gradient steps while updating $\Ubs_1$ and $\Ubs_2$, respectively are chosen as per equation~\eqref{eq:step_size_U}, and
    
    \item step size $\sigma^{(i)}$ for the update of $\Ebs$ is chosen as per equation~\eqref{eq:step_size_im},
\end{itemize}
then Algorithm~\ref{algo_polcal} is guaranteed to converge to a critical point $(\Ebs^{\prime}, \Ubs_1^{\prime}, \Ubs_2^{\prime})$ of the underlying objective function given in equation~\eqref{eq:global_min}. Additionally, iteration-by-iteration, the value of this objective function decreases.

\subsection{Computational complexity}
The computational complexity of Algorithm~\ref{algo_polcal} can be analyzed from its two sub-parts: Calibration and Imaging. In the former case, the update of matrix $\Ubs_1$ in steps~\ref{algo1:step:update_u1_fb_start}~-~\ref{algo1:step:update_u1_fb_end} (and $\Ubs_2$ in steps~\ref{algo1:step:update_u2_fb_start}~-~\ref{algo1:step:update_u2_fb_end}) is performed in parallel for $n_a$ antennas, using forward-backward (FB) iterations. In this case, the gradient evaluation of the data fidelity term in equation~\eqref{eq:data_fid_cal1} (and in~\eqref{eq:data_fid_cal2}) is required. It involves the degridding operation while generating the matrix $\mathcal{X}_{t, \beta, \alpha} \big( \,{\Fbs} \, \widetilde{\Zbs} \widetilde{\mathcal{L}}({\Sbs_0}+\Ebs) \big)$ for all $t \in \{1, \ldots, T\}$ and $\beta \in \{1, \ldots, n_a\}$, with $\beta \neq \alpha$, that turns out to be the most expensive step. However, it is to be noted that for each global calibration iteration, this matrix needs to be computed only once before the inner FB iterations (both for $\Ubs_1$ and $\Ubs_2$ updates).

Regarding the imaging step, the FB strategy presents the computationally most demanding part of the algorithm. In particular, in step~\ref{algo1:step:update_e}, the gradient computation (forward step) of the data fidelity term (equation~\eqref{eq:data_fid_im}) consists in performing the application of the operator $\Phi$, followed by its adjoint operator. Furthermore, the evaluation of the proximity operator (backward step) involves sub-iterations, performed either by Algorithm~\ref{algo:dualfb} or Algorithm~\ref{algo:pd_fb}. While this iterative process adds to the computational cost, the heaviest steps within either of the algorithms is the application of the wavelet transform operator $\Psib$ consisting of 8 wavelet bases and Dirac basis to impose sparsity. This can in turn be implemented in a parallel fashion, for each sparsity basis as well as for each of the underlying images.

%% file: sim_results.tex
\section{Simulations and results}
\label{sec:results}

%----------------------------------------------------
\begin{figure}
\centering
\begin{tabular}{@{}c@{}c@{}c@{}c@{}}
\includegraphics[trim ={0.2cm 0 0 0cm},clip,width=3.6cm]{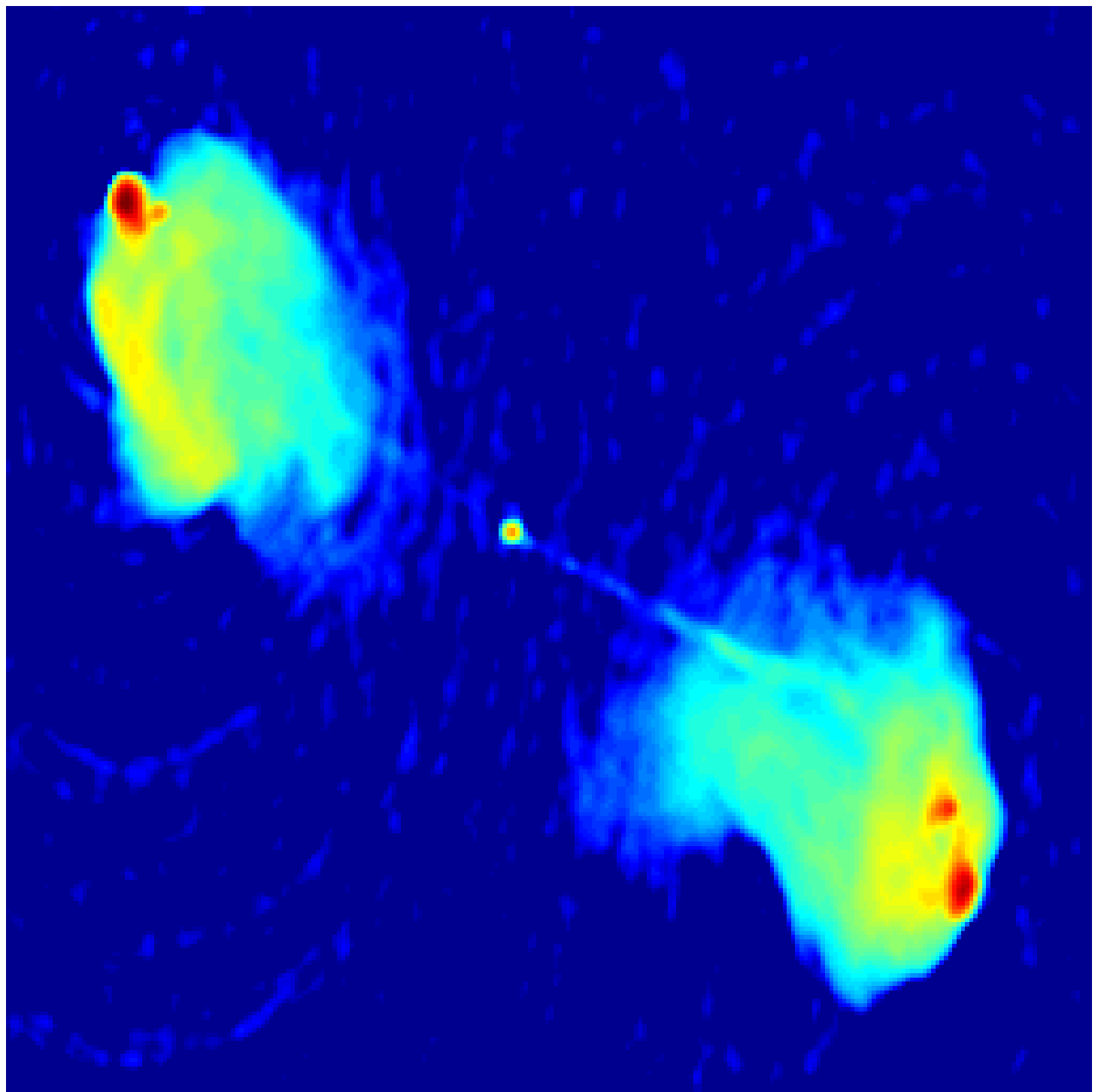} &
\hspace{-0.05cm}\includegraphics[trim ={16.2cm 0 0 0cm},clip,width=0.88cm]{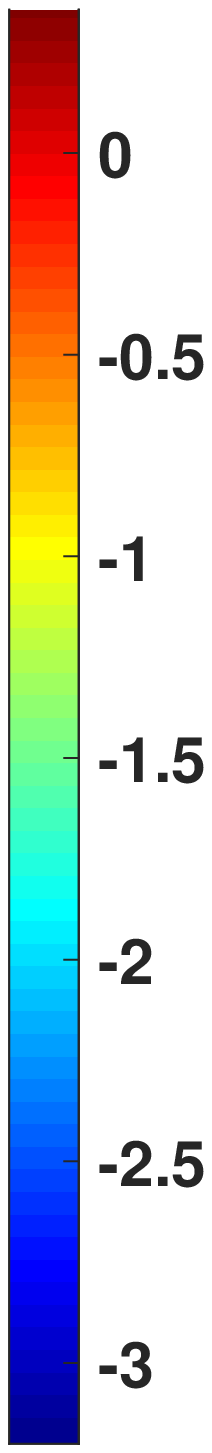} &
\hspace*{-0.2cm}\includegraphics[trim ={0.2cm 0 0 0cm},clip,width=3.6cm]{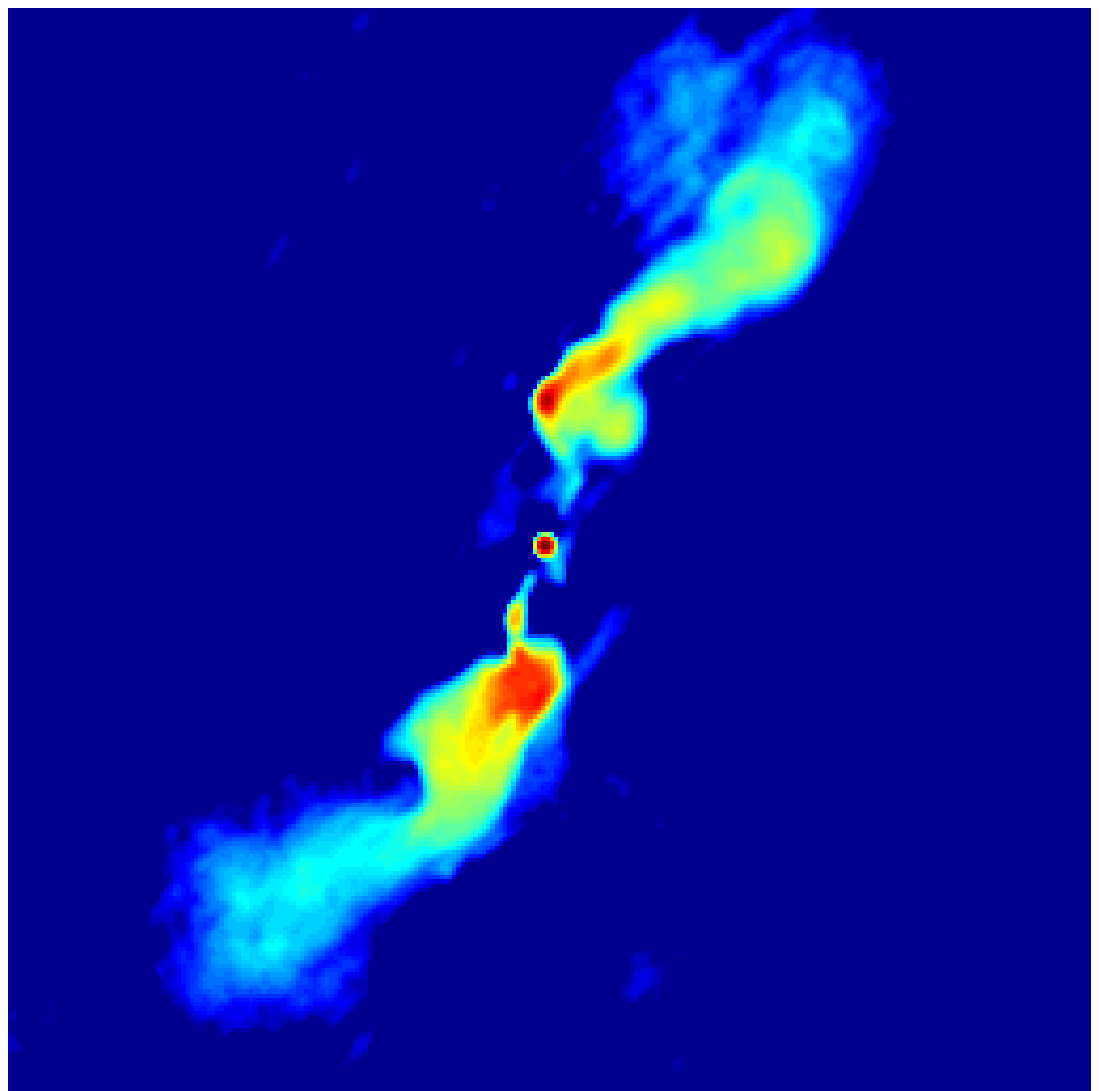} &
\hspace{-0.01cm}\includegraphics[trim ={16.2cm 0 0 0cm},clip,width=0.88cm]{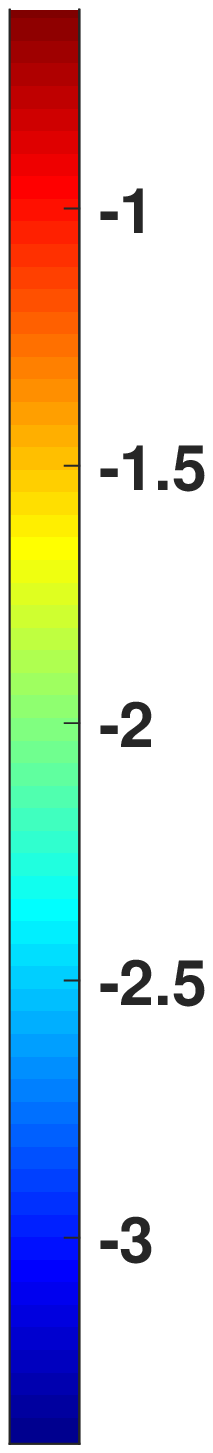} \\
\includegraphics[trim ={0.2cm 0 0 0cm},clip,width=3.6cm]{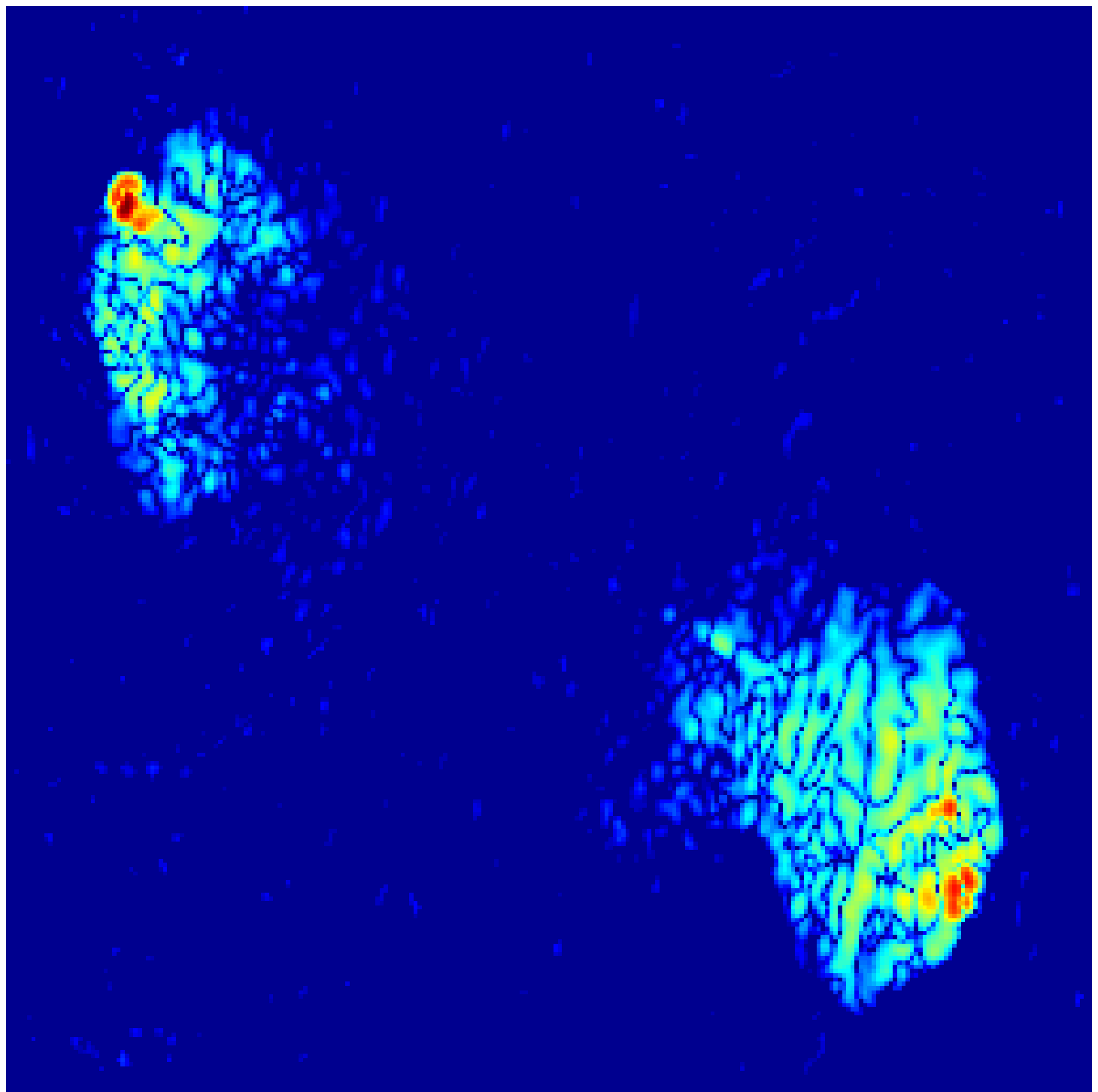} &
\hspace{-0.05cm}\includegraphics[trim ={16.2cm 0 0 0cm},clip,width=0.88cm]{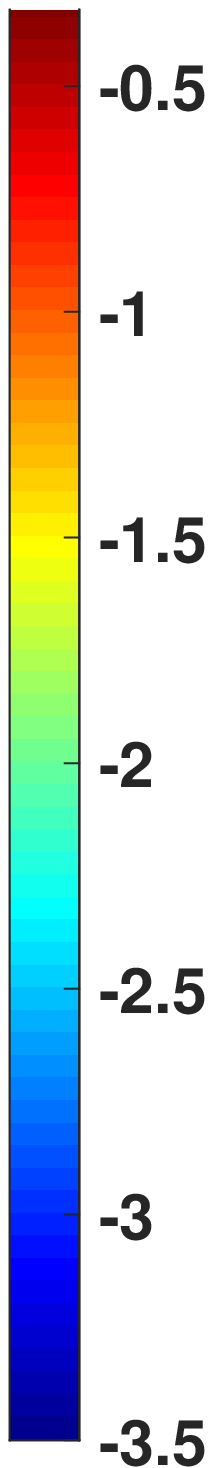} &
\hspace*{-0.2cm}\includegraphics[trim ={0.2cm 0 0 0cm},clip,width=3.6cm]{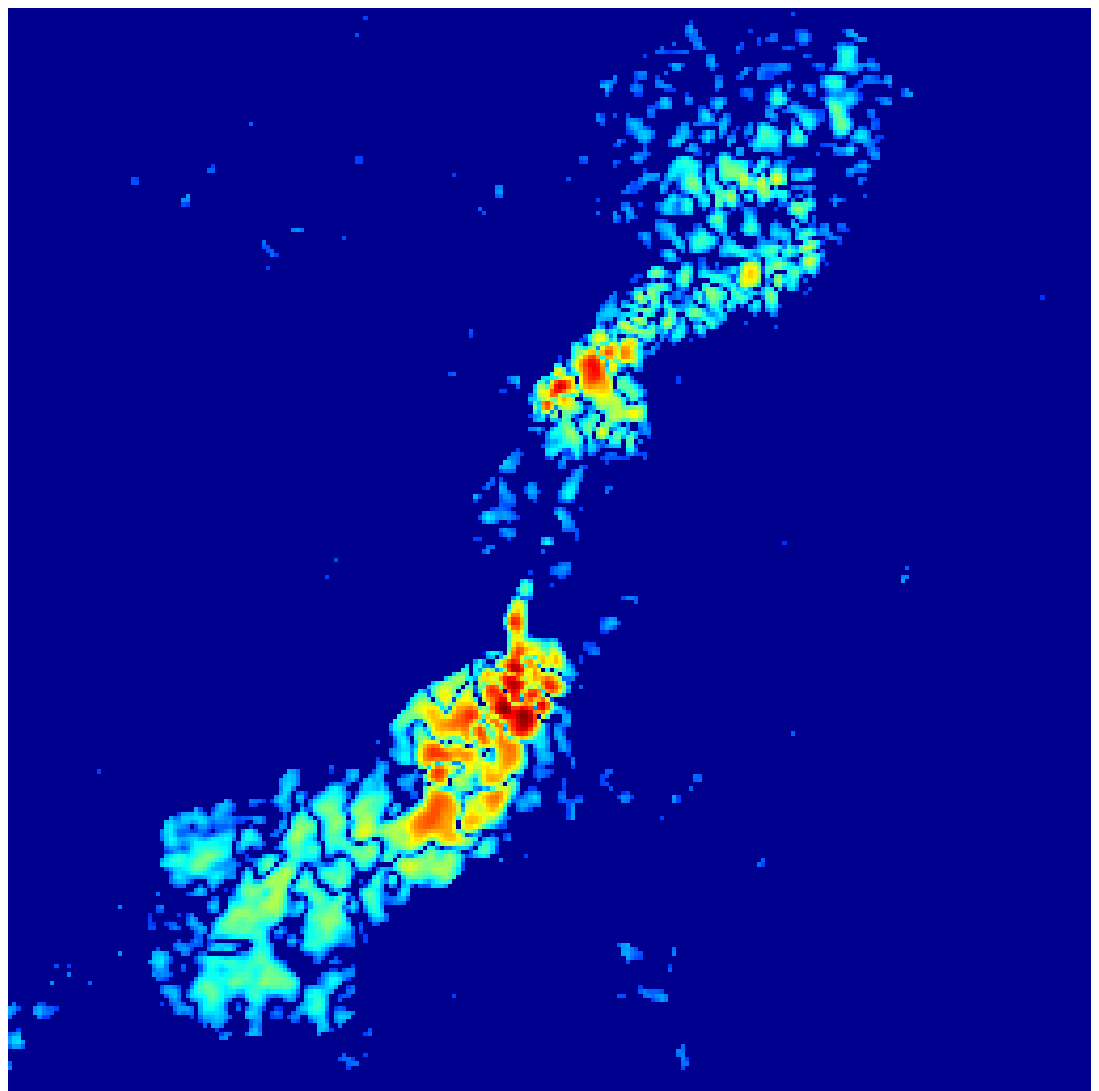} &
\hspace{-0.01cm}\includegraphics[trim ={16.2cm 0 0 0cm},clip,width=0.88cm]{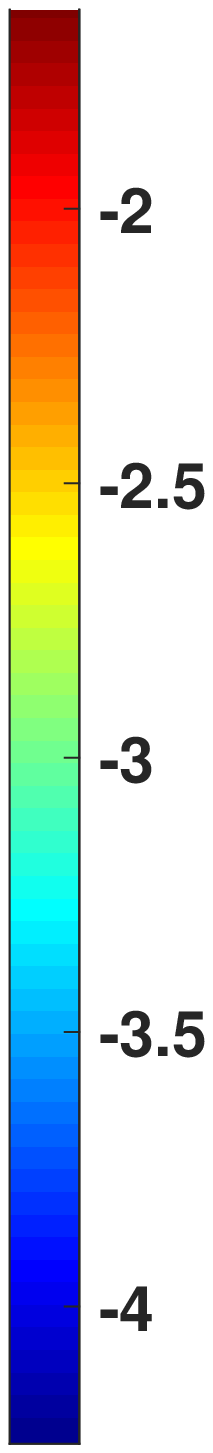} \\
\includegraphics[trim ={0.2cm 0 0 0cm},clip,width=3.6cm]{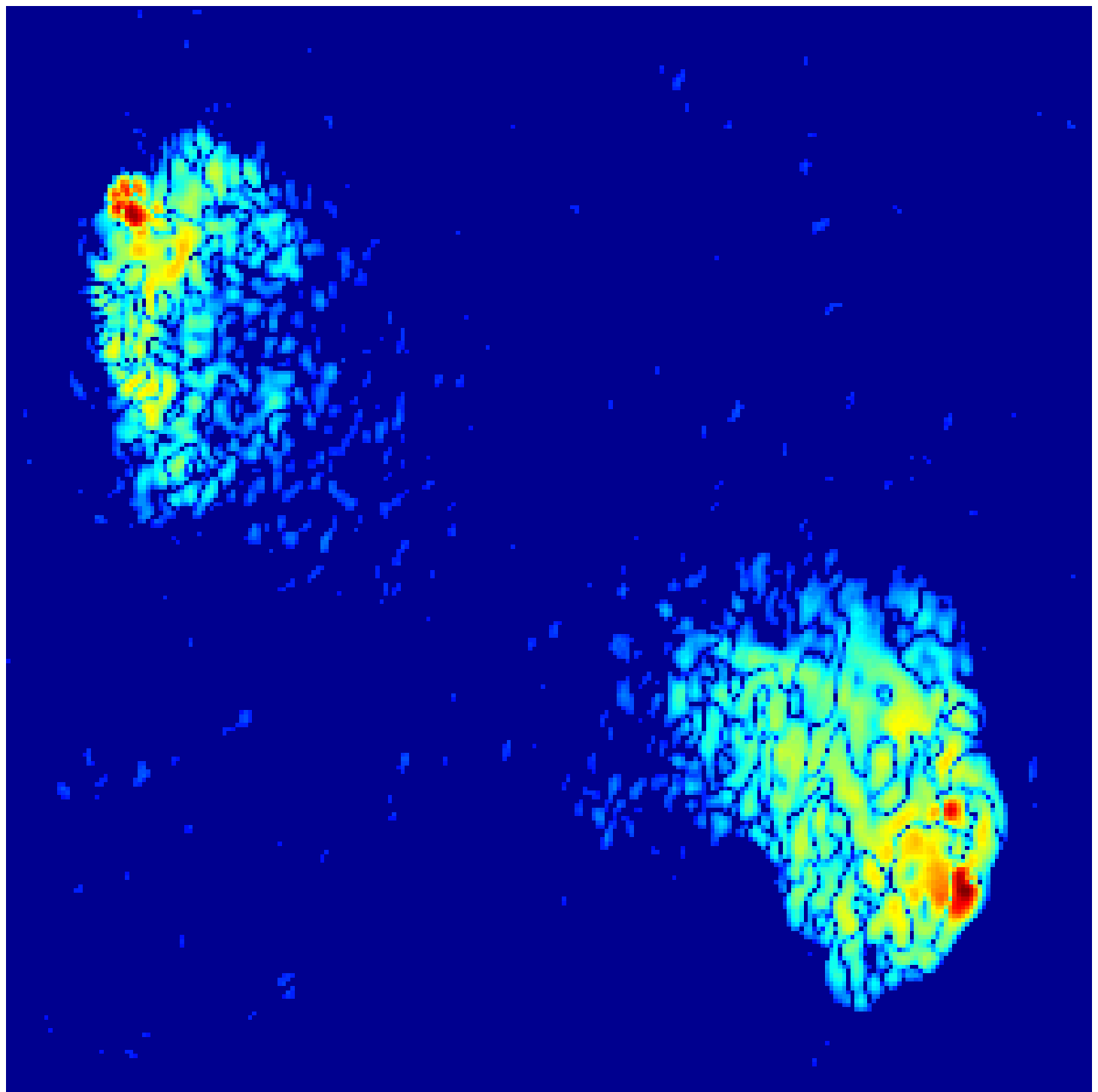} &
\hspace{-0.05cm}\includegraphics[trim ={16.2cm 0 0 0cm},clip,width=0.88cm]{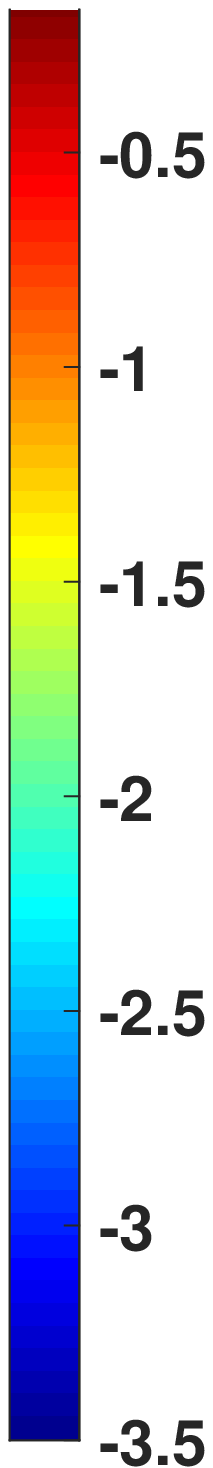} &
\hspace*{-0.2cm}\includegraphics[trim ={0.2cm 0 0 0cm},clip,width=3.6cm]{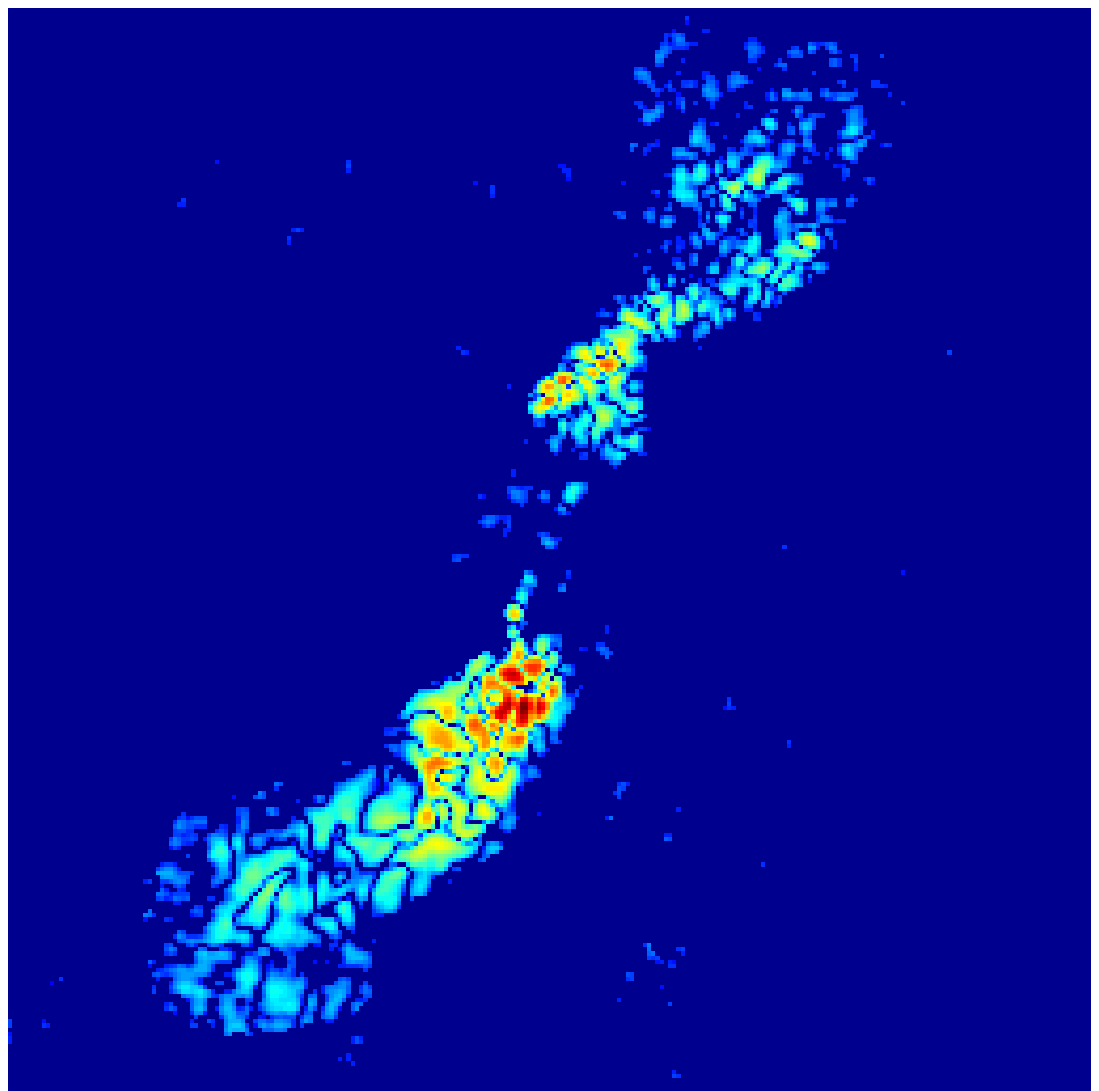} &
\hspace{-0.01cm}\includegraphics[trim ={16.2cm 0 0 0cm},clip,width=0.88cm]{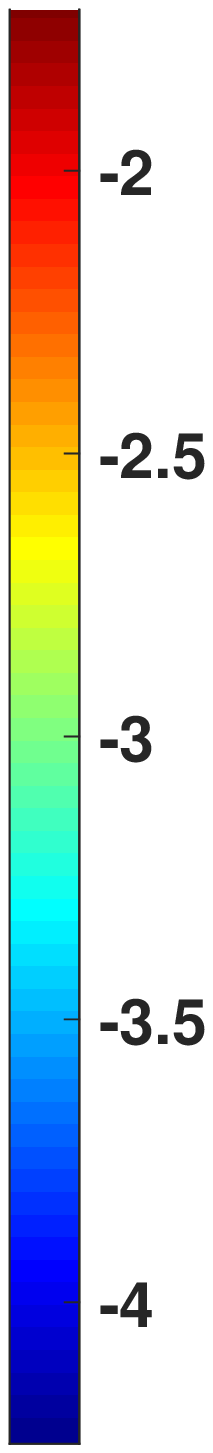} \\
\end{tabular}
\caption{Cygnus A (first column) and Hydra A (second column) ground truth images used for performing simulations. In both the columns, row-wise from top to bottom the following images are shown (all in log scale): Stokes $I$, Stokes $Q$ and Stokes $U$. }
\label{fig:true_images}
\end{figure}  
%----------------------------------------------------

In this section, we describe the various simulations conducted using a \textsc{matlab} implementation of the proposed algorithm to assess its performance.

To simulate the data sets, we consider VLA antenna's A configuration. It consists of $n_a = 27$ antennas with each antenna pair acquiring measurements at $T = 200$ snapshots, considering linear feeds. While performing the non-uniform Fourier transform at these sampled frequencies, we make use of the Kaiser-Bessel kernels of size $5 \times 5$ for interpolation \citep{Fessler2003}.
% The resultant $u-v$ coverage is shown in Fig.~\eqref{fig:uv_cov}.
Furthermore, we perform tests on two sets of model images of size $N = 256 \times 256$. The two sets correspond to the images of the radio galaxies Cygnus A and Hydra A, shown respectively in first and second columns of Fig.~\ref{fig:true_images}. For each set, Stokes $I$, $Q$ and $U$ images ($V = 0$ due to negligible circular polarization) are displayed from top to bottom. 

Regarding the DDE Fourier kernels, for every antenna $\alpha \in \{1,\ldots,n_a\}$ and at each time instant $t \in \{1, \ldots, T\}$, these kernels are generated randomly in the Fourier domain, with a spatial Fourier support of $K = 5 \times 5$ and a temporal Fourier support of size $P = 3 \times 3$. Furthermore, the values of the Fourier coefficients are chosen as per the discussion in Section~\ref{sec:cal_prob}. More precisely, for the diagonal terms in the Jones matrix, the central coefficients that correspond to the Fourier coefficients of the DIEs have their real and imaginary parts lying in the interval $[1-\theta_1, 1+ \theta_1]$  and  $[-\theta_1, \theta_1]$, respectively, whereas the other frequency coefficients (related to the DDEs Fourier coefficients) belong to $[-\theta_1, \theta_1]$ with $\theta_1 = 5 \times 10^{-2}$. 
On the other hand, the central and all the rest coefficients in the off-diagonal terms are considered belonging to $[-\theta_2, \theta_2]$ and $[-\theta_3, \theta_3]$, respectively. Here we choose $(\theta_2, \theta_3) = (5 \times 10^{-2}, 5 \times 10^{-4})$. The chosen values are in line with the VLA characteristics, wherein the leakage terms (i.e. the off-diagonal terms) are $\sim 10^{-2}$ lower in peak amplitude compared to the diagonal terms, representing a leakage of around $5 \%$ of the diagonal terms \citep{Bhatnagar2013}.

%%%%%%%%%%%%%%%%%%%%%%%%%%%%%%%%%%%%%%%%%%%%%%%%%%%%%
%----------------------------------------------
% Table style for Cygnus A: SNR
\begin{table*}
\begin{tabular}{c|c|c|c|c|c|c|c|c}
& \multicolumn{4}{|c|}{without polarization constraint}  &  \multicolumn{4}{c}{with polarization constraint} \\
\hline 
 & Normalized & DIE & DDE calibration & DDE calibration  & Normalized & DIE & DDE calibration & DDE calibration   \\
& DIEs & calibration  &  w/o off- & of full Jones & DIEs & calibration & w/o  off- & of full Jones \\
& & & diagonal terms & matrix & & & diagonal terms & matrix \\
\hline 
Stokes $I$ & 19.4 & 21.4 & 34 & 34.7 & 18.4 & 20.2 & 33.6 & 37.4 \\
Stokes $Q$ & 18.6 & 16.9 & 20.8 & 21.2 & 16.3 & 15.2 & 23.4 & 24.9 \\
Stokes $U$ & 16.9 & 18 & 19.9 & 22.5 & 15.6 & 16.7 & 21.3 & 25.2 \\
\hline 
\multicolumn{9}{c}{(a) SNR (in dB) values for Cygnus A} \vspace{0.2cm}\\
& \multicolumn{4}{|c|}{without polarization constraint}  &  \multicolumn{4}{c}{with polarization constraint} \\
\hline 
  & Normalized & DIE & DDE calibration & DDE calibration  & Normalized & DIE & DDE calibration & DDE calibration   \\
& DIEs & calibration  &  w/o off- & of full Jones & DIEs & calibration & w/o  off- & of full Jones \\
& & & diagonal terms & matrix & & & diagonal terms & matrix \\
\hline 
Stokes $I$ & 1.04 & 1.73 & 10.8 & 12.8 & 1.07 & 1.82 & 9.64 & 41.4 \\
Stokes $Q$ & 3.31 & 3.43 & 4.79 & 4.94 & 2.59 & 2.16 & 6.79 & 9.61 \\
Stokes $U$ & 2 & 2.14 & 2.62 & 3.38 & 0.55 & 0.87 & 1.39 & 19.8 \\
\hline
\multicolumn{9}{c}{(b) Dynamic range (DR) values for Cygnus A in units of $10^4$} \vspace{0.2cm} \\

& \multicolumn{4}{|c|}{without polarization constraint}  &  \multicolumn{4}{c}{with polarization constraint} \\
\hline 
   & Normalized & DIE & DDE calibration & DDE calibration  & Normalized & DIE & DDE calibration & DDE calibration   \\
& DIEs & calibration  &  w/o off- & of full Jones & DIEs & calibration & w/o  off- & of full Jones \\
& & & diagonal terms & matrix & & & diagonal terms & matrix \\
\hline 
Stokes $I$ & 18.3 & 20.6 & 24.9 & 25 & 18.3 & 20.7 & 31.2 & 33.2 \\
Stokes $Q$ & 17.8 & 18.6 & 11.3 & 11.4 & 17.9 & 18.8 & 19.7 & 20.7 \\
Stokes $U$ & 10.3 & 11.5 & 8.09 & 9.07 & 10.9 & 12.2 & 12.6 & 19.8 \\
\hline
\multicolumn{9}{c}{(c) SNR (in dB) values for Hydra A} \vspace{0.2cm}\\
& \multicolumn{4}{|c|}{without polarization constraint}  &  \multicolumn{4}{c}{with polarization constraint} \\
\hline 
   & Normalized & DIE & DDE calibration & DDE calibration  & Normalized & DIE & DDE calibration & DDE calibration   \\
& DIEs & calibration  &  w/o off- & of full Jones & DIEs & calibration & w/o  off- & of full Jones \\
& & & diagonal terms & matrix & & & diagonal terms & matrix \\
\hline 
Stokes $I$ & 0.82 & 1.54 & 2.35 & 2.45 & 0.82 & 1.49 & 6.43 & 15.9 \\
Stokes $Q$ & 2.01 & 2.21 & 0.23 & 0.23 & 0.68 & 0.93 & 1.01 & 1.19 \\
Stokes $U$ & 2.44 & 2.53 & 0.21 & 0.28 & 0.31 & 0.5 & 0.48 & 2.7 \\
\hline 
\multicolumn{9}{c}{(d) Dynamic range (DR) values for Hydra A in units of $10^4$} 
\end{tabular}
\caption{Cygnus A (top two tables) and Hydra A (bottom two tables) results: SNR and Dynamic range values for different considered cases. For each of the reconstructed Stokes image, the shown values are the mean computed over 5 performed simulations for the cases with following considerations- for calibration step: known normalized DIEs, DIE calibration, no off-diagonal Jones terms calibration, and full Jones matrix DDE calibration; and for imaging step: without and with the enforcement of the polarization constraint.}
\label{tab:results}
\end{table*}
%%%%%%%%%%%%%%%%%%%%%%%%%%%%%%%%%%%%%%%%%%%%%%%%%%%%%

\vspace{-0.4cm}
\subsection{Comparisons performed}
For an assessment of the proposed algorithm, we perform an extensive study on a varied number of cases based on choices made both on calibration and imaging fronts. From the calibration viewpoint, {we compare between different generations of calibration schemes, i.e.\ 1GC, 2GC and 3GC.} Particularly, we compare the results obtained by DDE calibration with those of considering only DIEs. Furthermore, to determine the importance of off-diagonal terms in the Jones matrices, we run tests with and without calibrating for these terms. On the other hand, from the imaging perspective, it is interesting to analyze the performance of different regularizations for the Stokes parameters. Thus, we consider the cases with and without enforcement of the polarization constraint. A blend of these approaches leads to the following list of tests:
\begin{itemize}[labelindent=0pt, labelwidth=\widthof{(\ref{algo:prox_step})}, itemindent=1em, leftmargin=!] \vspace{0.1cm}
%  \vspace{-0.2cm}
    \item[(1)] Imaging with normalized DIEs {(1GC)}: Working under the assumption of calibration transfer, the first step is to consider the Jones matrices as identity, without any directional dependency. It is important to mention here that this is the usual case considered by imaging algorithms in RI, either ignoring the calibration effects or relying on the data provided after the calibration transfer. In such a scenario, only imaging step is performed to obtain an estimation of the Stokes images $\Sbs$. Furthermore, in this imaging problem, we can consider two different regularizations and hence two sub-cases: Imaging with normalized DIEs (1.a) without and (1.b) with the enforcement of the polarization constraint.
   
   \vskip2pt
    
    \item[(2)] {Joint DIE calibration and imaging {(2GC)}: It involves using the proposed algorithm for calibration and imaging. While calibrating, only the DIEs, i.e. considering support size $K = 1$ for Fourier kernels, are solved for. A thresholded version of the images estimated from (1.a) and (1.b) are used respectively to initialize the problems while working in the absence and presence of the polarization constraint. This corresponds to two sub-cases: DIE calibration and imaging (2.a) without and (2.b) with the polarization constraint.}
     
    \vskip2pt

    \item[(3)] {Joint DDE calibration and imaging {(3GC-polarized)}: This approach follows the proposed algorithm for calibrating the DDEs in conjunction with the imaging step. Similar to (2), the images are initialized from thresholded versions of (1.a) and (1.b), resulting into four cases:}
    \begin{itemize}[labelindent=0pt, labelwidth=\widthof{(\ref{algo:prox_step})}, itemindent=1em, leftmargin=!]
     \vspace{-0.1cm}
     \item[(3.a)] DDE calibration for the Jones matrices, excluding the off-diagonal terms \& imaging without the polarization constraint: It corresponds to the case when the calibration steps are applied only to update the diagonal terms of the Jones matrices, without accounting for the polarization constraint in the imaging step.
      \vspace{0.05cm}
     
     \item[(3.b)] DDE calibration for the Jones matrices, excluding the off-diagonal terms \& imaging with the polarization constraint: Here, the calibration strategy is the same as (3.a), whereas the polarization constraint is considered in the imaging step.
  \vspace{0.05cm}
     
     \item[(3.c)] DDE calibration for the whole Jones matrix, including the off-diagonal terms \& imaging without the polarization constraint: It consists in calibrating for the full Jones matrix, but without enforcing the polarization constraint in the imaging step.
      \vspace{0.05cm}
     
     \item[(3.d)] DDE calibration for the whole Jones matrix, including the off-diagonal terms \& imaging with the polarization constraint: This approach comprises of full Jones matrix calibration and imposing the polarization constraint in the imaging step.

    \end{itemize}
\end{itemize}

\vspace{-0.5cm}
\subsection{Simulation settings}

As previously mentioned, one of the key points for convergence of the proposed algorithm is to update the variables at least once in every finite number of given iterations. To ensure this, within the global loop in Algorithm~\ref{algo_polcal}, we perform $L_{\text{cyc}} = 2$ global iterations for the DDEs (i.e. steps~\ref{algo1:step:update_u_start} to~\ref{algo1:step:update_u_end} are iterated twice) followed by an update of the Stokes images (steps~\ref{algo1:step:update_e_start} to~\ref{algo1:step:update_e_end}).
Nevertheless, after a certain number of global loops, it may happen that the DDE updates stabilize in fewer than $L_{\text{cyc}}$ iterations. Thus, to avoid unnecessary computation, we define a stopping criterion for the DDE updates as the relative variation between the consecutive estimates of DDEs to be less than a threshold {$\varepsilon_\Us$}, that is
\begin{equation}
    \underset{q \in \{1,2\}}{\operatorname{max}} \left( \|\Ubs_q^{(i+1)} - \Ubs_q^{(i)} \|_F / \| \Ubs_q^{(i)} \|_F \right ) \leq \varepsilon_\Us.
\end{equation}
where we set $\varepsilon_\Us = 10^{-5}$. In other words, if this criterion is met, then no more DDE iterations are performed, and the algorithm leaps to the imaging step using the estimated DDEs.
The estimation of the Stokes images adopts a similar strategy wherein iterations in the inner imaging loop are stopped and the algorithm resorts to the calibration step once the following stopping criterion is satisfied:
\begin{equation}
    \underset{q \in \{1,2,3\}}{\operatorname{max}} \left( \|\Ebs_{\colon, q}^{(i+1)} - \Ebs_{\colon, q}^{(i)} \|_2 / \| \Ebs_{\colon, q}^{(i)} \|_2 \right ) \leq \varepsilon_\Es.
\end{equation}
with {$\varepsilon_\Es = 10^{-5}$}.
Furthermore,
in order to ensure stopping of the global algorithm at convergence, we define a stopping criterion as the relative variation between the values of the objective function at consecutive iterates to be less than a threshold $\varepsilon_0$, that is
\begin{equation}
    \|\varphi^{(i+1)} - \varphi^{(i)} \|_2 / \| \varphi^{(i)} \|_2  \leq  \varepsilon_0 ,
\end{equation}
where $\varphi^{(i)}$ denotes the objective function value for the $i^{\text{th}}$ iteration computed using the updated values $(\Ebs^{(i+1)}, \Ubs_1^{(i+1)}, \Ubs_2^{(i+1)})$ and {$\varepsilon_0$ is fixed to $ 2 \times 10^{-2}$}. 
Regarding the choice of number of iterations $L^{(i)}$ and $J^{(i)}$ for calibration and imaging inner loops, respectively, we choose $L^{(i)} = 5$ when DDE updates need to be performed and $J ^{(i)}= 100$ when imaging step is carried out. Furthermore, for each of the above mentioned cases, we run 5 simulations varying the DIEs/DDEs and noise realizations. 

%
%%%%%%%  CYGNUS-A: Stokes I  %%%%%%%%%%%%%%%%%
\begin{figure*}
\centering
\begin{tabular}{@{}c@{}c@{}c@{}c@{}}
\includegraphics[trim ={0.2cm 0 0 0cm},clip,width=4cm]{true_I_cyg_a.eps} &
\hspace{-3cm}\includegraphics[trim ={16.2cm 0 0 0cm},clip,width=0.97cm]{I_cyg_a_colorbar.eps}
& & \\
Recovered images w/o &  Absolute error images &  Recovered images with & Absolute error images \\
polarization constraint & &polarization constraint & \\
\includegraphics[trim ={0.2cm 0 0 0cm},clip,width=4cm,align=c]{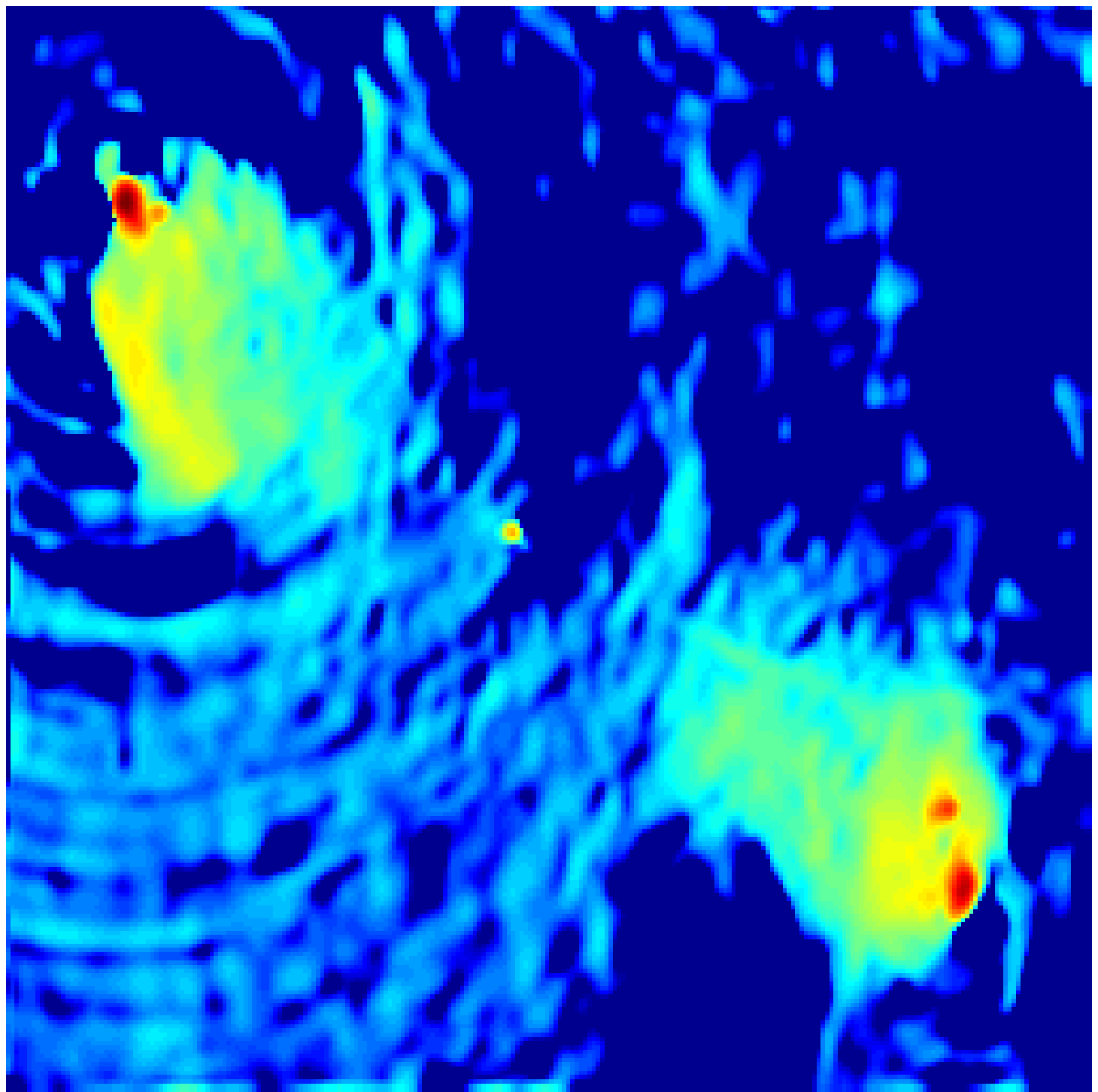} &
\hspace*{0.3cm}\includegraphics[trim ={0.2cm 0 0 0cm},clip,width=4cm,align=c]{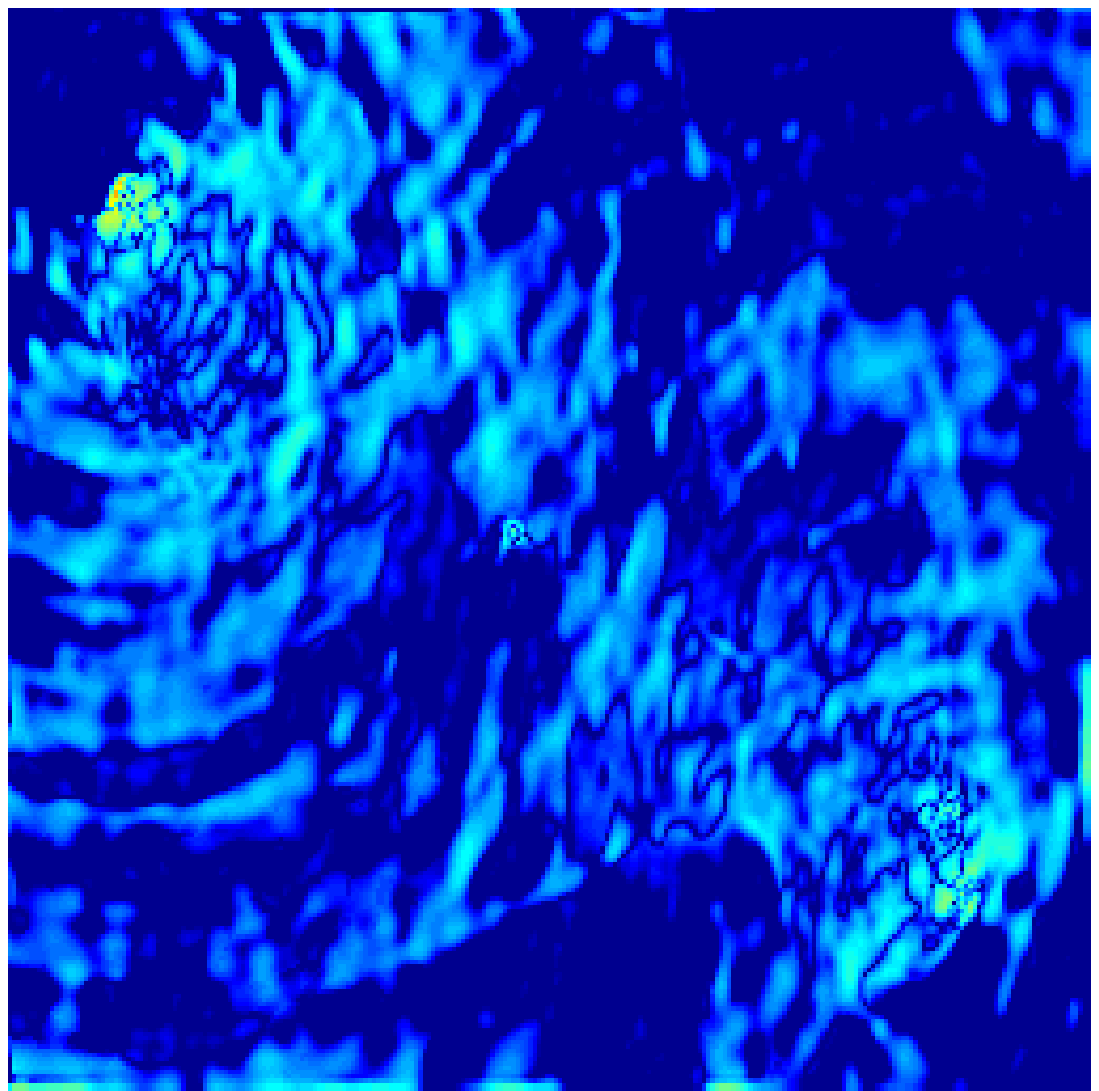} &
\hspace*{0.3cm}\includegraphics[trim ={0.2cm 0 0 0cm},clip,width=4cm,align=c]{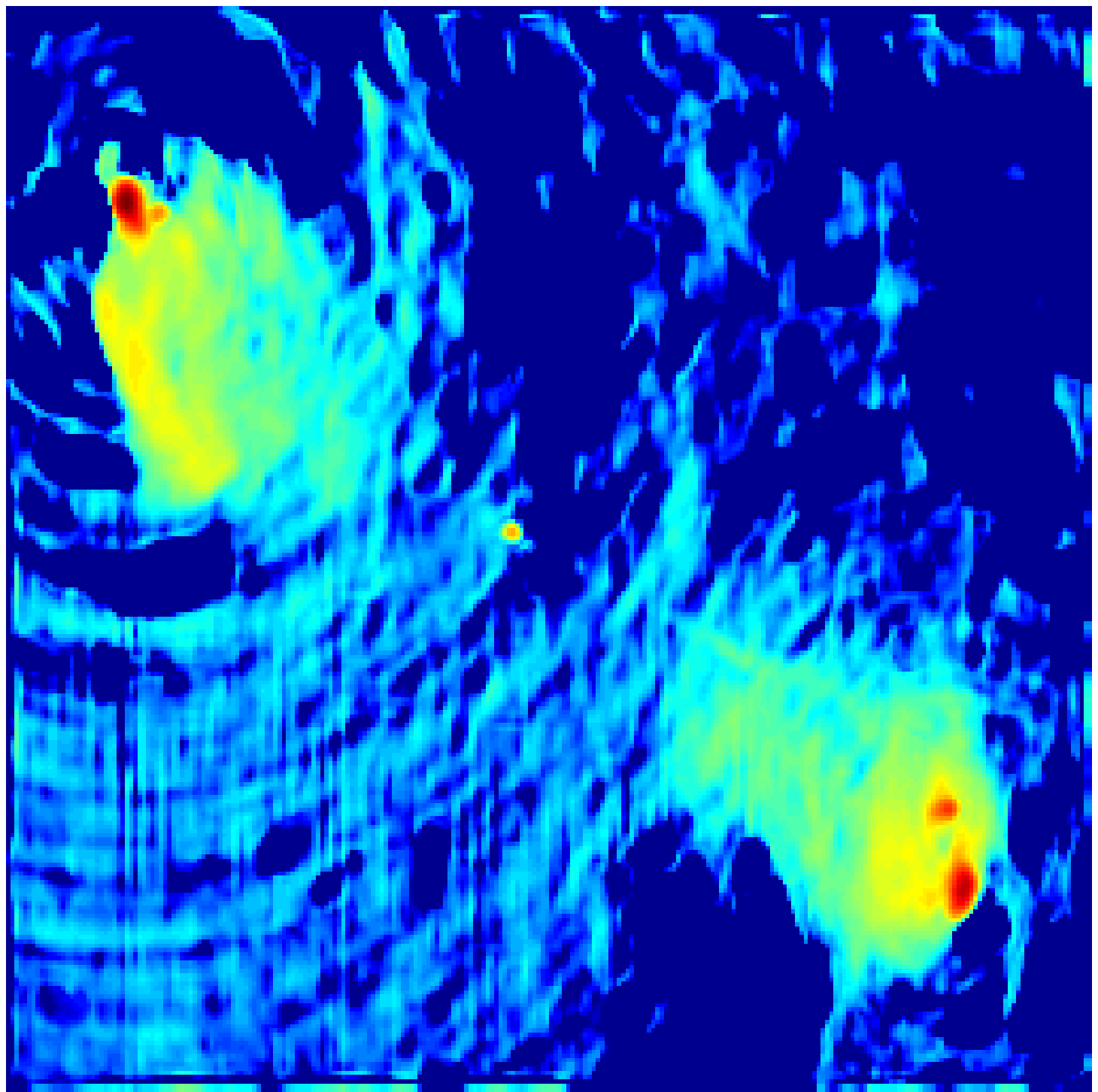} &
\hspace*{0.3cm}\includegraphics[trim ={0.2cm 0 0 0cm},clip,width=4cm,align=c]{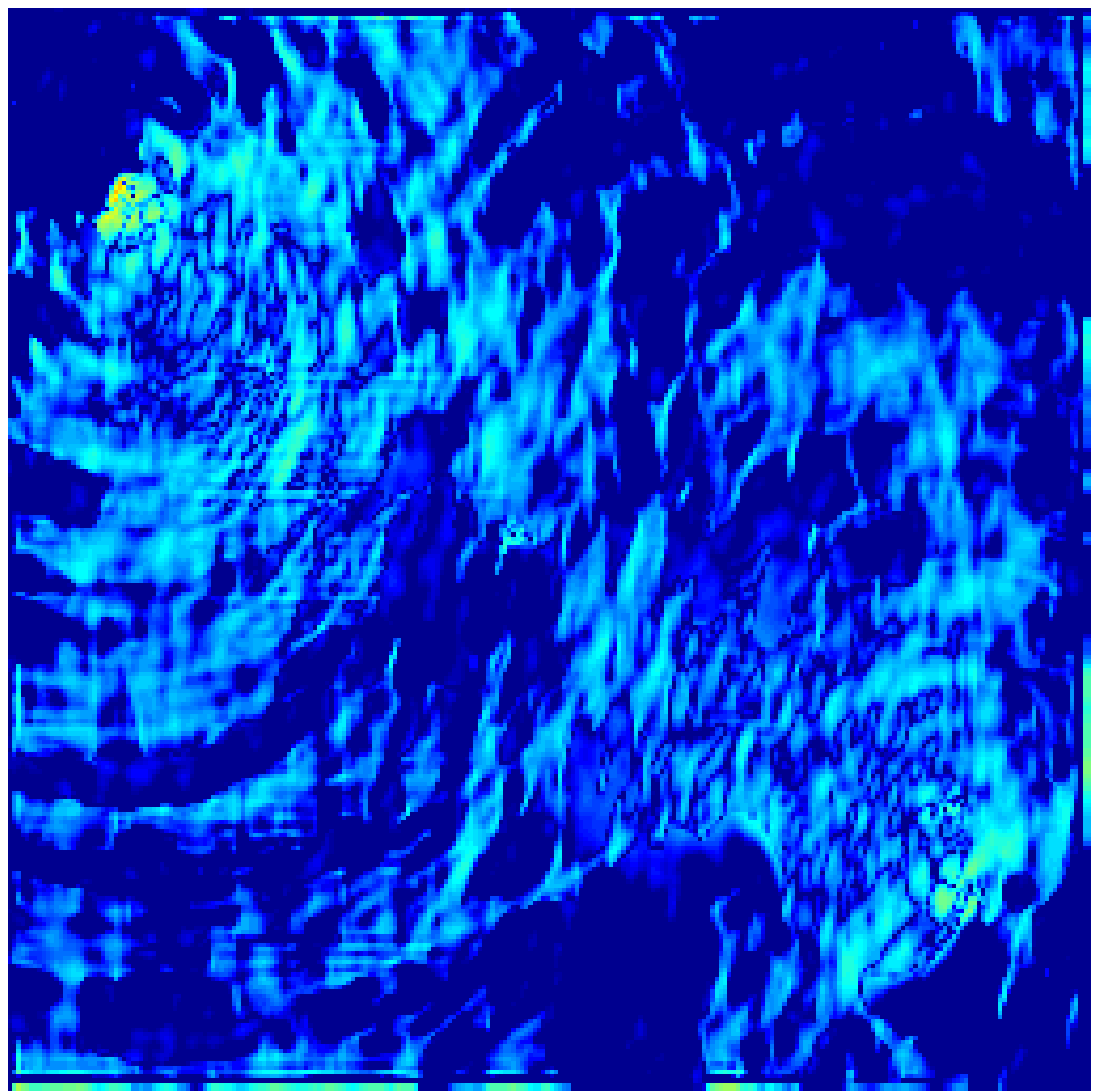} 
\vspace{0.25cm}
\\
%-----------------------------------------------
{\includegraphics[trim ={0.2cm 0 0 0cm},clip,width=4cm,align=c]{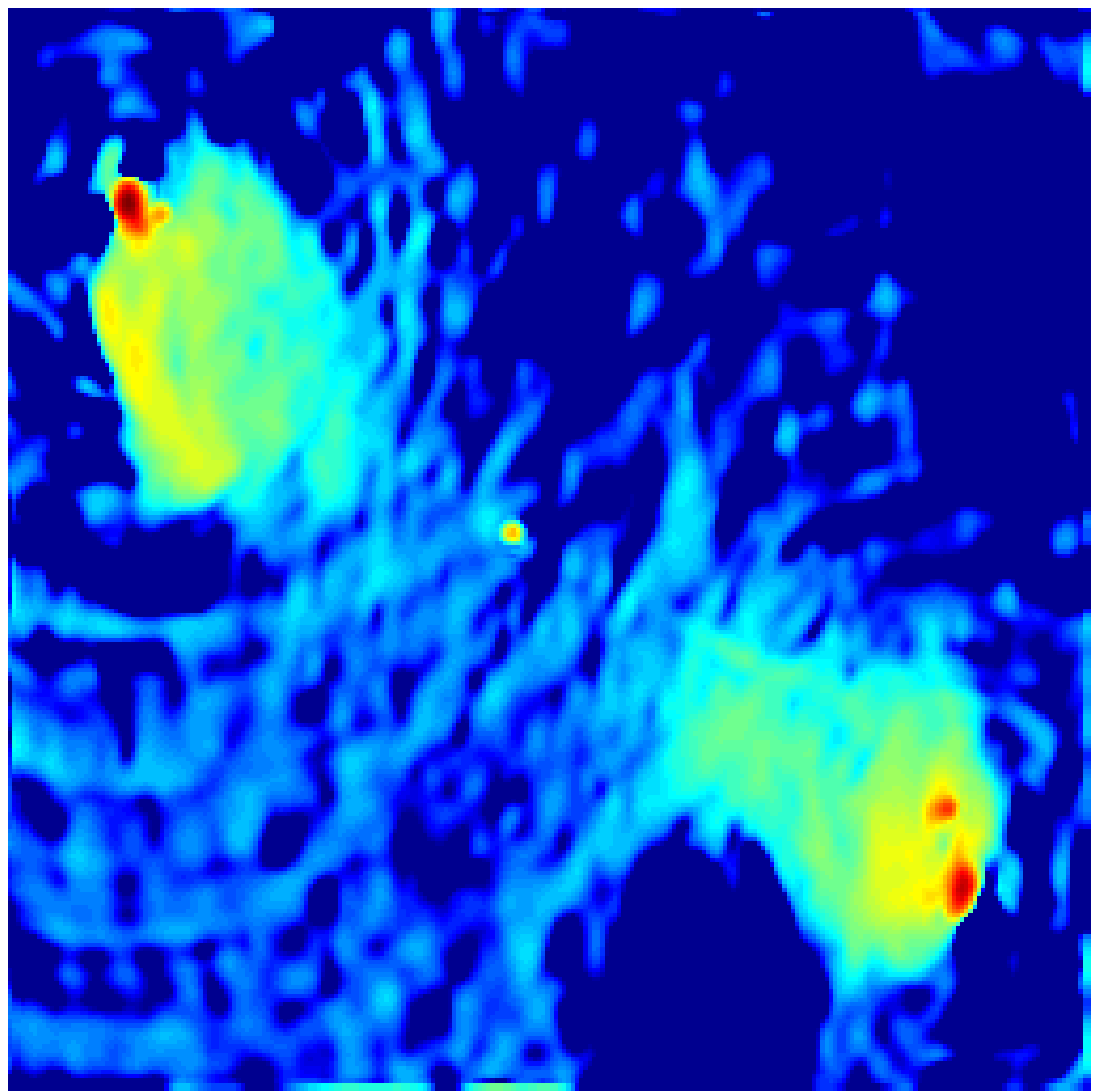}} &
\hspace*{0.3cm}\includegraphics[trim ={0.2cm 0 0 0cm},clip,width=4cm,align=c]{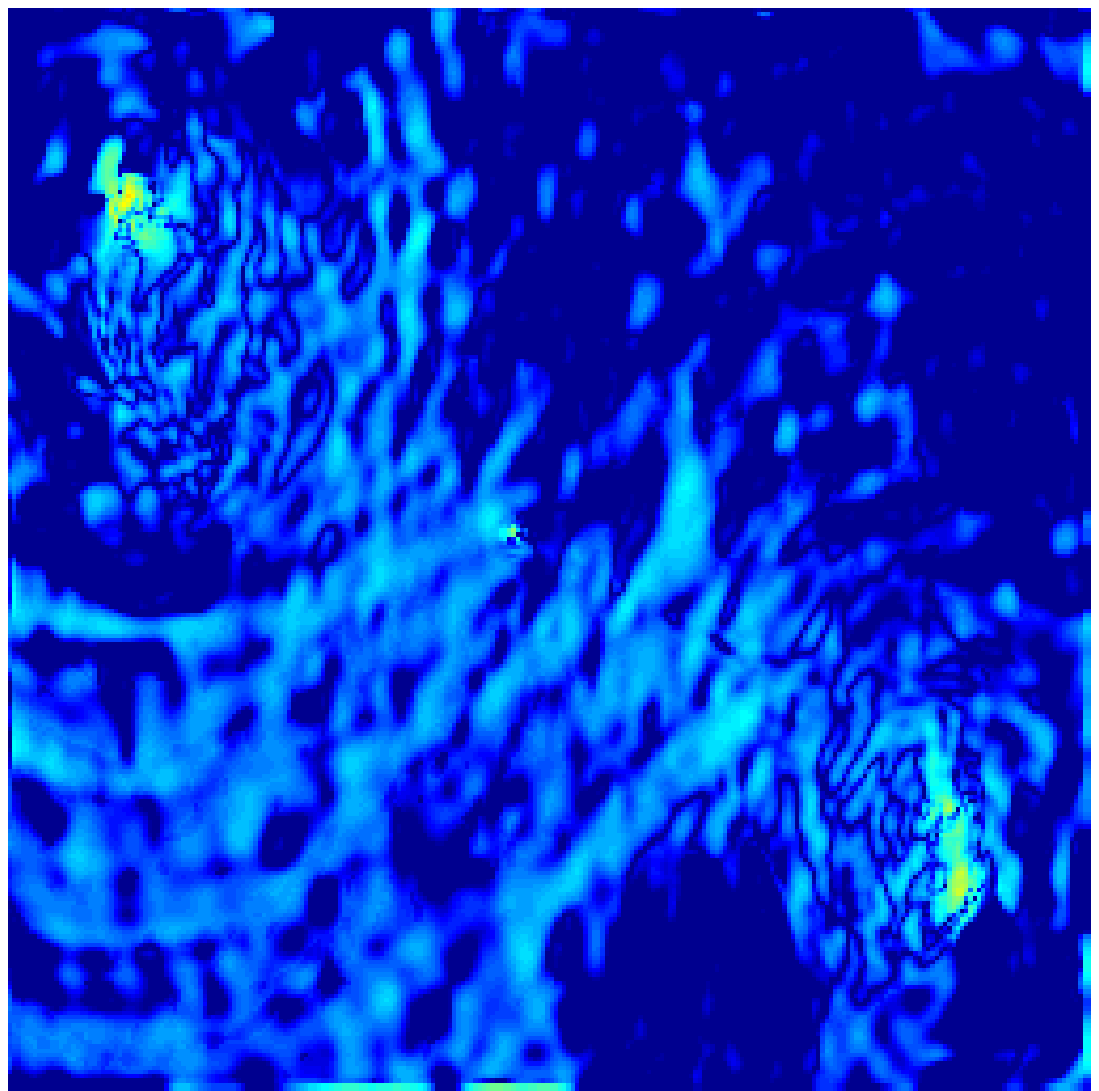} &
\hspace*{0.3cm}\includegraphics[trim ={0.2cm 0 0 0cm},clip,width=4cm,align=c]{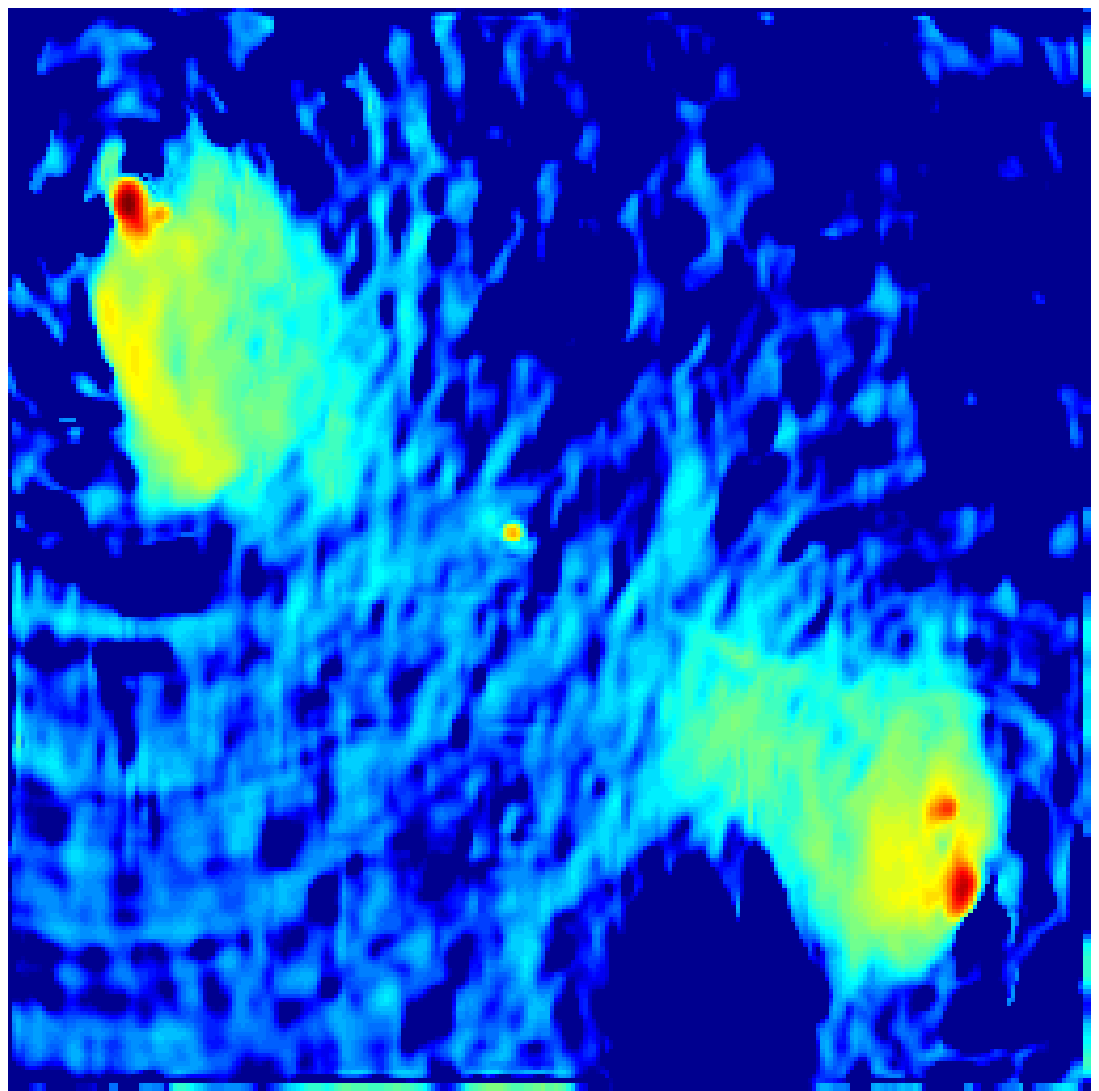} &
\hspace*{0.3cm}\includegraphics[trim ={0.2cm 0 0 0cm},clip,width=4cm,align=c]{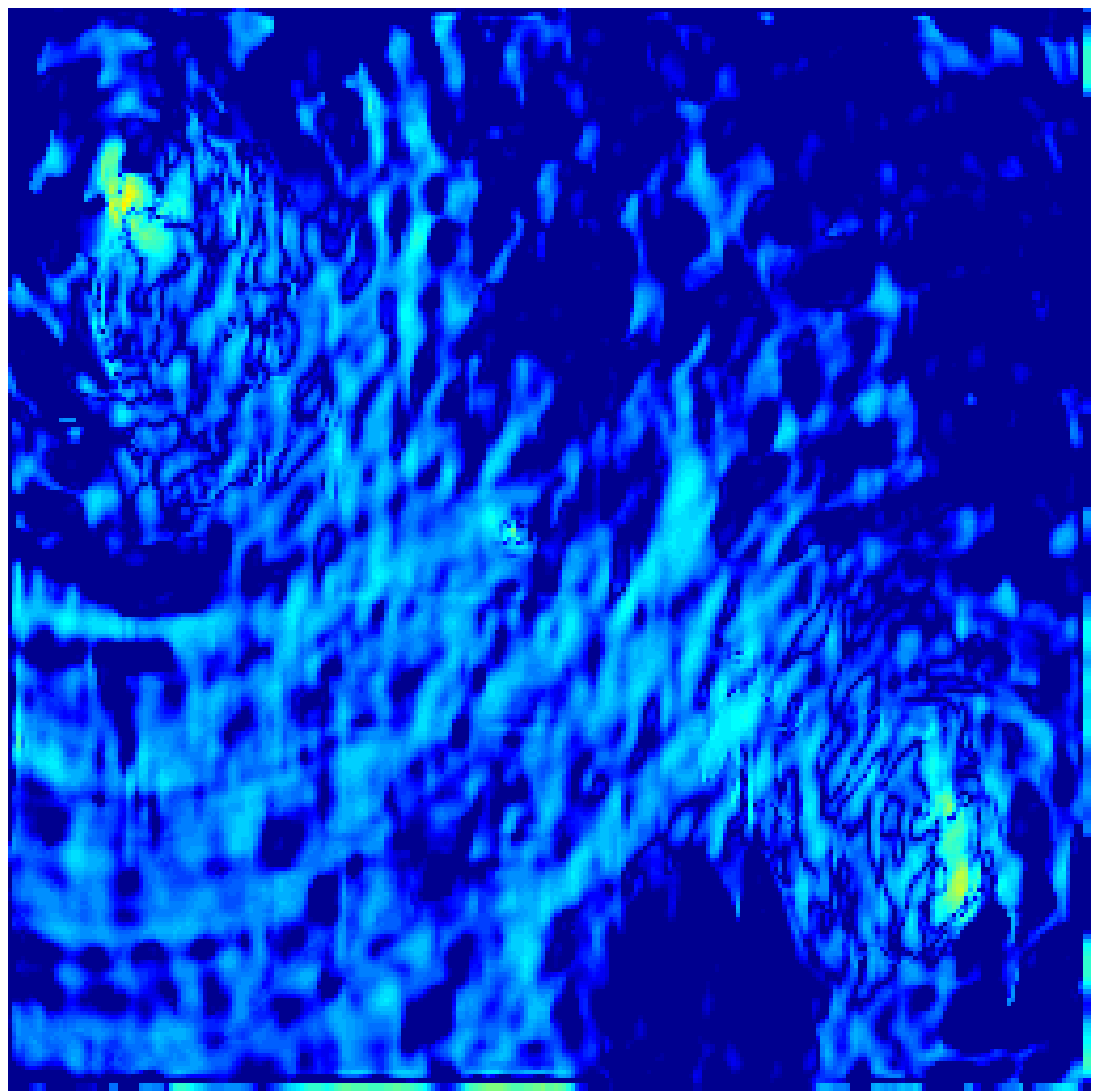} 
\vspace{0.25cm}
\\
%-----------------------------------------------
{\includegraphics[trim ={0.2cm 0 0 0cm},clip,width=4cm,align=c]{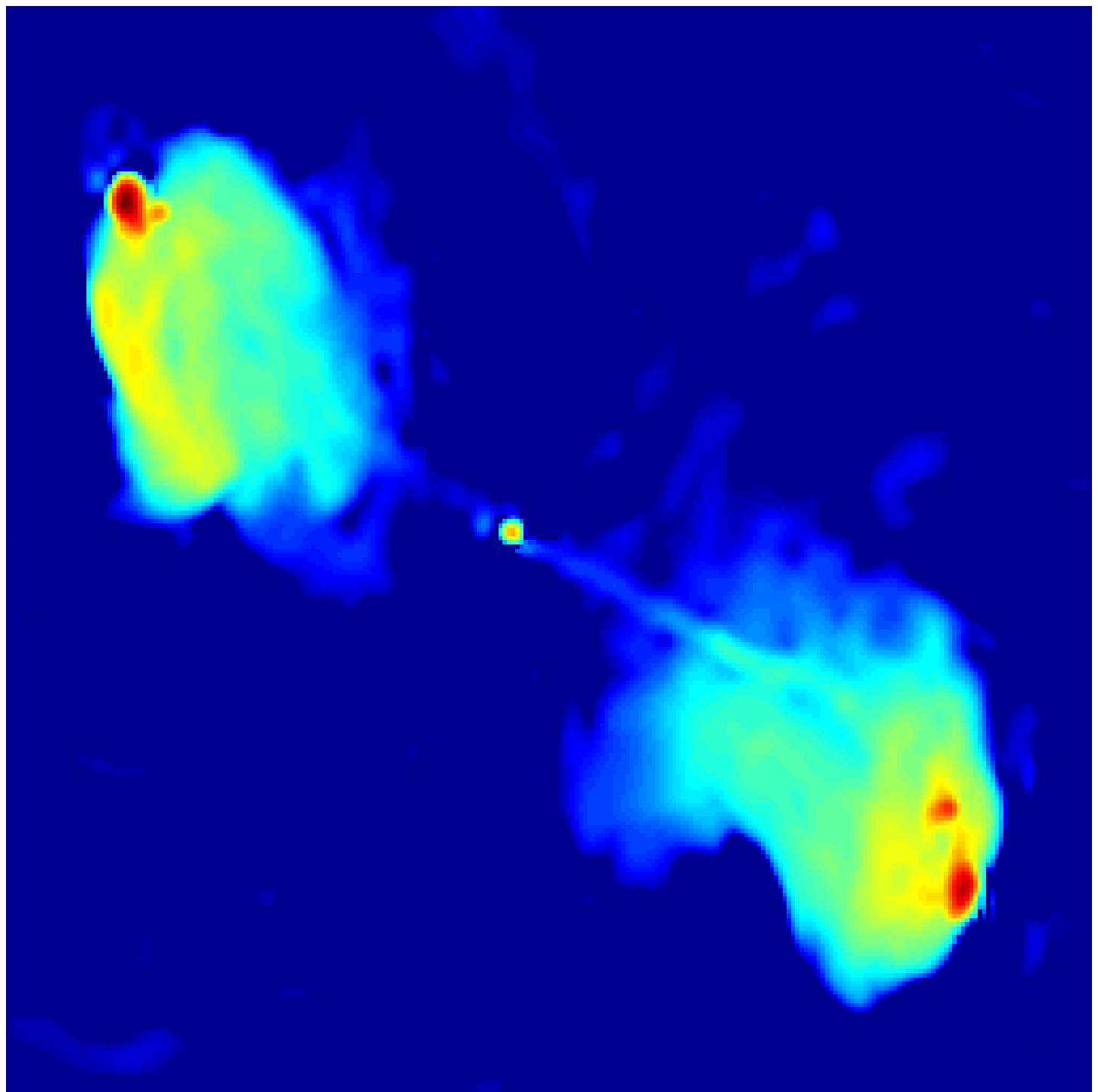}} &
\hspace*{0.3cm}\includegraphics[trim ={0.2cm 0 0 0cm},clip,width=4cm,align=c]{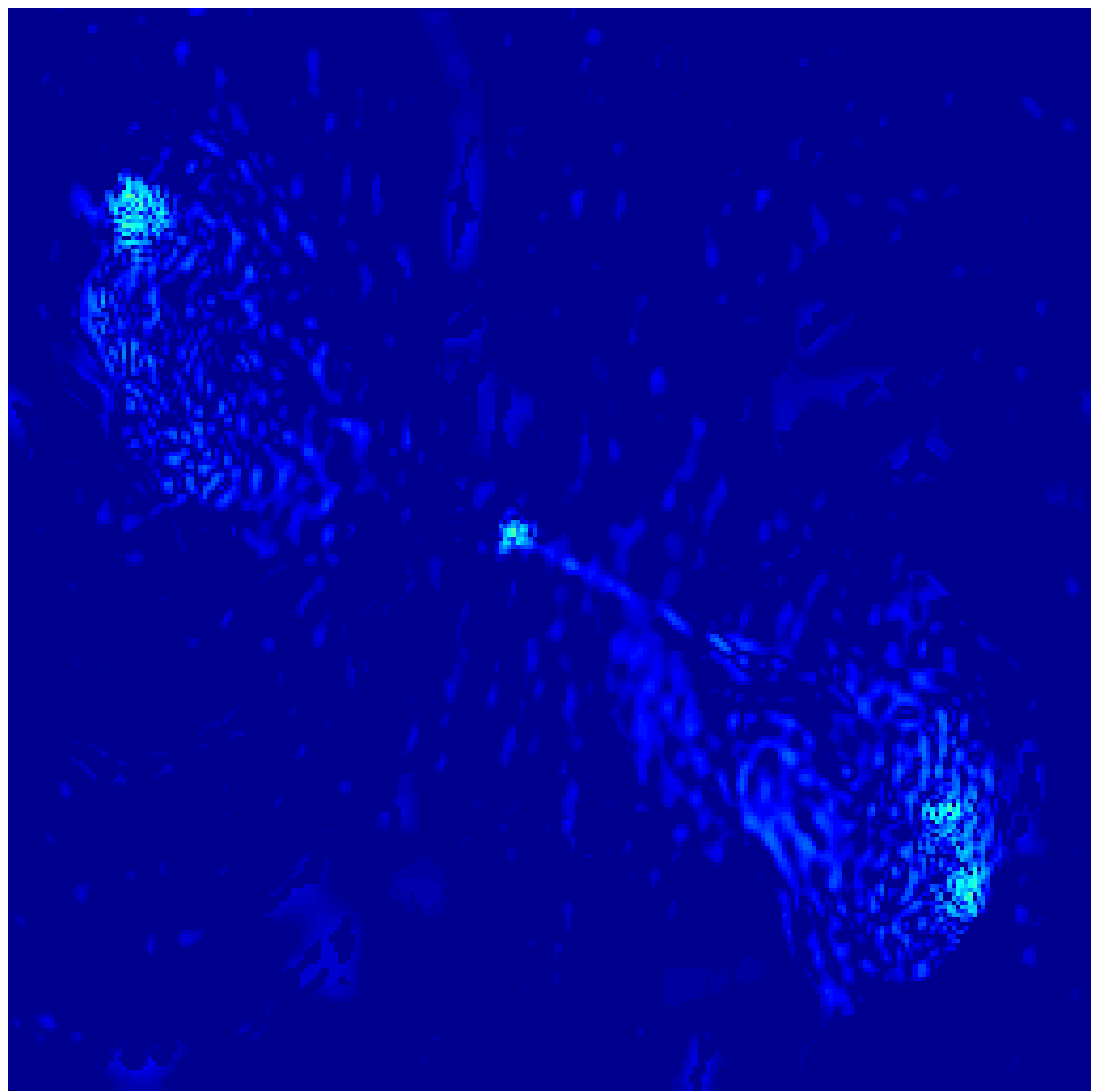} &
\hspace*{0.3cm}\includegraphics[trim ={0.2cm 0 0 0cm},clip,width=4cm,align=c]{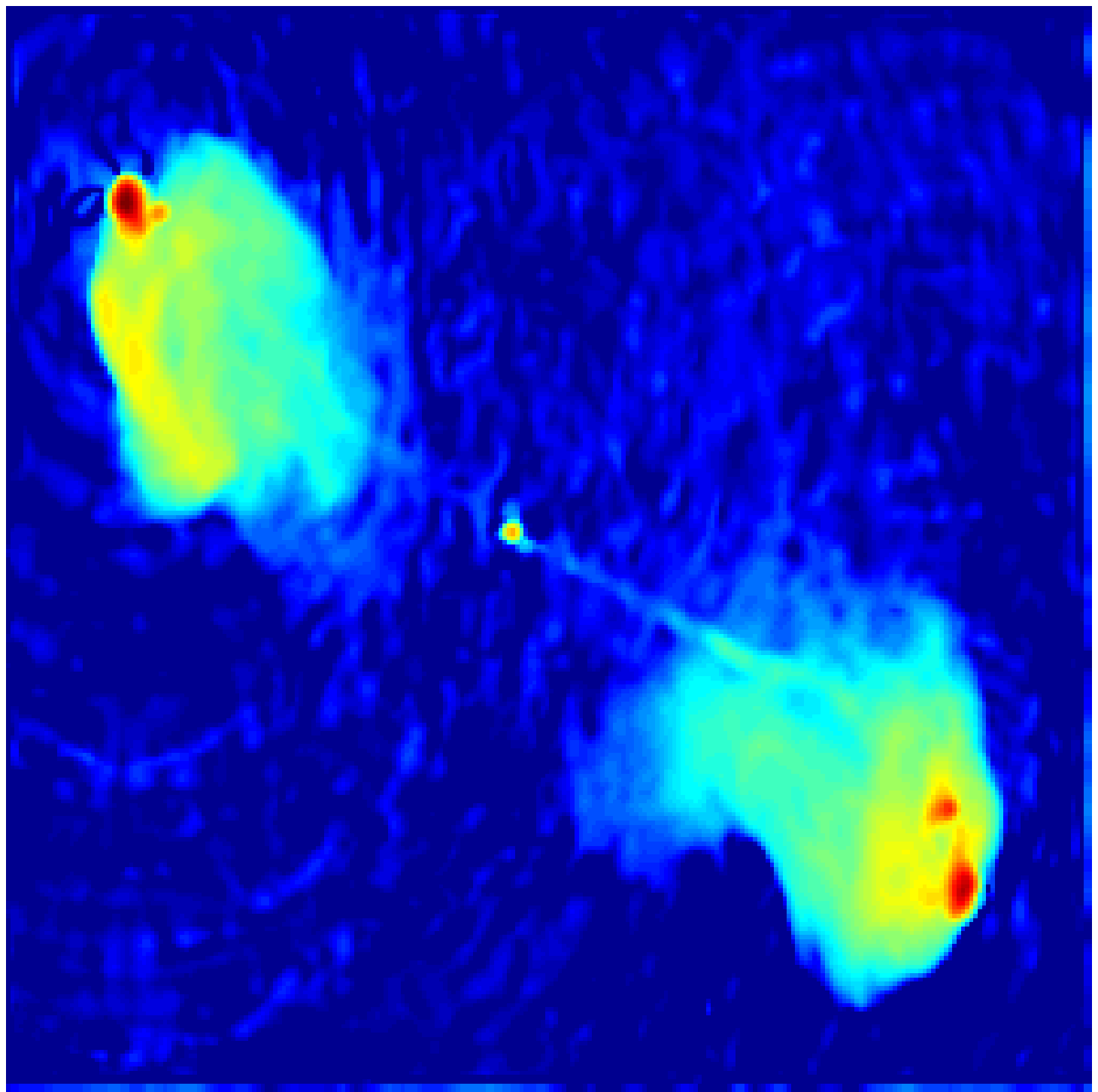} &
\hspace*{0.3cm}\includegraphics[trim ={0.2cm 0 0 0cm},clip,width=4cm,align=c]{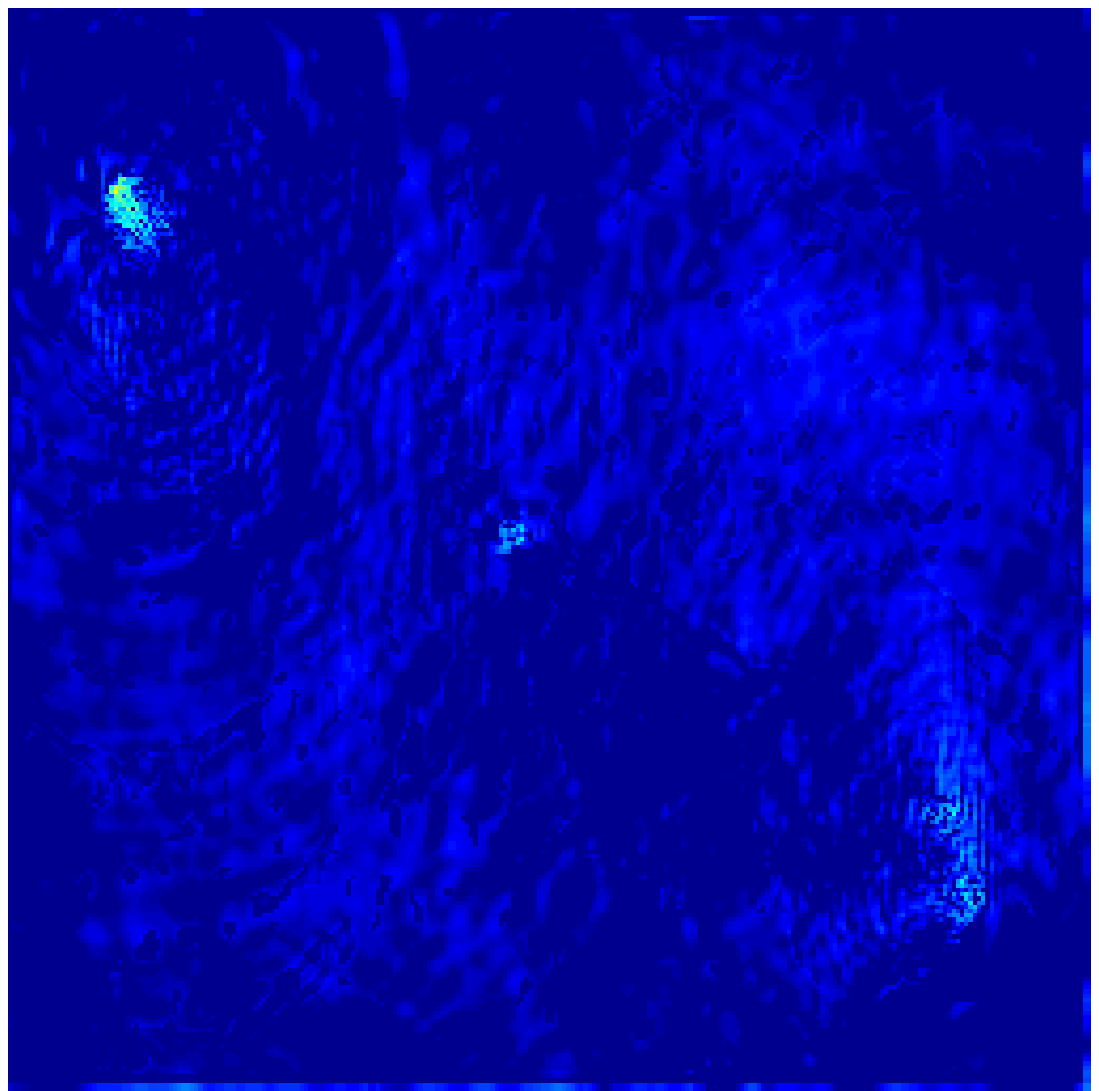} 
\vspace{0.25cm}
\\
%-----------------------------------------------
\includegraphics[trim ={0.2cm 0 0 0cm},clip,width=4cm,align=c]{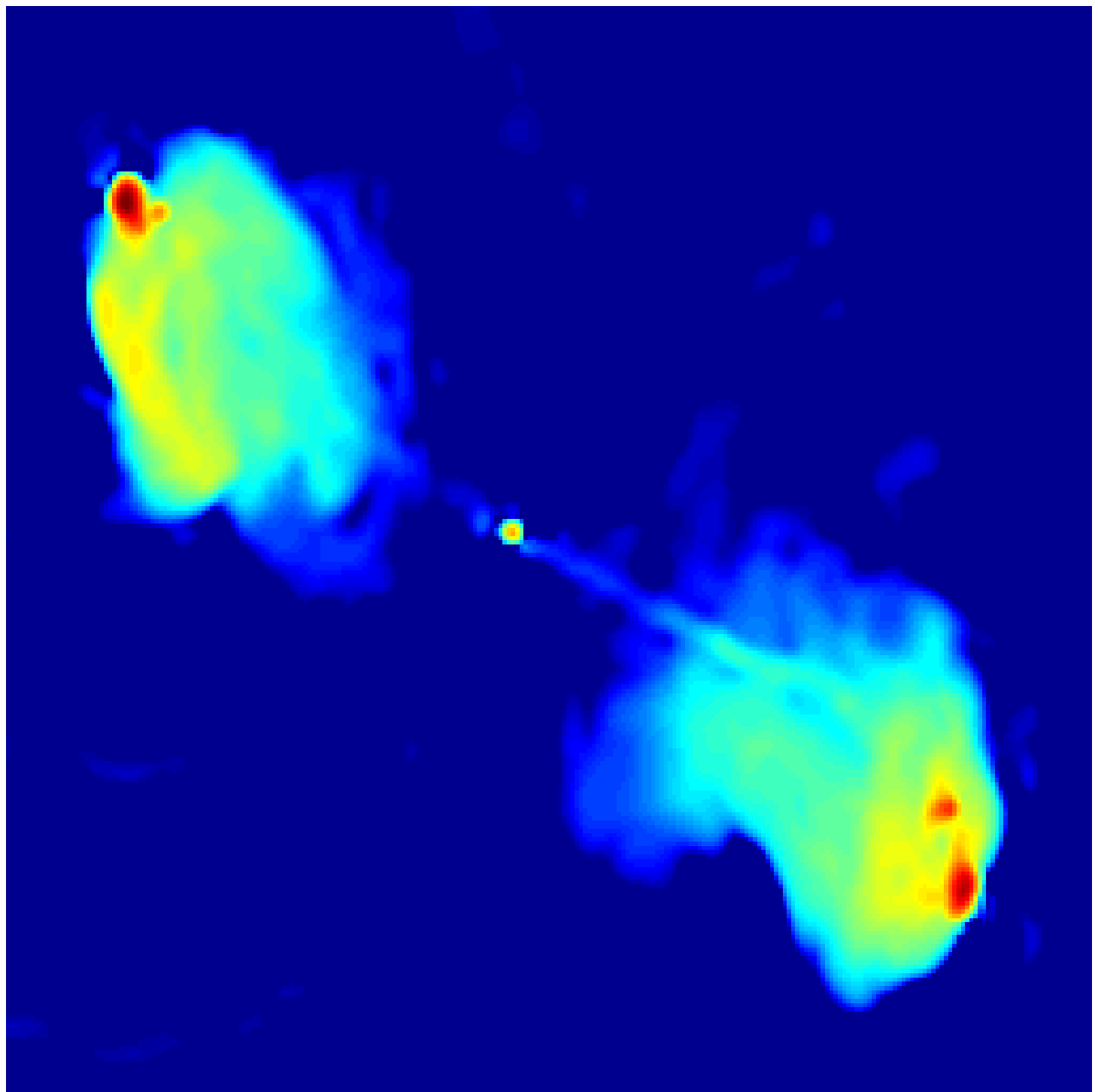} &
\hspace*{0.3cm}\includegraphics[trim ={0.2cm 0 0 0cm},clip,width=4cm,align=c]{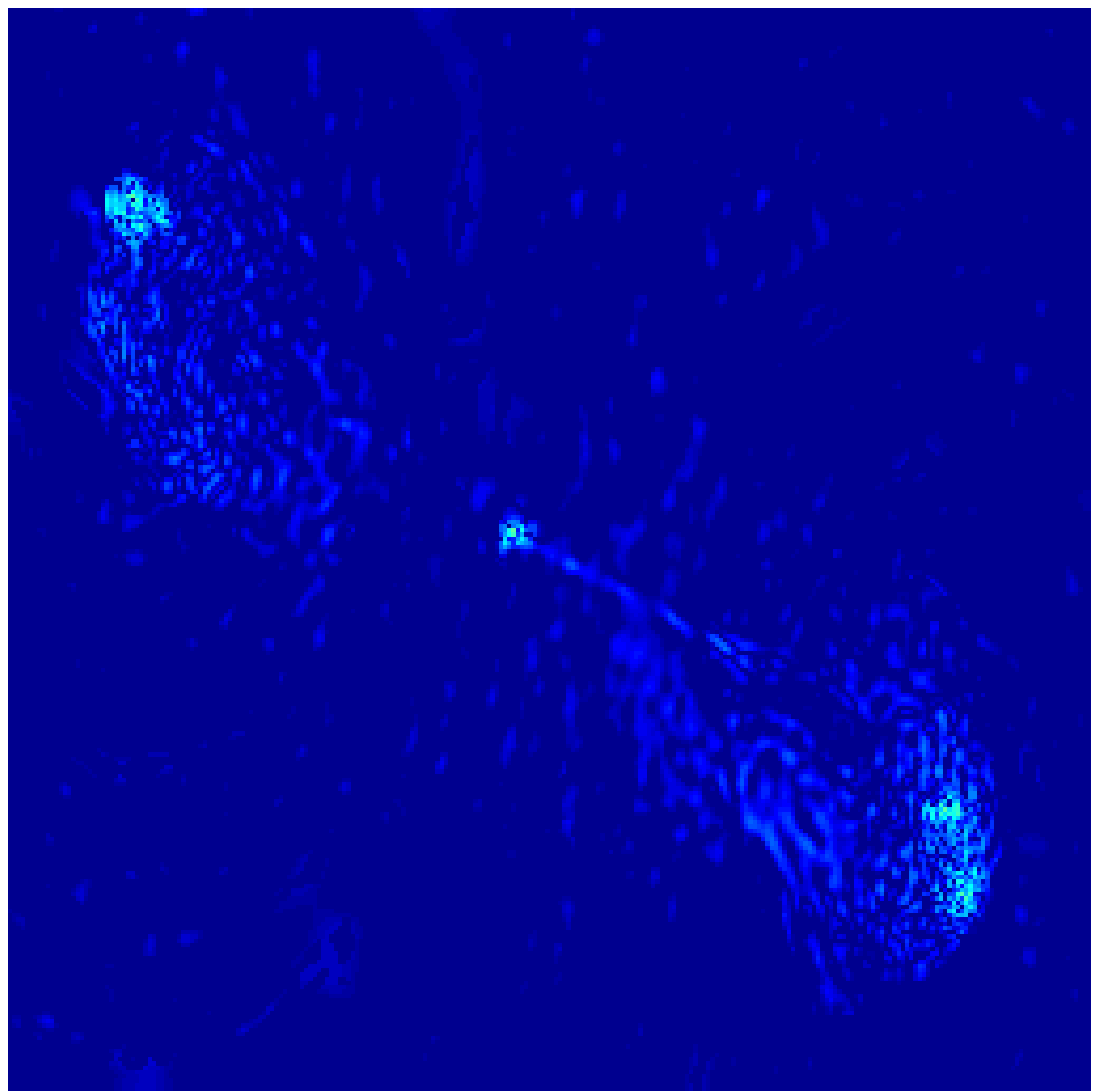} &
\hspace*{0.3cm}\includegraphics[trim ={0.2cm 0 0 0cm},clip,width=4cm,align=c]{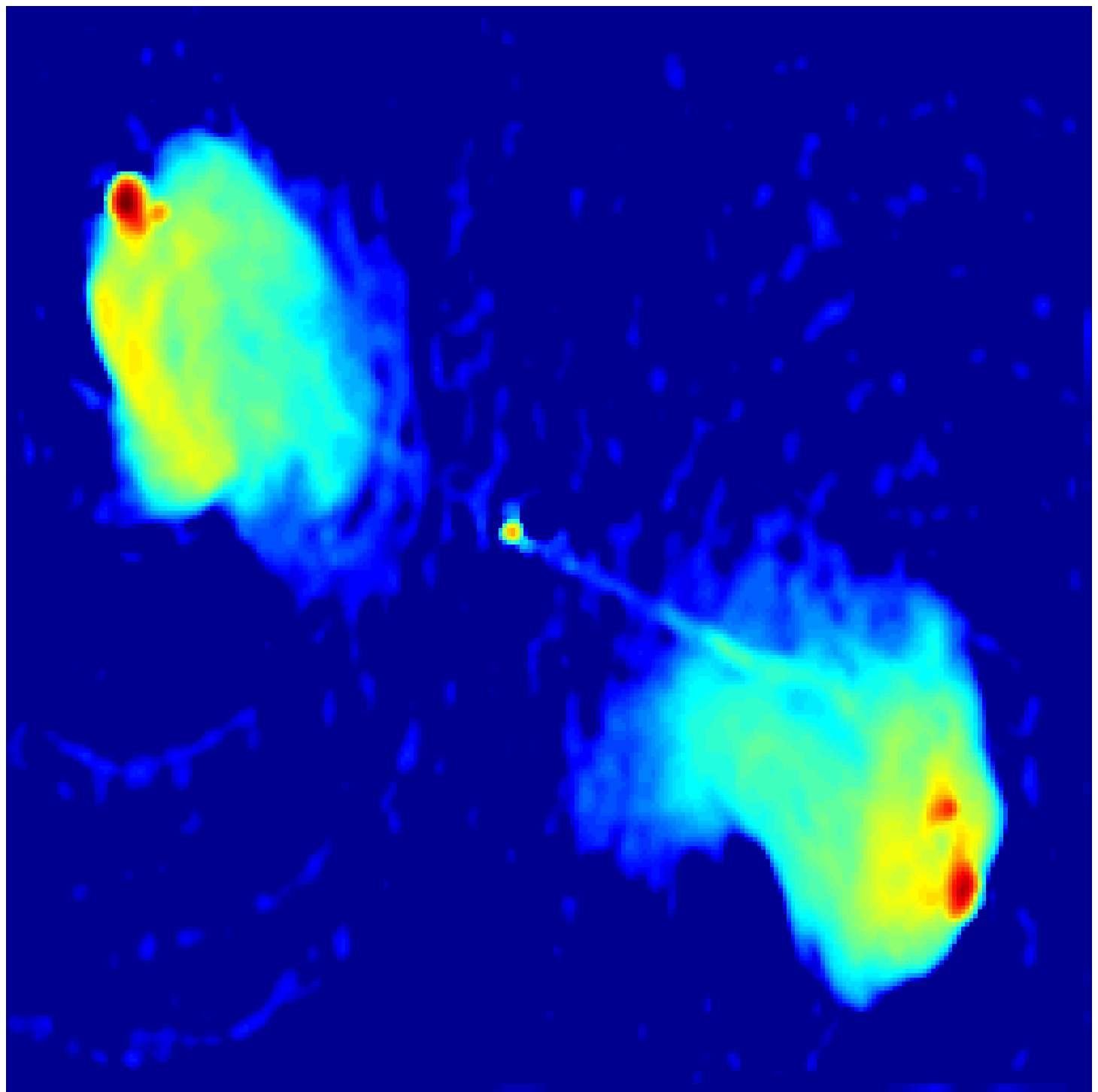} &
\hspace*{0.3cm}\includegraphics[trim ={0.2cm 0 0 0cm},clip,width=4cm,align=c]{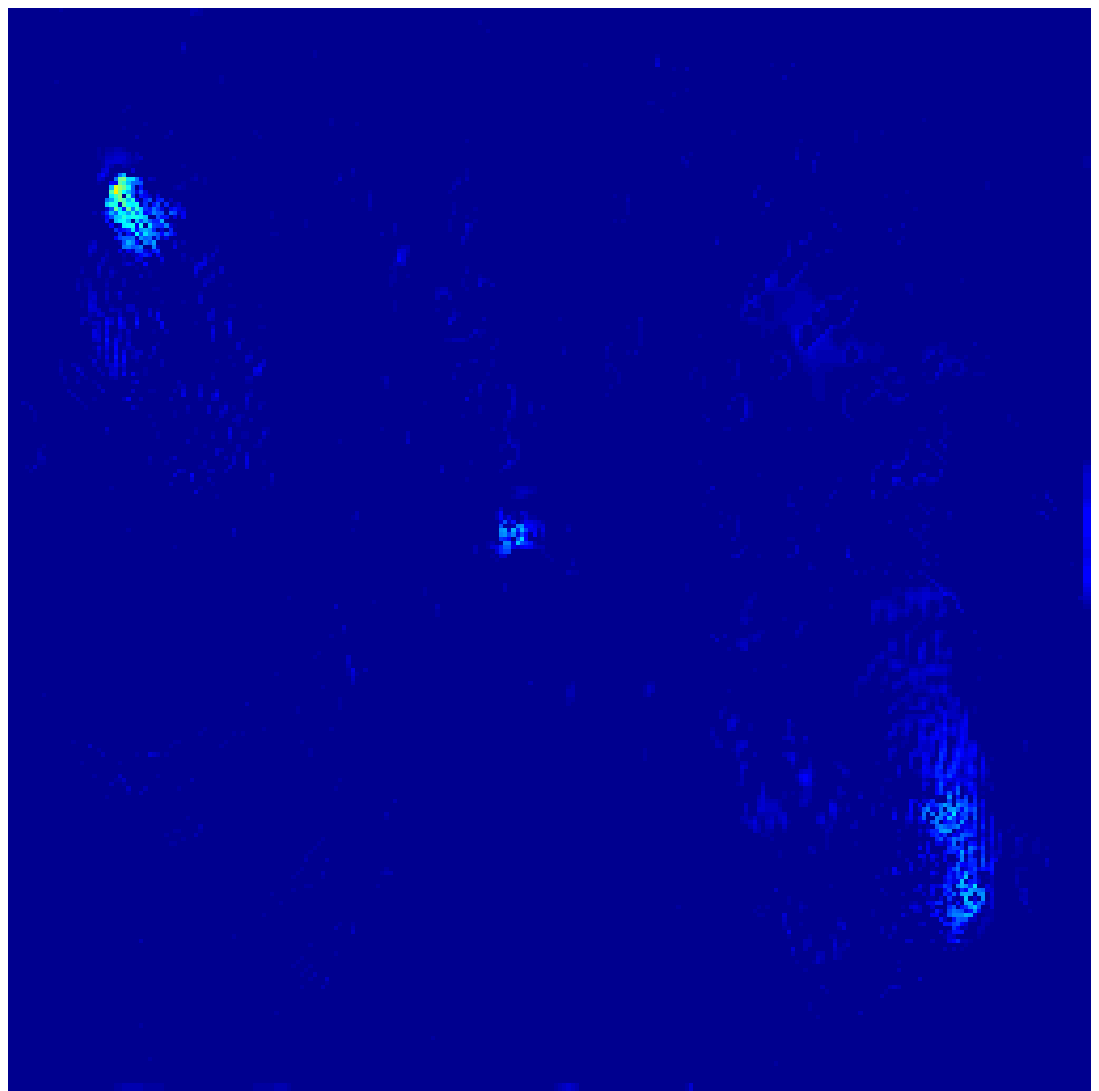} 
\end{tabular}
\caption{Cygnus A Stokes $I$ true image in first row and reconstructed images (best ones over 5 performed simulations for each case) in other rows for the cases: Imaging with normalized DIEs (second row), {Joint DIE calibration and imaging (third row)}, Joint DDE calibration and imaging excluding the off-diagonal terms (fourth row), and considering full Jones matrix (fifth row). In each case, column-wise recovered images followed by their corresponding error images are displayed when imaging is performed without polarization constraint (first two columns) and with polarization constraint (last two columns). All the images are shown in log scale, with the same color range corresponding to the colorbar given in first row.}
\label{fig:cyg_a_I}
\end{figure*}  

%%%%%%%%%%%%%%%%%%%%%%%%
%%%%%%%  CYGNUS-A: Stokes Q  %%%%%%%%%%%%%%%%%

\begin{figure*}
\centering
\begin{tabular}{@{}c@{}c@{}c@{}c@{}}
\includegraphics[trim ={0.2cm 0 0 0cm},clip,width=4cm]{true_Q_cyg_a.eps} &
\hspace{-3cm}\includegraphics[trim ={16.2cm 0 0 0cm},clip,width=0.97cm]{Q_cyg_a_colorbar.eps}
& & \\
Recovered images w/o &  Absolute error images &  Recovered images with & Absolute error images \\
polarization constraint & &polarization constraint & \\
\includegraphics[trim ={0.2cm 0 0 0cm},clip,width=4cm,align=c]{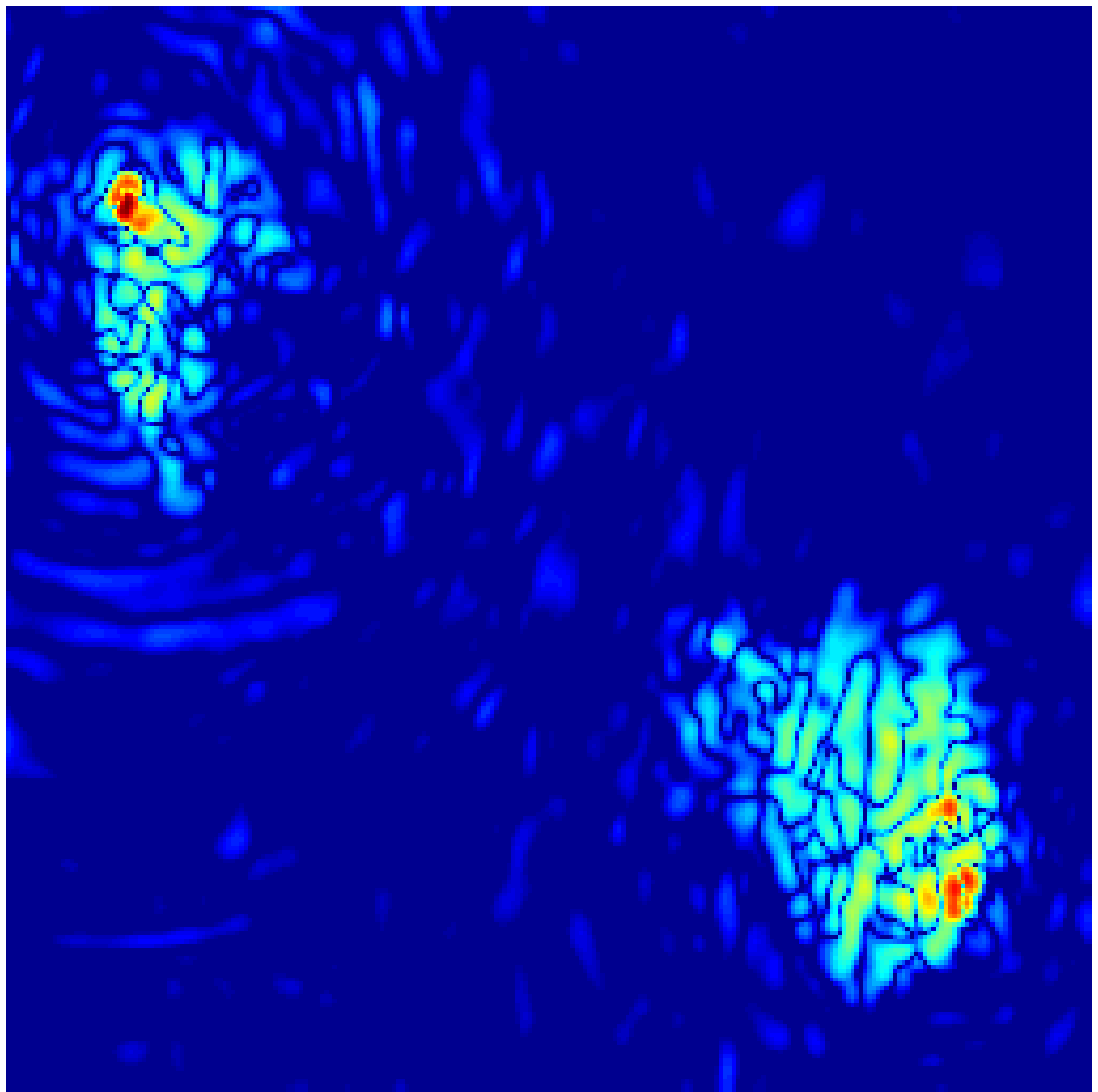} &
\hspace*{0.3cm}\includegraphics[trim ={0.2cm 0 0 0cm},clip,width=4cm,align=c]{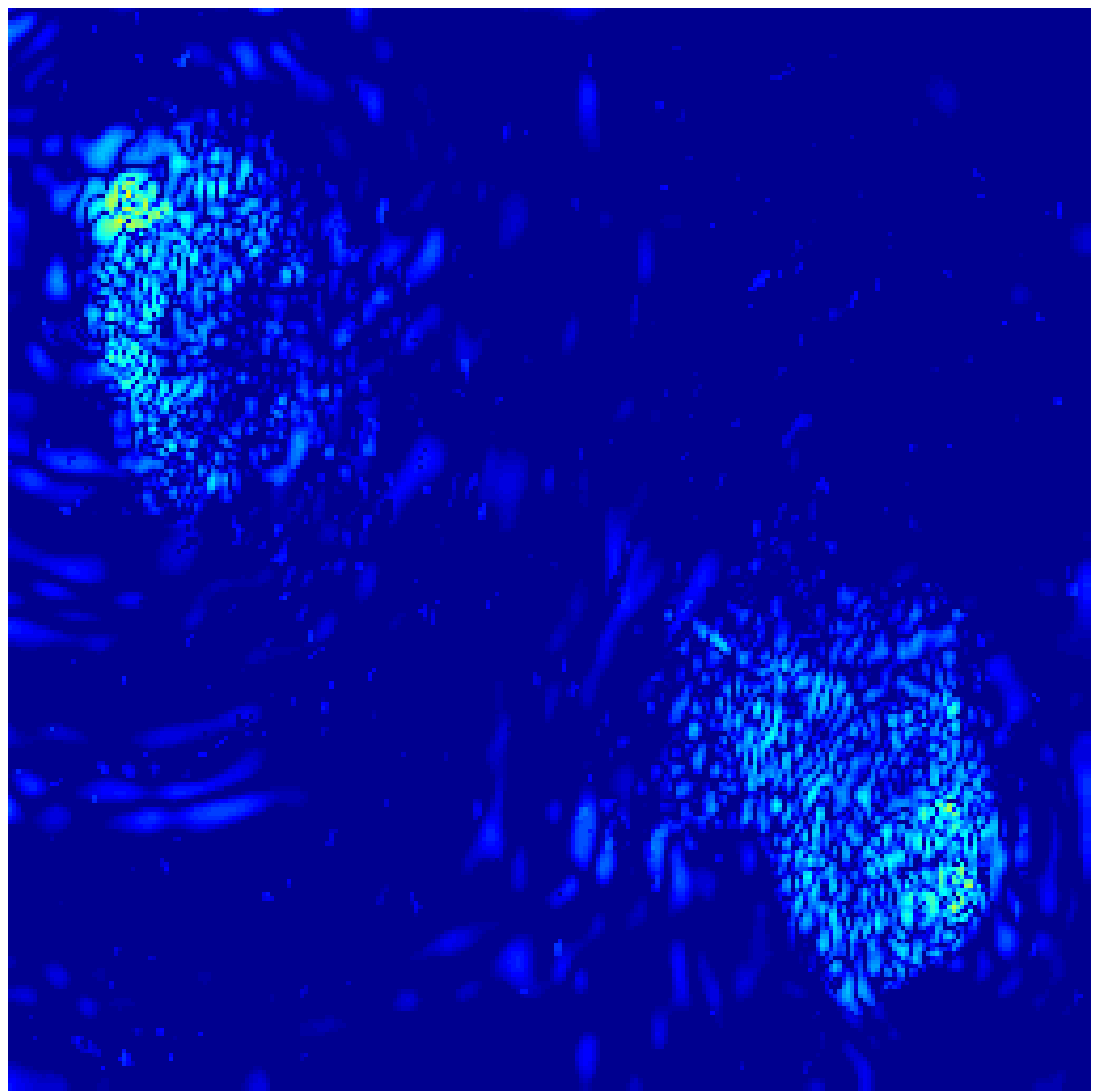} &
\hspace*{0.3cm}\includegraphics[trim ={0.2cm 0 0 0cm},clip,width=4cm,align=c]{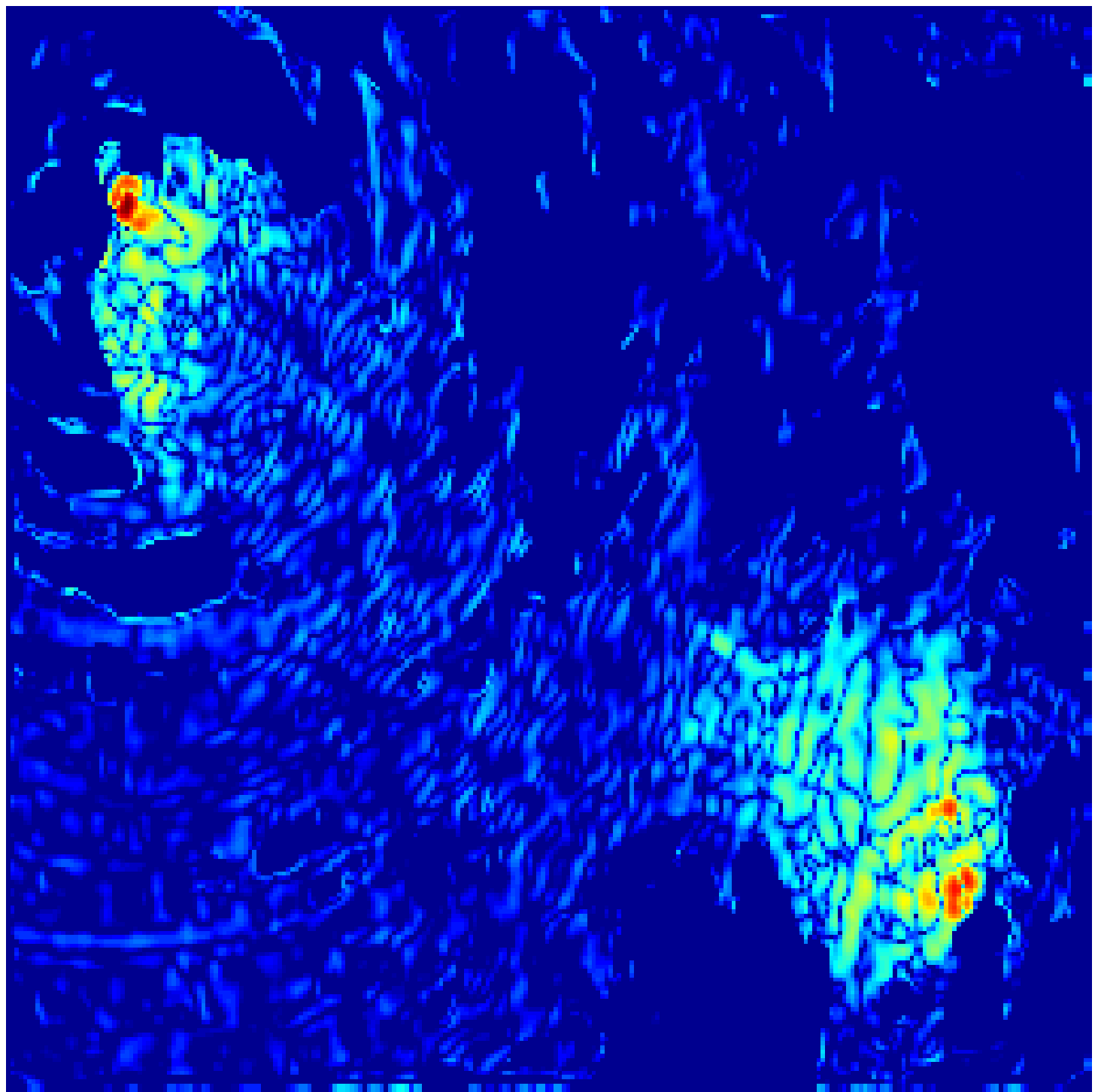} &
\hspace*{0.3cm}\includegraphics[trim ={0.2cm 0 0 0cm},clip,width=4cm,align=c]{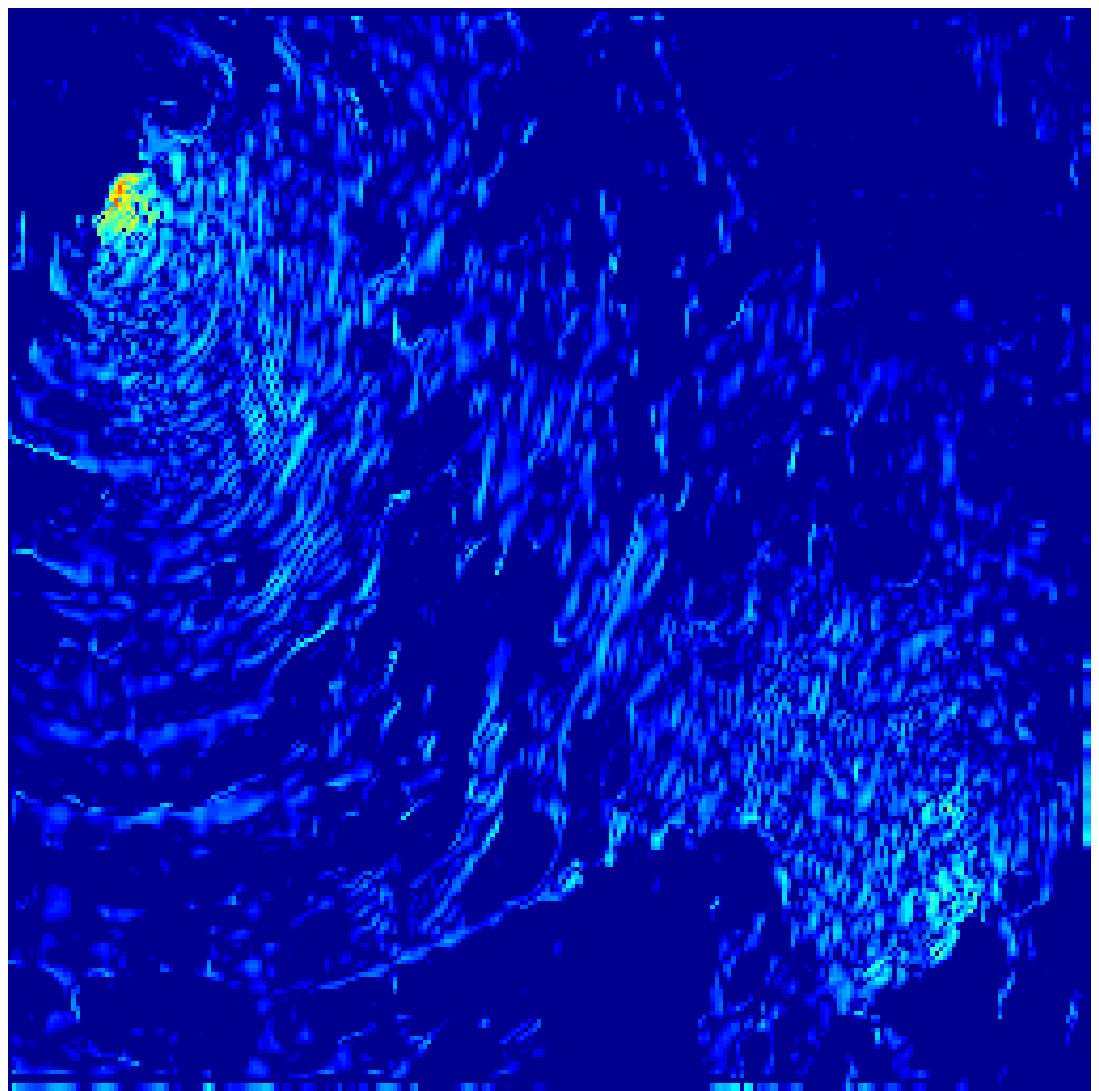} 
\vspace{0.25cm}
\\
%-----------------------------------------------
{\includegraphics[trim ={0.2cm 0 0 0cm},clip,width=4cm,align=c]{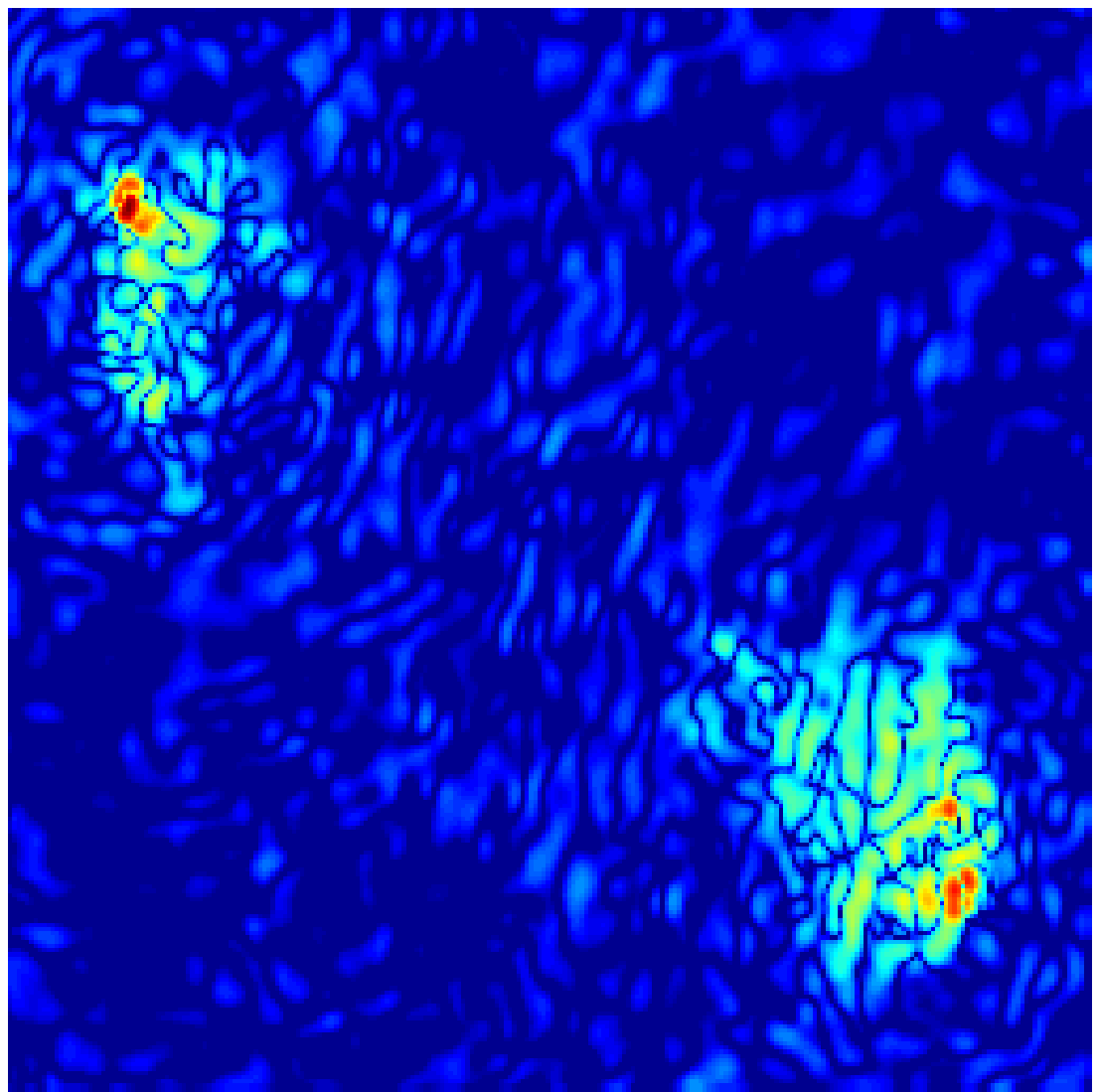}} &
\hspace*{0.3cm}\includegraphics[trim ={0.2cm 0 0 0cm},clip,width=4cm,align=c]{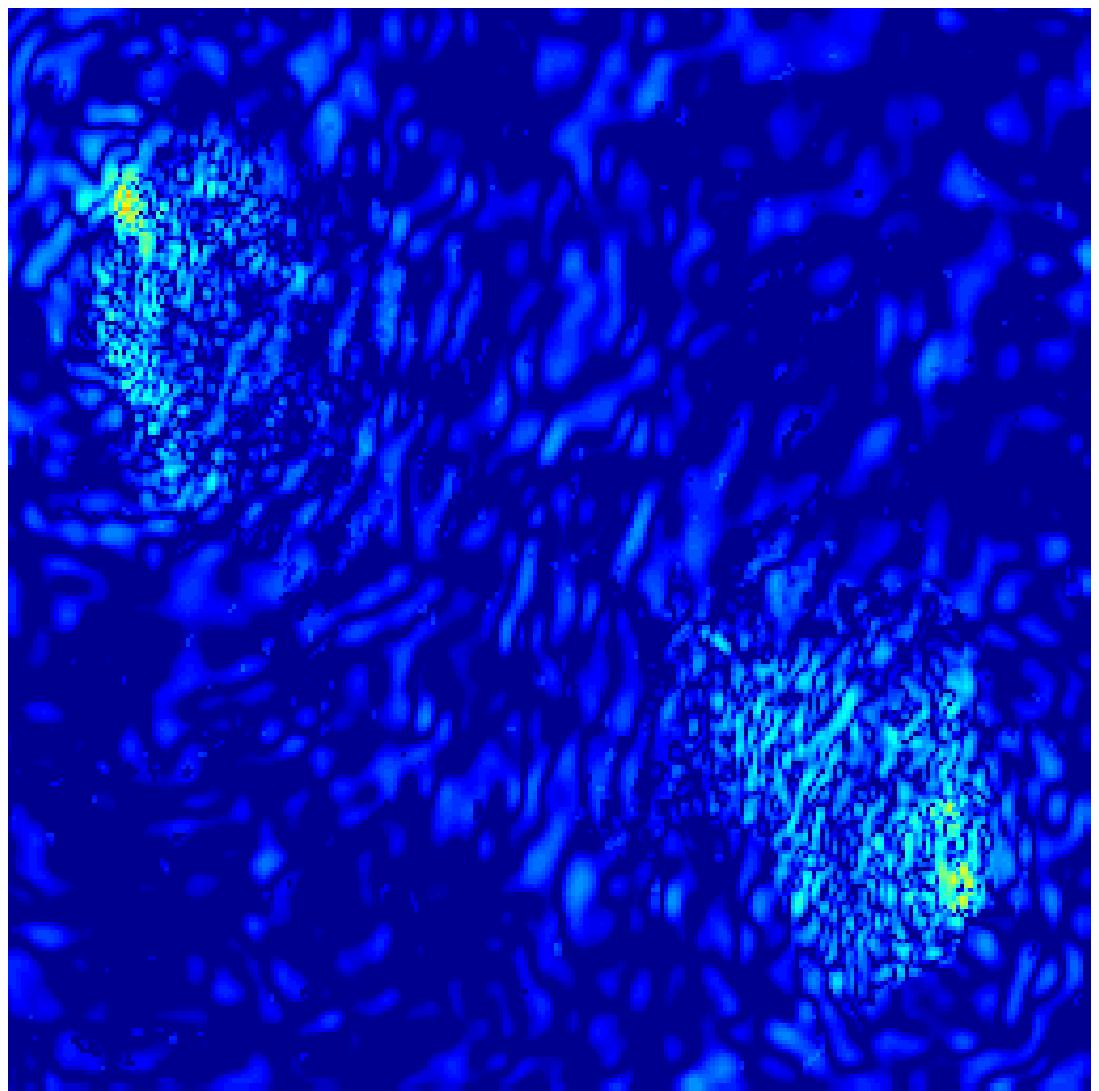} &
\hspace*{0.3cm}\includegraphics[trim ={0.2cm 0 0 0cm},clip,width=4cm,align=c]{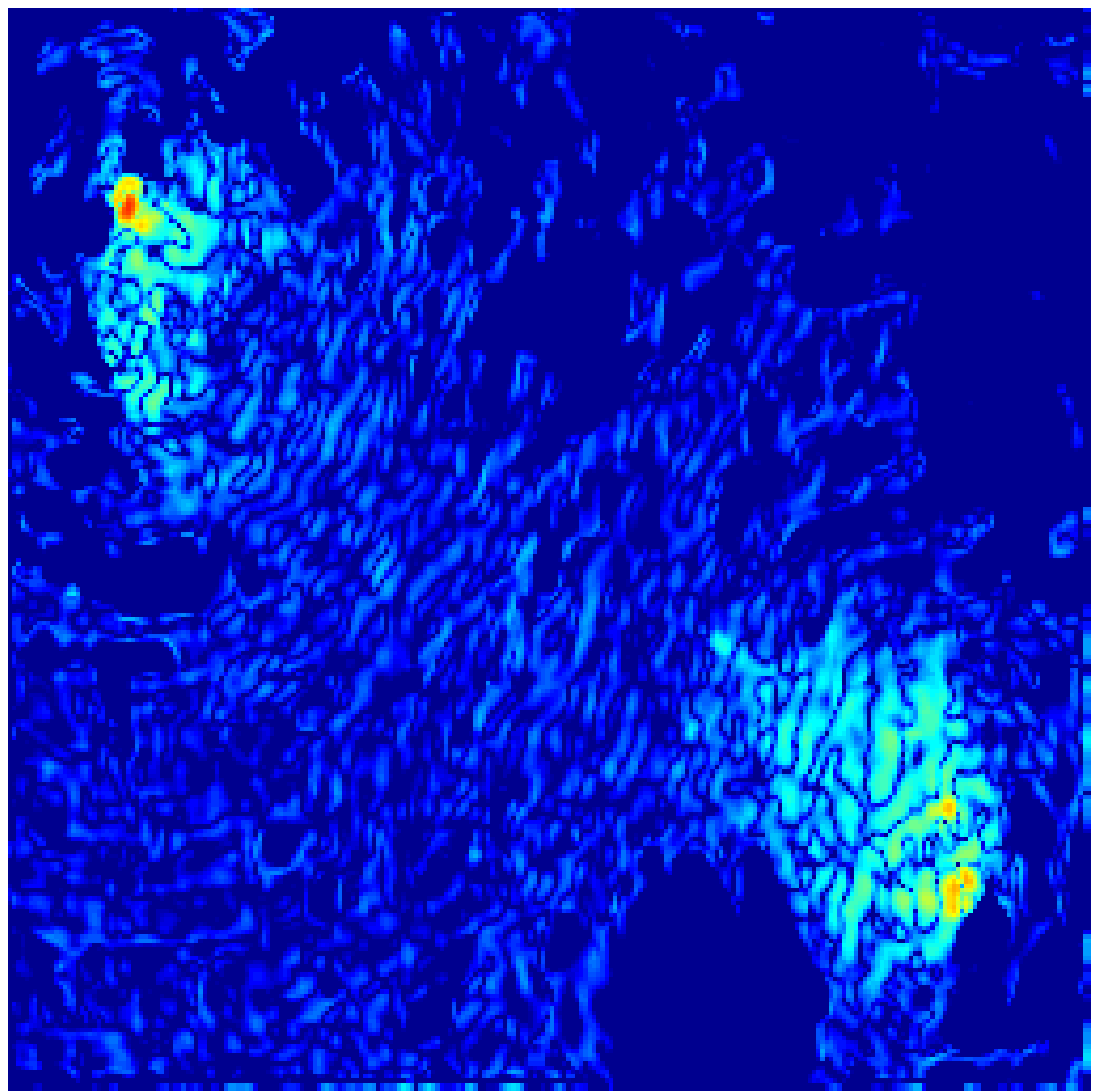} &
\hspace*{0.3cm}\includegraphics[trim ={0.2cm 0 0 0cm},clip,width=4cm,align=c]{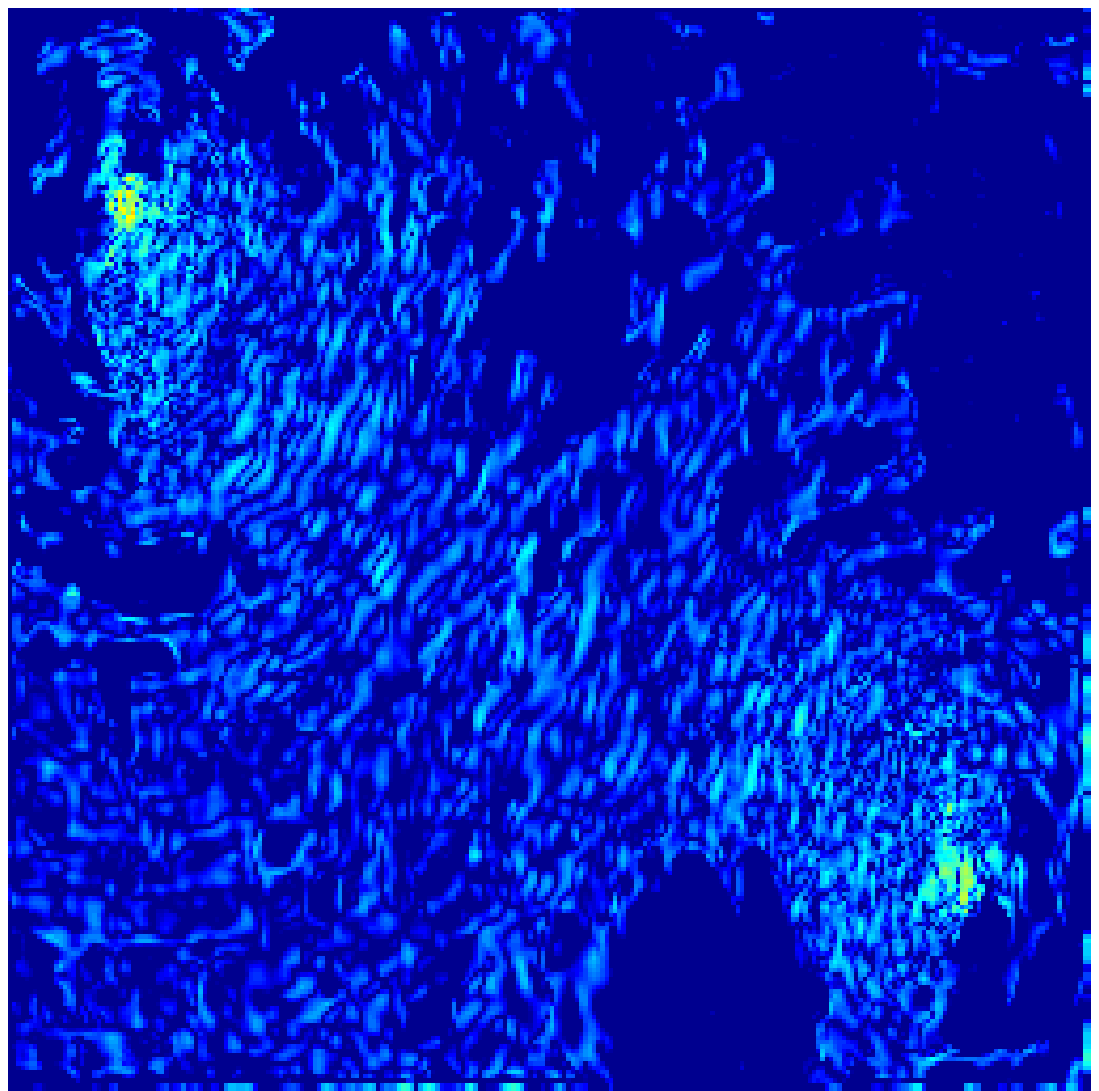} 
\vspace{0.25cm}
\\
%-----------------------------------------------
{\includegraphics[trim ={0.2cm 0 0 0cm},clip,width=4cm,align=c]{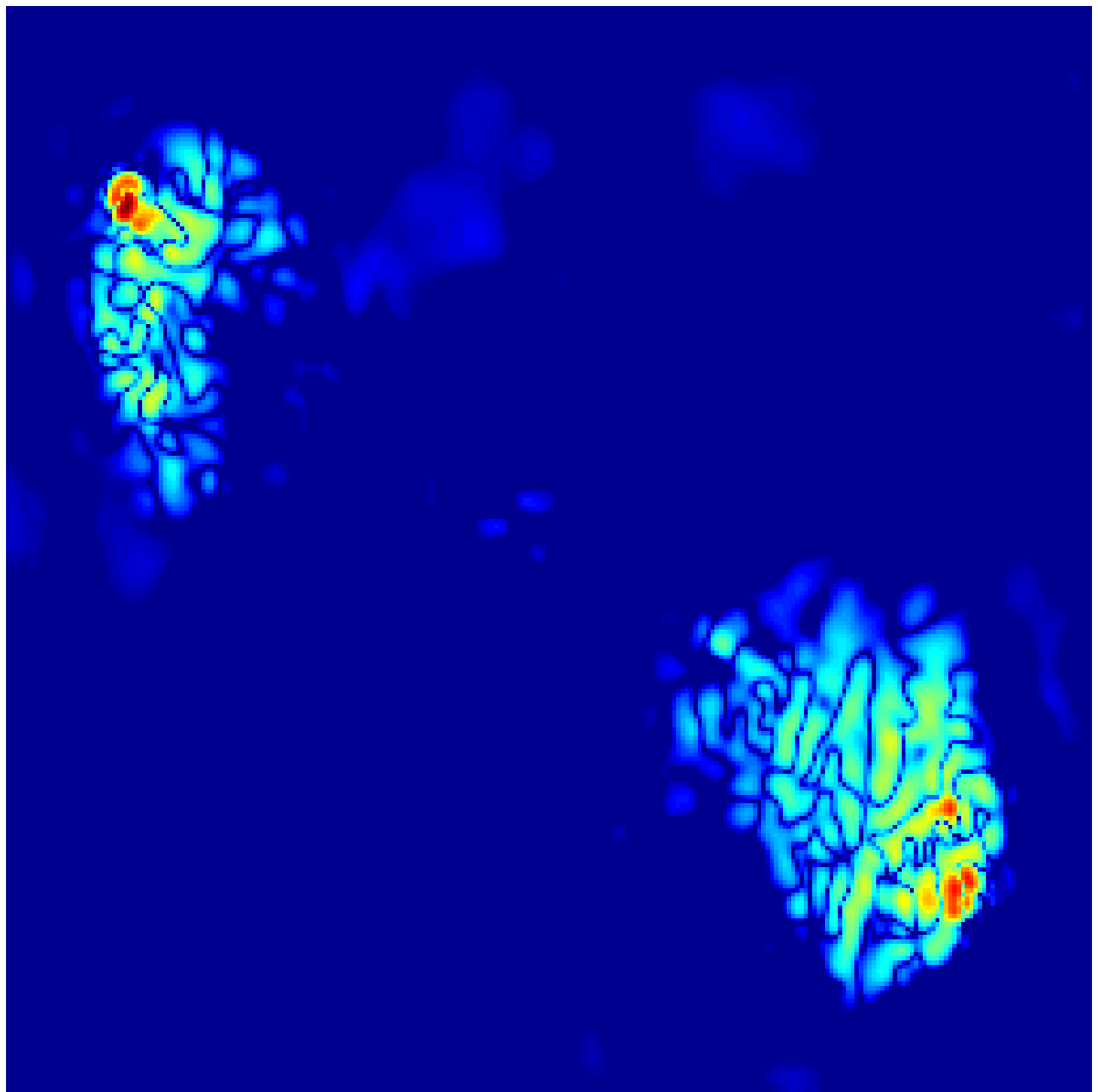}} &
\hspace*{0.3cm}\includegraphics[trim ={0.2cm 0 0 0cm},clip,width=4cm,align=c]{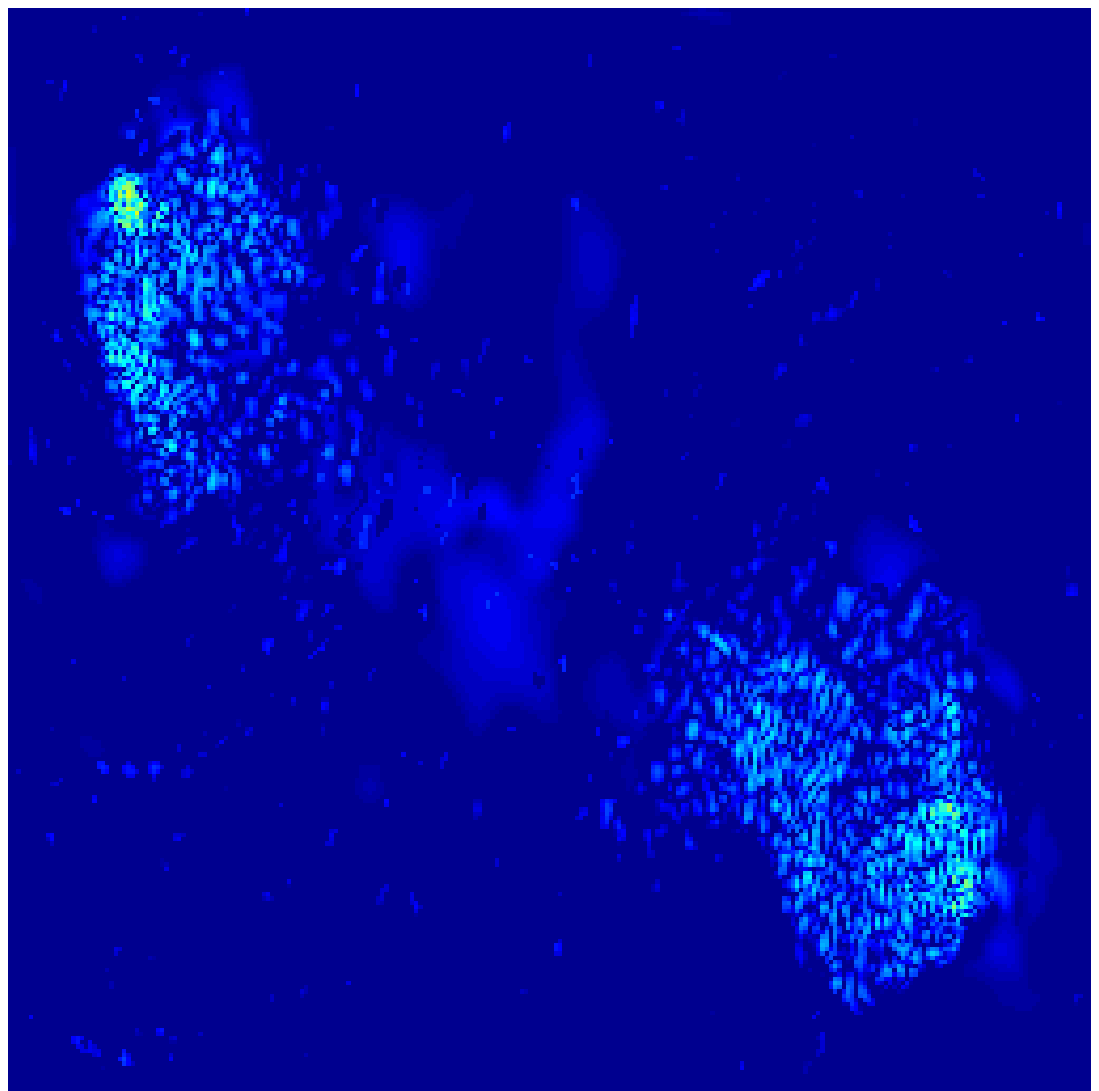} &
\hspace*{0.3cm}\includegraphics[trim ={0.2cm 0 0 0cm},clip,width=4cm,align=c]{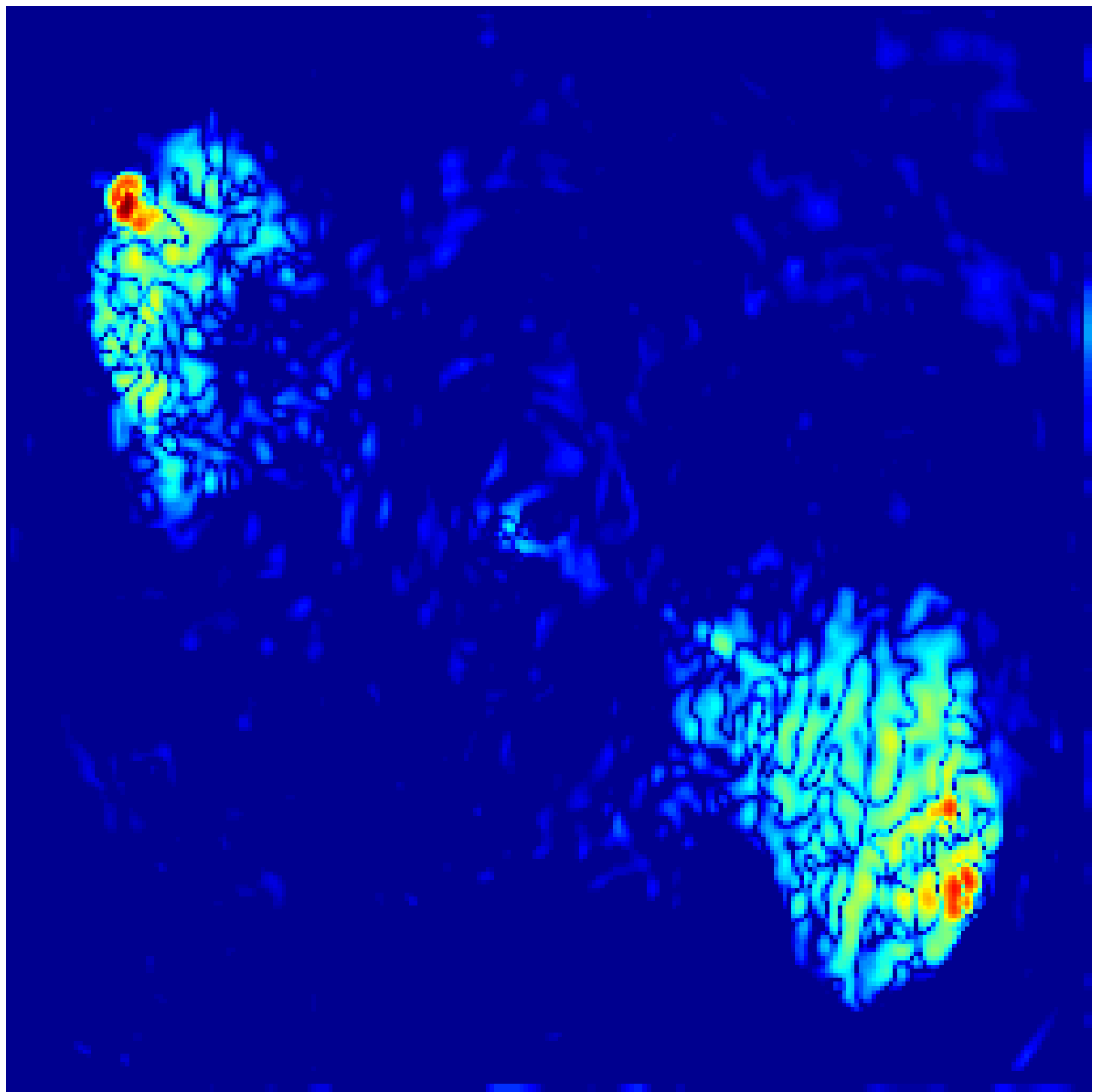} &
\hspace*{0.3cm}\includegraphics[trim ={0.2cm 0 0 0cm},clip,width=4cm,align=c]{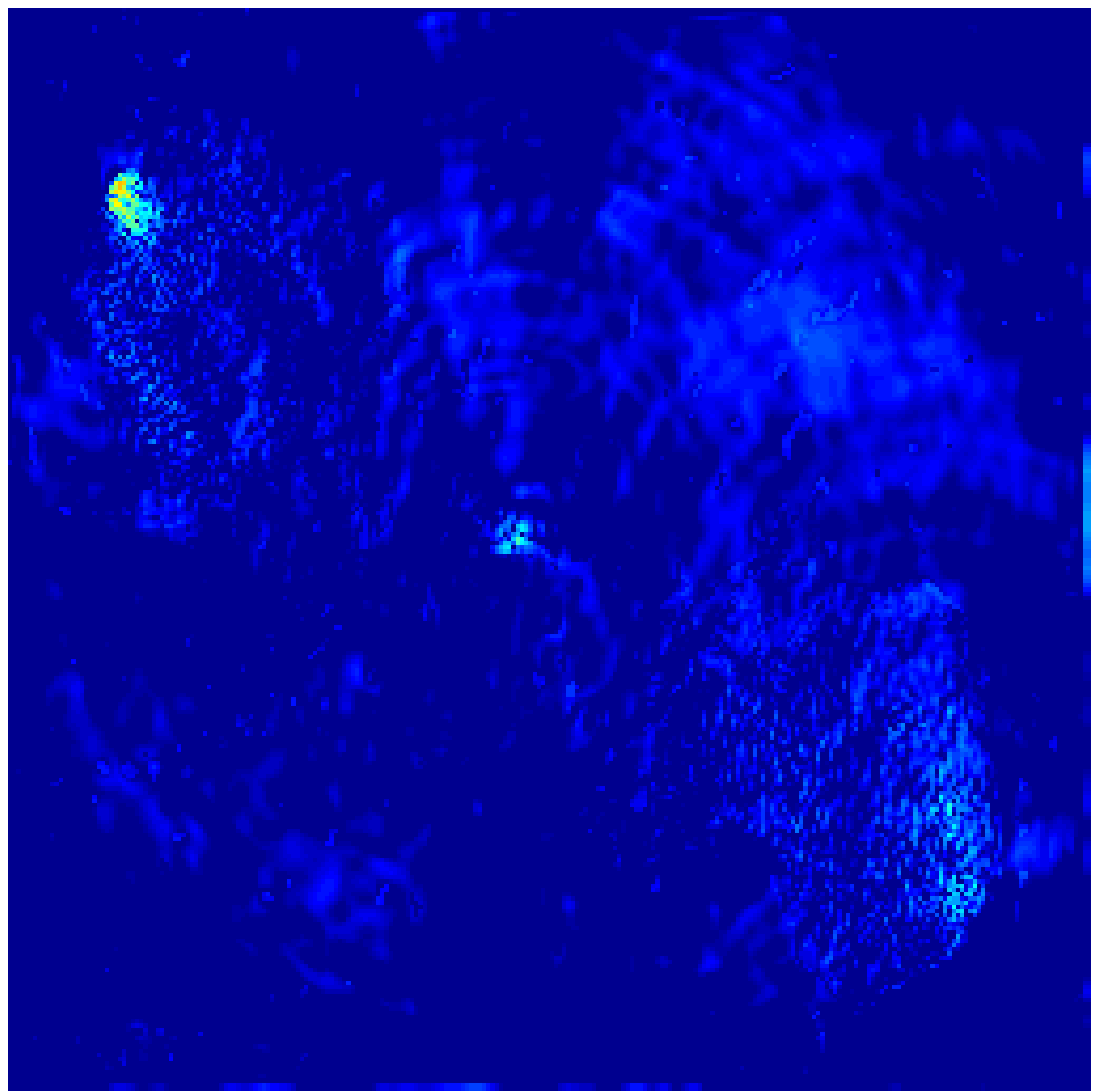} 
\vspace{0.25cm}
\\
%-----------------------------------------------
\includegraphics[trim ={0.2cm 0 0 0cm},clip,width=4cm,align=c]{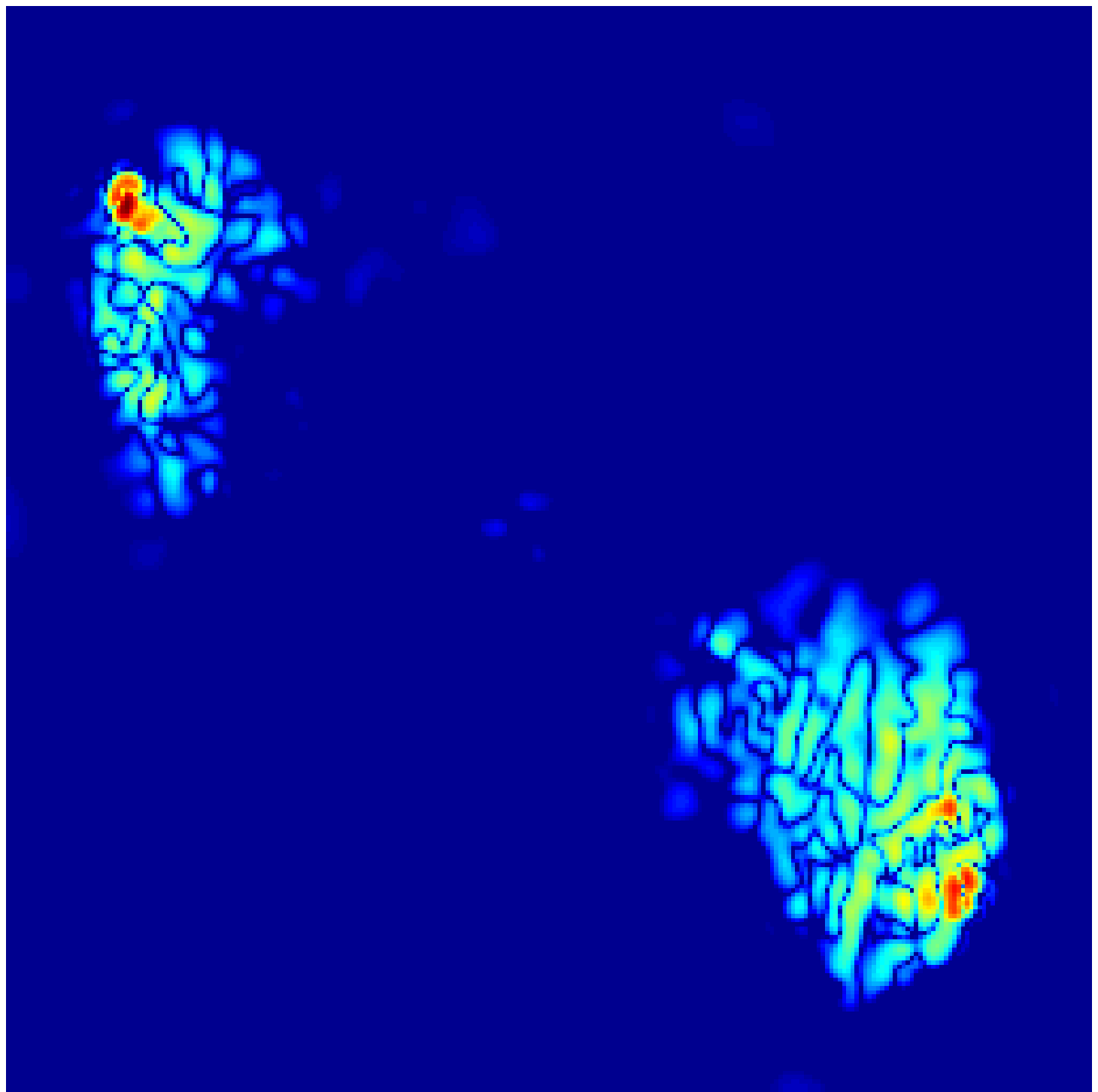} &
\hspace*{0.3cm}\includegraphics[trim ={0.2cm 0 0 0cm},clip,width=4cm,align=c]{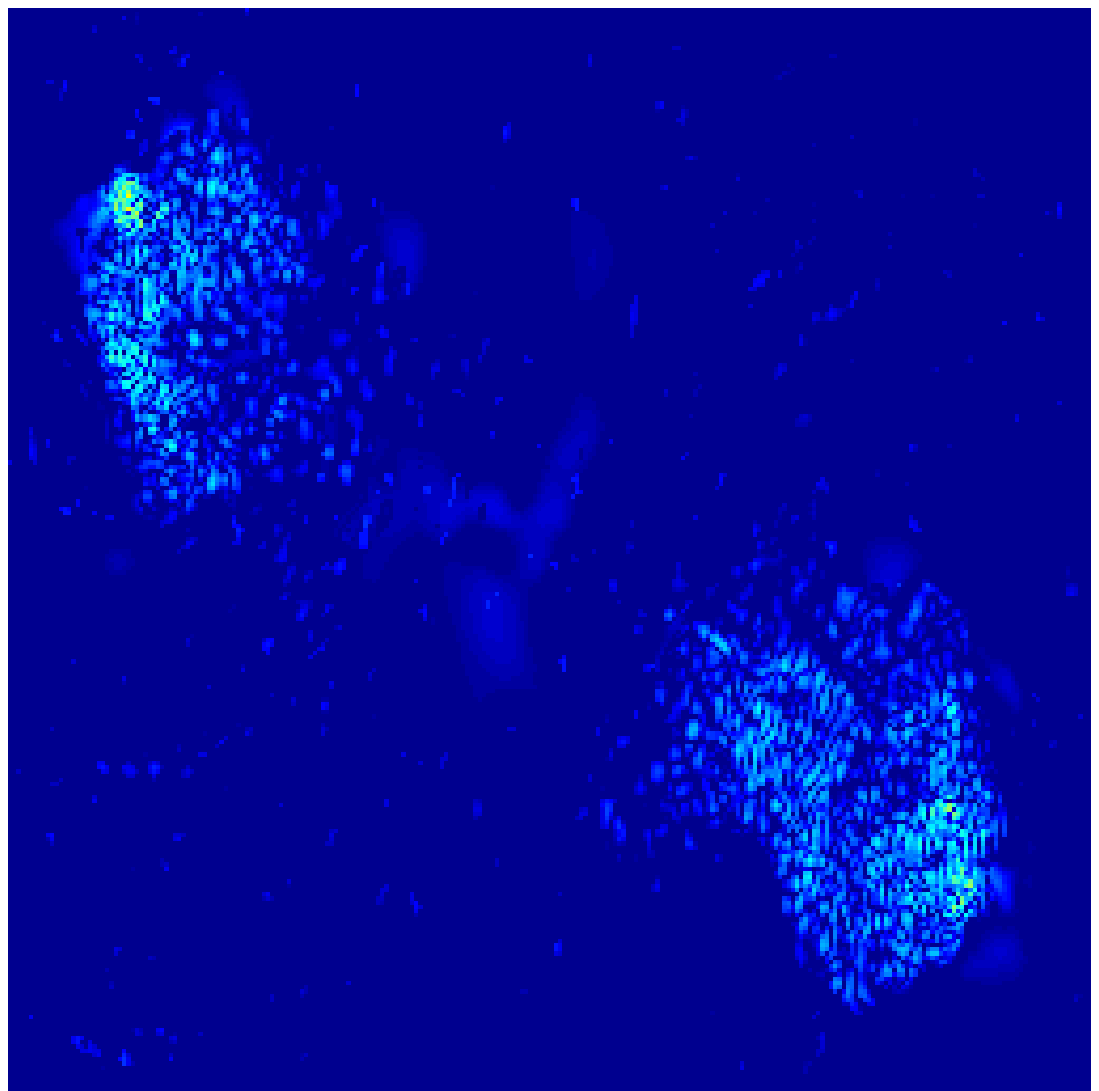} &
\hspace*{0.3cm}\includegraphics[trim ={0.2cm 0 0 0cm},clip,width=4cm,align=c]{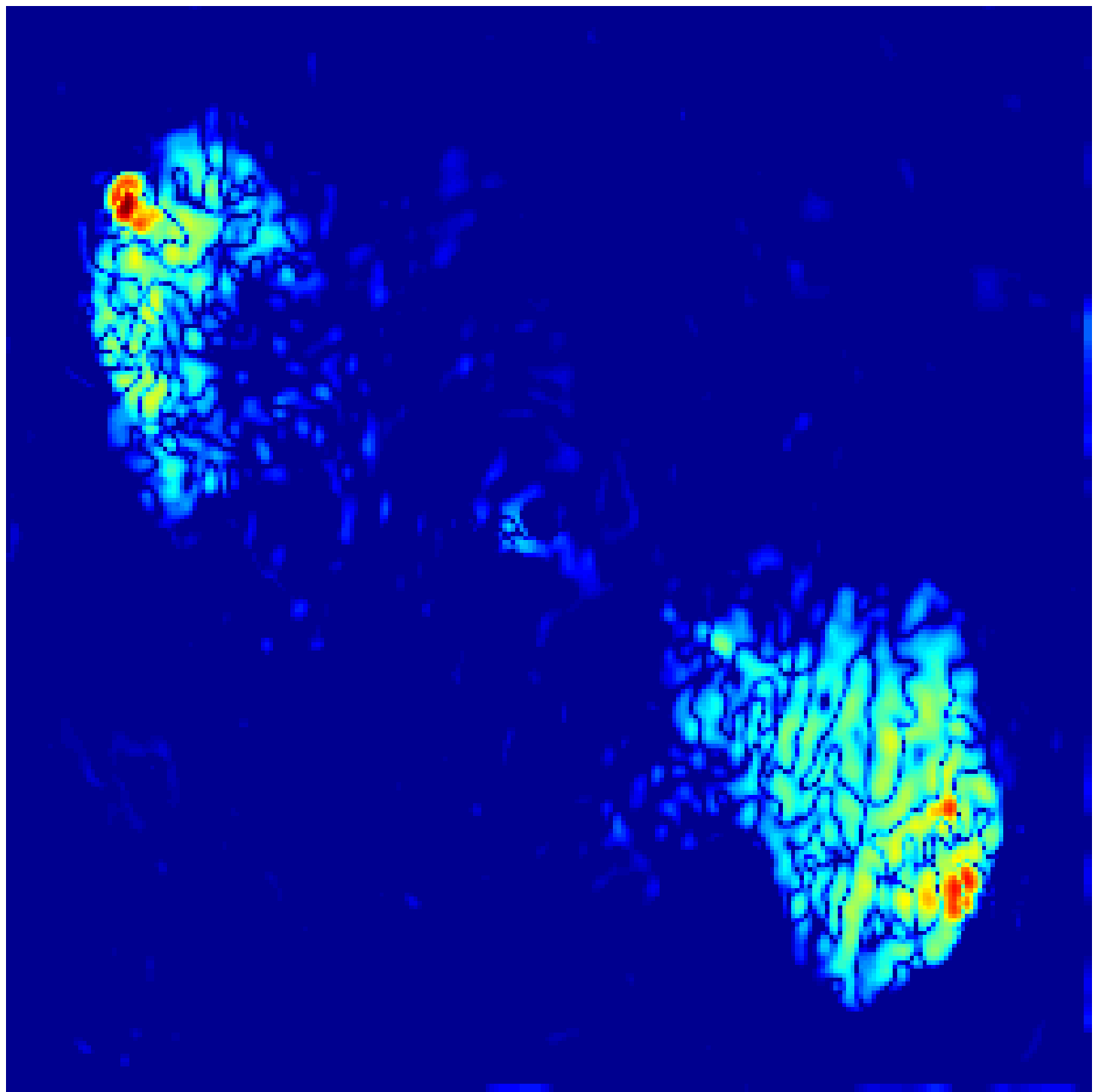} &
\hspace*{0.3cm}\includegraphics[trim ={0.2cm 0 0 0cm},clip,width=4cm,align=c]{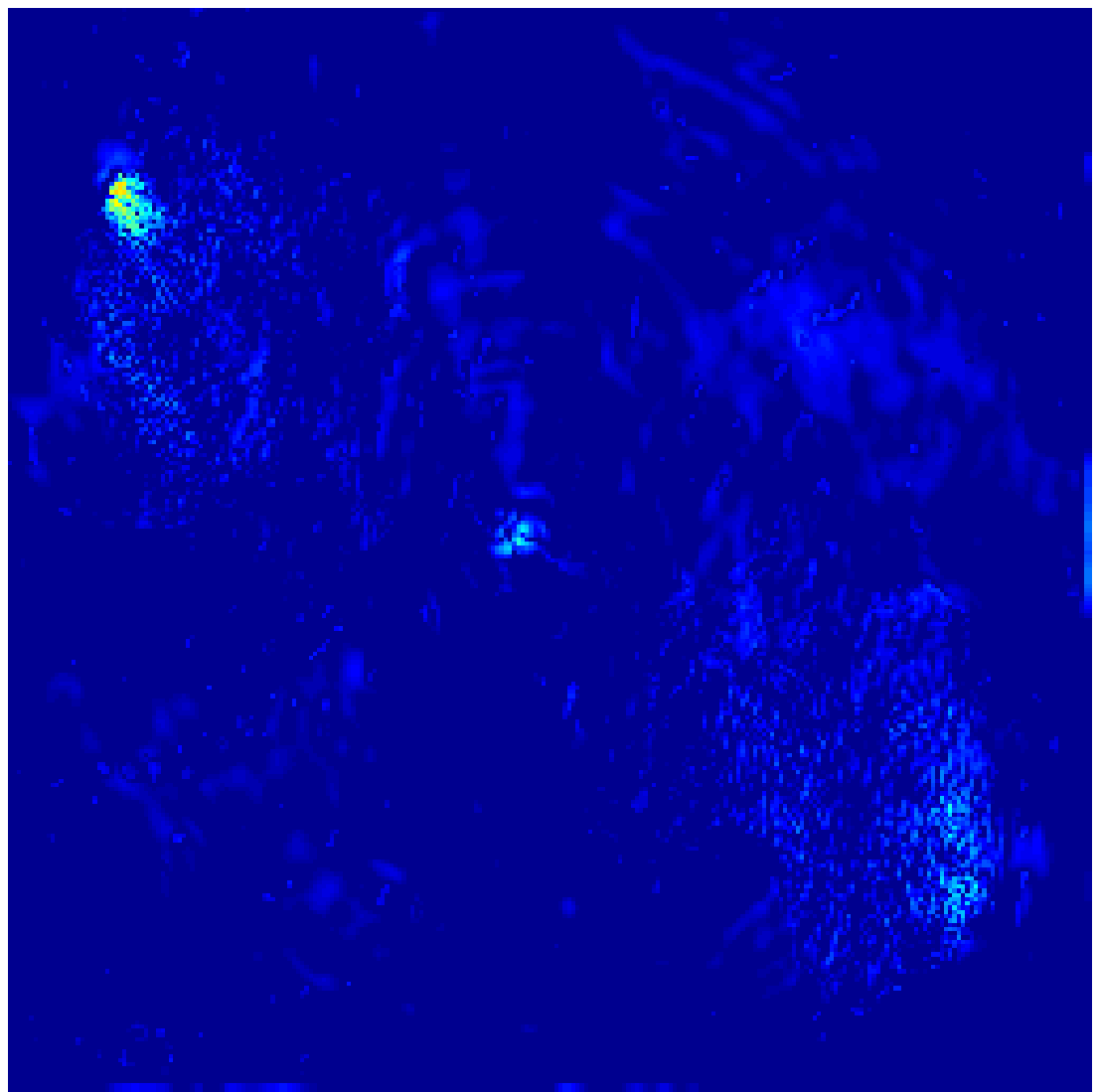} 
\end{tabular}
\caption{Cygnus A Stokes $Q$ true image in first row and reconstructed images (best ones over 5 performed simulations for each case) in other rows for the cases: Imaging with normalized DIEs (second row), {Joint DIE calibration and imaging (third row)}, Joint DDE calibration and imaging excluding the off-diagonal terms (fourth row), and considering full Jones matrix (fifth row). In each case, column-wise recovered images followed by their corresponding error images are displayed when imaging is performed without polarization constraint (first two columns) and with polarization constraint (last two columns). All the images are shown in log scale, with the same color range corresponding to the colorbar given in first row.}
\label{fig:cyg_a_Q}
\end{figure*}  

%%%%%%%%%%%%%%%%%%%%%%%%

%%%%%%%%%%%%%%%%%%%%%%%%
%%%%%%%  CYGNUS-A: Stokes U  %%%%%%%%%%%%%%%%%

\begin{figure*}
\centering
\begin{tabular}{@{}c@{}c@{}c@{}c@{}}
\includegraphics[trim ={0.2cm 0 0 0cm},clip,width=4cm]{true_U_cyg_a.eps} &
\hspace{-3cm}\includegraphics[trim ={16.2cm 0 0 0cm},clip,width=0.97cm]{U_cyg_a_colorbar.eps}
& & \\
Recovered images w/o &  Absolute error images &  Recovered images with & Absolute error images \\
polarization constraint & &polarization constraint & \\
\includegraphics[trim ={0.2cm 0 0 0cm},clip,width=4cm,align=c]{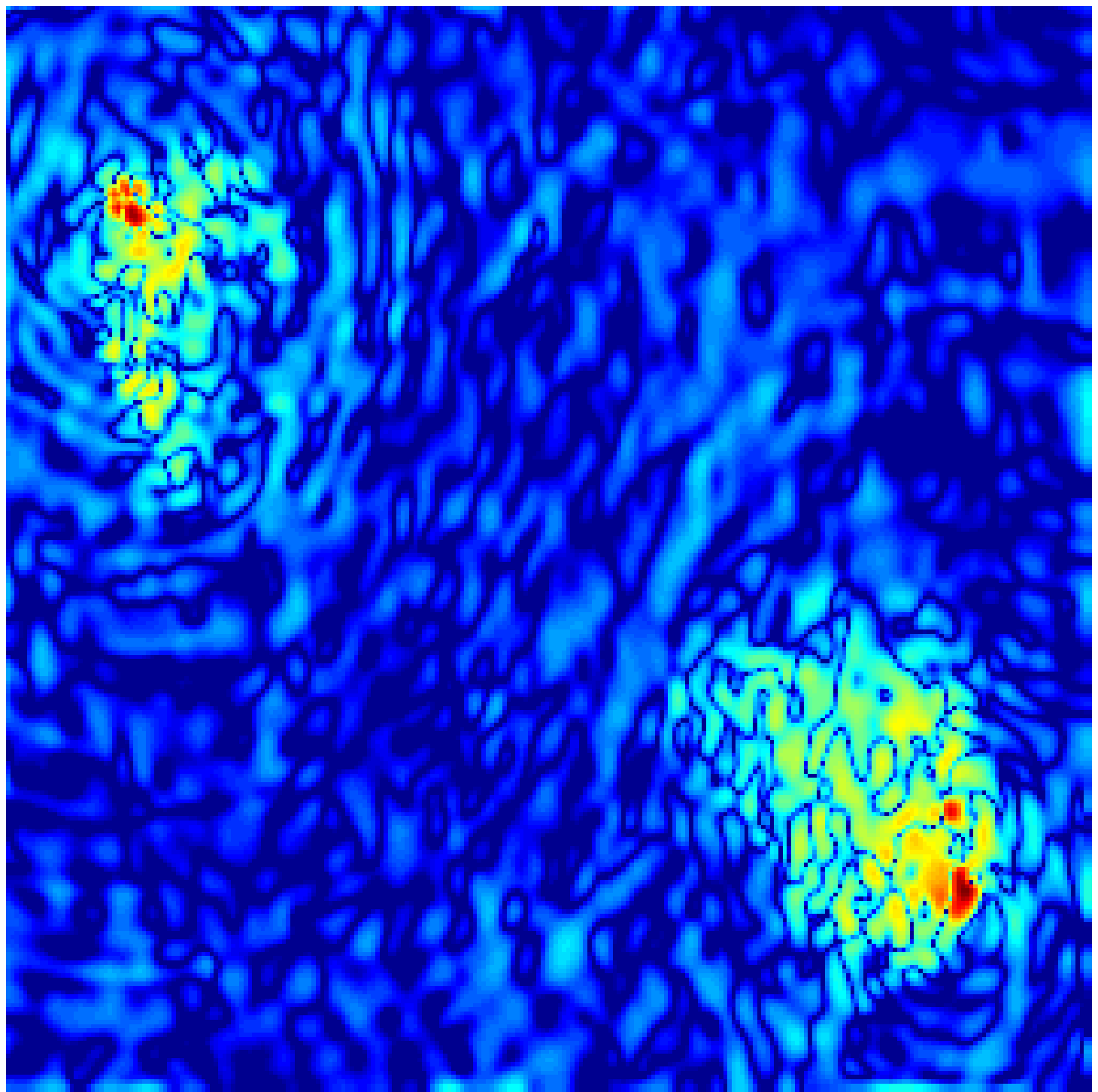} &
\hspace*{0.3cm}\includegraphics[trim ={0.2cm 0 0 0cm},clip,width=4cm,align=c]{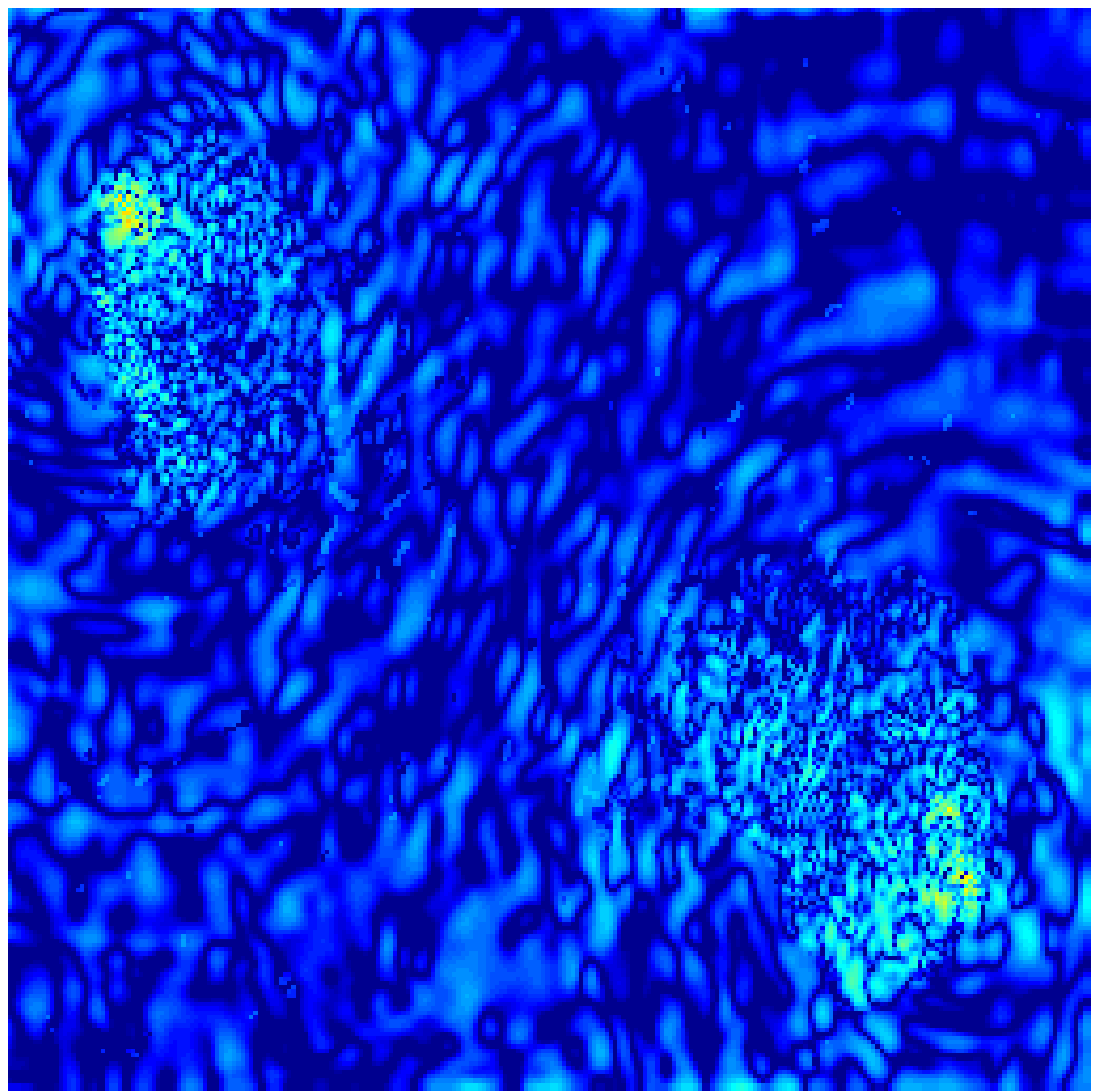} &
\hspace*{0.3cm}\includegraphics[trim ={0.2cm 0 0 0cm},clip,width=4cm,align=c]{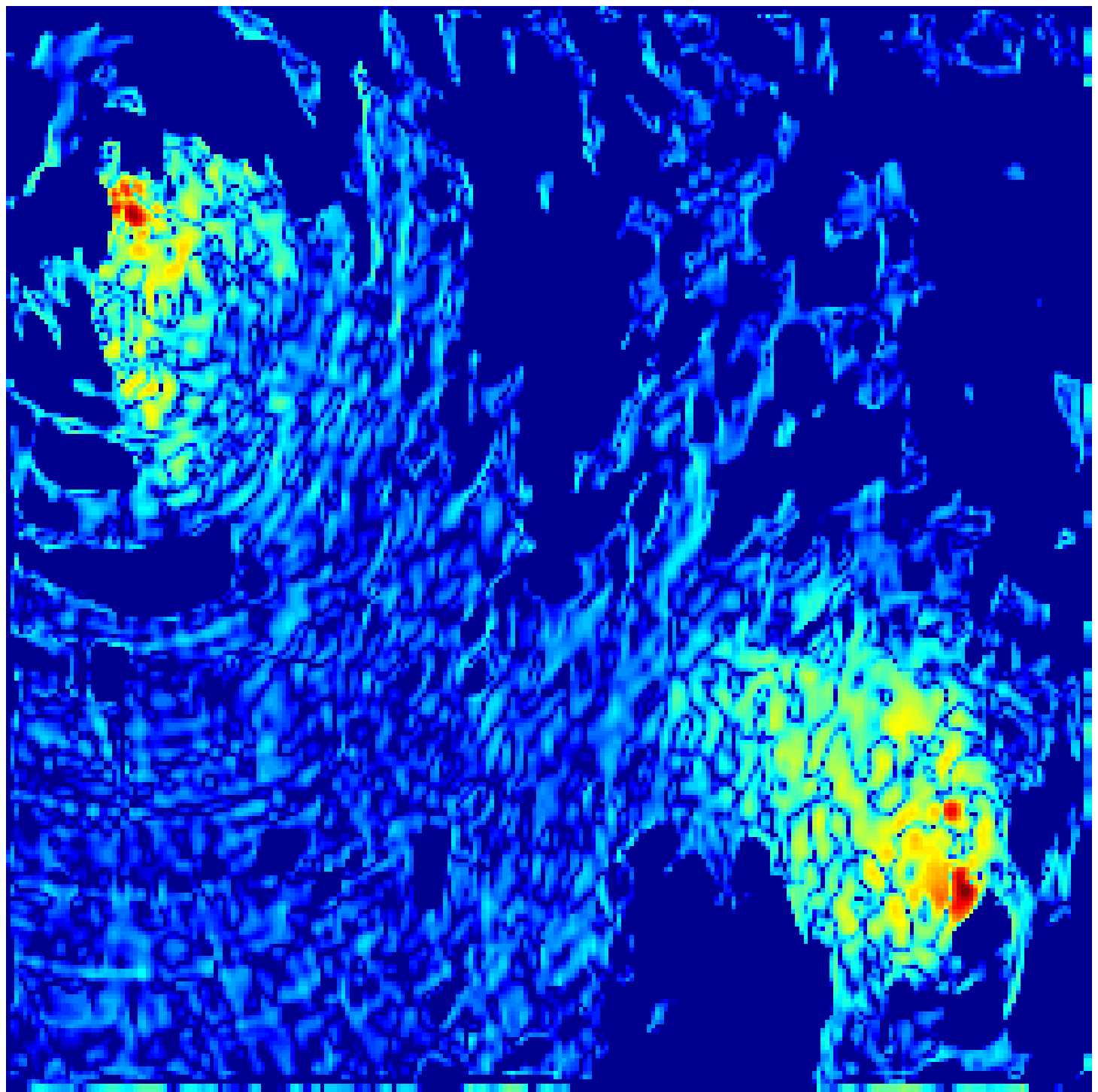} &
\hspace*{0.3cm}\includegraphics[trim ={0.2cm 0 0 0cm},clip,width=4cm,align=c]{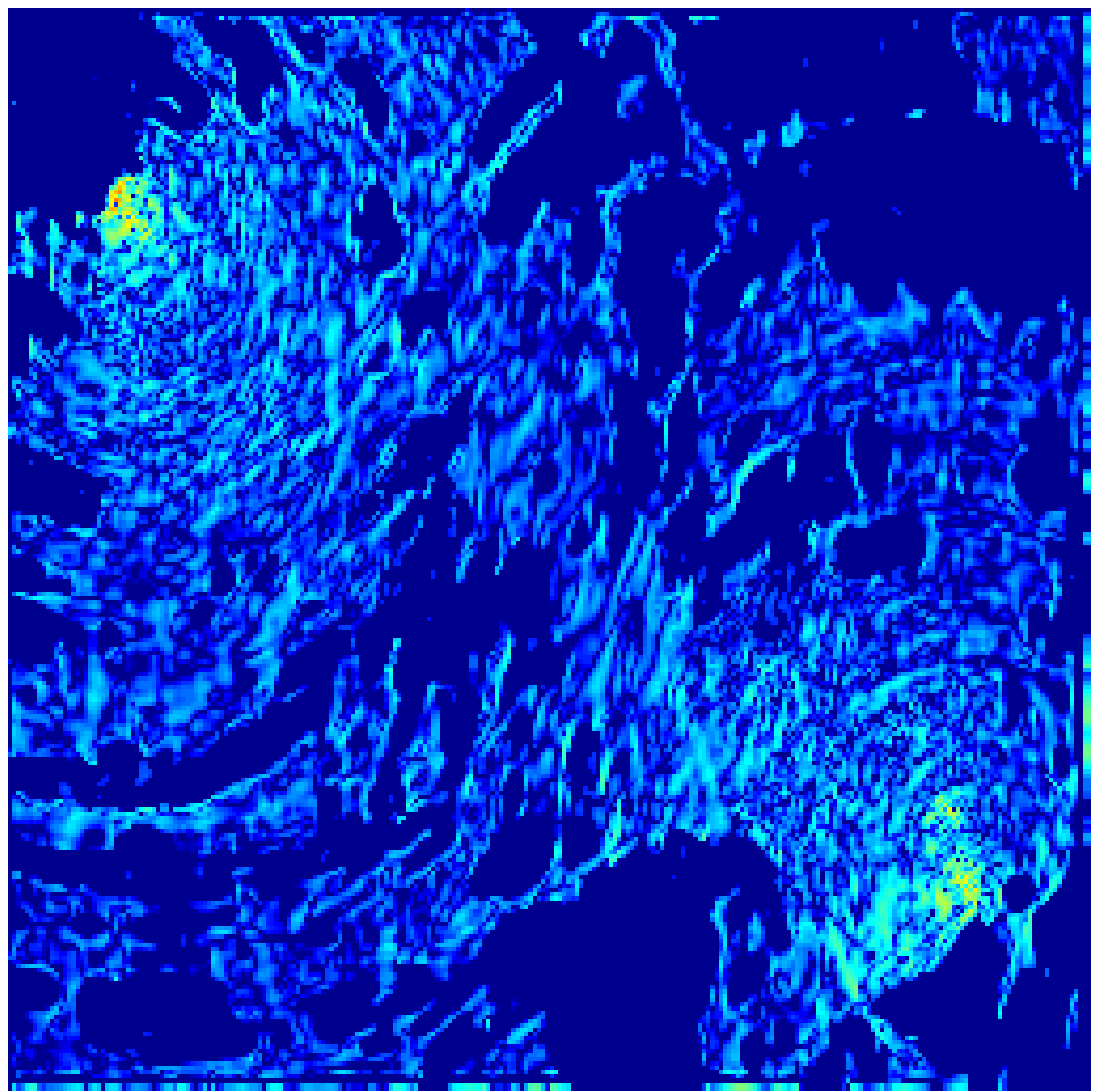} 
\vspace{0.25cm}
\\
%-----------------------------------------------
{\includegraphics[trim ={0.2cm 0 0 0cm},clip,width=4cm,align=c]{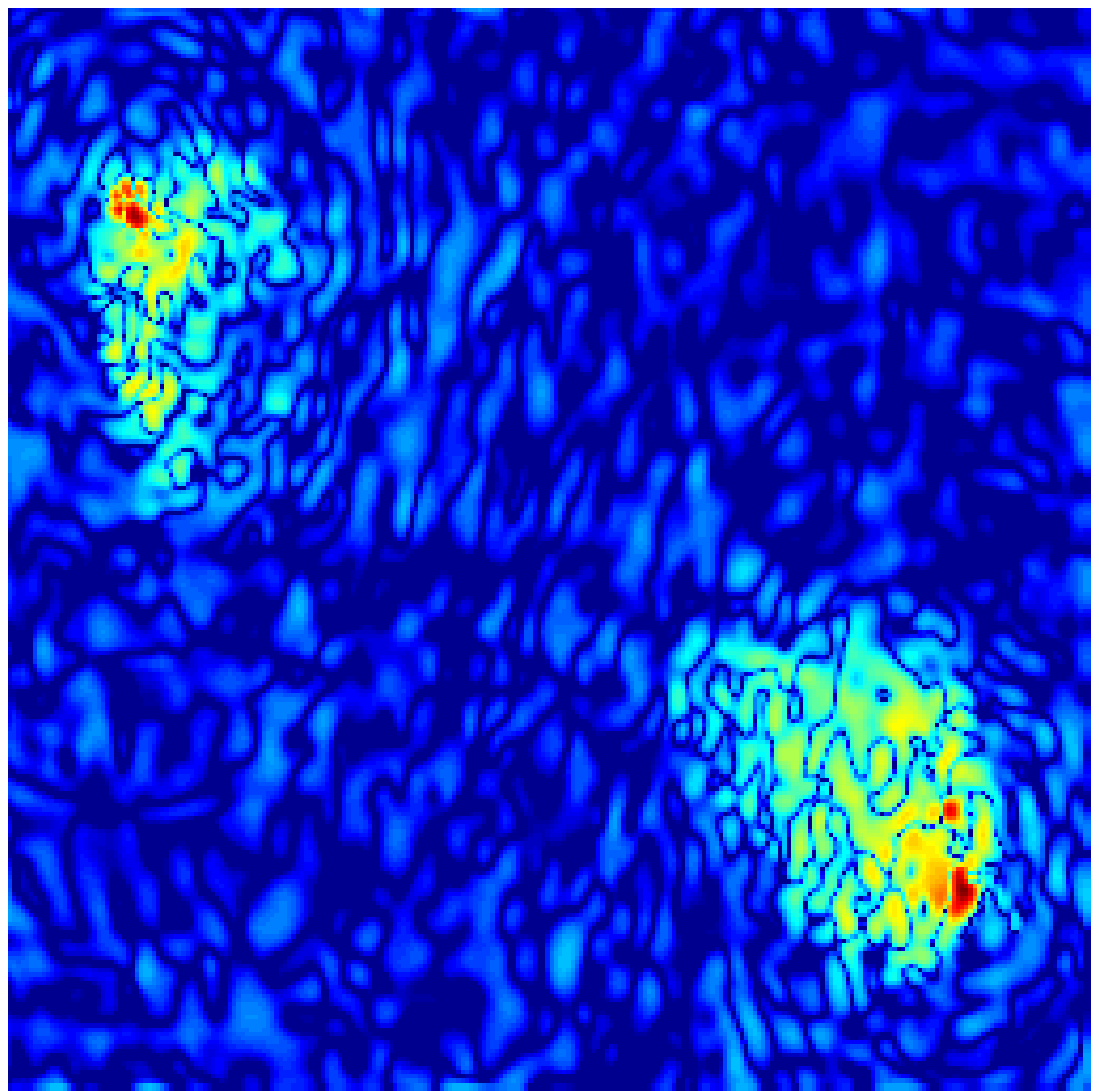}} &
\hspace*{0.3cm}\includegraphics[trim ={0.2cm 0 0 0cm},clip,width=4cm,align=c]{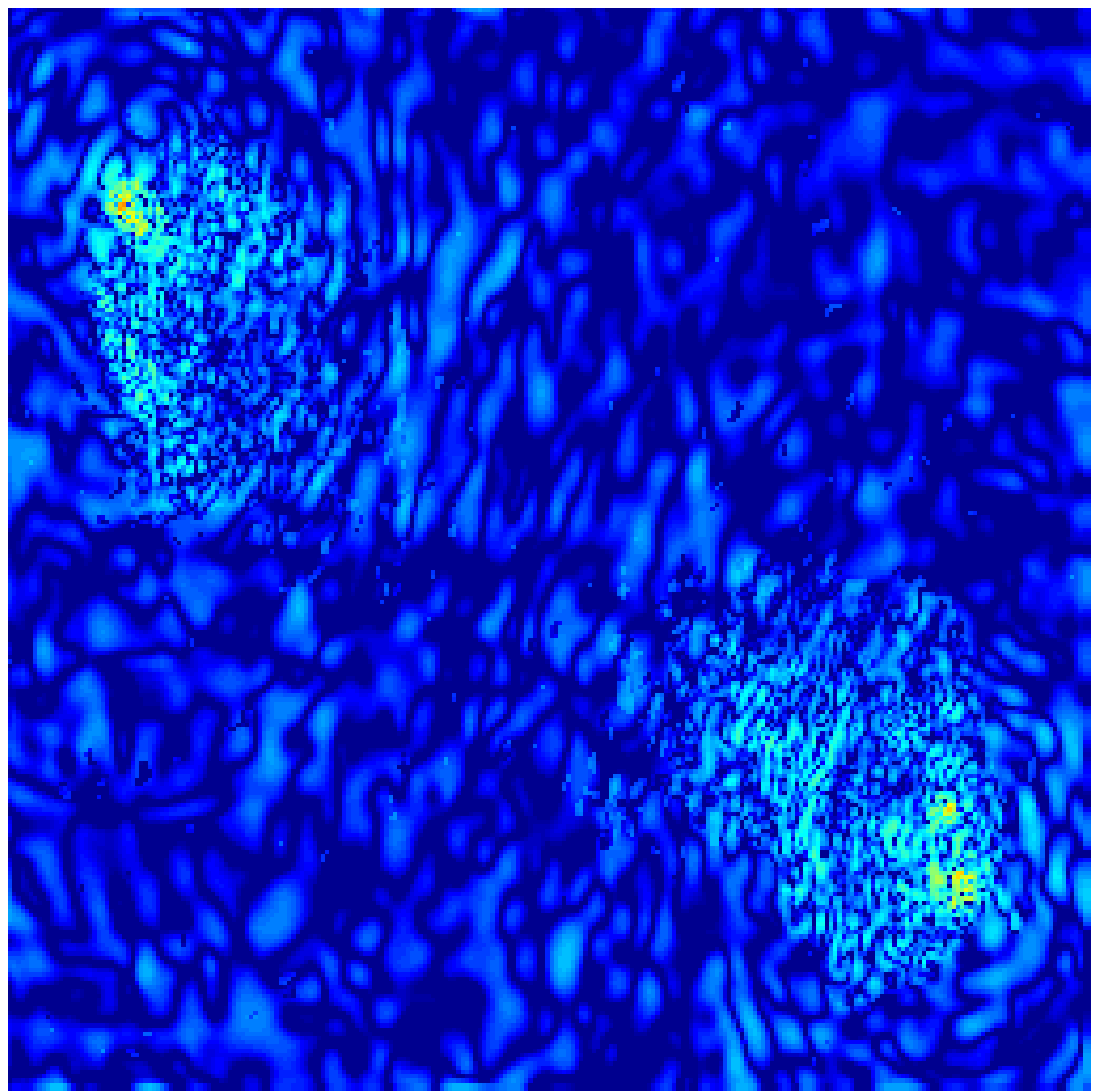} &
\hspace*{0.3cm}\includegraphics[trim ={0.2cm 0 0 0cm},clip,width=4cm,align=c]{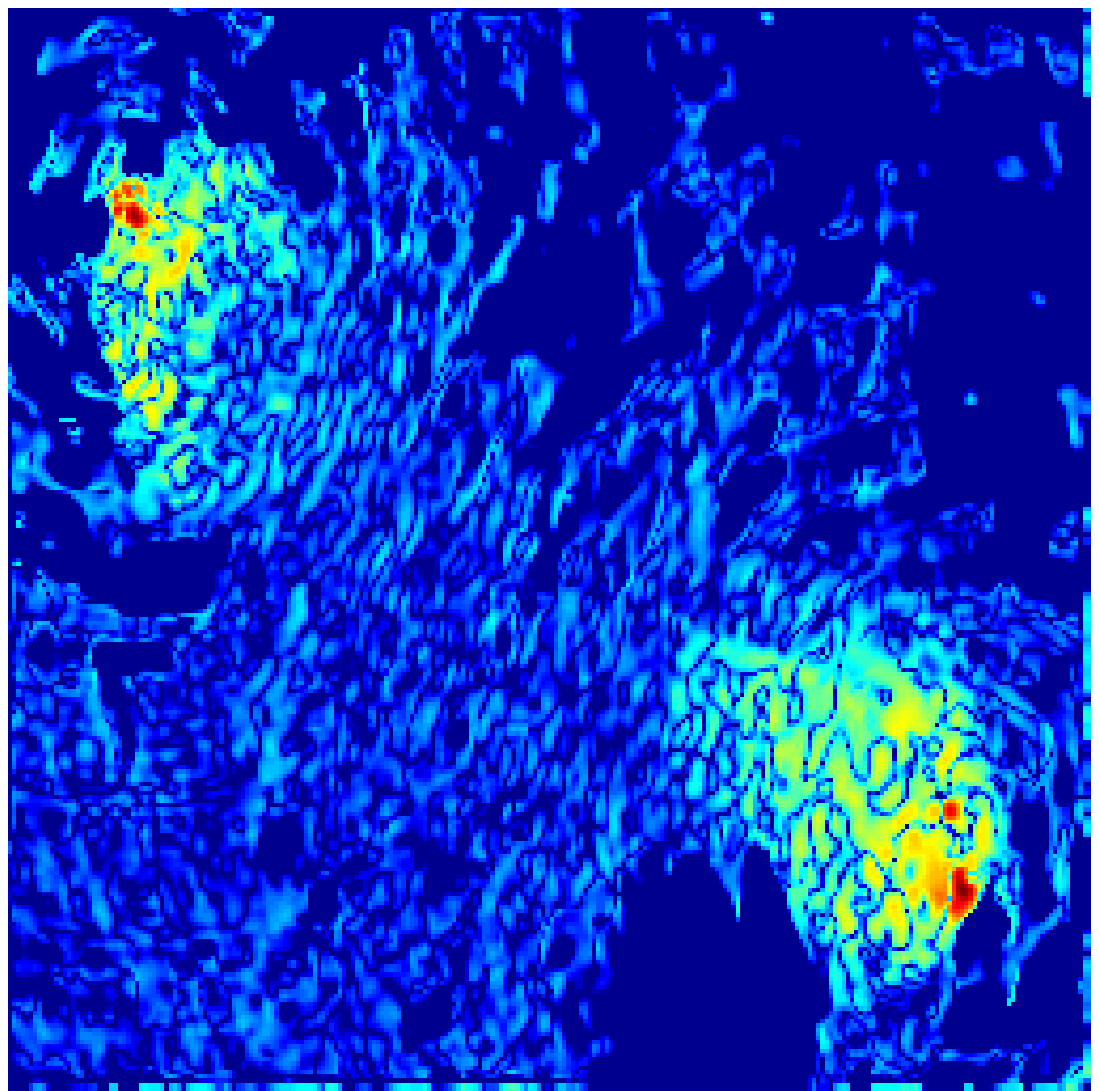} &
\hspace*{0.3cm}\includegraphics[trim ={0.2cm 0 0 0cm},clip,width=4cm,align=c]{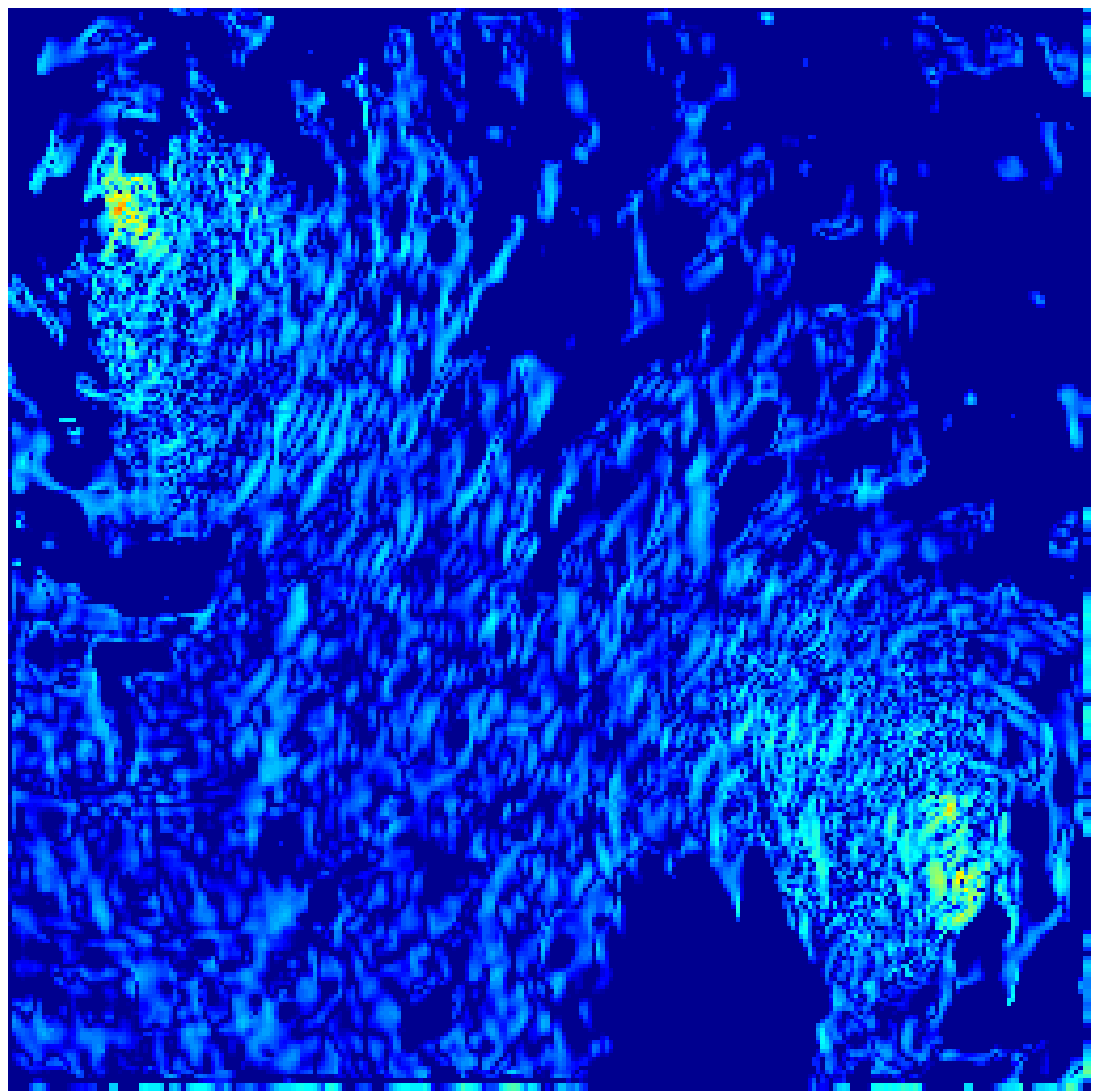} 
\vspace{0.25cm}
\\

%-----------------------------------------------
{\includegraphics[trim ={0.2cm 0 0 0cm},clip,width=4cm,align=c]{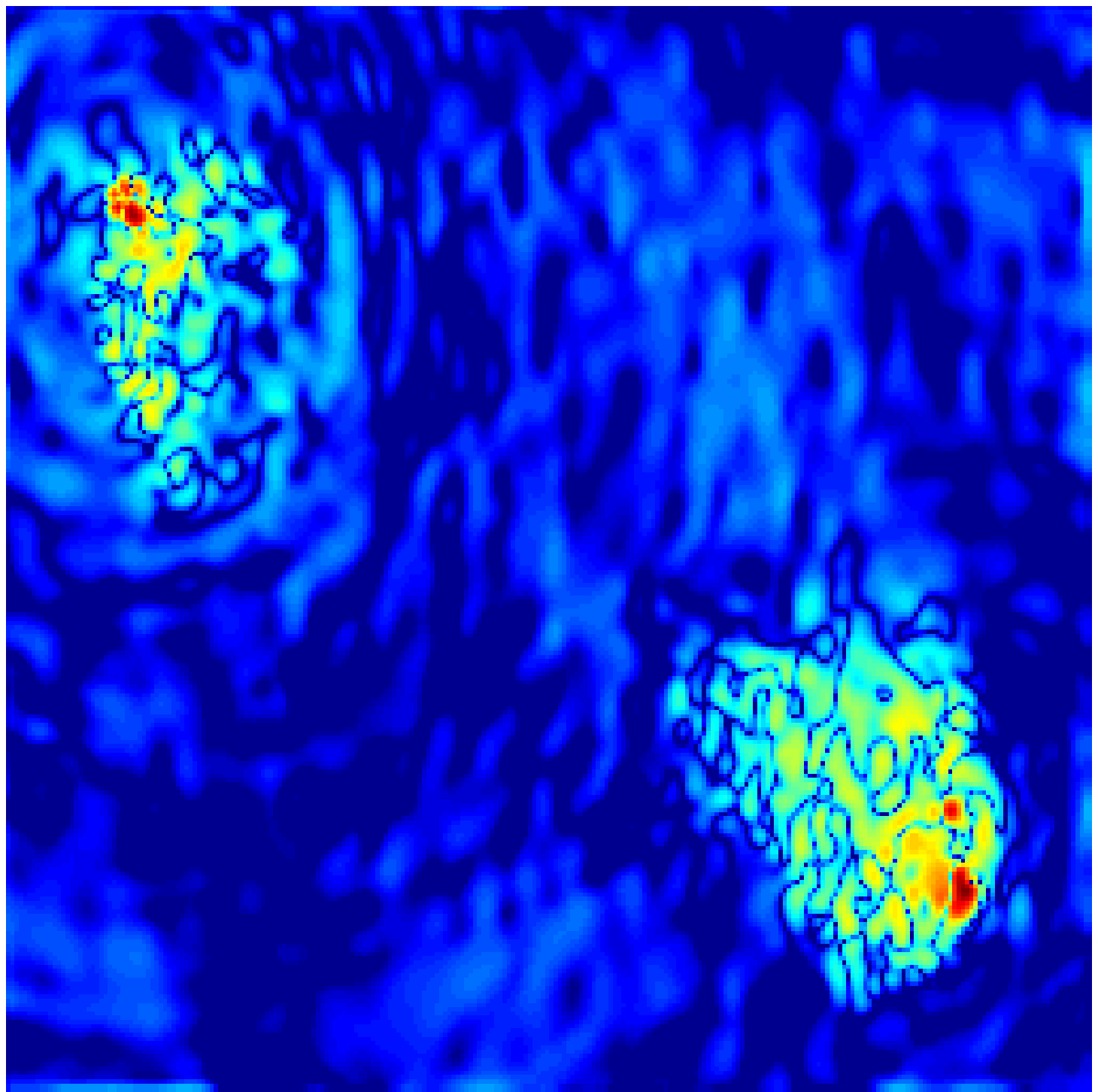}} &
\hspace*{0.3cm}\includegraphics[trim ={0.2cm 0 0 0cm},clip,width=4cm,align=c]{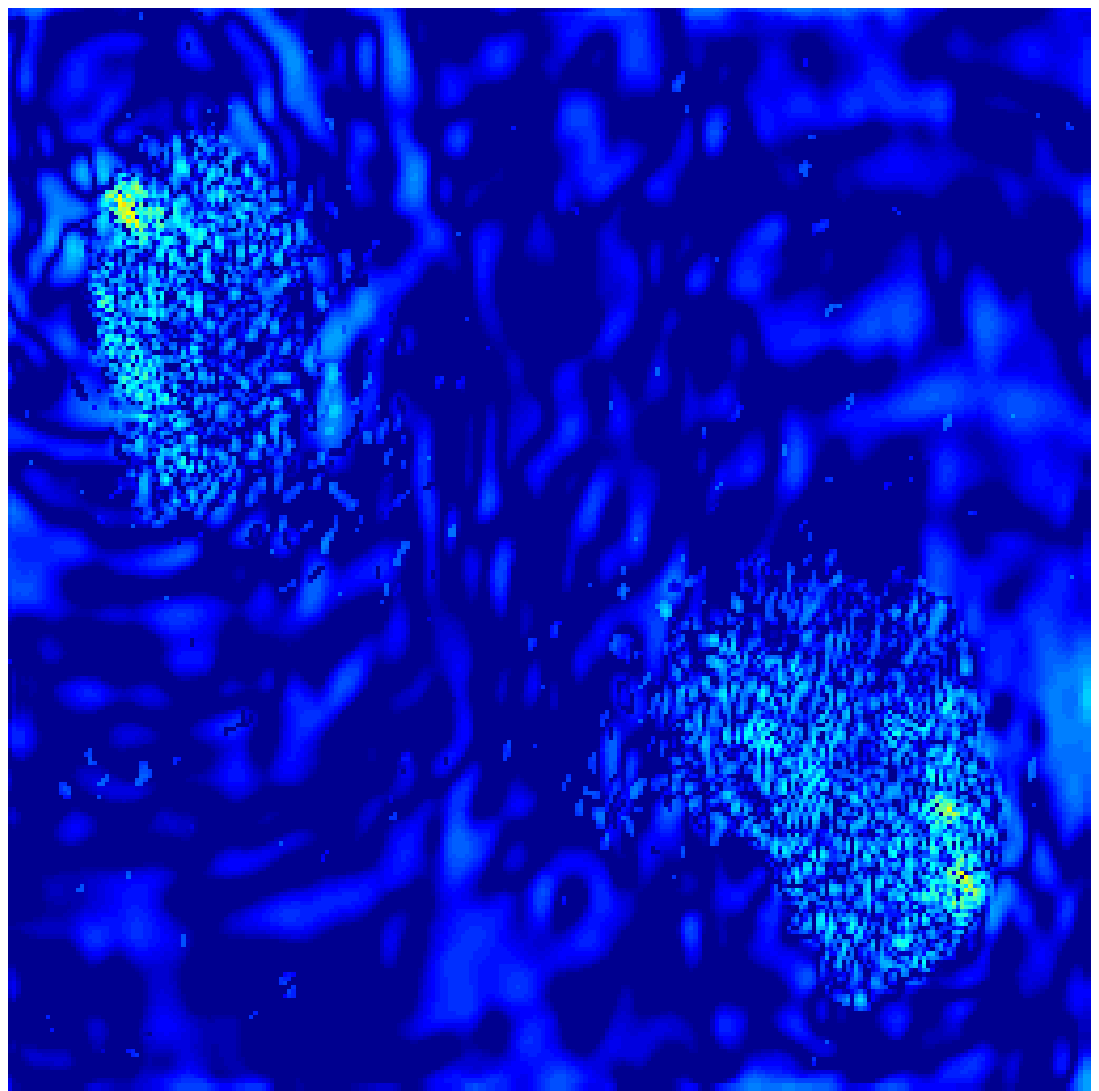} &
\hspace*{0.3cm}\includegraphics[trim ={0.2cm 0 0 0cm},clip,width=4cm,align=c]{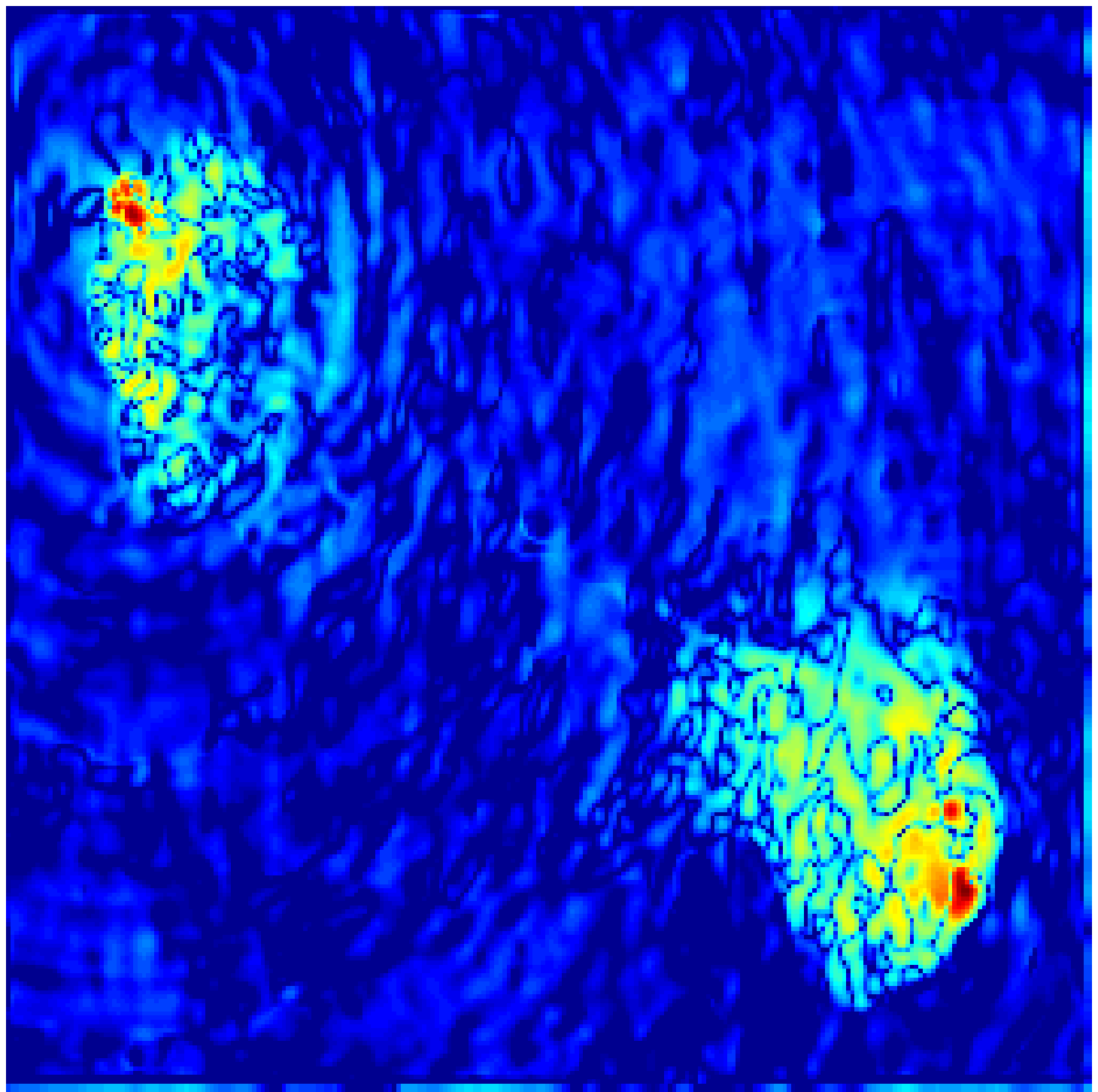} &
\hspace*{0.3cm}\includegraphics[trim ={0.2cm 0 0 0cm},clip,width=4cm,align=c]{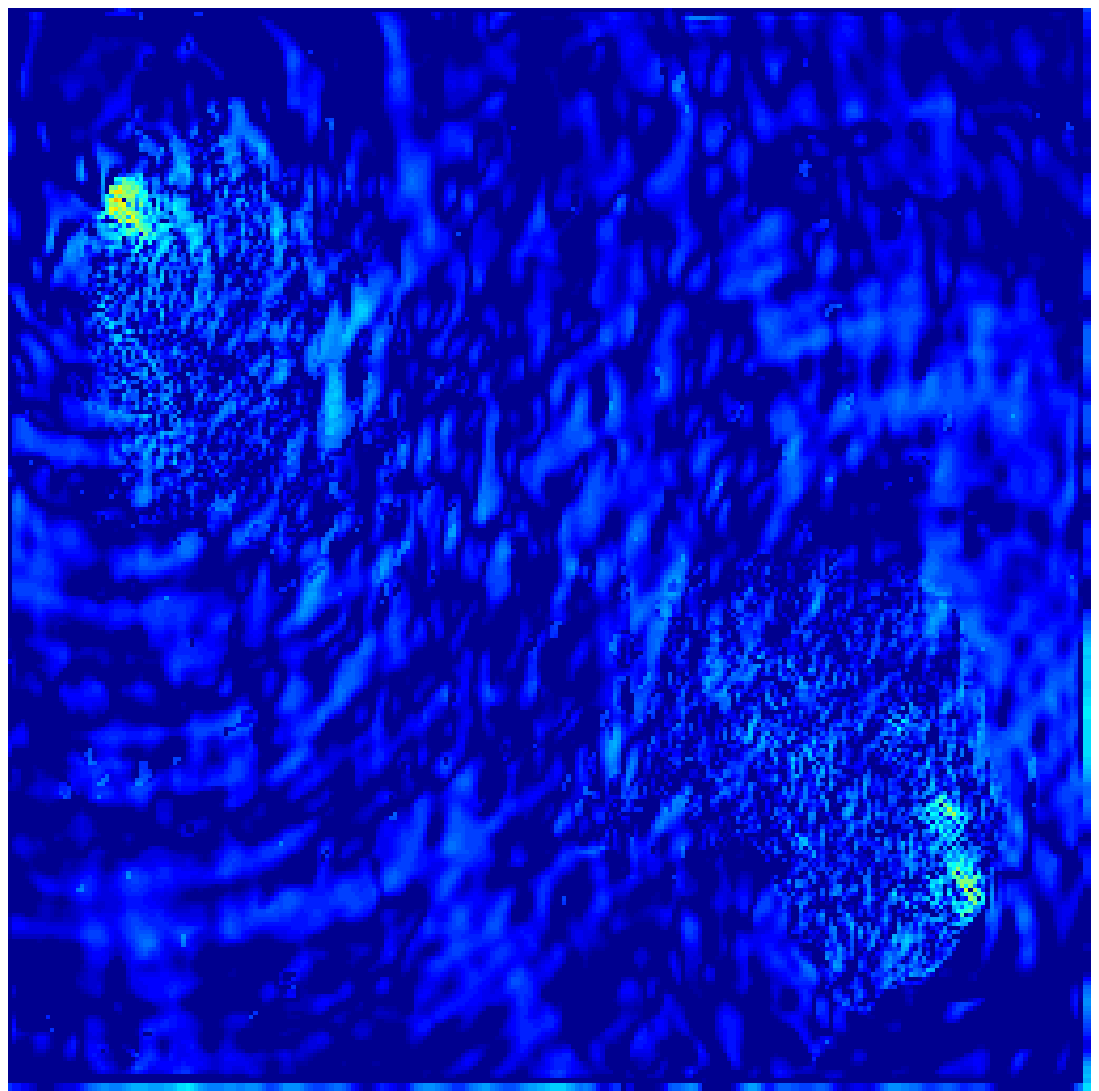} 
\vspace{0.25cm}
\\
%-----------------------------------------------
\includegraphics[trim ={0.2cm 0 0 0cm},clip,width=4cm,align=c]{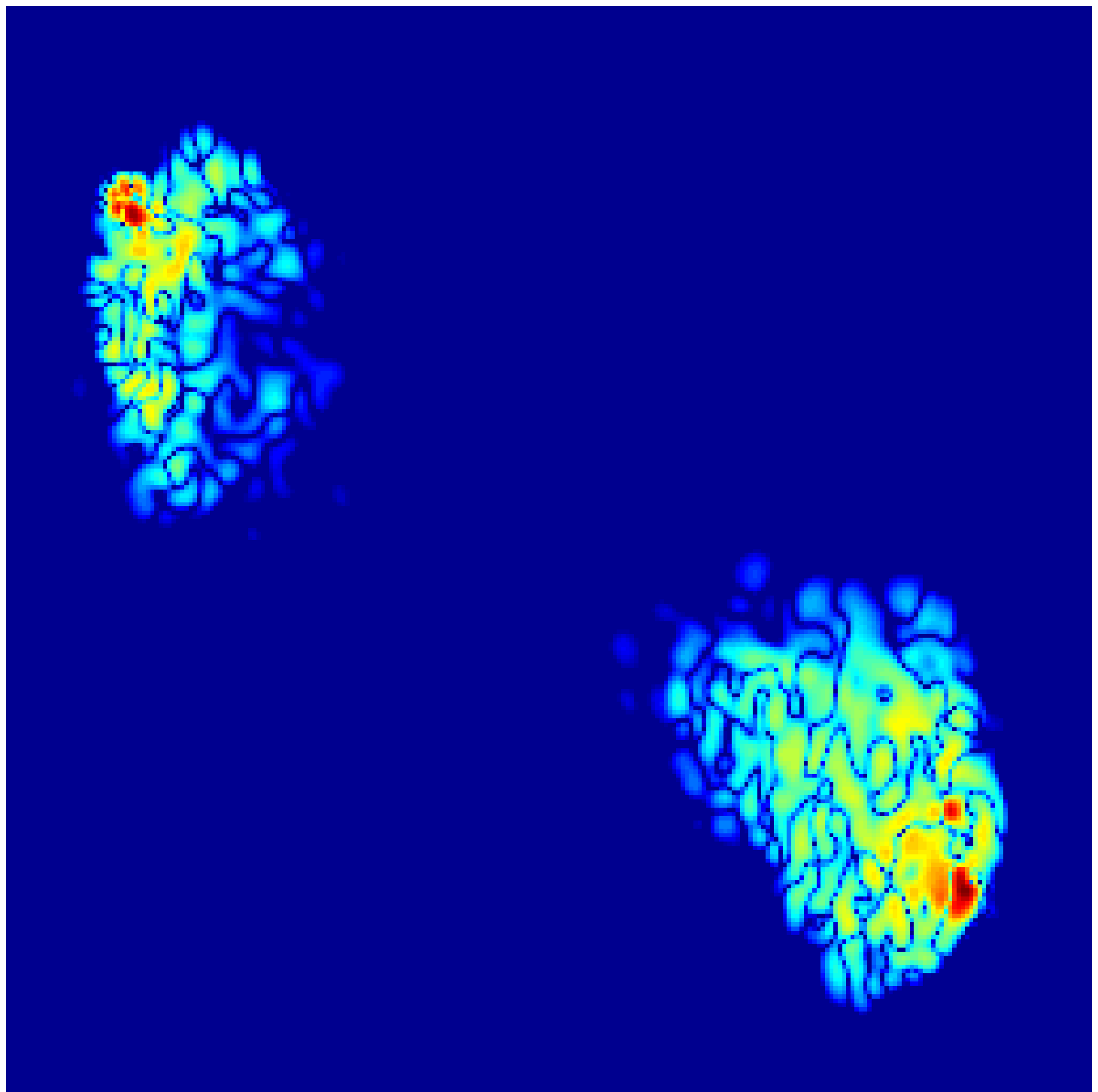} &
\hspace*{0.3cm}\includegraphics[trim ={0.2cm 0 0 0cm},clip,width=4cm,align=c]{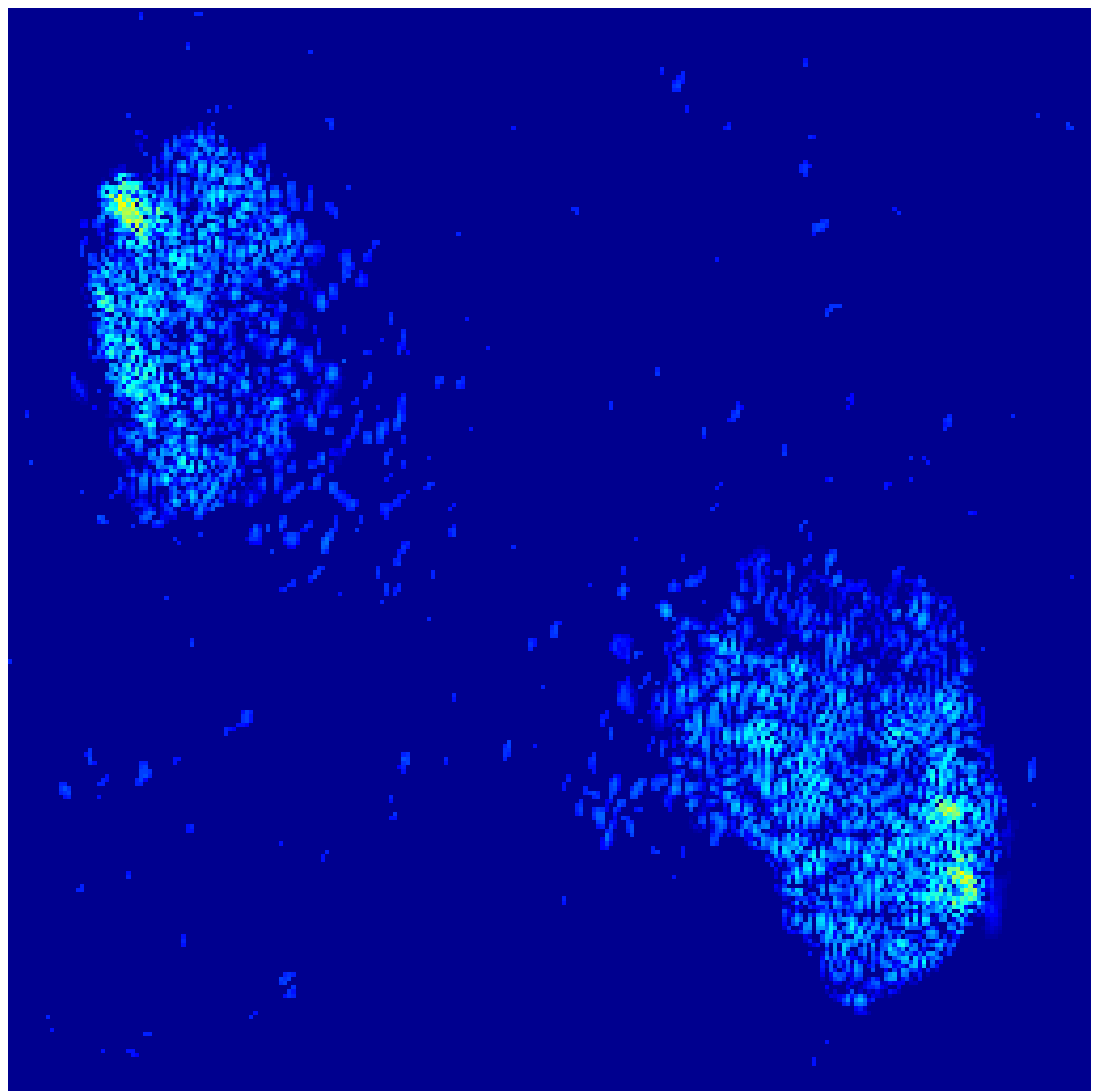} &
\hspace*{0.3cm}\includegraphics[trim ={0.2cm 0 0 0cm},clip,width=4cm,align=c]{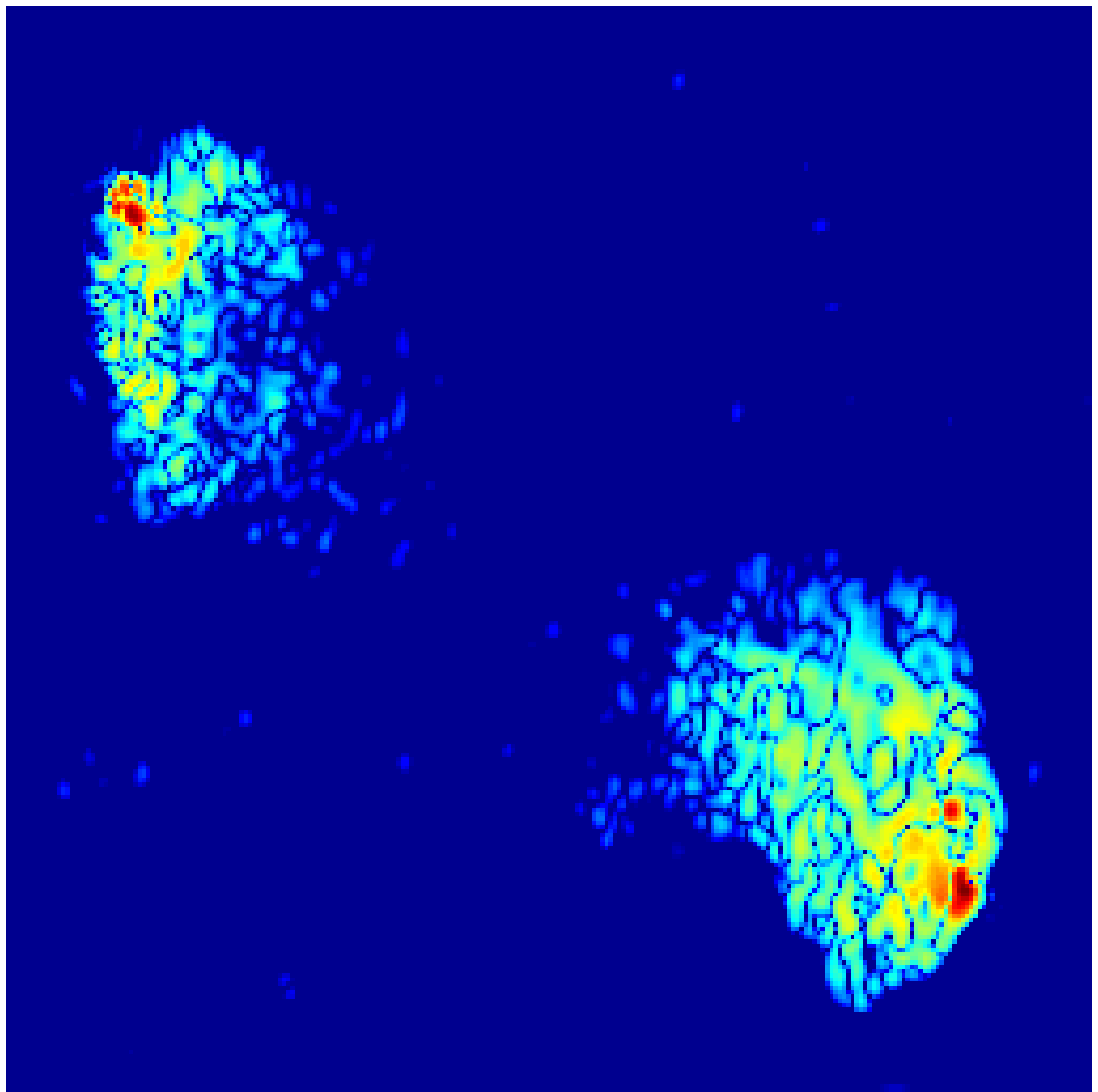} &
\hspace*{0.3cm}\includegraphics[trim ={0.2cm 0 0 0cm},clip,width=4cm,align=c]{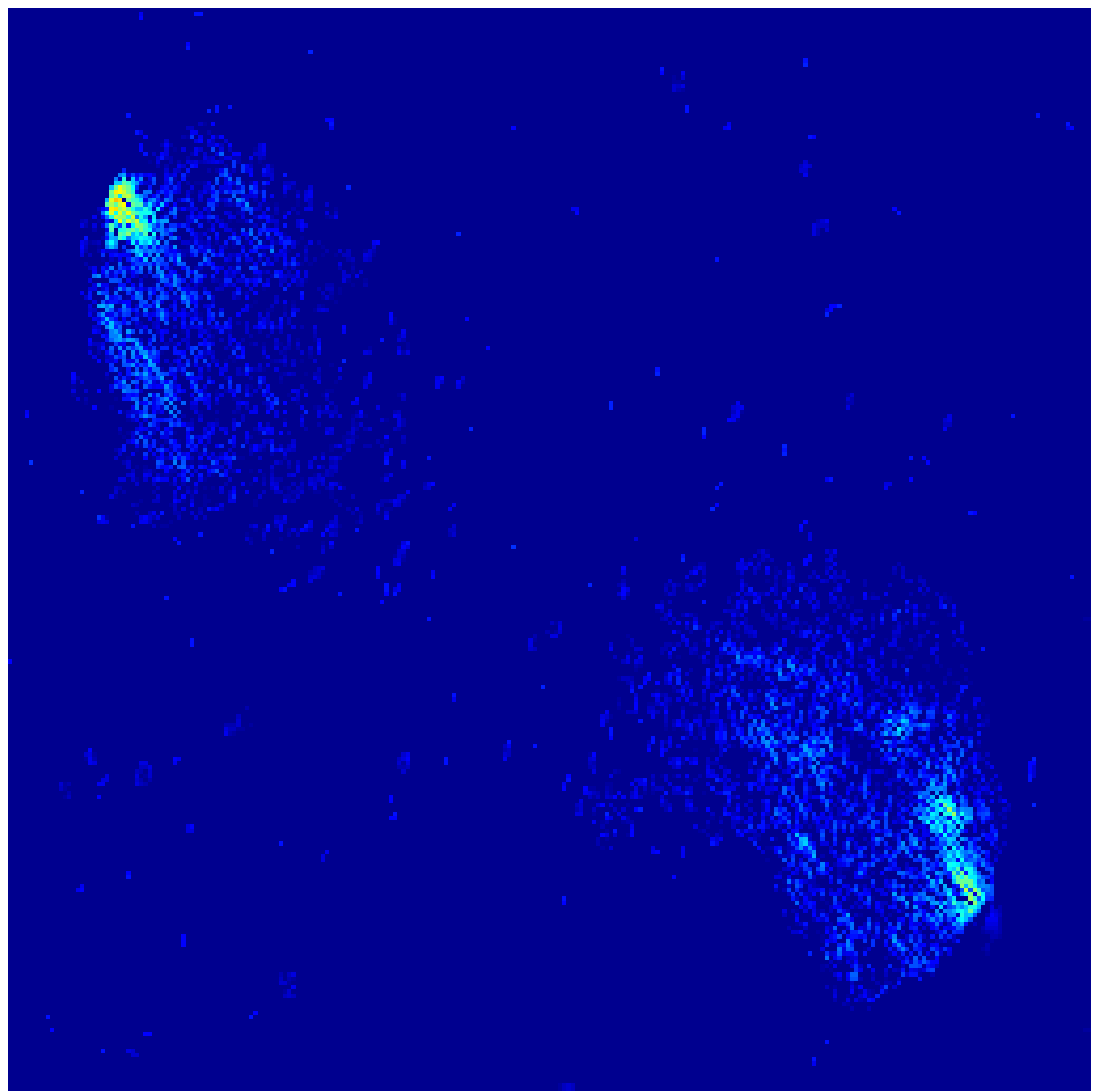} 
\end{tabular}
\caption{Cygnus A Stokes $U$ true image in first row and reconstructed images (best ones over 5 performed simulations for each case) in other rows for the cases: Imaging with normalized DIEs (second row), {Joint DIE calibration and imaging (third row)}, Joint DDE calibration and imaging excluding the off-diagonal terms (fourth row), and considering full Jones matrix (fifth row). In each case, column-wise recovered images followed by their corresponding error images are displayed when imaging is performed without polarization constraint (first two columns) and with polarization constraint (last two columns). All the images are shown in log scale, with the same color range corresponding to the colorbar given in first row.}
\label{fig:cyg_a_U}
\end{figure*}  

For quantitative comparison of the results obtained by different tests, we use signal-to-noise ratio (SNR) as a metric. Given an original image $\overline{\bm{s}} \in \eR^{N}$, the SNR of the reconstructed image $\bm{s}^\prime$ with  respect to $\overline{\bm{s}}$ is defined as
\vspace{-0.15cm}
\begin{equation}
    \text{SNR} = 20 \operatorname{log_{10}} \left( \frac{\|\overline{\bm{s}}\|_2}{\|\upsilon^\prime \bm{s}^\prime -\overline{\bm{s}}\|_2 } \right),
\end{equation}
where $\upsilon^\prime = \underset{\upsilon > 0}{\operatorname{argmin}} \, \, \|\upsilon \bm{s}^\prime -\overline{\bm{s}}\|_2^2$ accounts for the ambiguity problem in the underlying blind deconvolution problem. Furthermore, the values of the regularization parameters $\bm{\eta}=[\eta_1, \eta_2, \eta_3]$ and $\gamma$ are chosen to maximize the SNR. 

We also report the dynamic range (DR) obtained for the reconstructed images. For every image contained in the estimated Stokes matrix $\Sbs^{\prime} = \Sbs_0 + \Ebs^{\prime}$, with $(i,j) \in \{1,2\}^2$ it is defined as follows:
\begin{equation}
    \text{DR}_{ij} = \frac{\sqrt{N} \, \|\Phi \|_{2}^2}{\big\| \big[\Phi^\dagger \big( \bm{y} - \Phi(\Sbs^\prime) \big)\big]_{ij} \big\|_2} \underset{n}{\operatorname{max}} \, { [\Sbs^\prime(n)]_{ij}}.
\end{equation}

\vspace{-0.5cm}

\subsection{Results and analysis}
We now present the results obtained by conducting tests for the various cases mentioned before. In terms of quantitative comparison between the different cases performed, Table~\ref{tab:results} provides the obtained SNR and dynamic range values for both the sets of images. In each case, the mean values evaluated over the 5 performed simulations are shown. From the calibration front, it can be noticed that joint DDE calibration and imaging leads to a significant improvement in the reconstruction quality, both in terms of SNR and DR, in comparison with either of the cases {when only imaging step is performed considering normalized DIEs or jointly calibrating for DIEs and imaging.}
In particular, the case of normalized DIEs considers Jones matrices to be identity, whereas the DIE calibration scheme also accounts for the DIEs in the off-diagonal Jones terms and is thus further affected by the estimation of these terms. In fact, these terms lead to flux leakage from one Stokes parameter to others and if not perfectly calibrated, can lead to error propagation. It particularly affects the low amplitude Stokes $Q$ and $U$ images. This can be seen from the smaller SNR values of Stokes $Q$ image than that obtained by the case of normalized DIEs (Table~\ref{tab:results}(a)).
Another observation that can be made for these two cases is regarding the higher DR values obtained for Stokes $Q$ and $U$ images than that of Stokes $I$ image. This non-physical feature can be attributed to the usage of incomplete measurement model (i.e. without the incorporation of the DDEs), severely affecting the underlying non-convex minimization problem giving inaccurate solutions.
The aforementioned observations for these cases further indicate the necessity of taking direction-dependent terms into account and performing their calibration along with imaging for a better reconstruction quality.
Second, comparing the results obtained with and without calibrating for the off-diagonal terms in the Jones matrices shows 
better SNR and DR values are achieved in the former case, thereby demonstrating
the significance of calibrating for the off-diagonal Jones terms to recover high quality, high dynamic range images.

On the other hand, on the imaging front, it is interesting to note that when the image recovery is performed {either using only normalized DIEs or calibrating for DIEs, i.e. having inexact knowledge of the Jones matrices and neglecting the DDEs}, regularizing the underlying problem with the polarization constraint may not be very effective. Indeed, in such a case, not enforcing the constraint can produce slightly better results, as seen for Cygnus A images.
Nevertheless, when the DDE calibration scheme is incorporated, the result analysis highlights the superior performance of this constraint, recovering images with much higher SNR and DR compared to the case where it is not imposed.
Especially in the case of model images for Hydra A, the enforcement of the polarization constraint leads to an appreciable improvement of $\sim 9 - 10$ dBs in SNR of Stokes $Q$ and $U$ images (Table~\ref{tab:results}(c)) and around one order of magnitude in DR (Table~\ref{tab:results}(d)). It should be noted that Hydra A images have lesser amplitude than the first set of images (Cygnus A). This is true specifically for Stokes $Q$ and $U$ images which are around one order of magnitude lower in amplitude than the corresponding Stokes $I$ image and hence, difficult to be recovered. In this respect, the obtained results emphasize the crucial role played by suitably chosen regularization prior in enhancing the reconstruction quality. {In fact, if the problem is not well regularized (i.e. the polarization constraint is not enforced), the solution of the underlying non-convex joint DDE calibration and imaging problem may stuck in a local minimum leading to poor reconstruction quality.} The good performance of the polarization constraint to recover these images shows suitability of this prior for full polarization imaging. This validates the findings of \cite{Birdi2018b} and further extends them to the case of joint calibration and imaging.

\begin{figure*}
\centering
\begin{tabular}{@{}c@{}c@{}c@{}c@{}}
\includegraphics[trim ={0.2cm 0 0 0cm},clip,width=4cm]{true_I_hydra5.eps} &
\hspace{-3cm}\includegraphics[trim ={16.2cm 0 0 0cm},clip,width=0.97cm]{I_hydra_colorbar.eps}
& & \\
Recovered images w/o &  Absolute error images &  Recovered images with & Absolute error images \\
polarization constraint & &polarization constraint & \\
\includegraphics[trim ={0.2cm 0 0 0cm},clip,width=4cm,align=c]{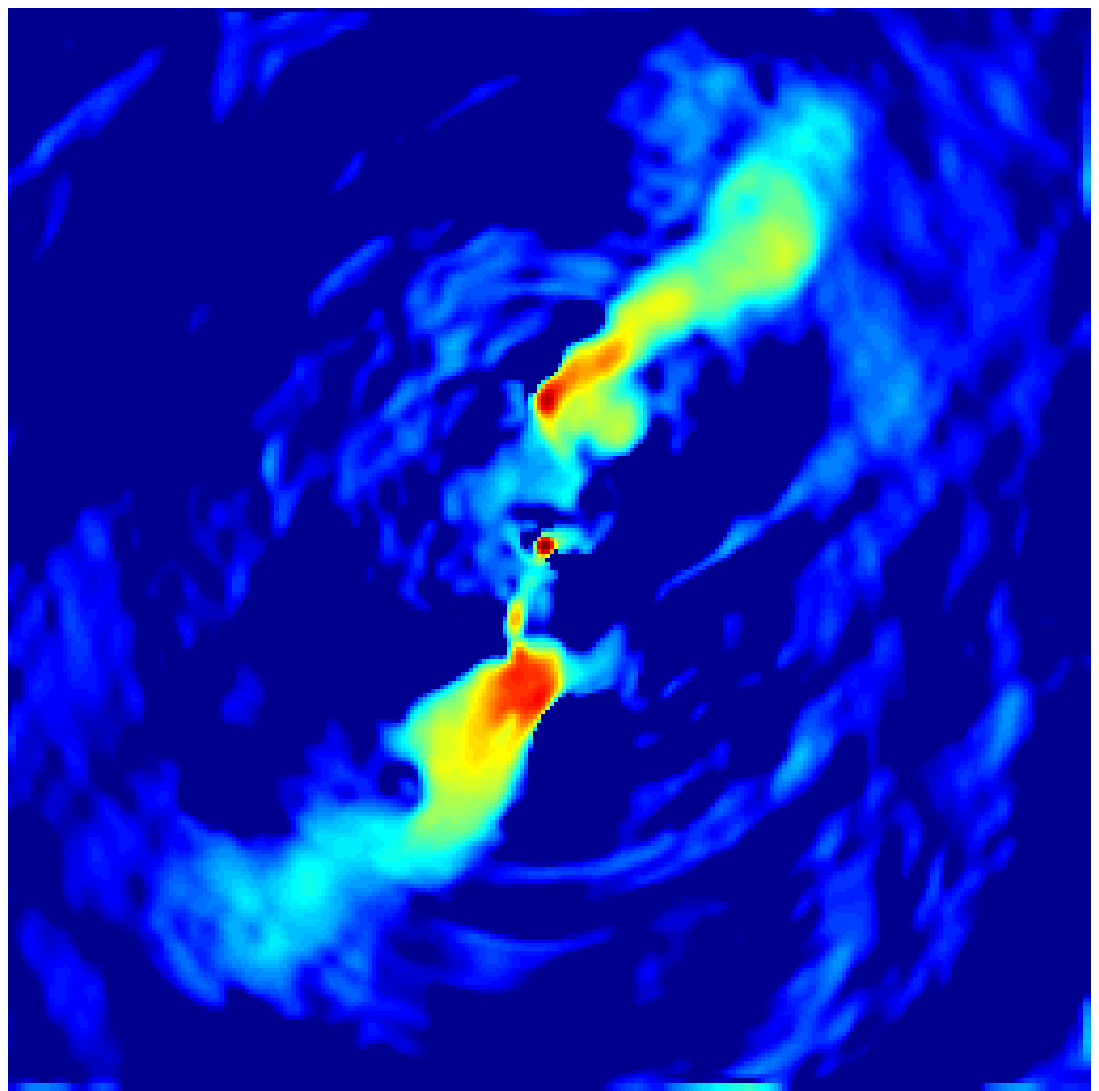} &
\hspace*{0.3cm}\includegraphics[trim ={0.2cm 0 0 0cm},clip,width=4cm,align=c]{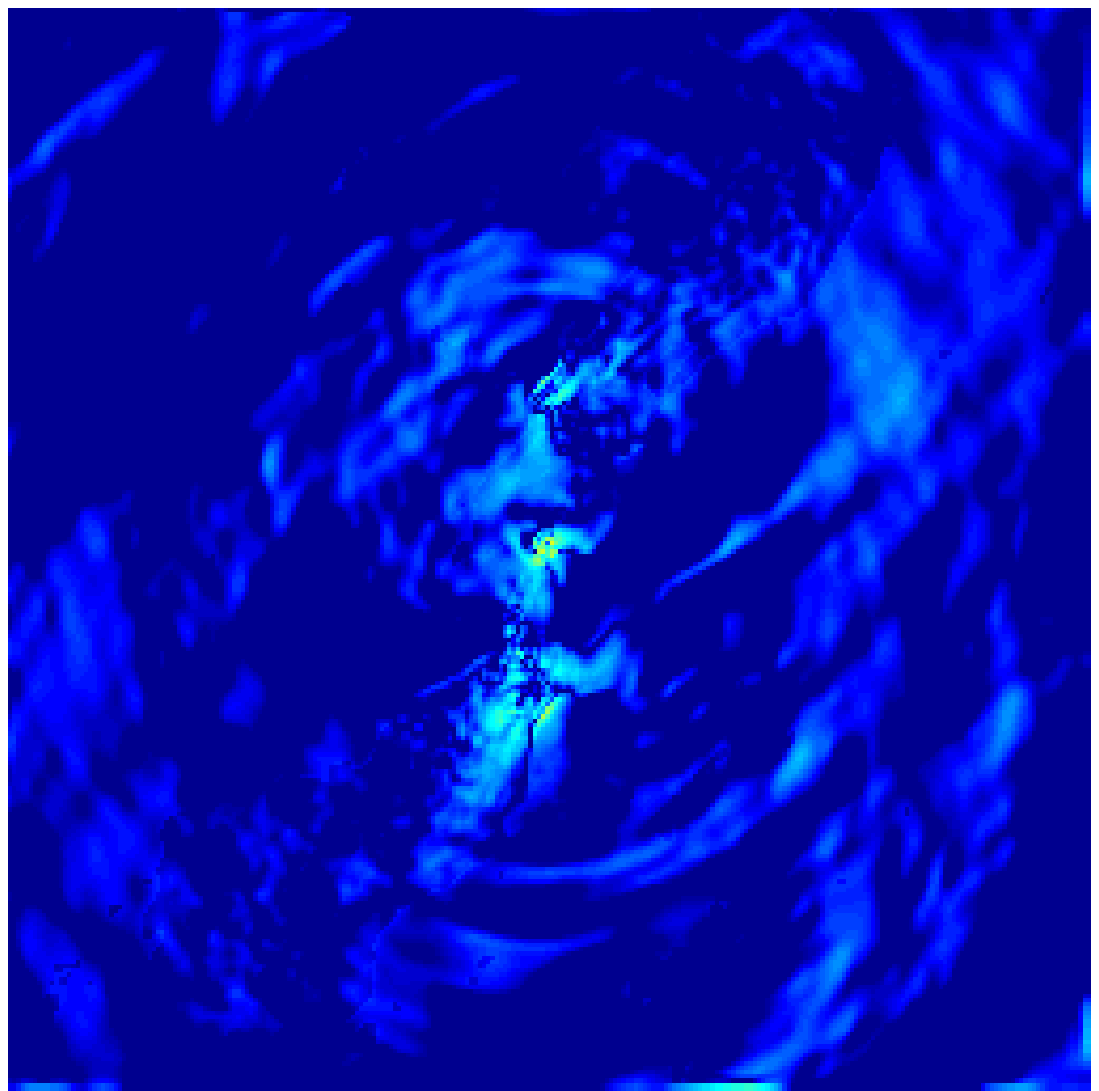} &
\hspace*{0.3cm}\includegraphics[trim ={0.2cm 0 0 0cm},clip,width=4cm,align=c]{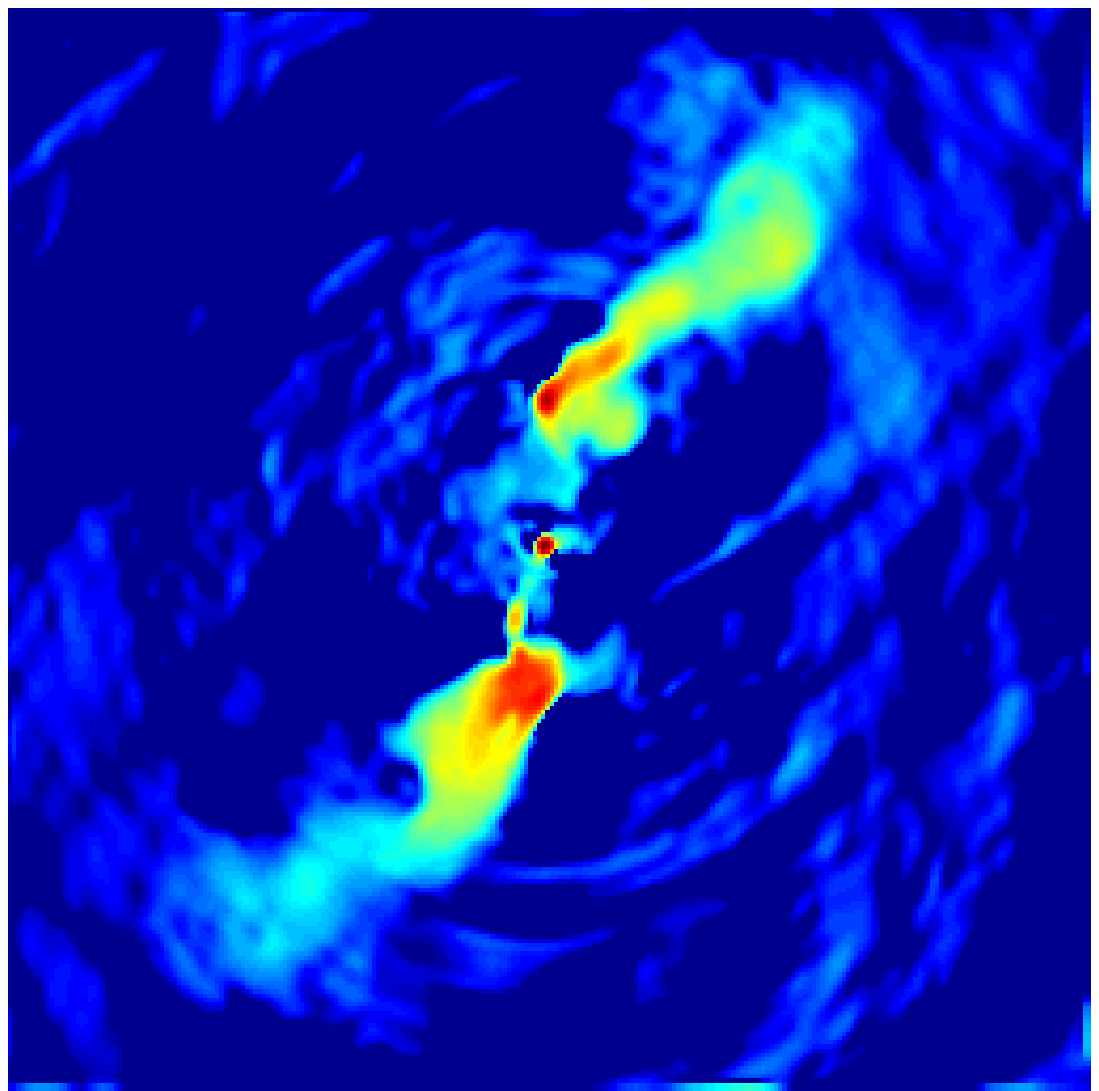} &
\hspace*{0.3cm}\includegraphics[trim ={0.2cm 0 0 0cm},clip,width=4cm,align=c]{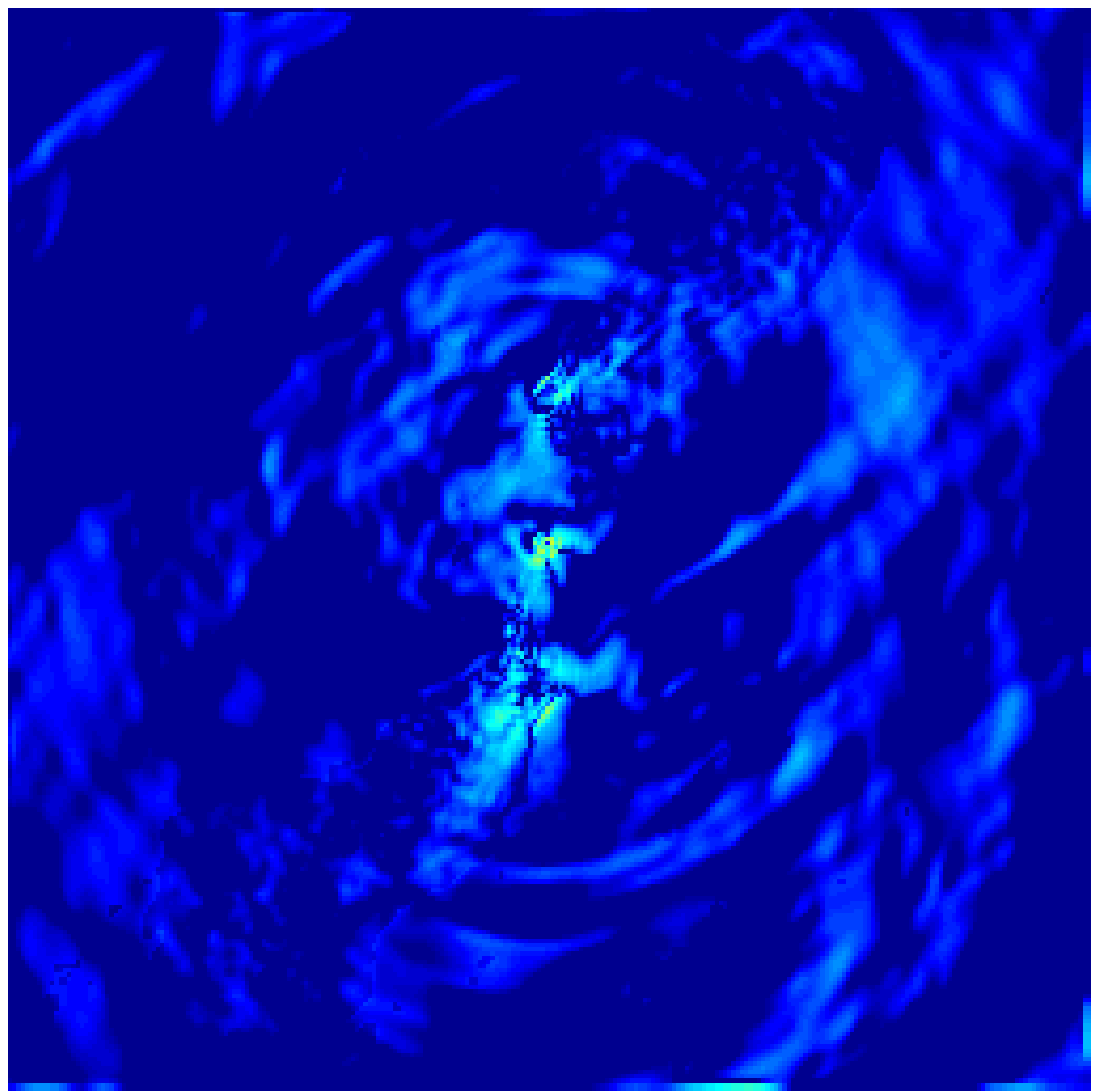} 
\vspace{0.25cm}
\\
%-----------------------------------------------
{\includegraphics[trim ={0.2cm 0 0 0cm},clip,width=4cm,align=c]{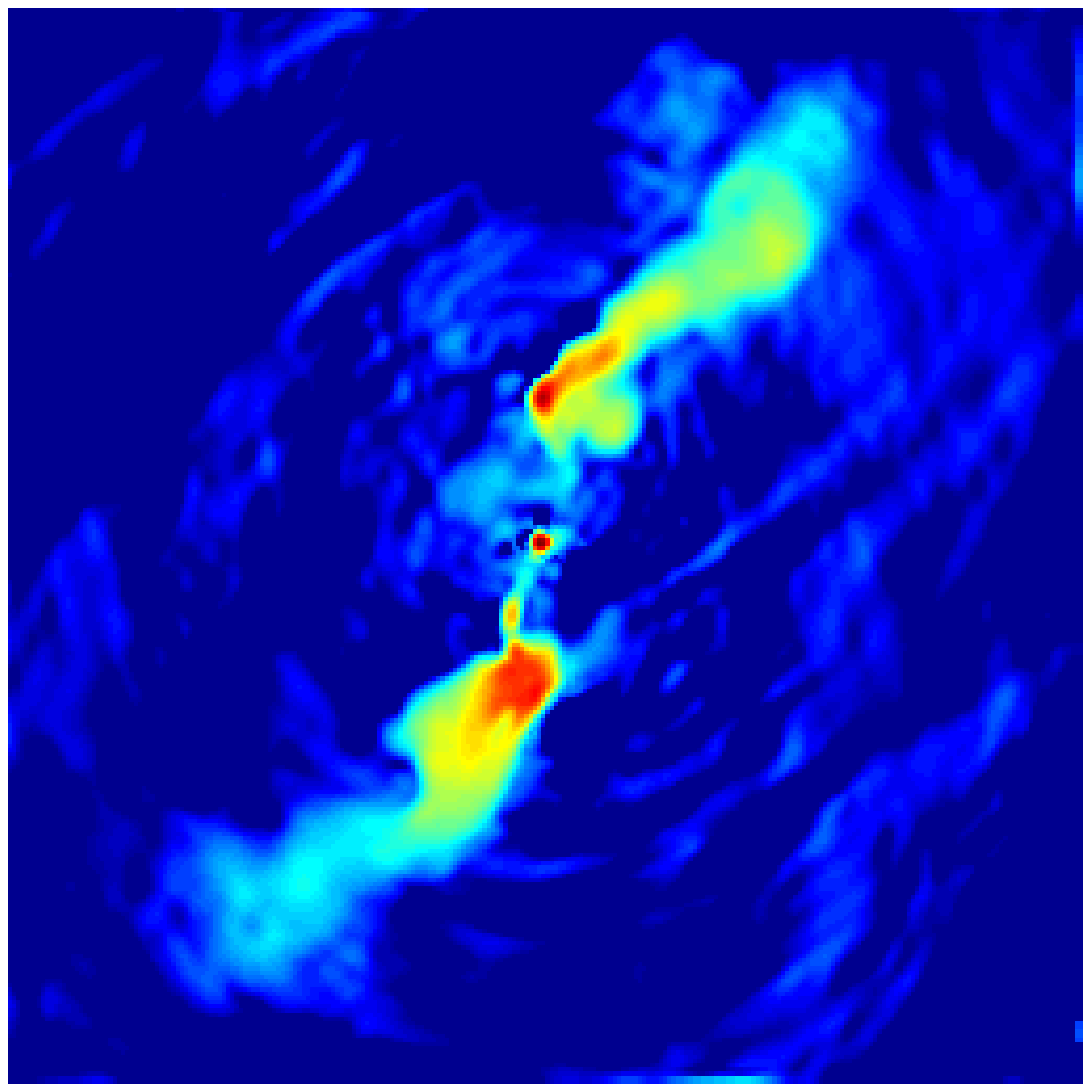}} &
\hspace*{0.3cm}\includegraphics[trim ={0.2cm 0 0 0cm},clip,width=4cm,align=c]{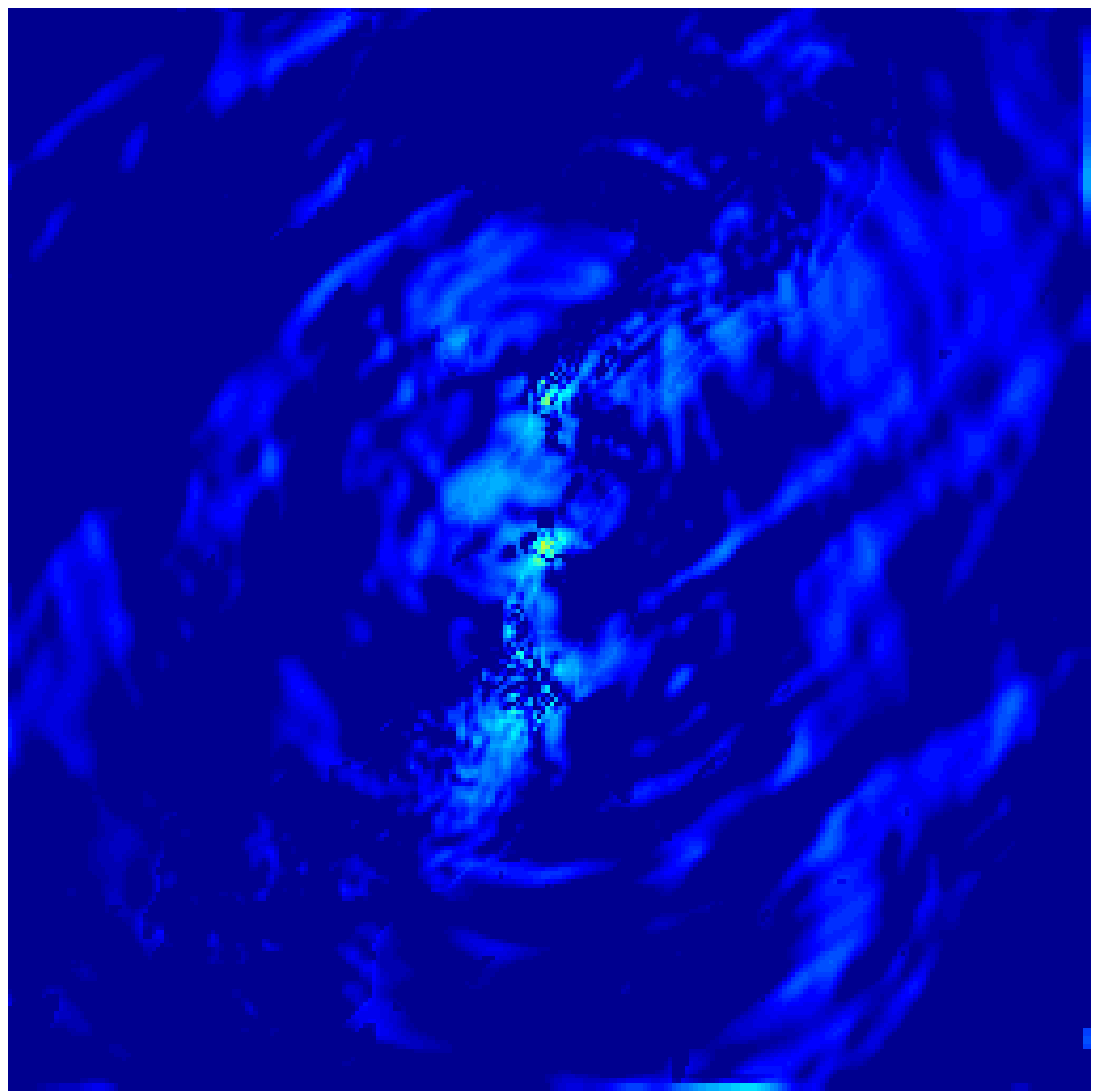} &
\hspace*{0.3cm}\includegraphics[trim ={0.2cm 0 0 0cm},clip,width=4cm,align=c]{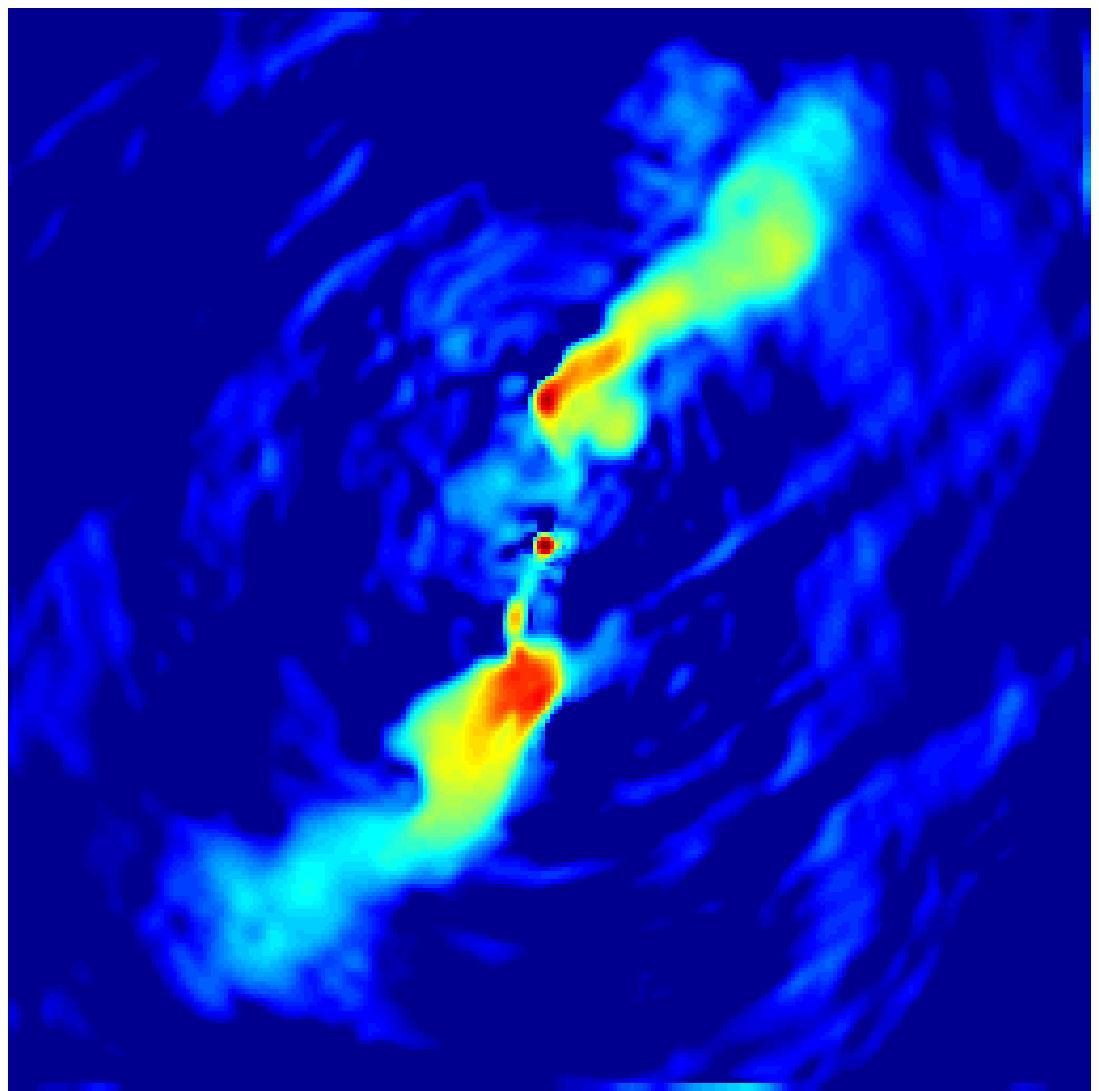} &
\hspace*{0.3cm}\includegraphics[trim ={0.2cm 0 0 0cm},clip,width=4cm,align=c]{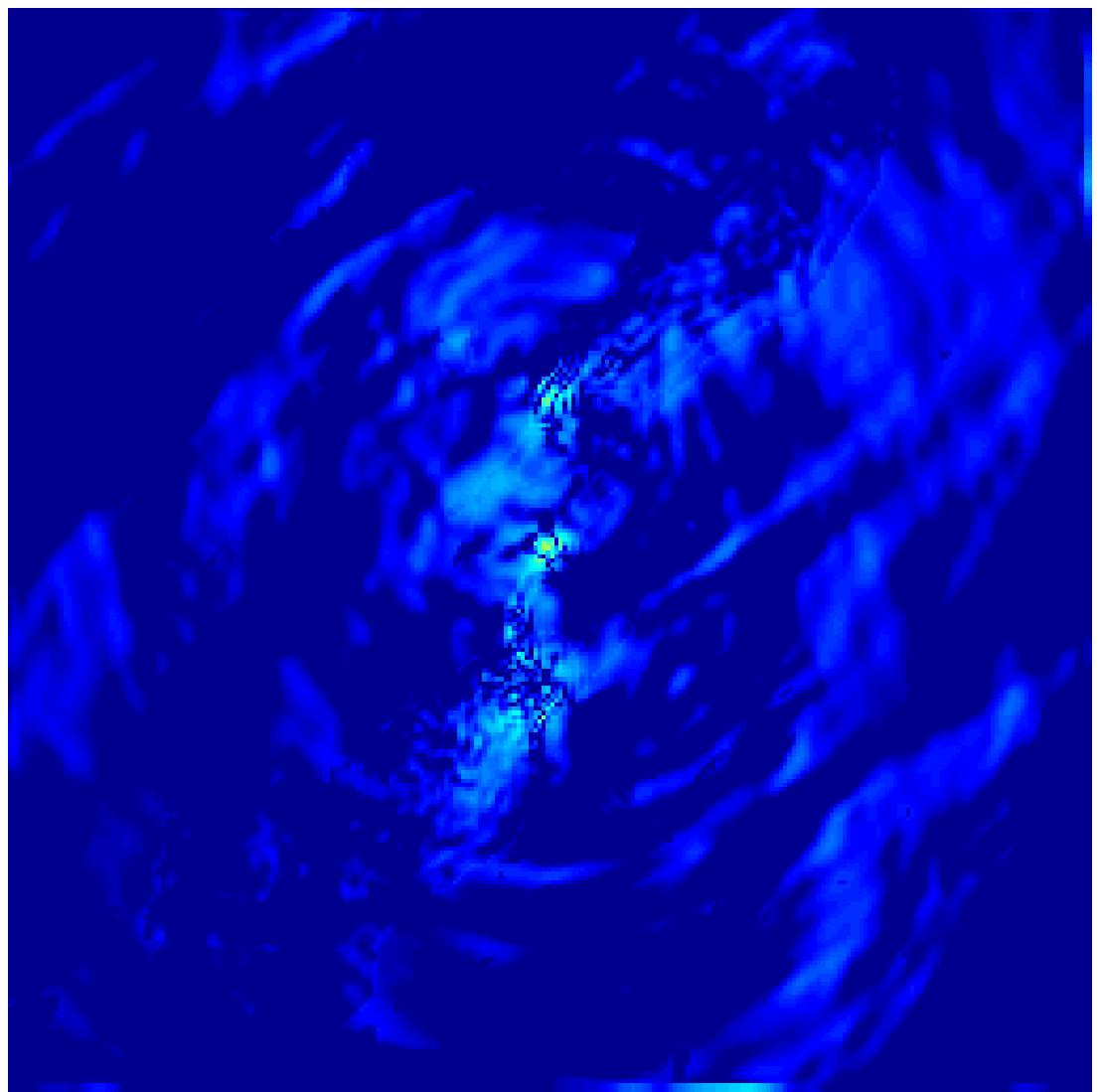} 
\vspace{0.25cm}
\\

%-----------------------------------------------
{\includegraphics[trim ={0.2cm 0 0 0cm},clip,width=4cm,align=c]{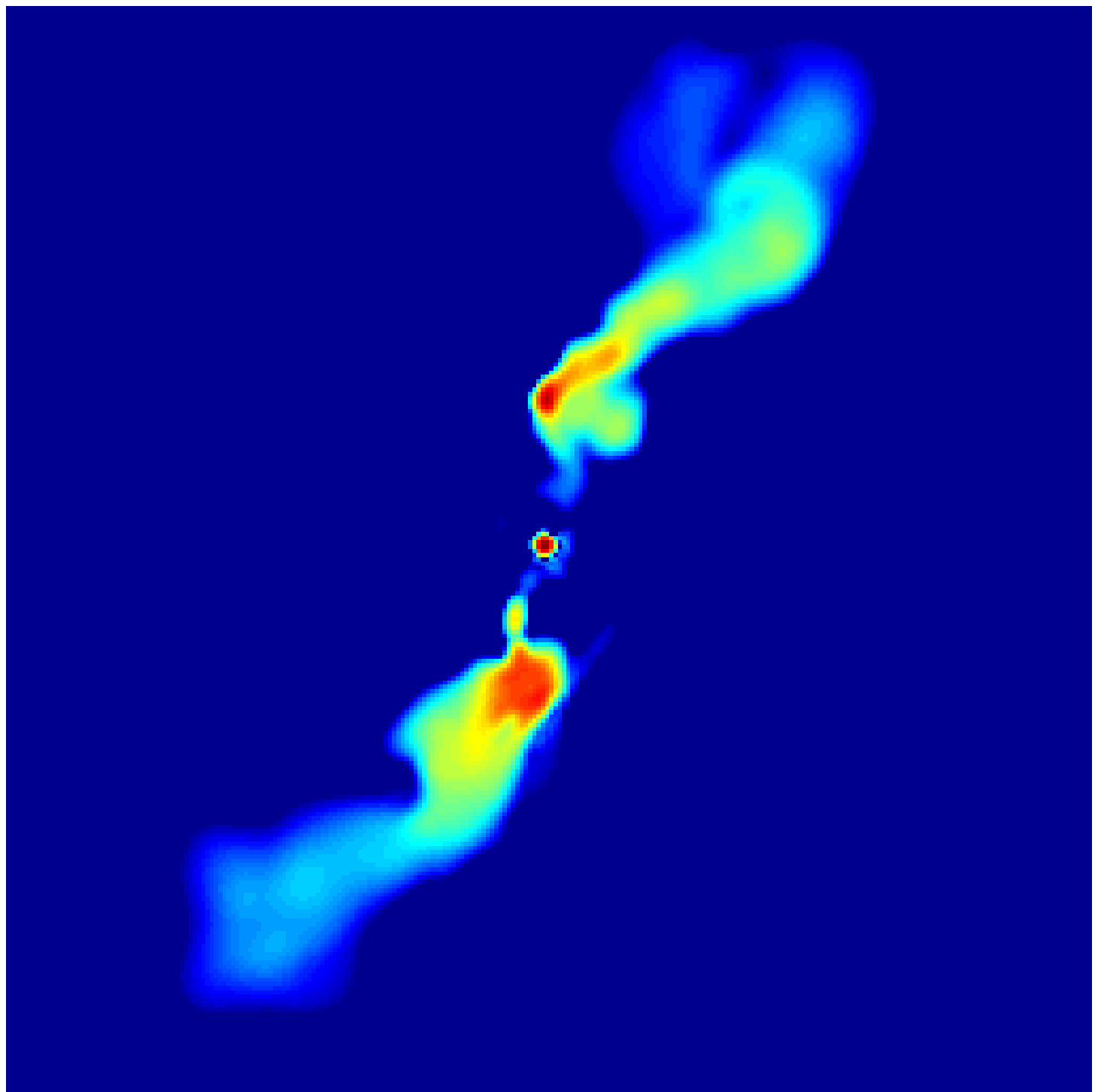}} &
\hspace*{0.3cm}\includegraphics[trim ={0.2cm 0 0 0cm},clip,width=4cm,align=c]{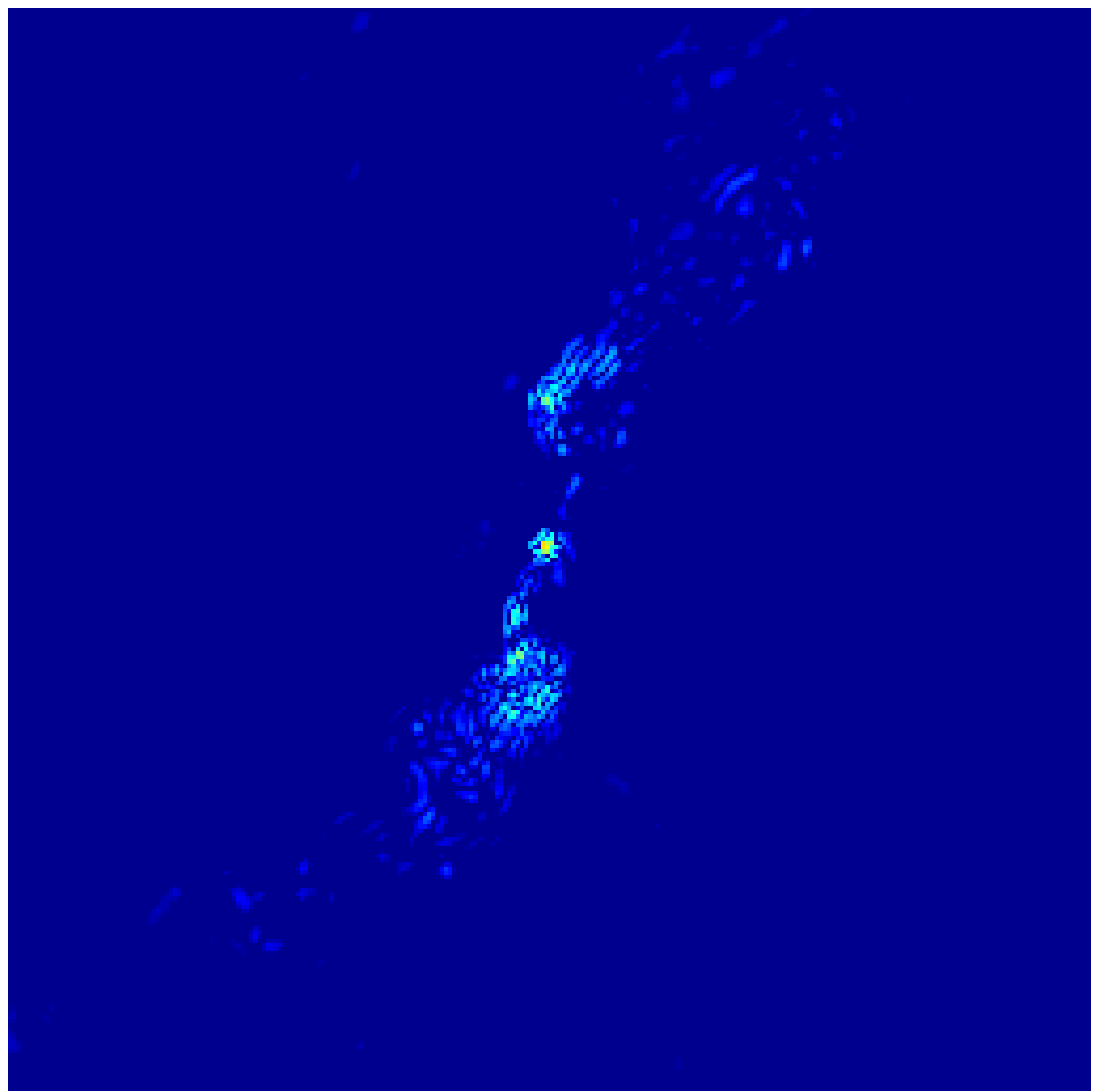} &
\hspace*{0.3cm}\includegraphics[trim ={0.2cm 0 0 0cm},clip,width=4cm,align=c]{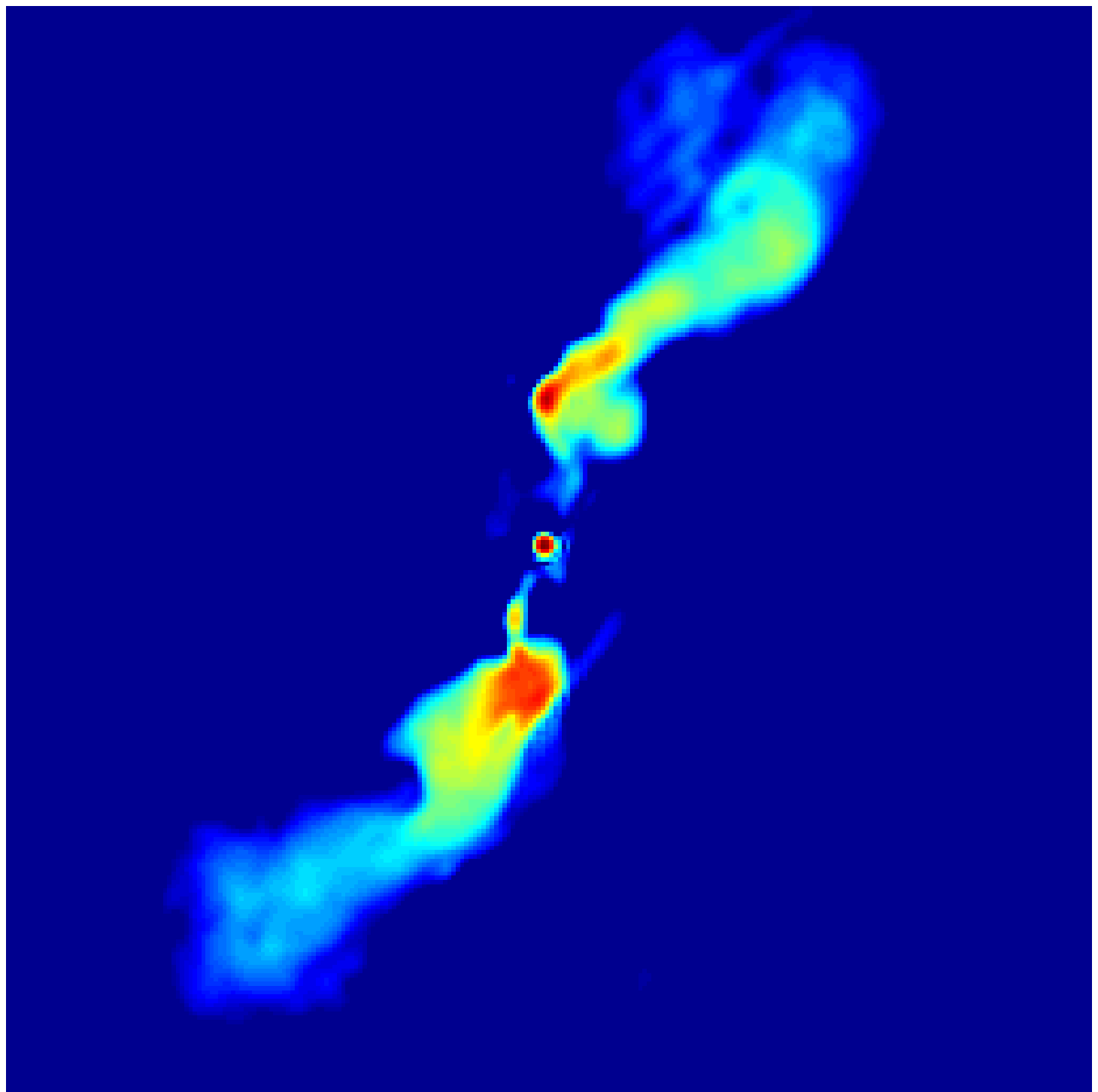} &
\hspace*{0.3cm}\includegraphics[trim ={0.2cm 0 0 0cm},clip,width=4cm,align=c]{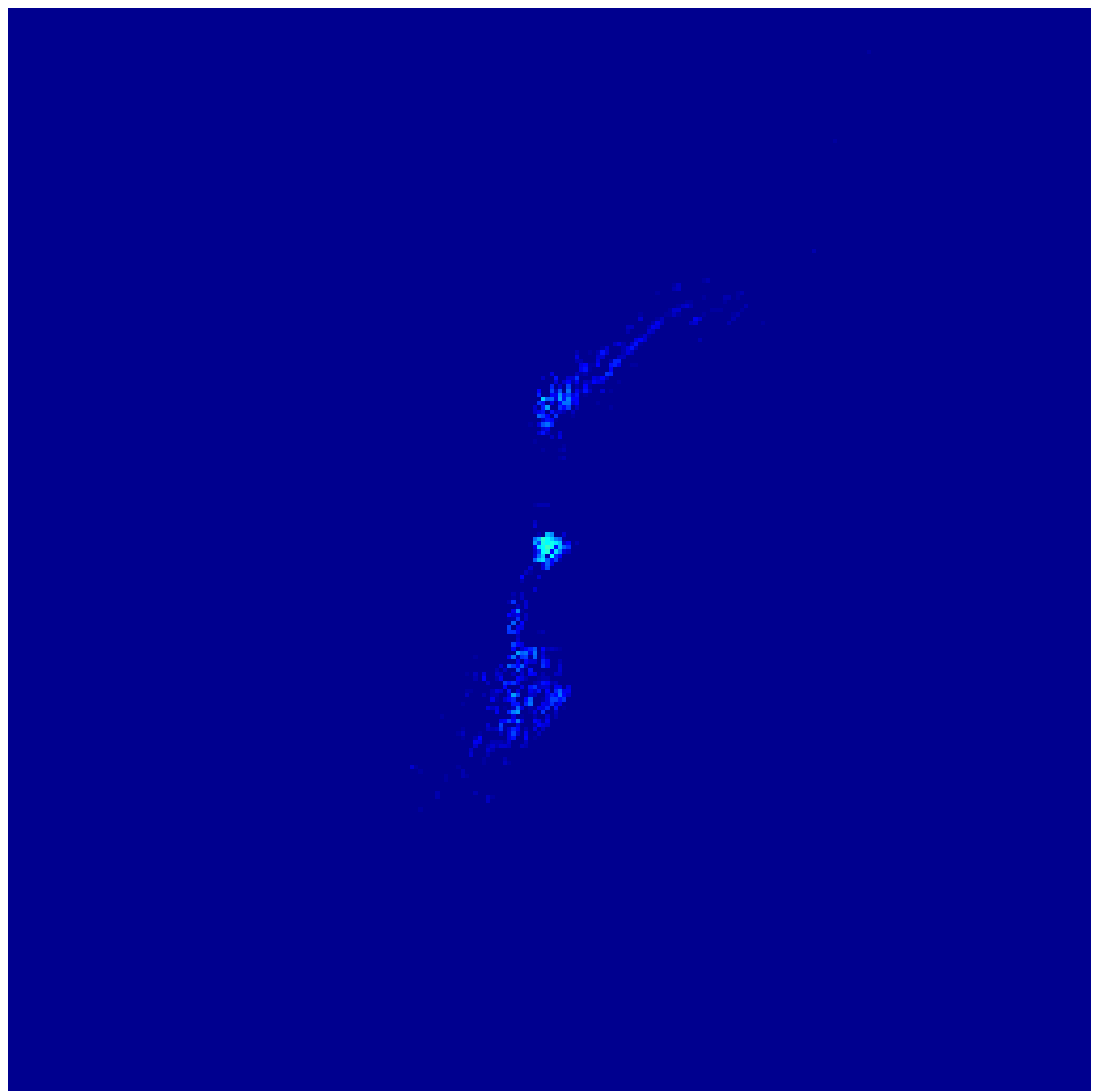} 
\vspace{0.25cm}
\\
%-----------------------------------------------
\includegraphics[trim ={0.2cm 0 0 0cm},clip,width=4cm,align=c]{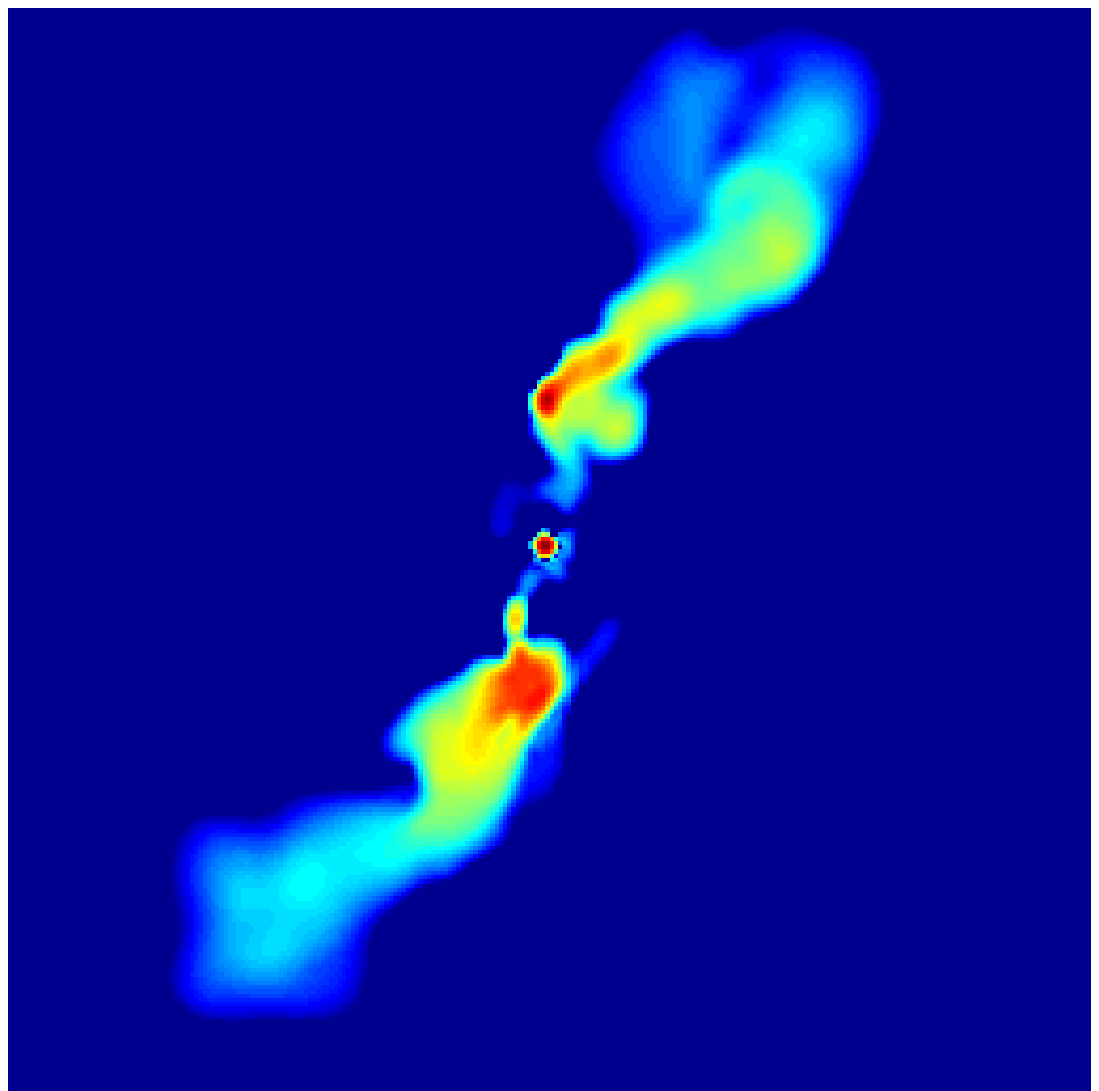} &
\hspace*{0.3cm}\includegraphics[trim ={0.2cm 0 0 0cm},clip,width=4cm,align=c]{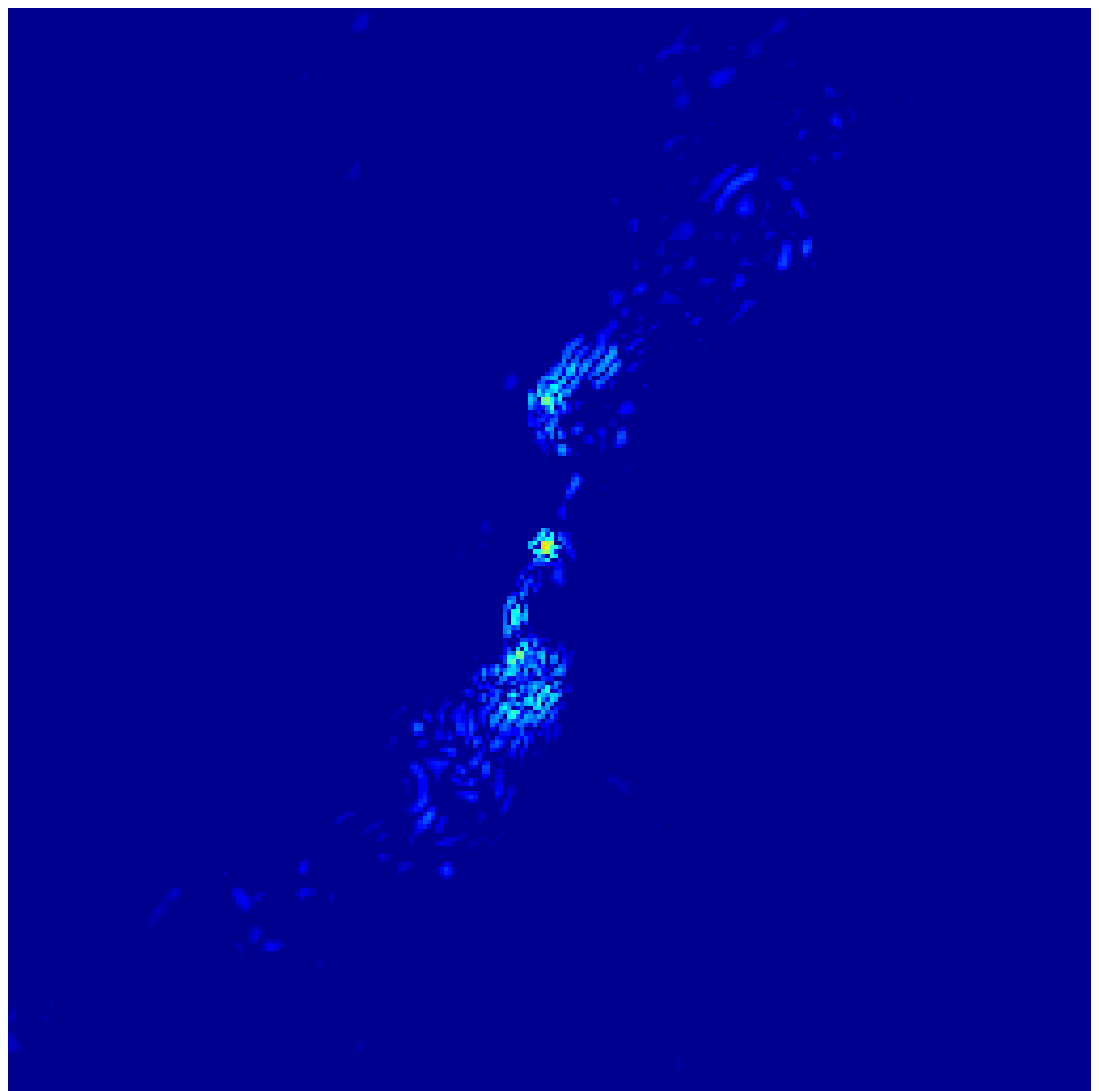} &
\hspace*{0.3cm}\includegraphics[trim ={0.2cm 0 0 0cm},clip,width=4cm,align=c]{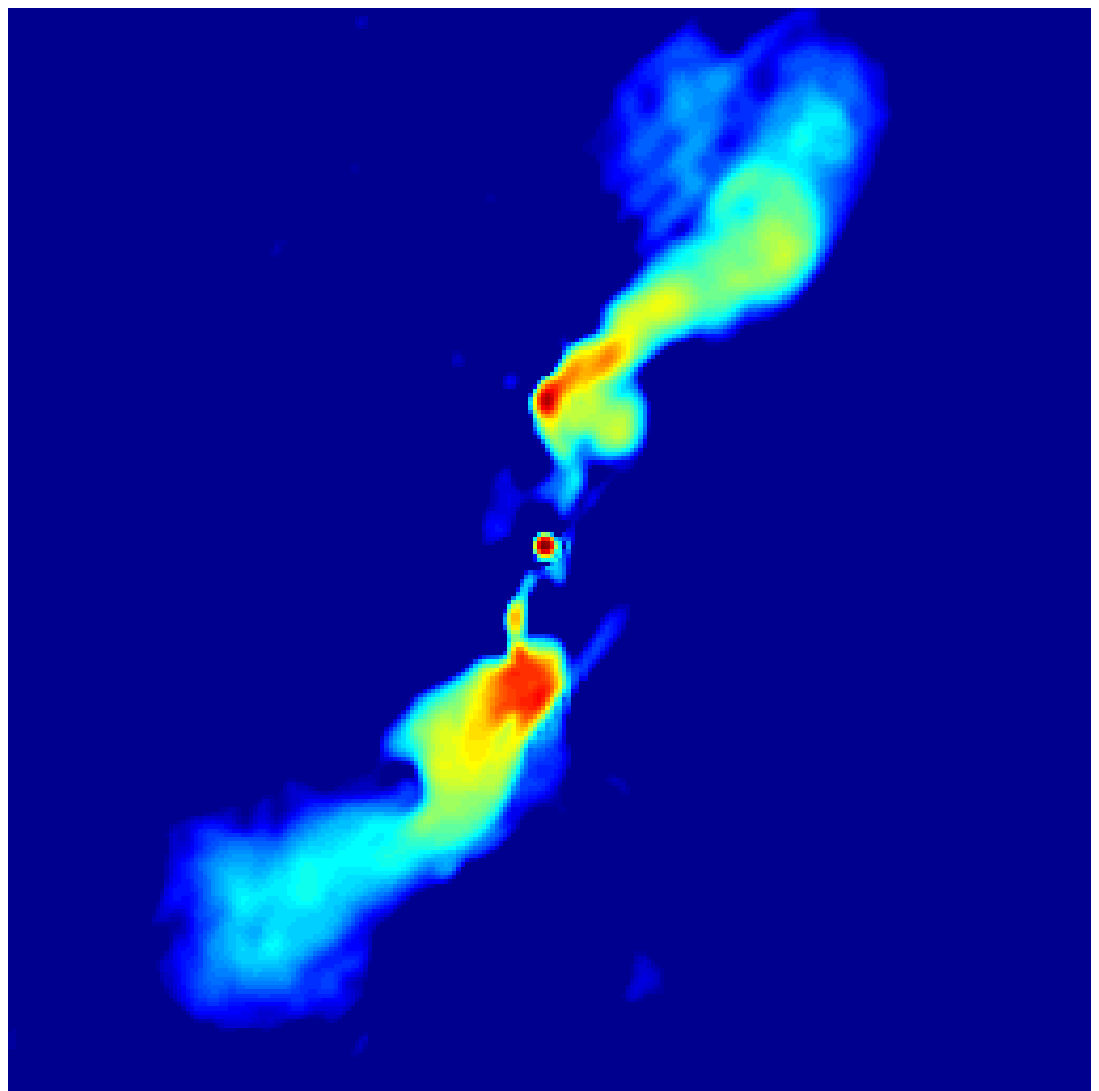} &
\hspace*{0.3cm}\includegraphics[trim ={0.2cm 0 0 0cm},clip,width=4cm,align=c]{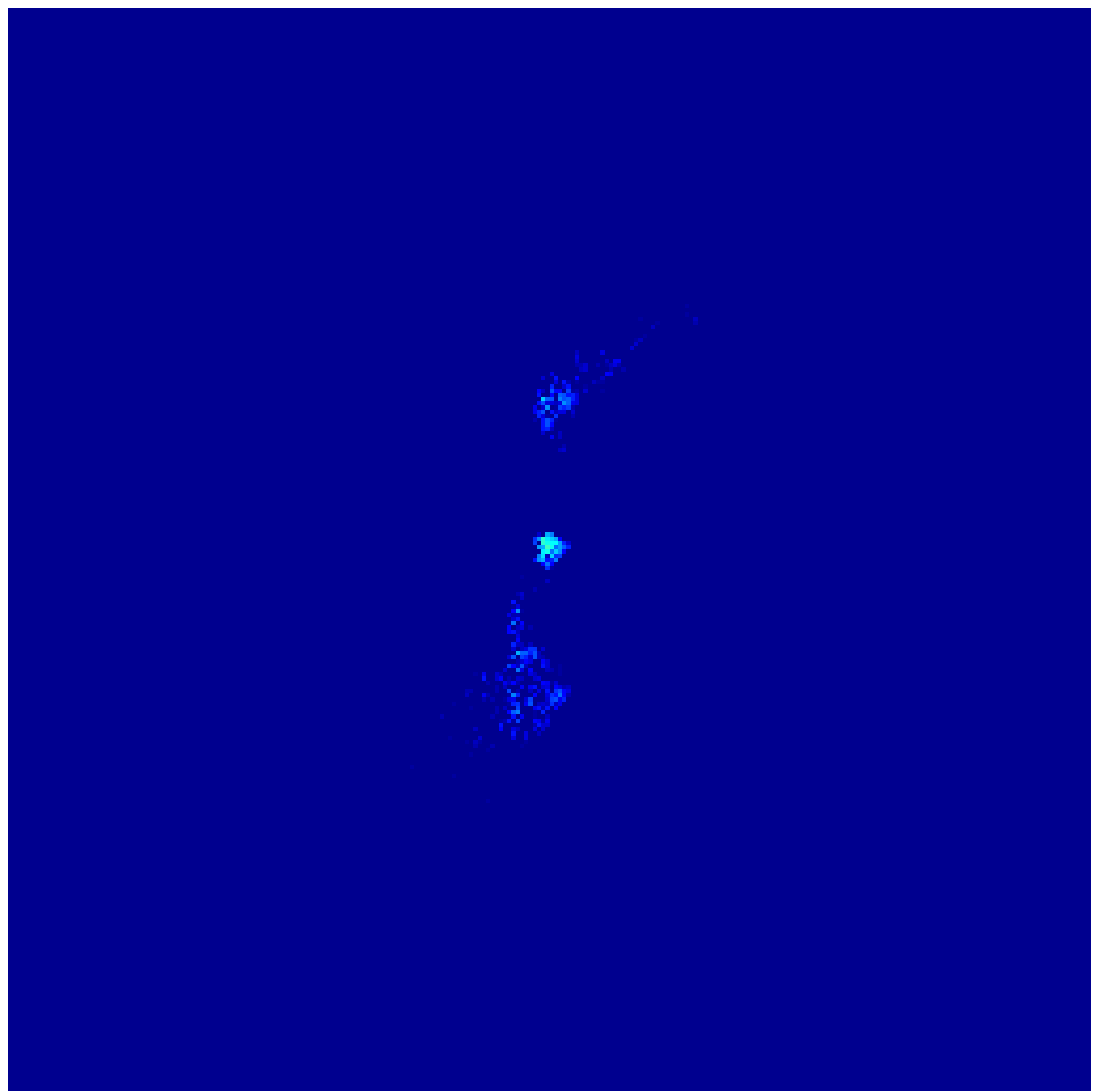} 
\end{tabular}
\caption{Hydra A Stokes $I$ true image in first row and reconstructed images (best ones over 5 performed simulations for each case) in other rows for the cases: Imaging with normalized DIEs (second row), {Joint DIE calibration and imaging (third row)}, Joint DDE calibration and imaging excluding the off-diagonal terms (fourth row), and considering full Jones matrix (fifth row). In each case, column-wise recovered images followed by their corresponding error images are displayed when imaging is performed without polarization constraint (first two columns) and with polarization constraint (last two columns). All the images are shown in log scale, with the same color range corresponding to the colorbar given in first row.}
\label{fig:hydra5_I}
\end{figure*}

\begin{figure*}
\centering
\begin{tabular}{@{}c@{}c@{}c@{}c@{}}
\includegraphics[trim ={0.2cm 0 0 0cm},clip,width=4cm]{true_Q_hydra5.eps} &
\hspace{-3cm}\includegraphics[trim ={16.2cm 0 0 0cm},clip,width=0.97cm]{Q_hydra_colorbar.eps}
& & \\
Recovered images w/o &  Absolute error images &  Recovered images with & Absolute error images \\
polarization constraint & &polarization constraint & \\
\includegraphics[trim ={0.2cm 0 0 0cm},clip,width=4cm,align=c]{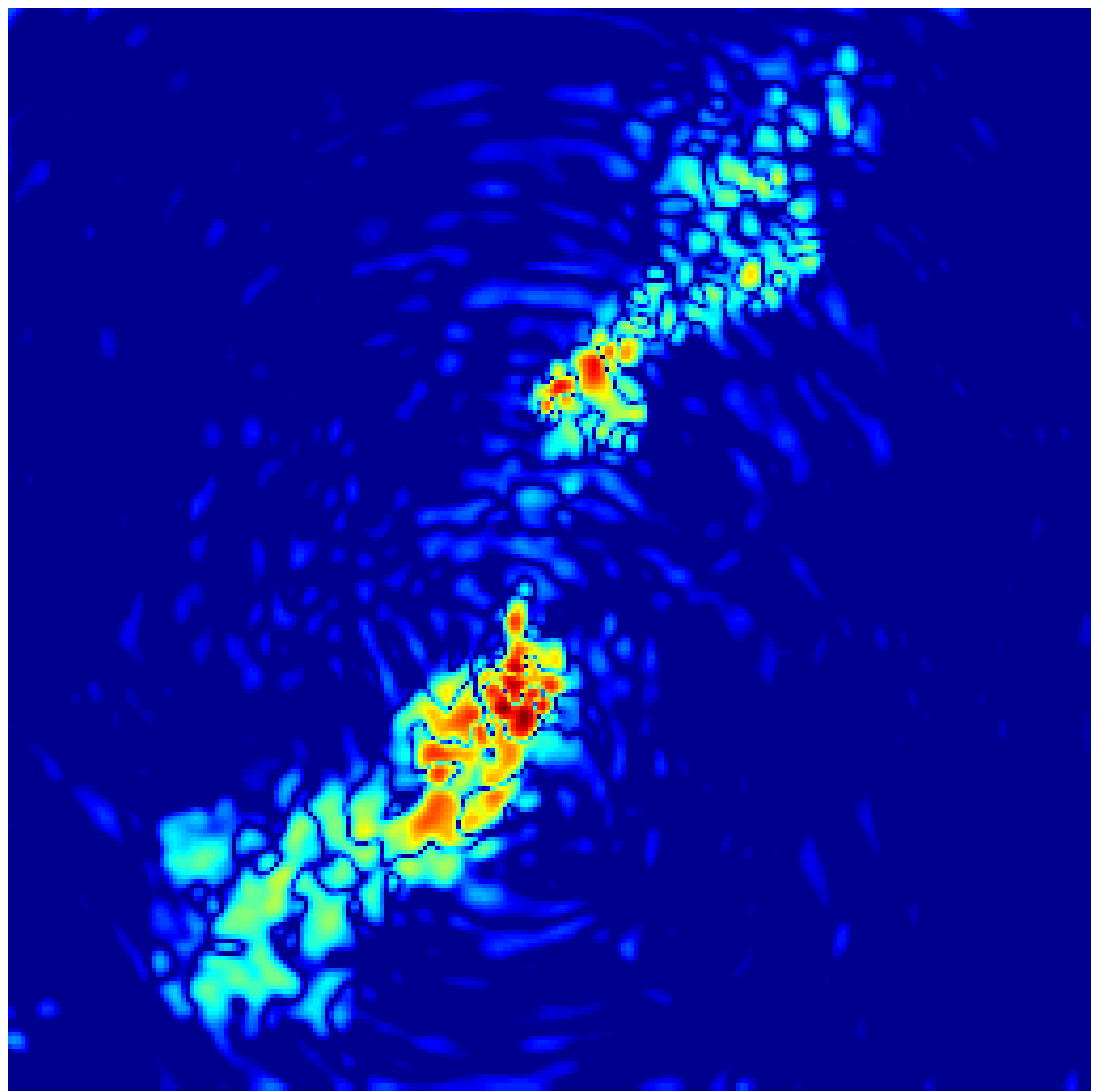} &
\hspace*{0.3cm}\includegraphics[trim ={0.2cm 0 0 0cm},clip,width=4cm,align=c]{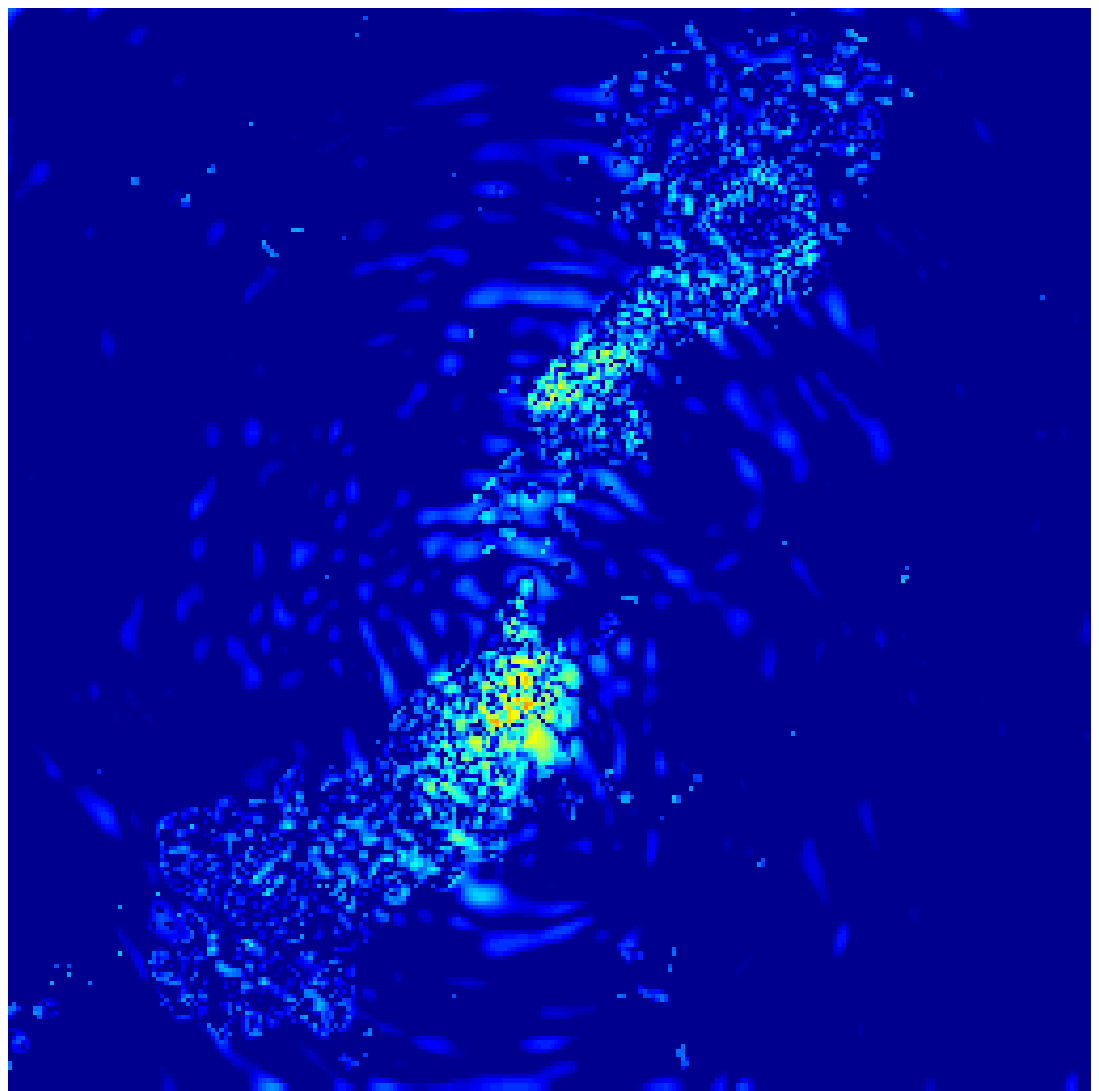} &
\hspace*{0.3cm}\includegraphics[trim ={0.2cm 0 0 0cm},clip,width=4cm,align=c]{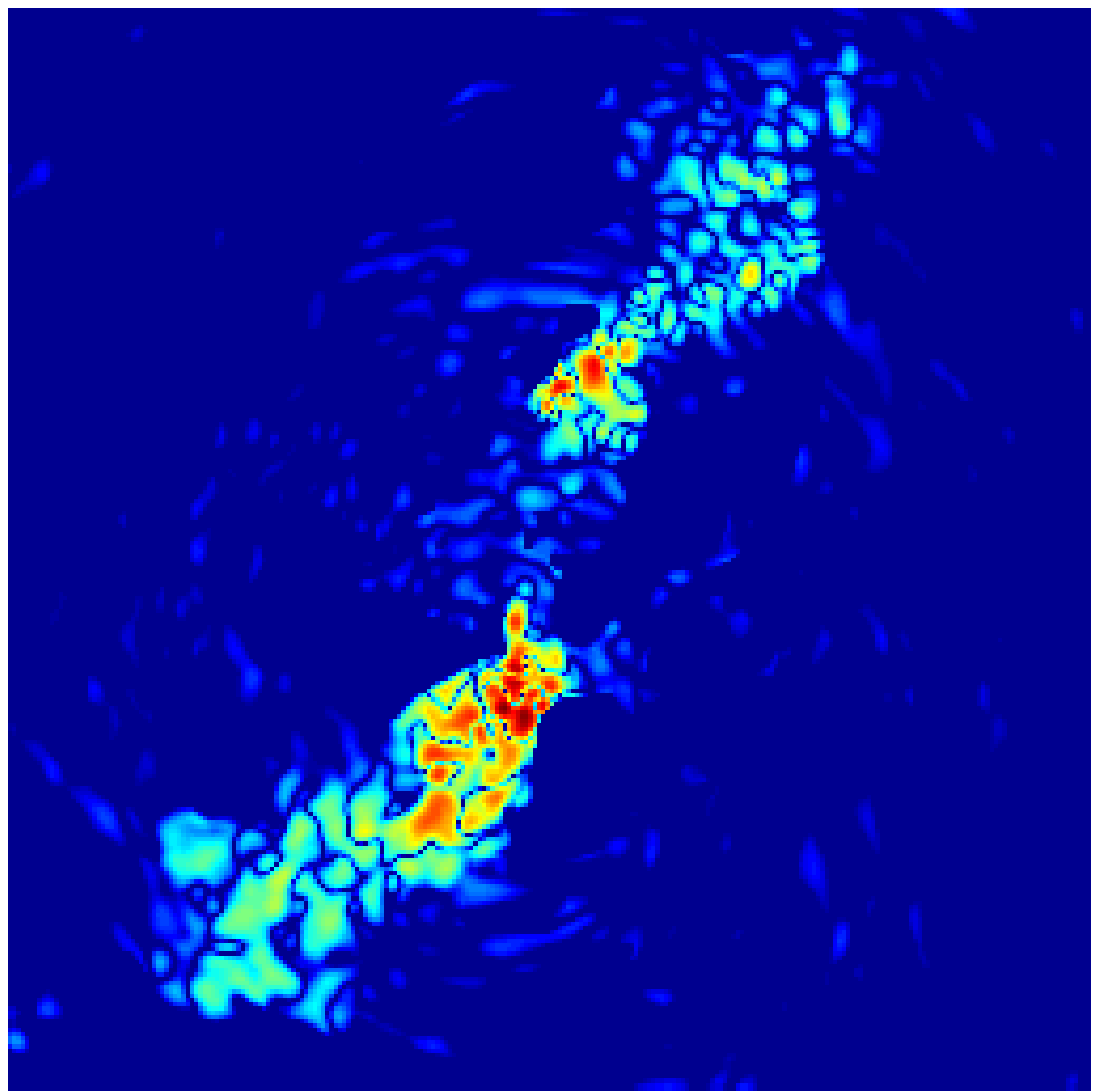} &
\hspace*{0.3cm}\includegraphics[trim ={0.2cm 0 0 0cm},clip,width=4cm,align=c]{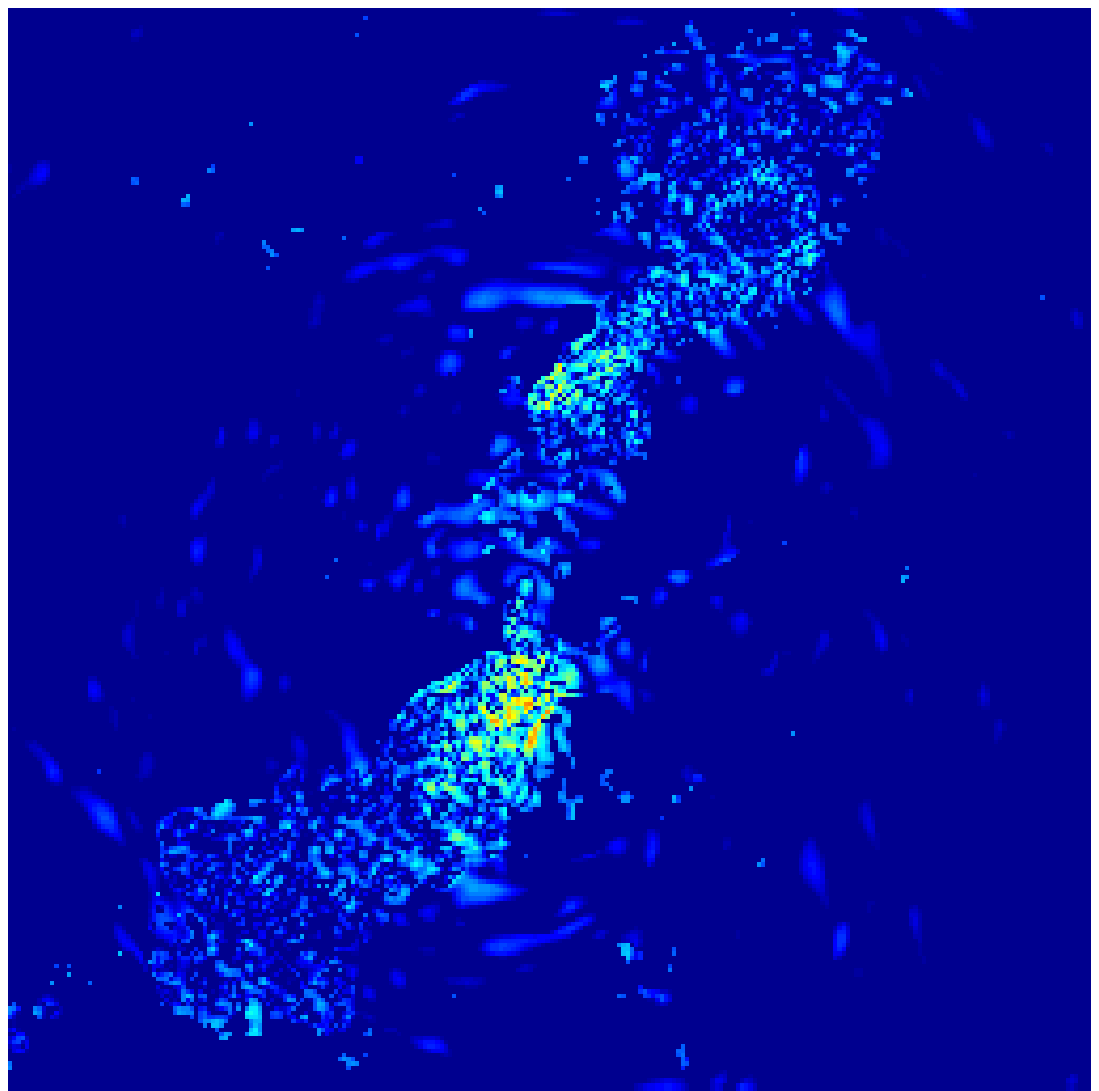} 
\vspace{0.25cm}
\\
%-----------------------------------------------
{\includegraphics[trim ={0.2cm 0 0 0cm},clip,width=4cm,align=c]{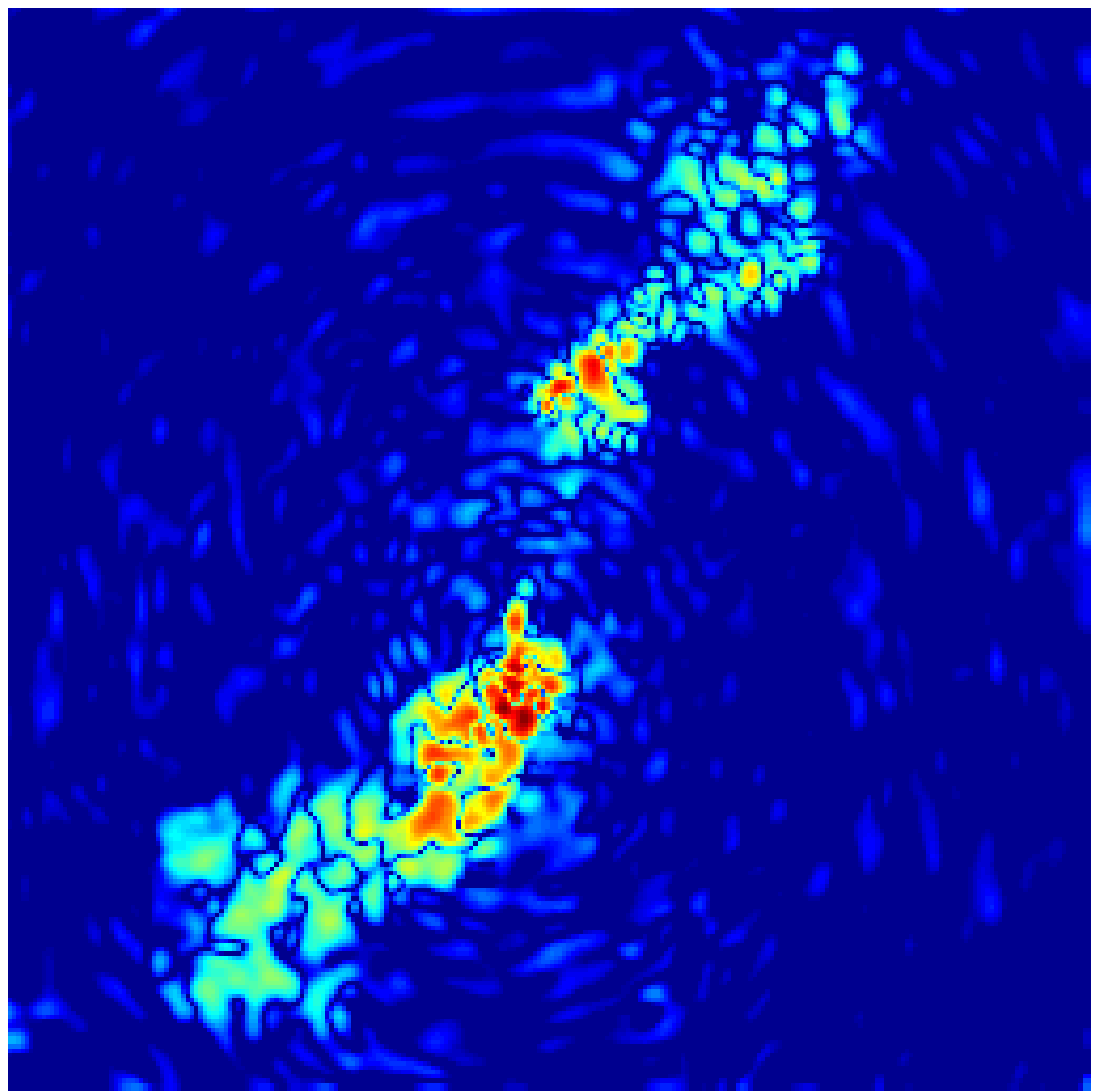}} &
\hspace*{0.3cm}\includegraphics[trim ={0.2cm 0 0 0cm},clip,width=4cm,align=c]{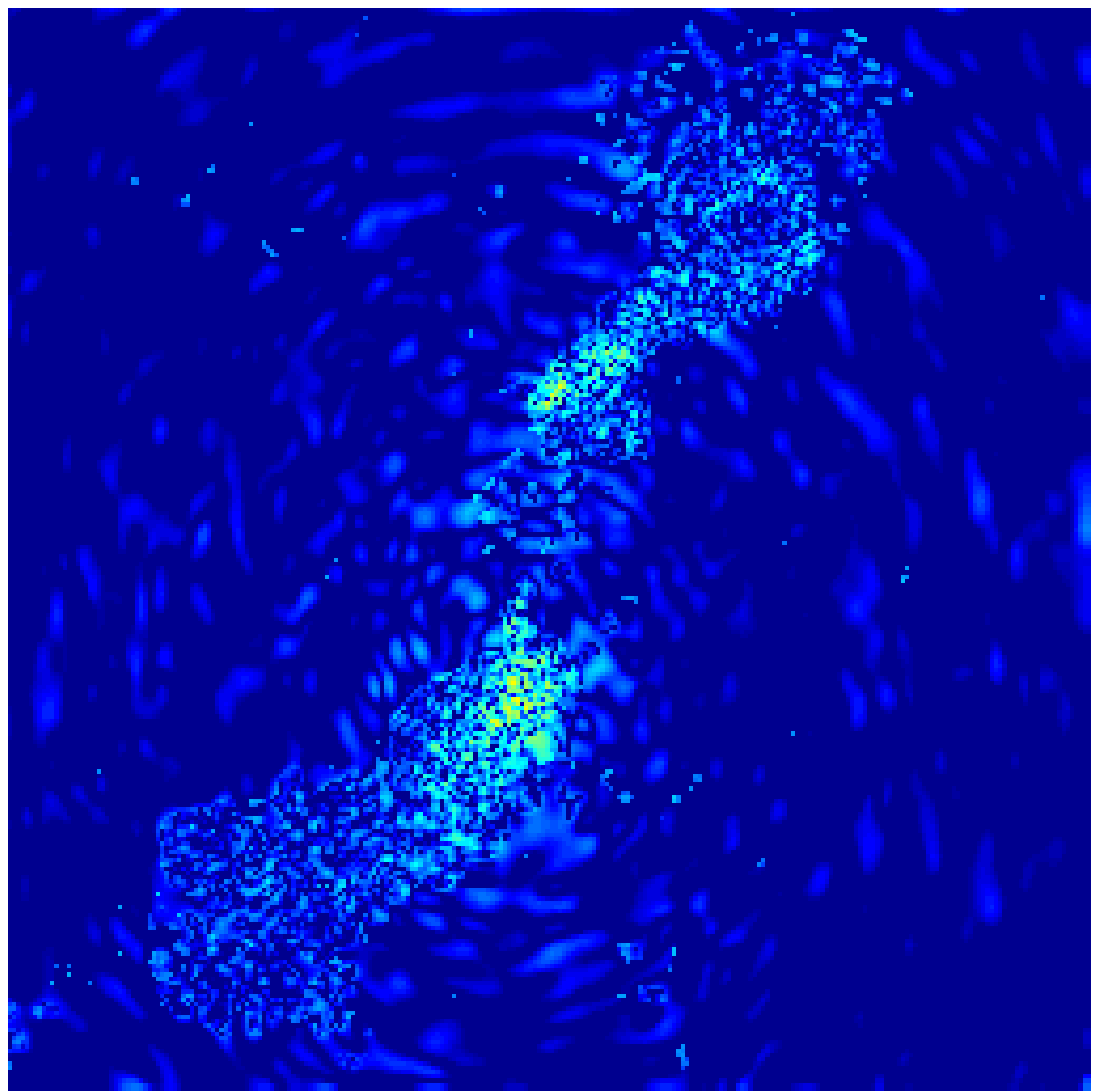} &
\hspace*{0.3cm}\includegraphics[trim ={0.2cm 0 0 0cm},clip,width=4cm,align=c]{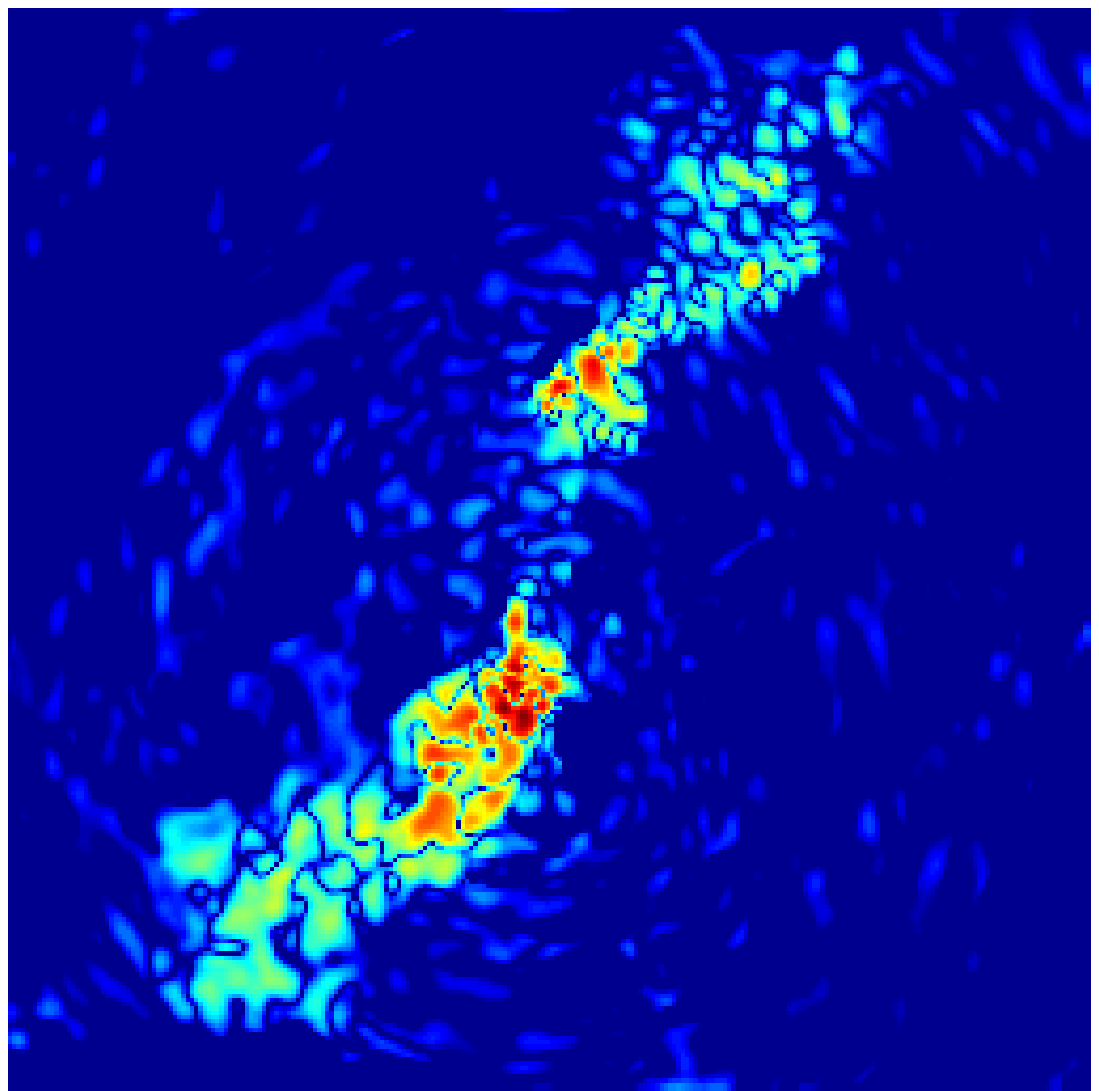} &
\hspace*{0.3cm}\includegraphics[trim ={0.2cm 0 0 0cm},clip,width=4cm,align=c]{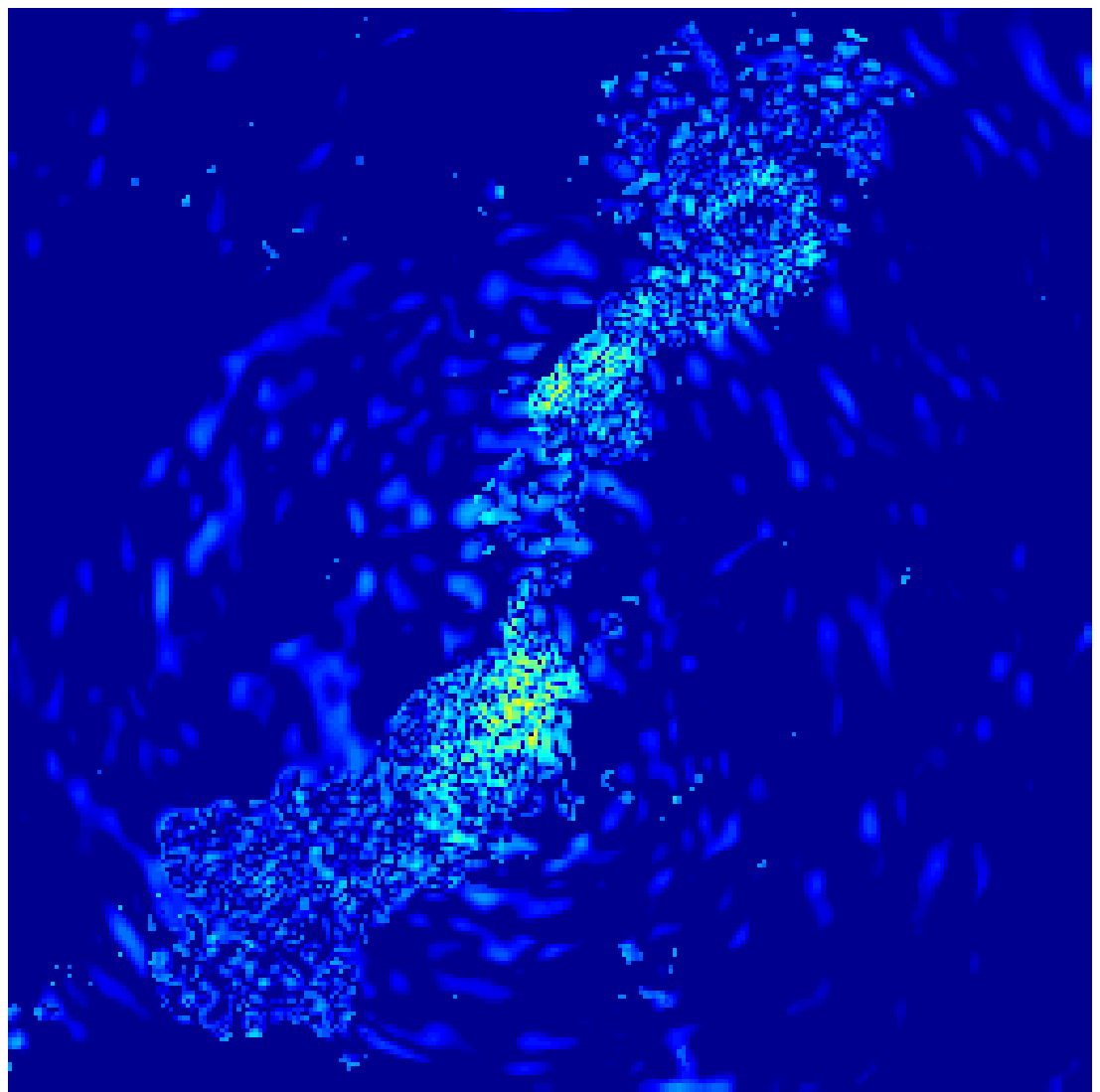} 
\vspace{0.25cm}
\\
%-----------------------------------------------
{\includegraphics[trim ={0.2cm 0 0 0cm},clip,width=4cm,align=c]{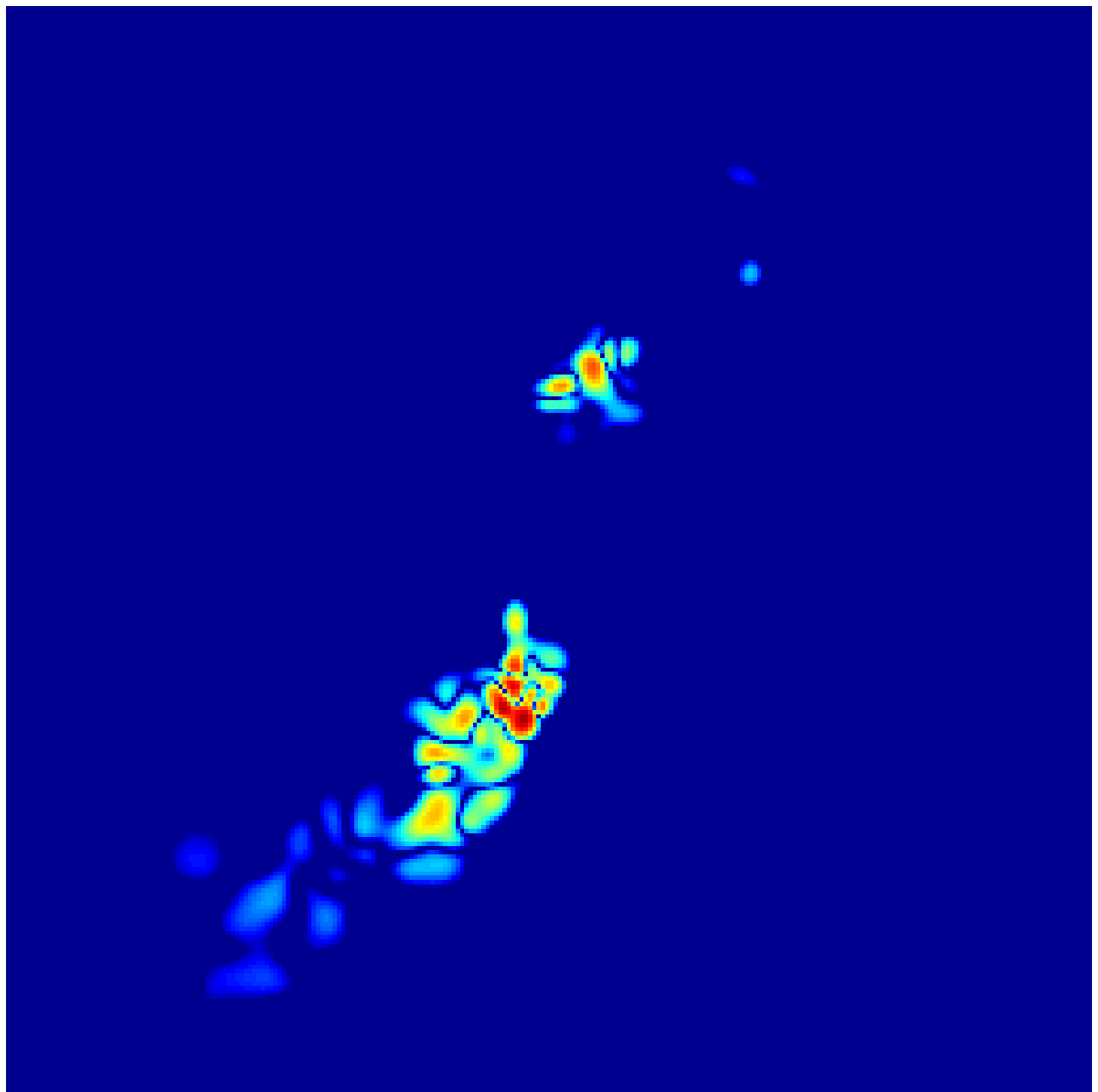}} &
\hspace*{0.3cm}\includegraphics[trim ={0.2cm 0 0 0cm},clip,width=4cm,align=c]{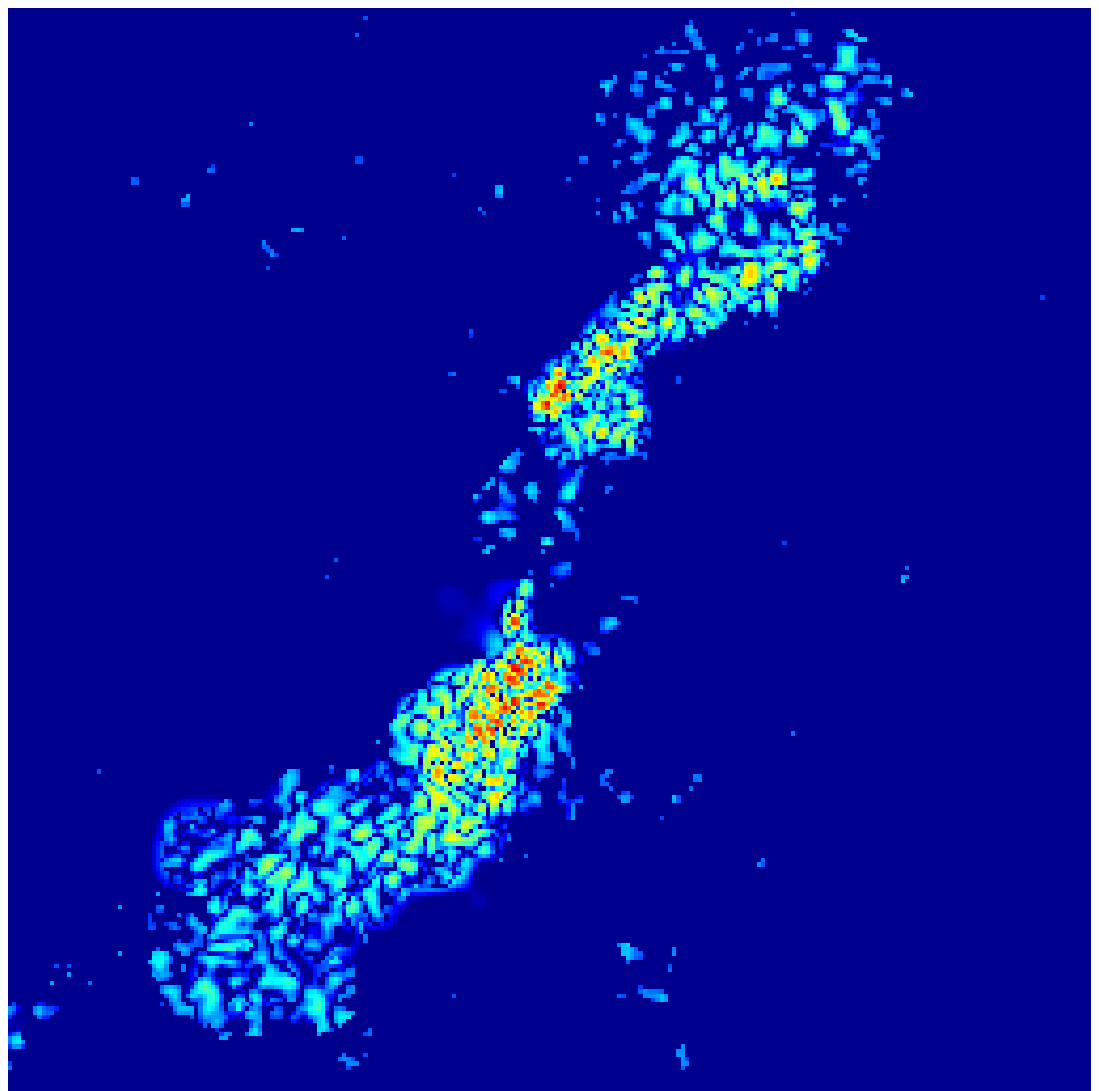} &
\hspace*{0.3cm}\includegraphics[trim ={0.2cm 0 0 0cm},clip,width=4cm,align=c]{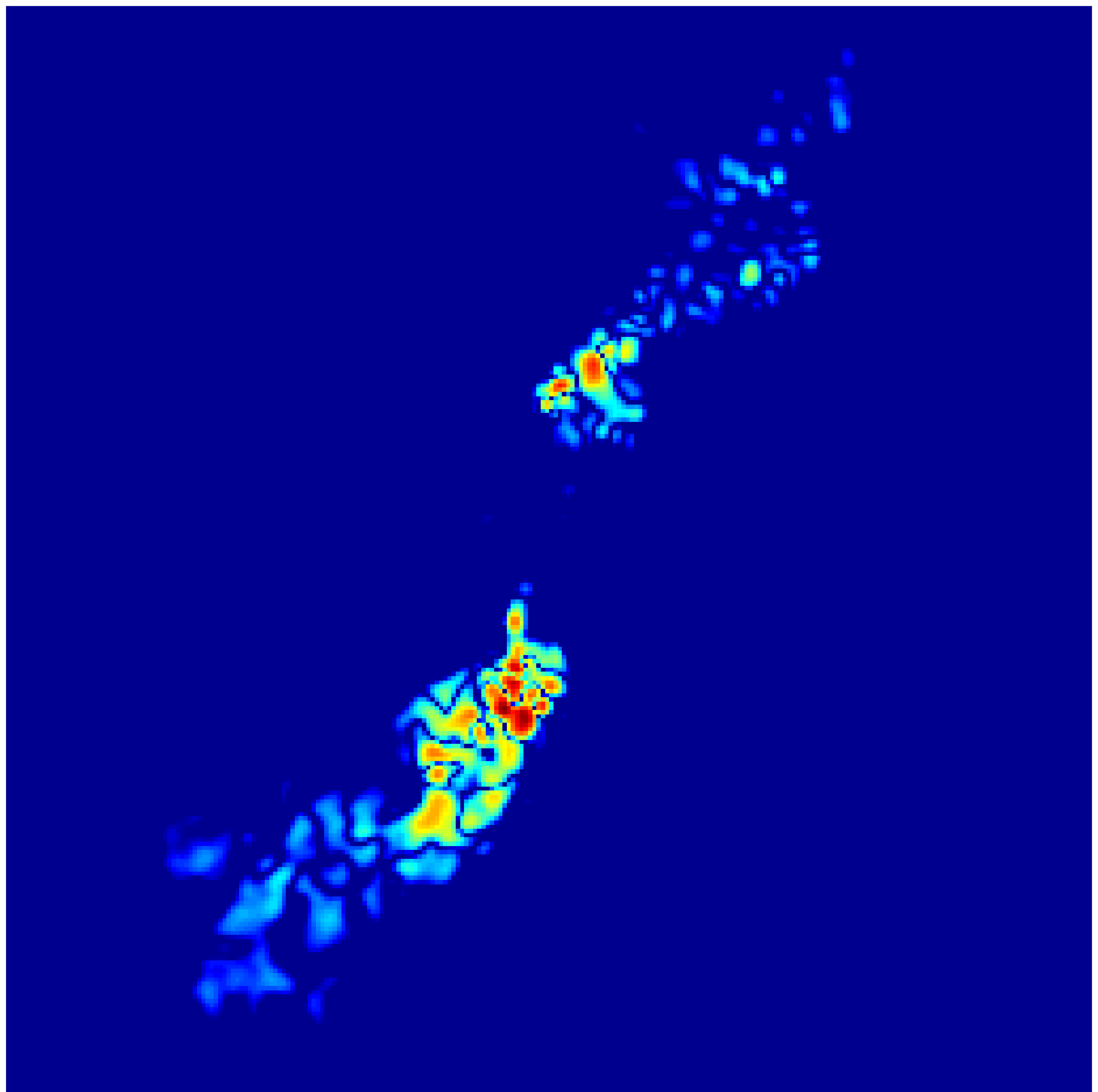} &
\hspace*{0.3cm}\includegraphics[trim ={0.2cm 0 0 0cm},clip,width=4cm,align=c]{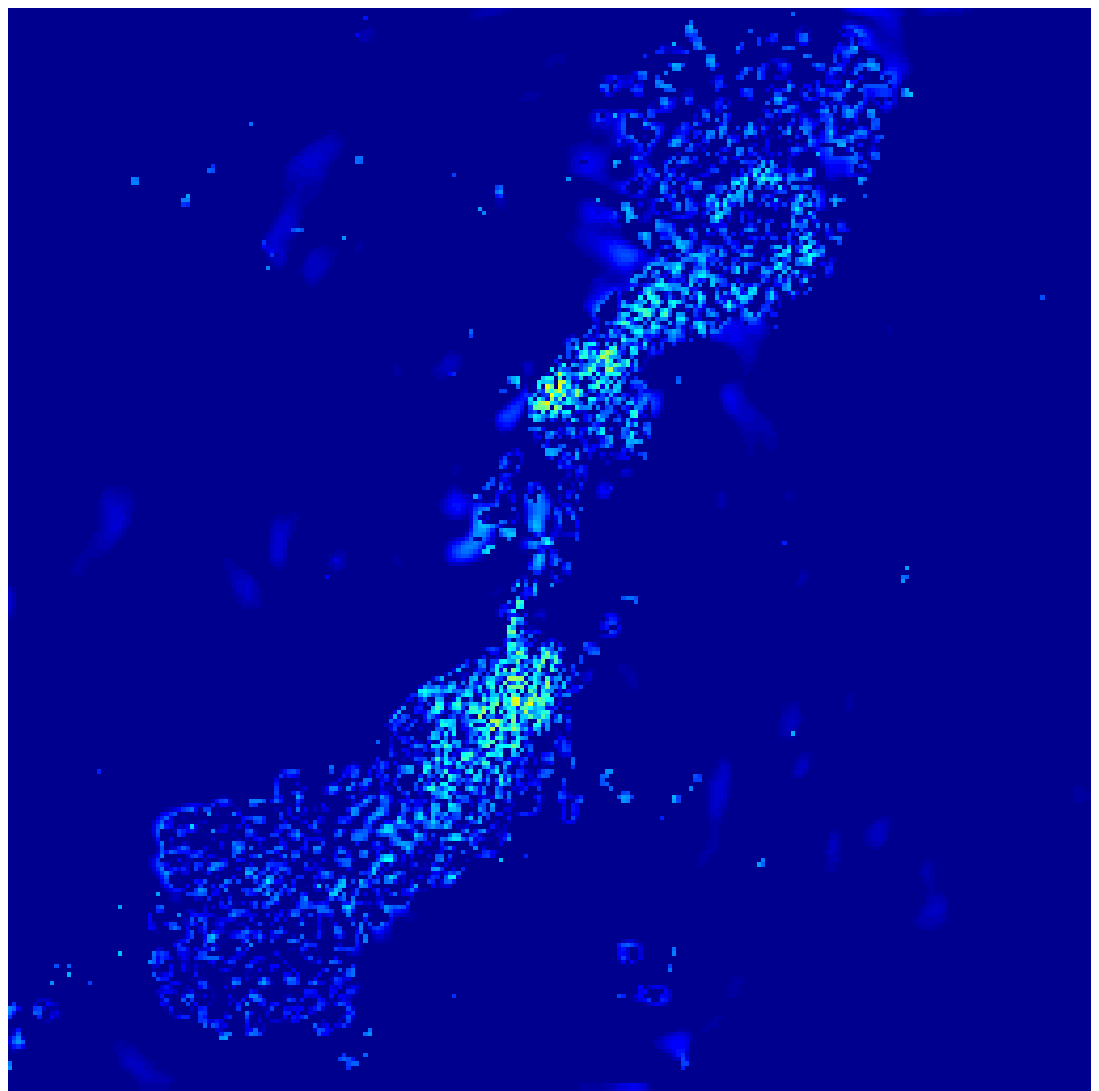} 
\vspace{0.25cm}
\\
%-----------------------------------------------
\includegraphics[trim ={0.2cm 0 0 0cm},clip,width=4cm,align=c]{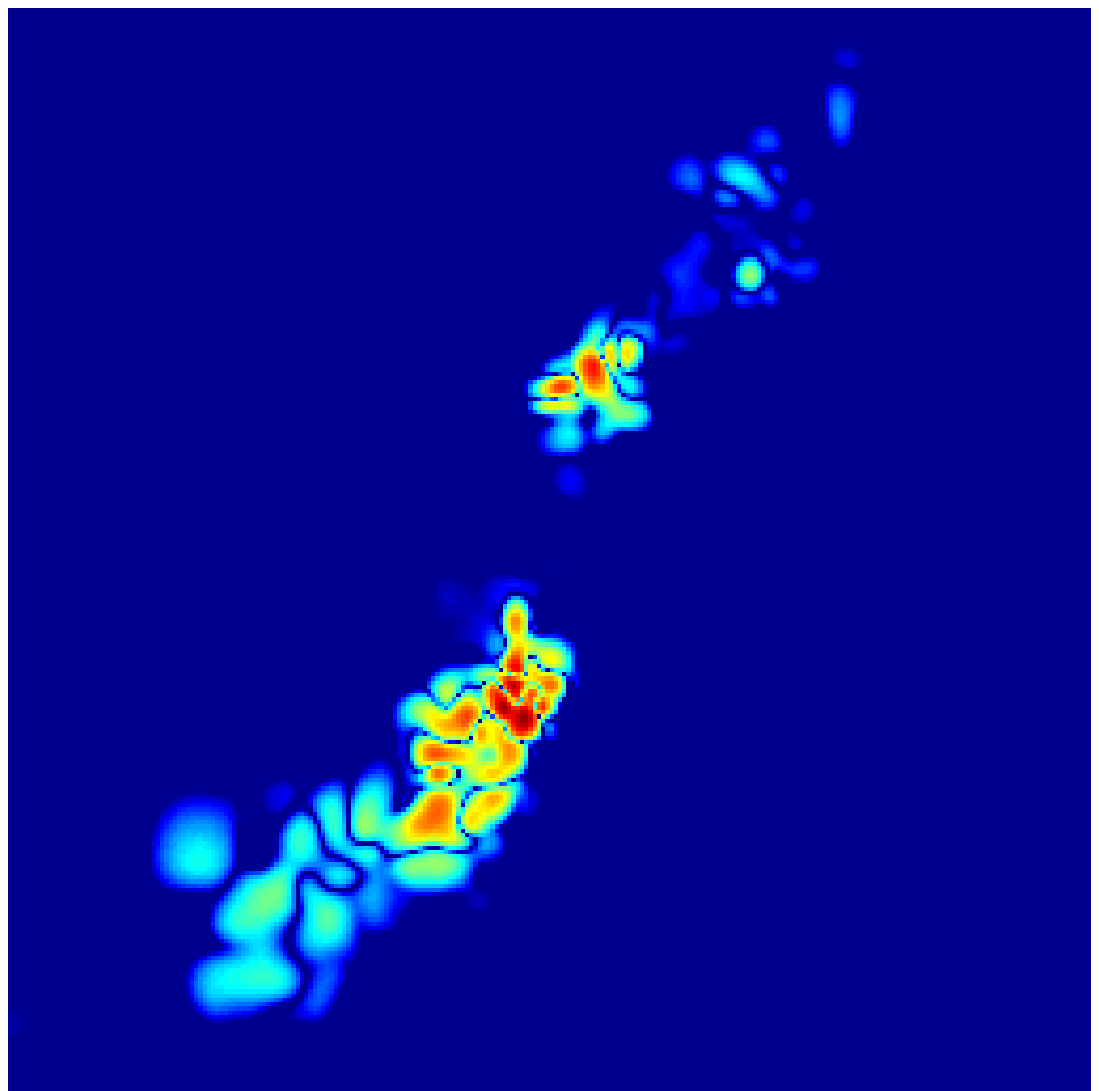} &
\hspace*{0.3cm}\includegraphics[trim ={0.2cm 0 0 0cm},clip,width=4cm,align=c]{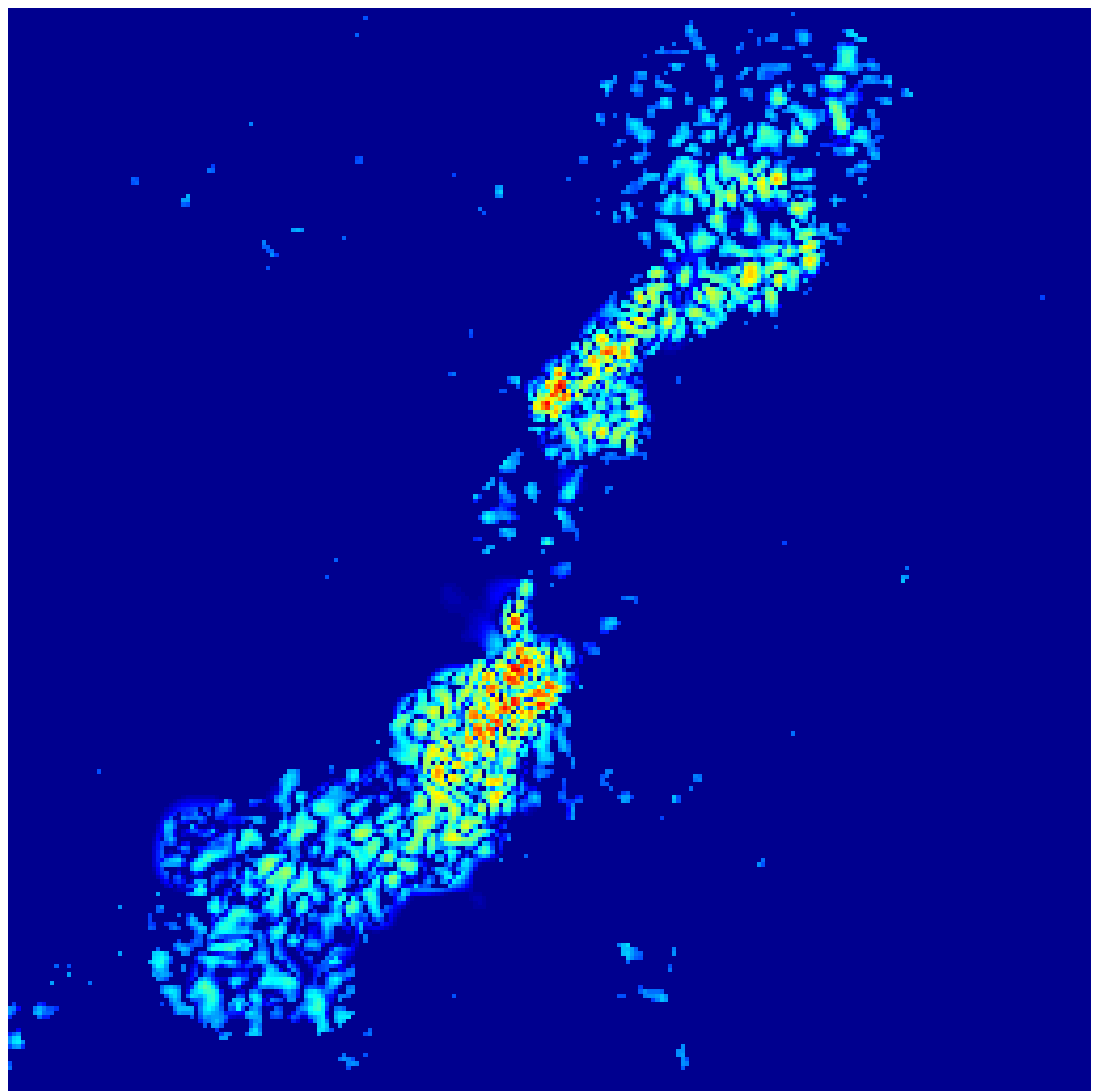} &
\hspace*{0.3cm}\includegraphics[trim ={0.2cm 0 0 0cm},clip,width=4cm,align=c]{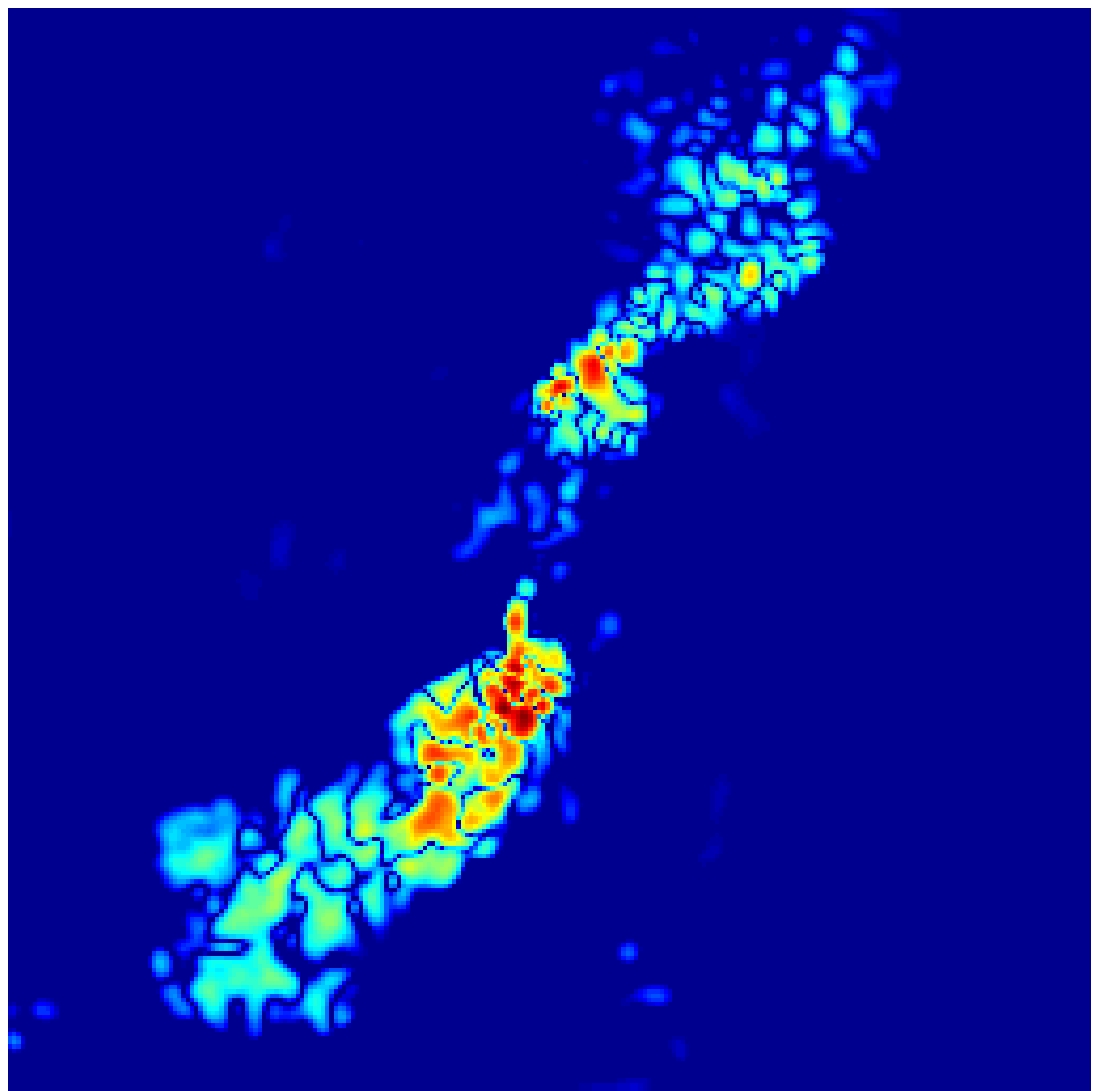} &
\hspace*{0.3cm}\includegraphics[trim ={0.2cm 0 0 0cm},clip,width=4cm,align=c]{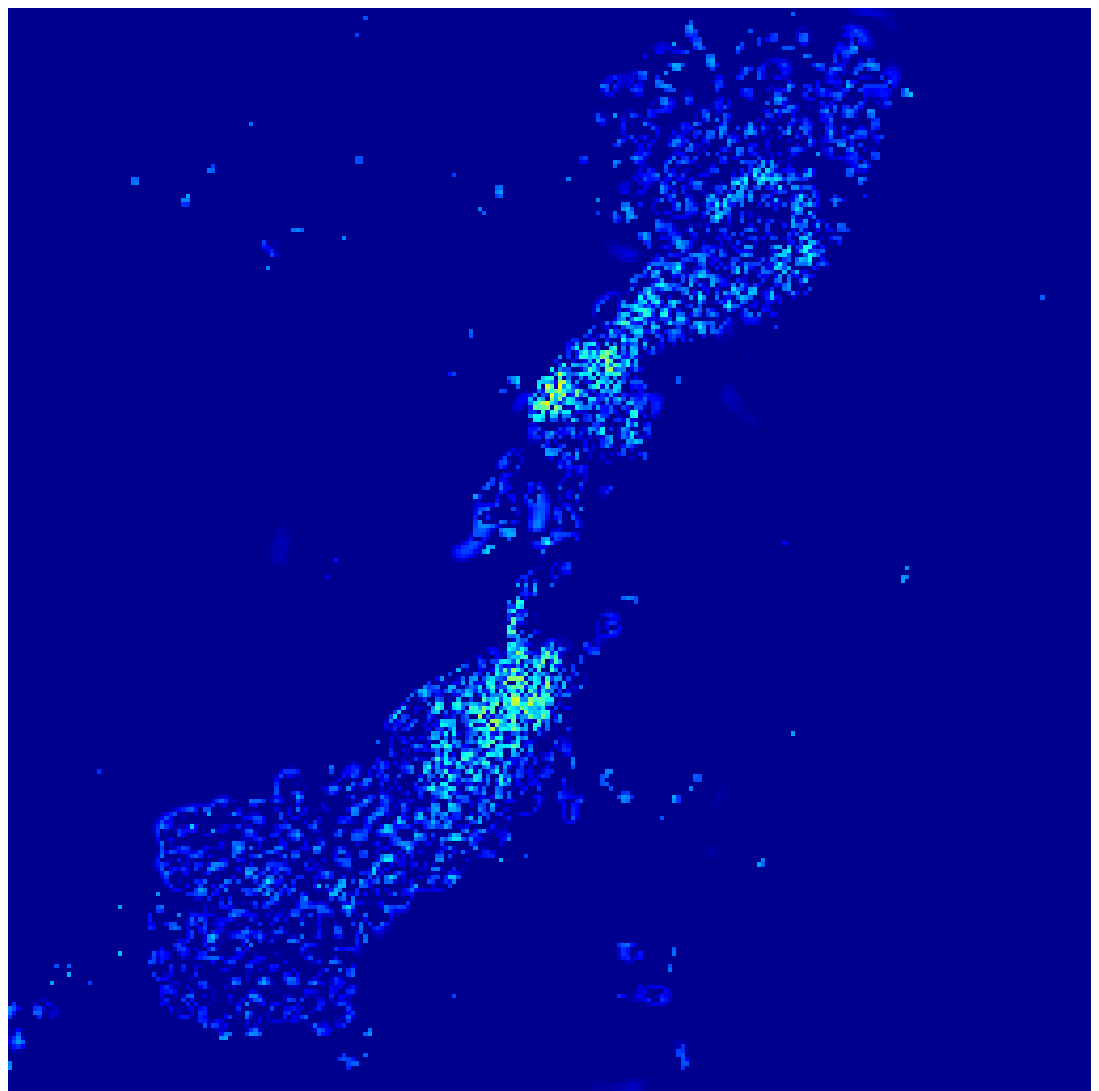} 
\end{tabular}
\caption{Hydra A Stokes $Q$ true image in first row and reconstructed images (best ones over 5 performed simulations for each case) in other rows for the cases: Imaging with normalized DIEs (second row), {Joint DIE calibration and imaging (third row)}, Joint DDE calibration and imaging excluding the off-diagonal terms (fourth row), and considering full Jones matrix (fifth row). In each case, column-wise recovered images followed by their corresponding error images are displayed when imaging is performed without polarization constraint (first two columns) and with polarization constraint (last two columns). All the images are shown in log scale, with the same color range corresponding to the colorbar given in first row.}
\label{fig:hydra5_Q}
\end{figure*}

\begin{figure*}
\centering
\begin{tabular}{@{}c@{}c@{}c@{}c@{}}
\includegraphics[trim ={0.2cm 0 0 0cm},clip,width=4cm]{true_U_hydra5.eps} &
\hspace{-3cm}\includegraphics[trim ={16.2cm 0 0 0cm},clip,width=0.97cm]{U_hydra_colorbar.eps}
& & \\
Recovered images w/o &  Absolute error images &  Recovered images with & Absolute error images \\
polarization constraint & &polarization constraint & \\
\includegraphics[trim ={0.2cm 0 0 0cm},clip,width=4cm,align=c]{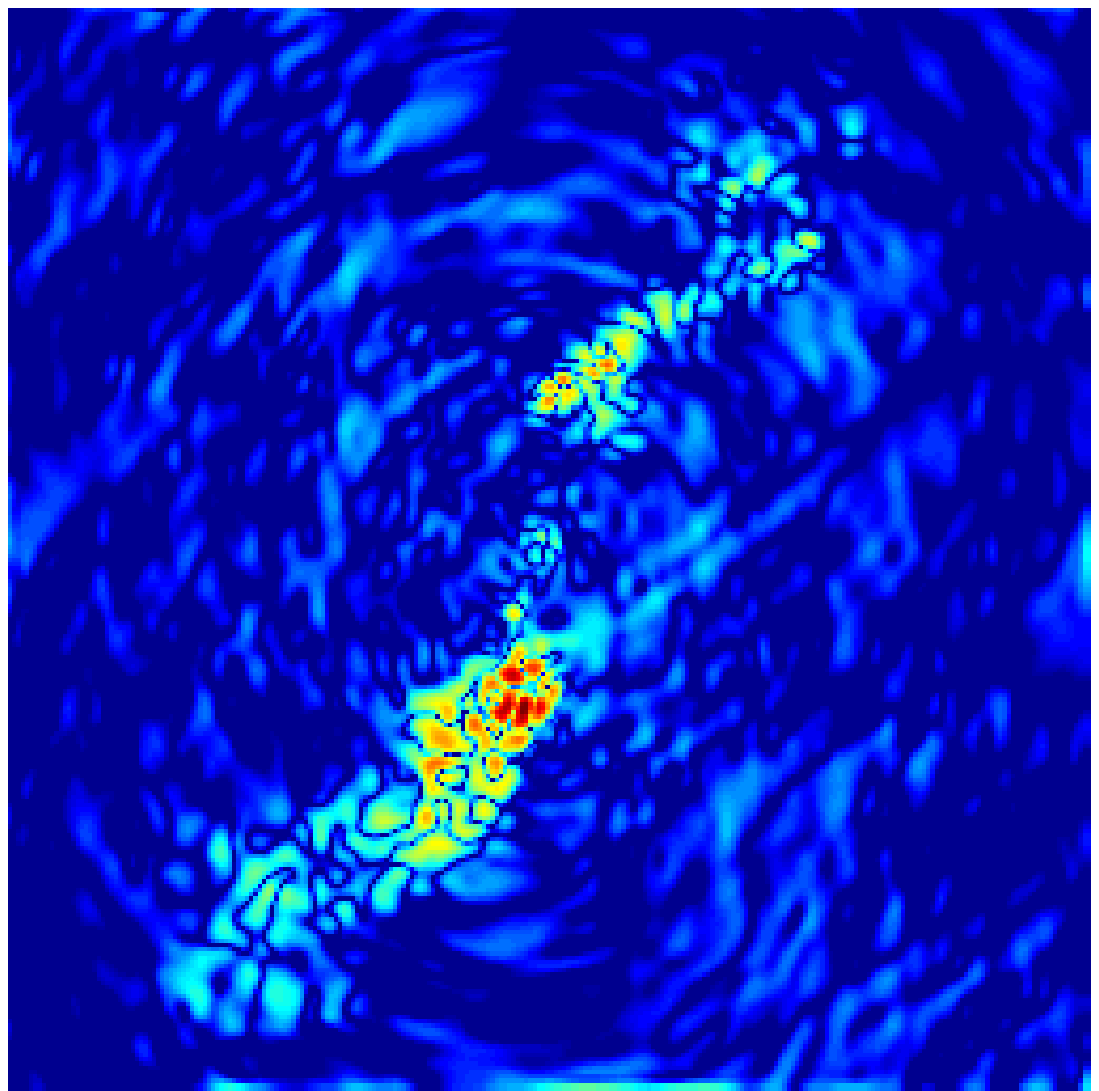} &
\hspace*{0.3cm}\includegraphics[trim ={0.2cm 0 0 0cm},clip,width=4cm,align=c]{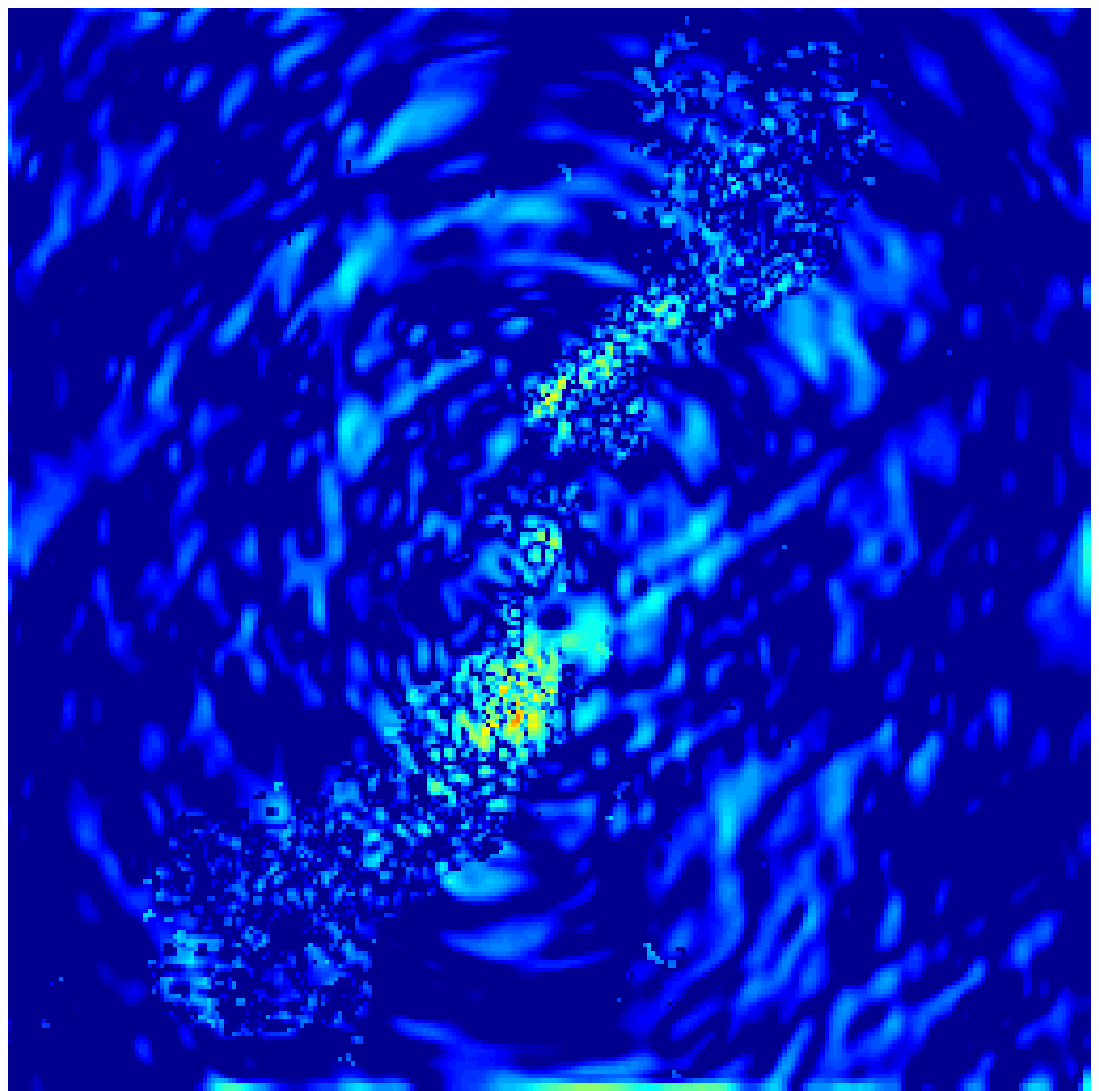} &
\hspace*{0.3cm}\includegraphics[trim ={0.2cm 0 0 0cm},clip,width=4cm,align=c]{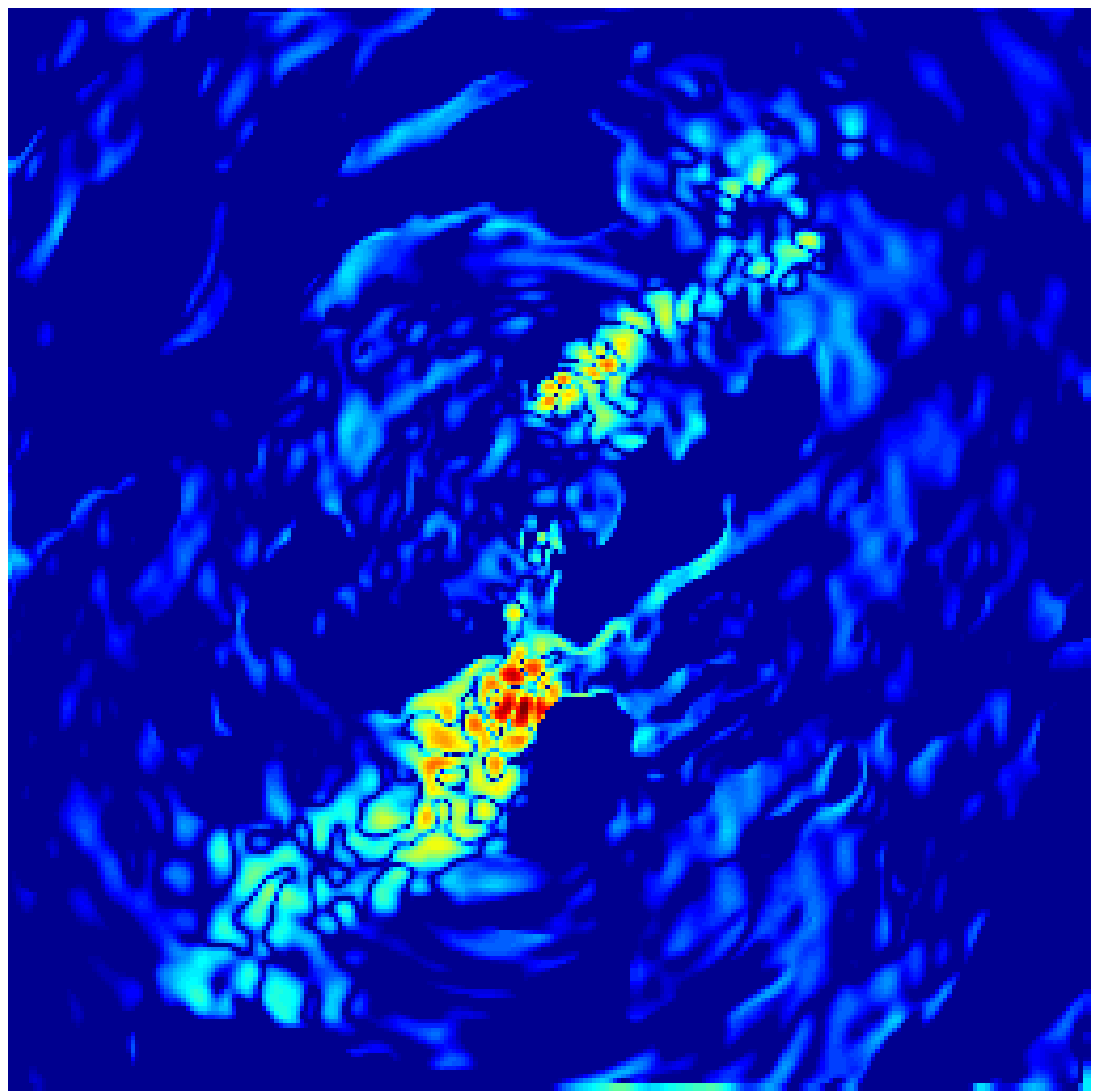} &
\hspace*{0.3cm}\includegraphics[trim ={0.2cm 0 0 0cm},clip,width=4cm,align=c]{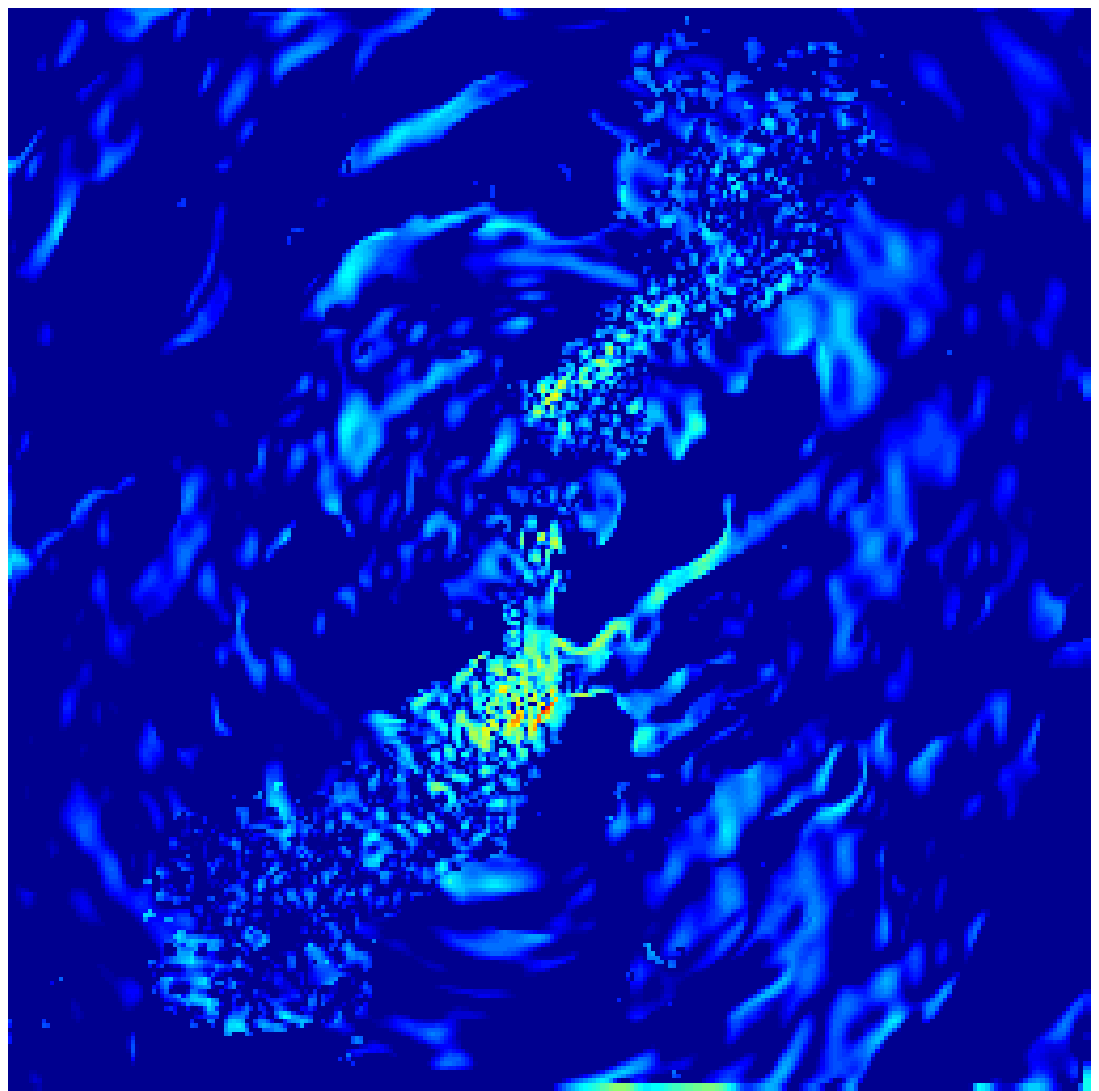} 
\vspace{0.25cm}
\\
%-----------------------------------------------
{\includegraphics[trim ={0.2cm 0 0 0cm},clip,width=4cm,align=c]{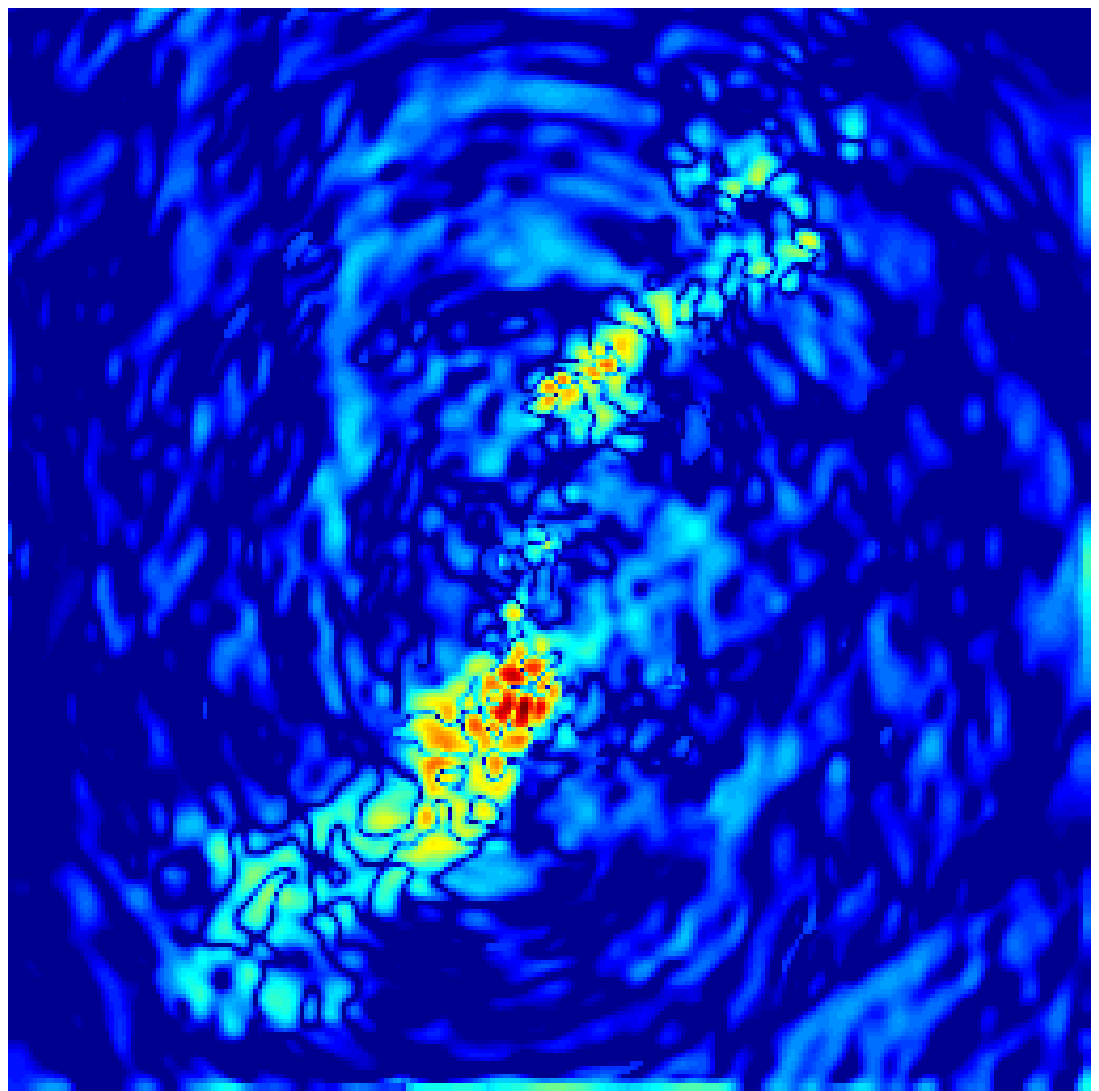}} &
\hspace*{0.3cm}\includegraphics[trim ={0.2cm 0 0 0cm},clip,width=4cm,align=c]{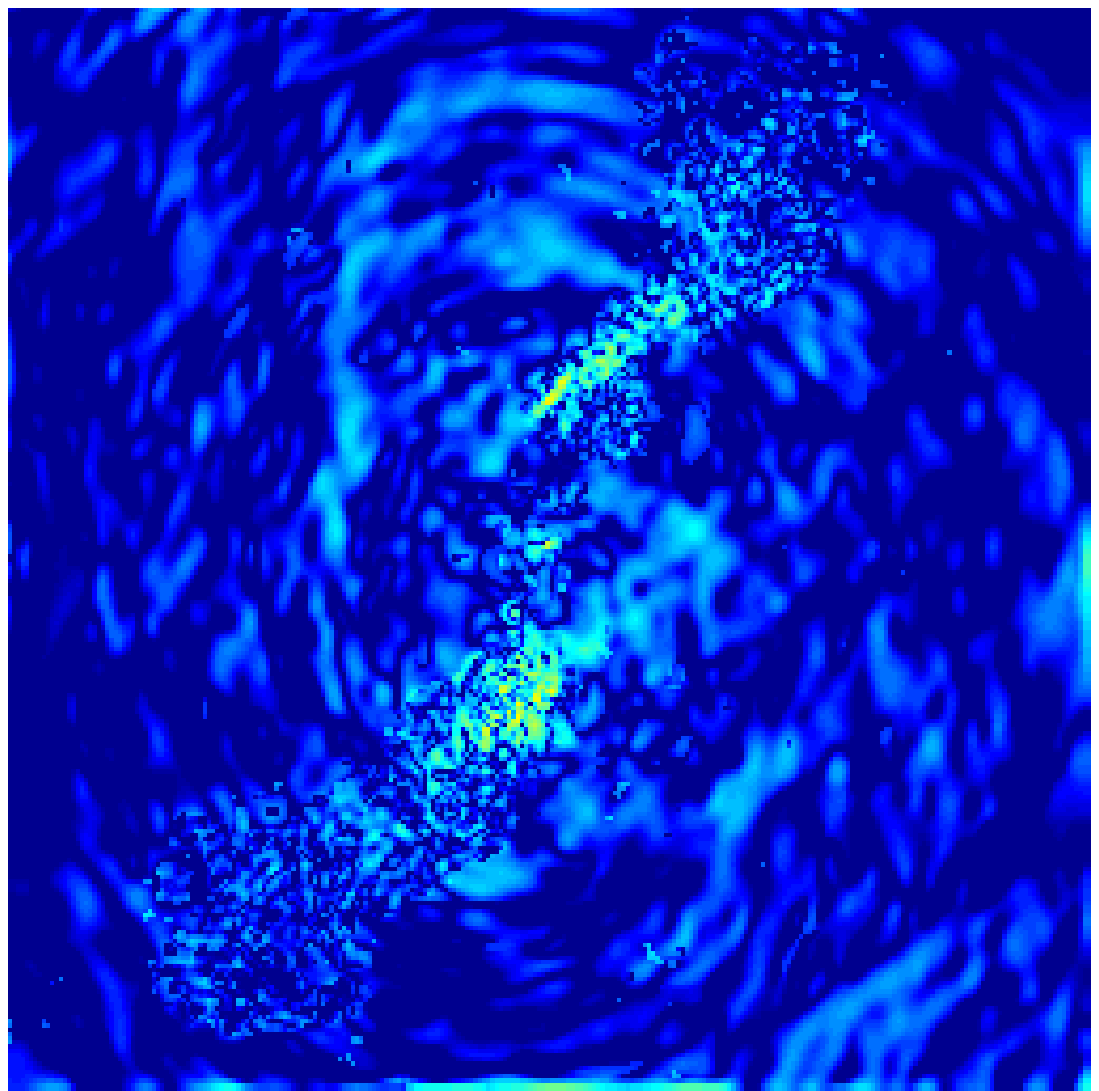} &
\hspace*{0.3cm}\includegraphics[trim ={0.2cm 0 0 0cm},clip,width=4cm,align=c]{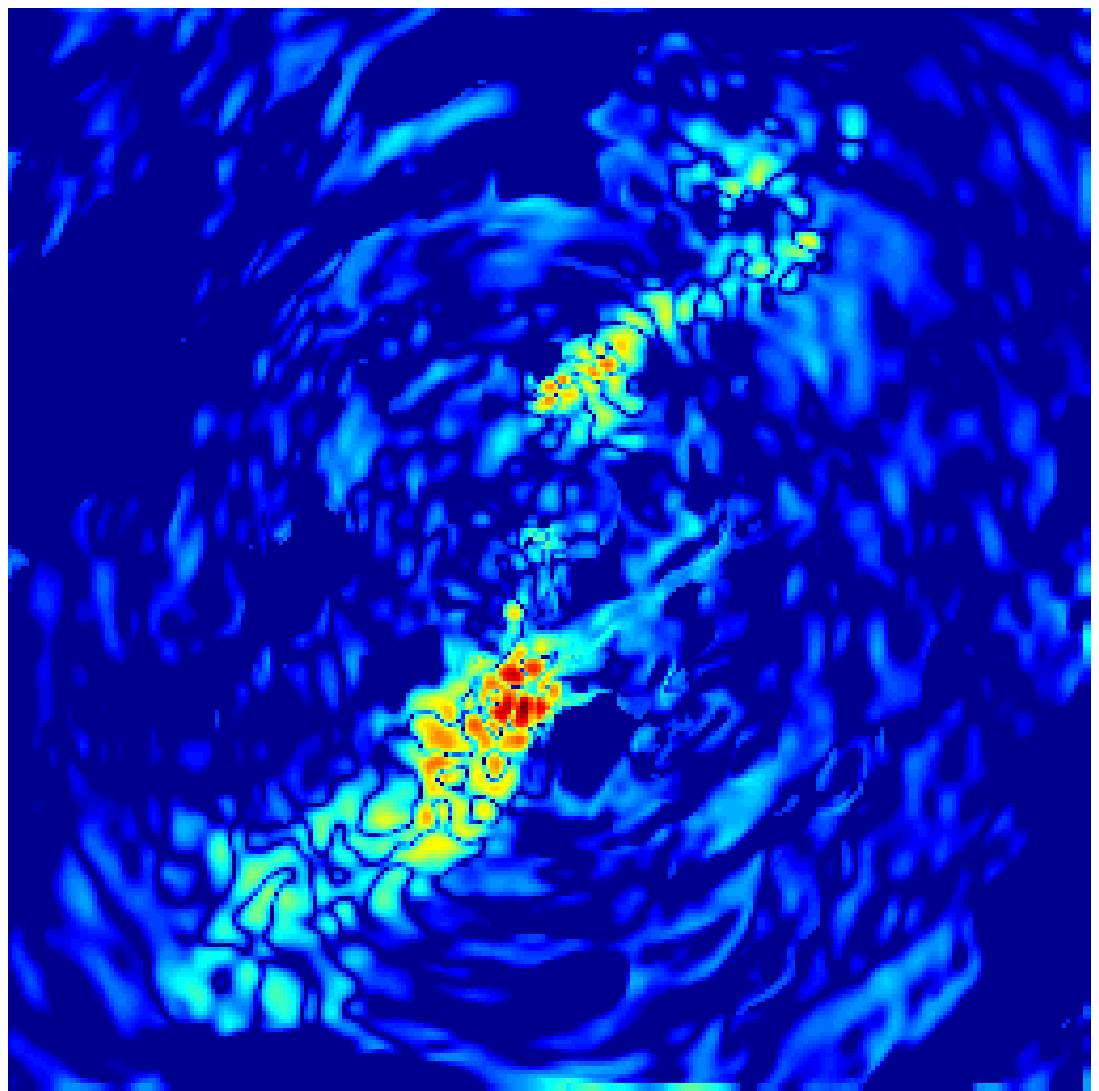} &
\hspace*{0.3cm}\includegraphics[trim ={0.2cm 0 0 0cm},clip,width=4cm,align=c]{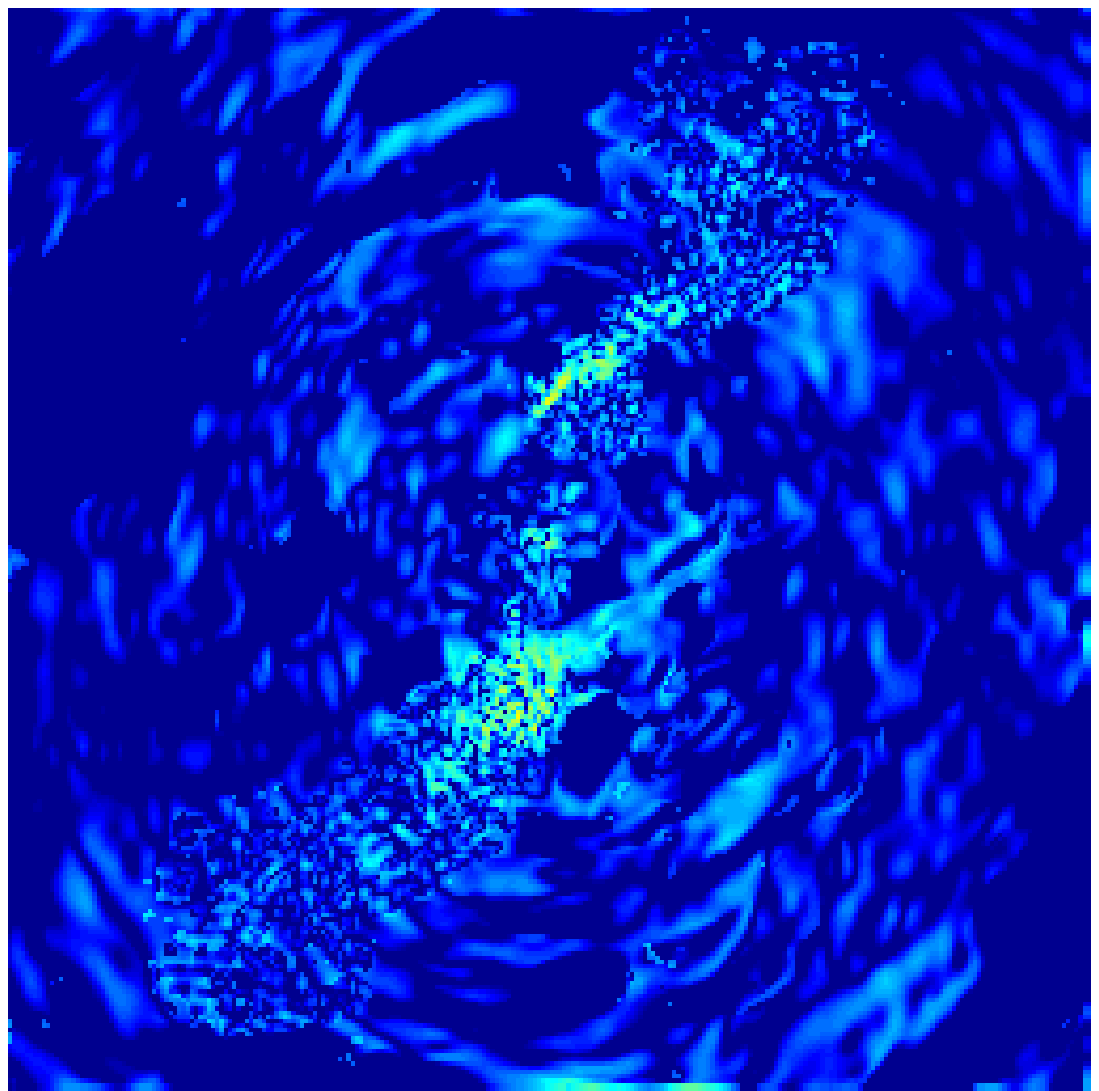} 
\vspace{0.25cm}
\\
%-----------------------------------------------
{\includegraphics[trim ={0.2cm 0 0 0cm},clip,width=4cm,align=c]{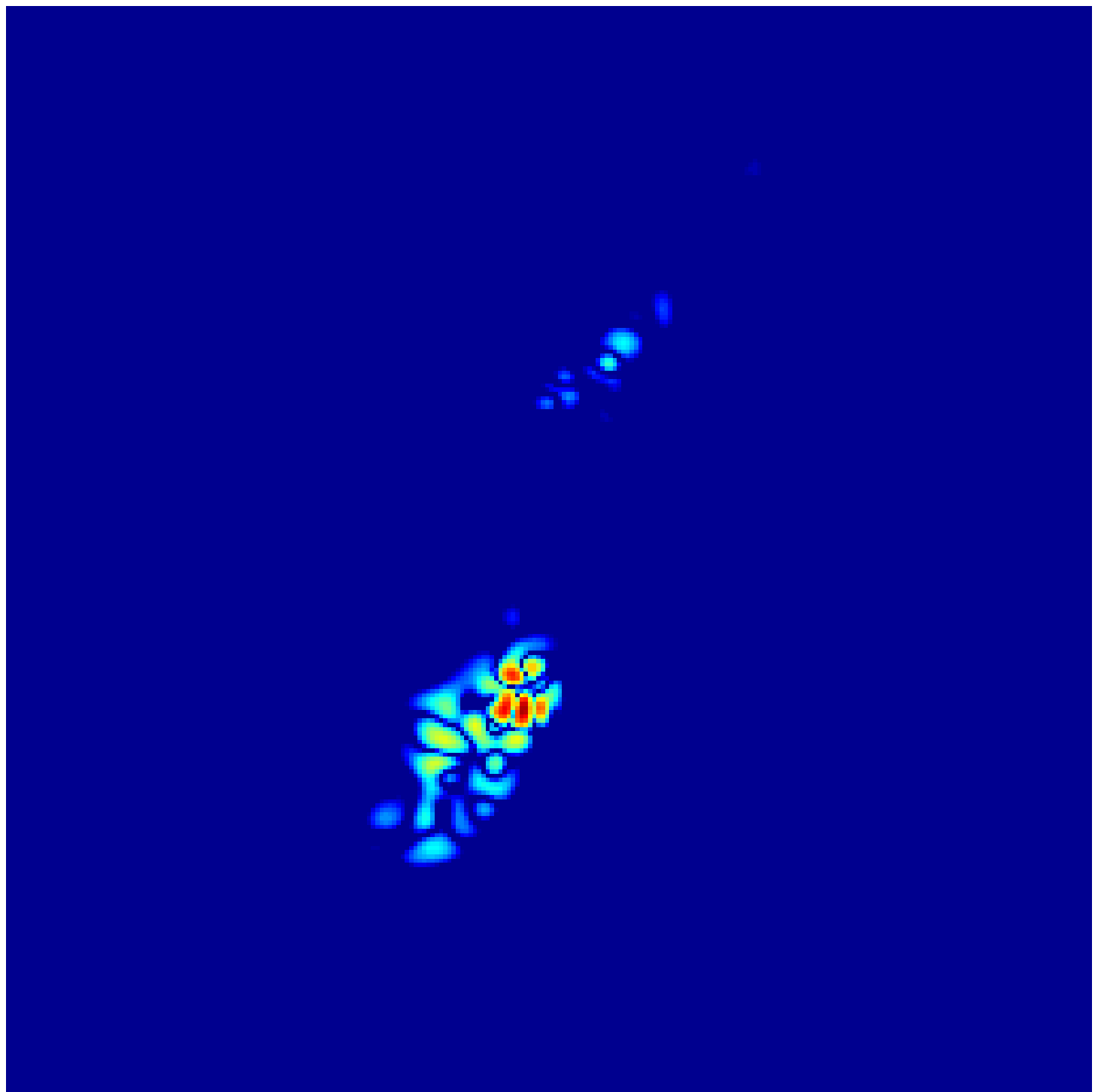}} &
\hspace*{0.3cm}\includegraphics[trim ={0.2cm 0 0 0cm},clip,width=4cm,align=c]{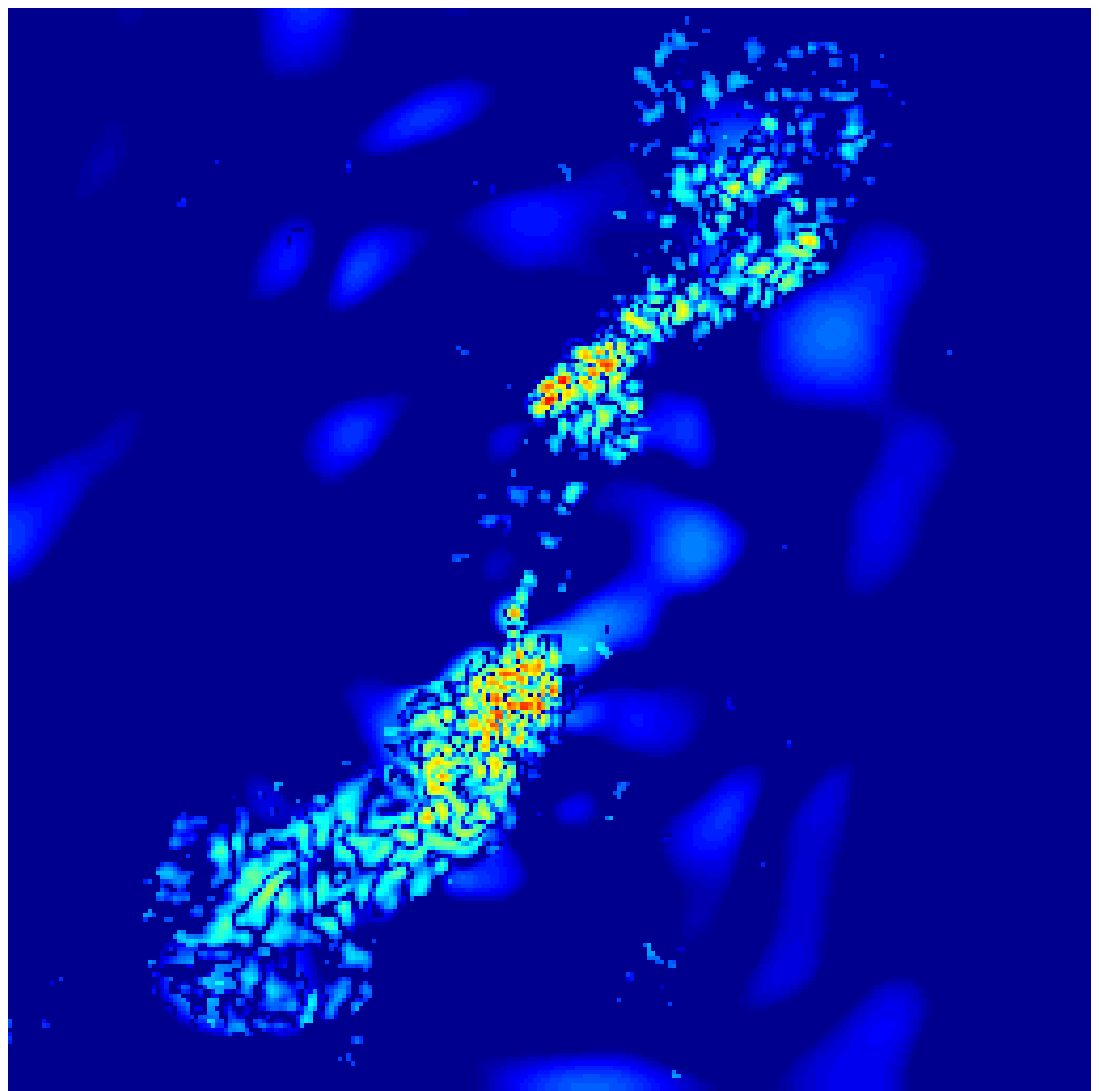} &
\hspace*{0.3cm}\includegraphics[trim ={0.2cm 0 0 0cm},clip,width=4cm,align=c]{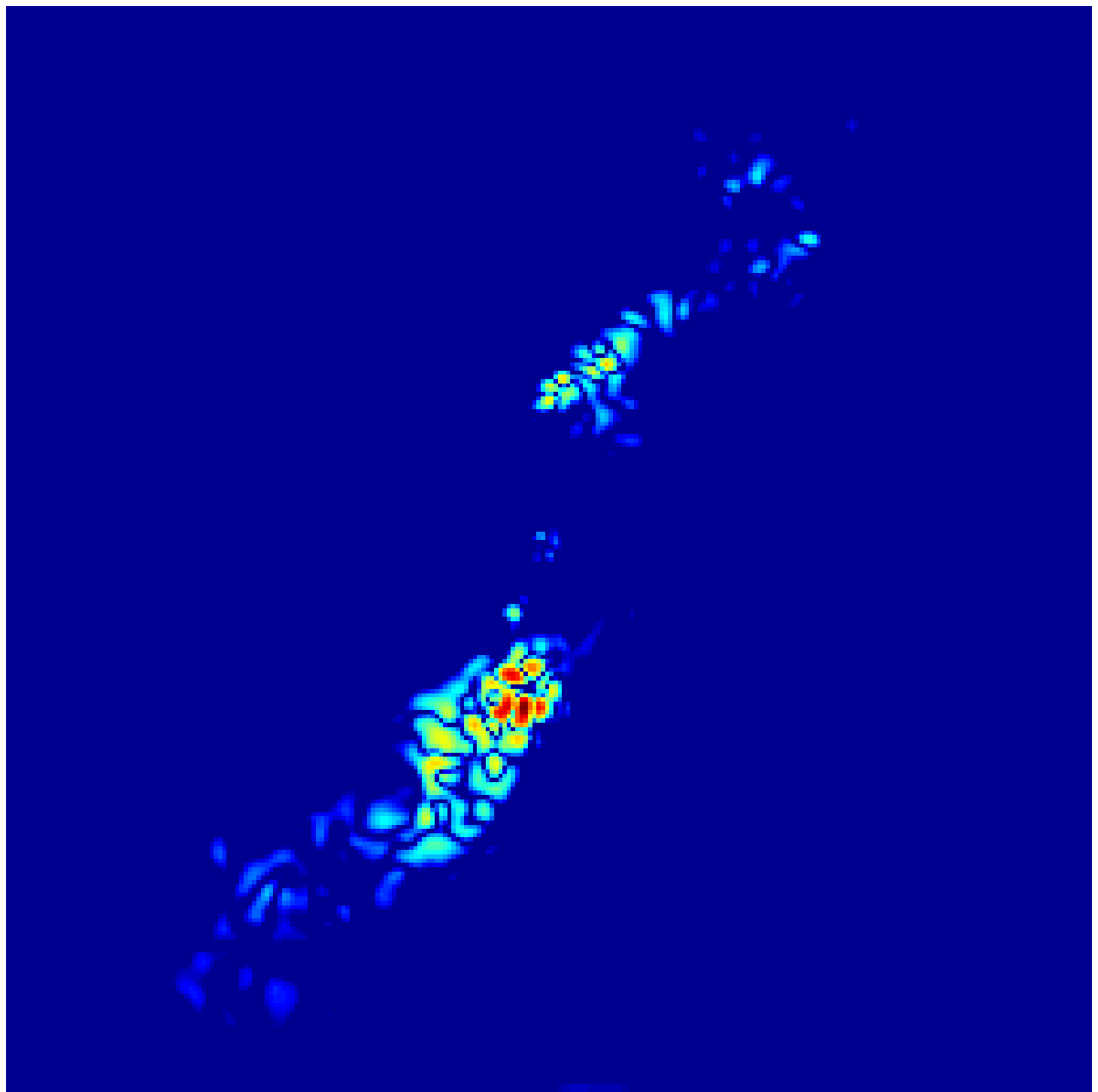} &
\hspace*{0.3cm}\includegraphics[trim ={0.2cm 0 0 0cm},clip,width=4cm,align=c]{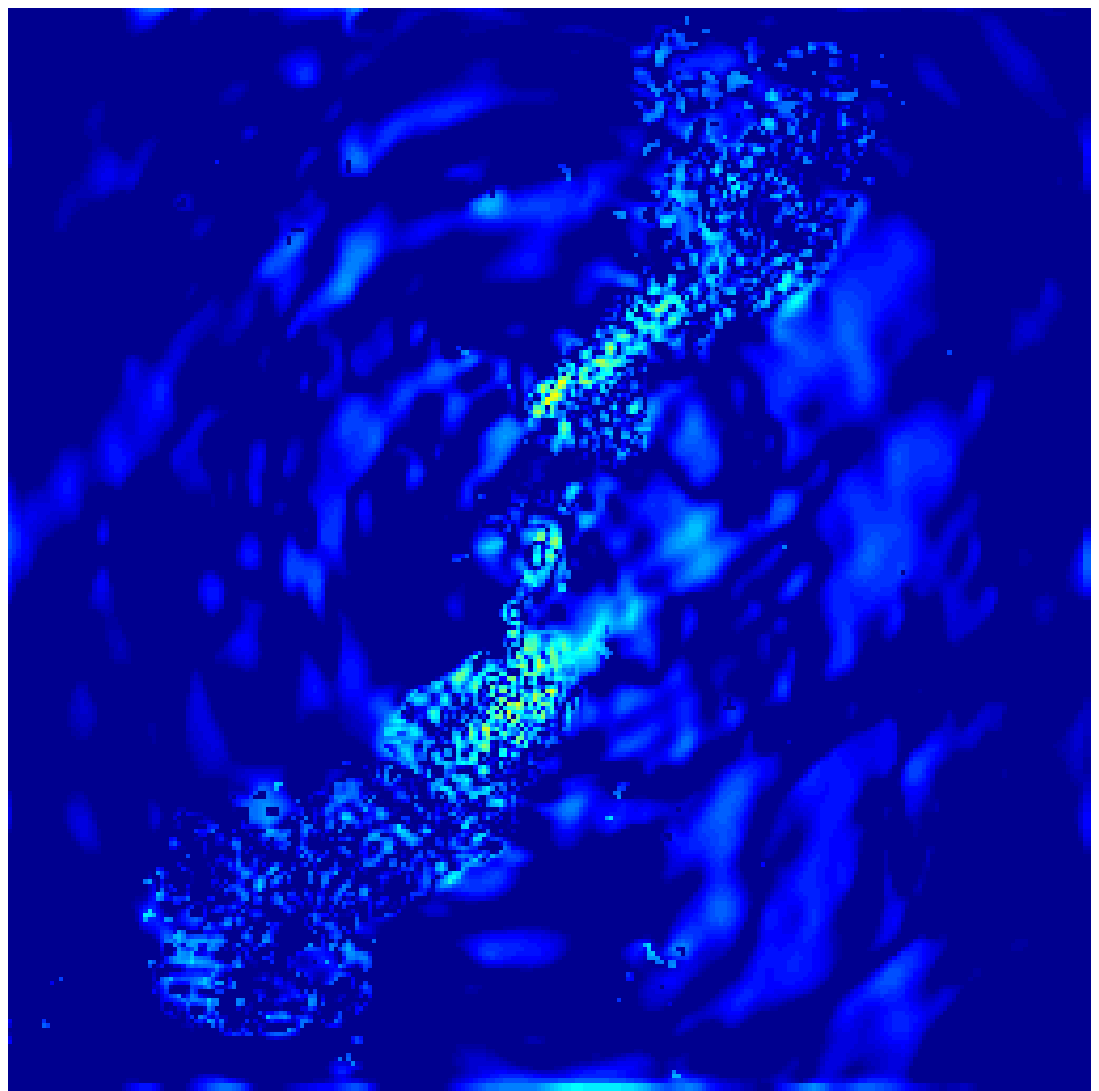} 
\vspace{0.25cm}
\\
%-----------------------------------------------
\includegraphics[trim ={0.2cm 0 0 0cm},clip,width=4cm,align=c]{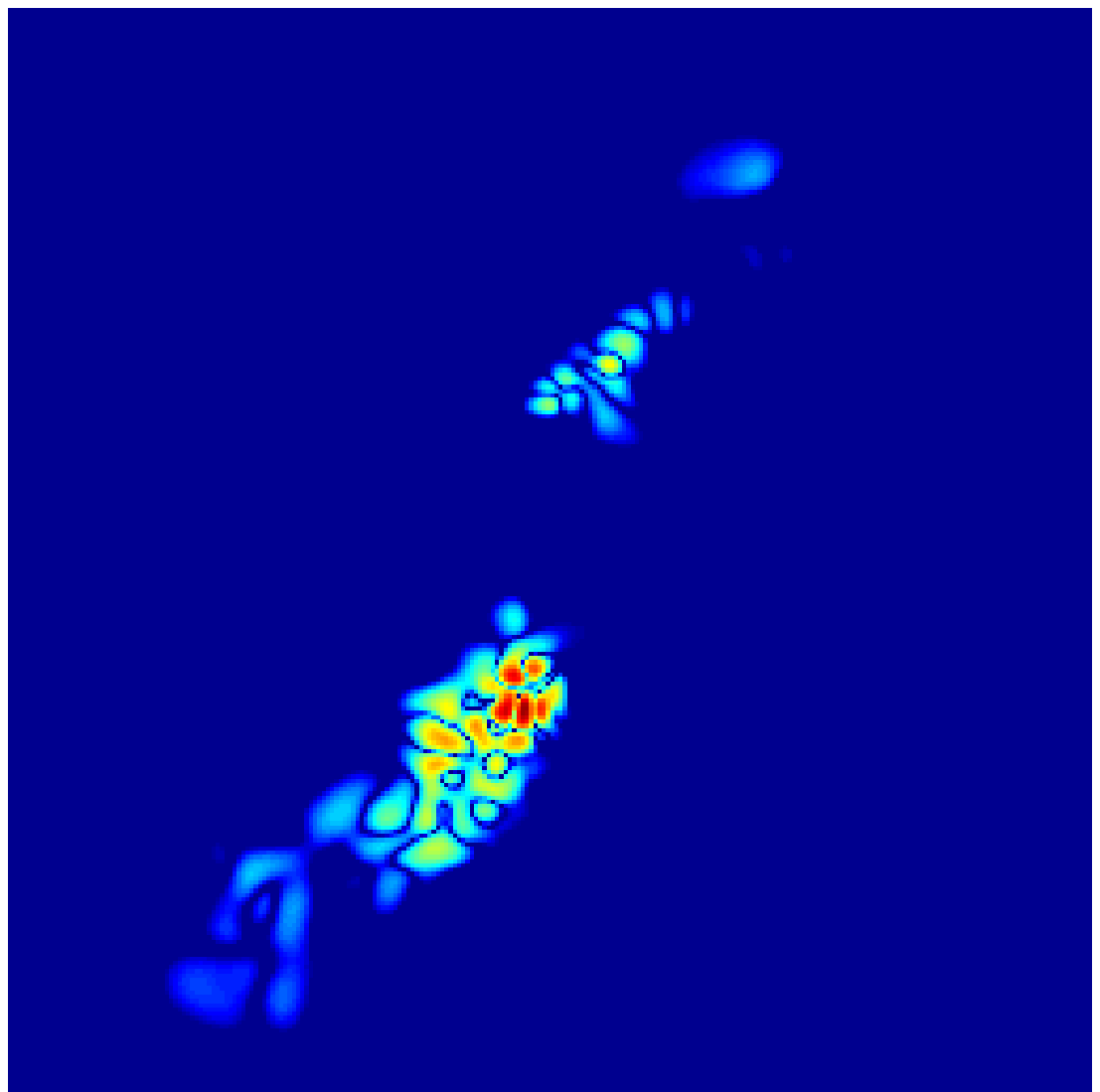} &
\hspace*{0.3cm}\includegraphics[trim ={0.2cm 0 0 0cm},clip,width=4cm,align=c]{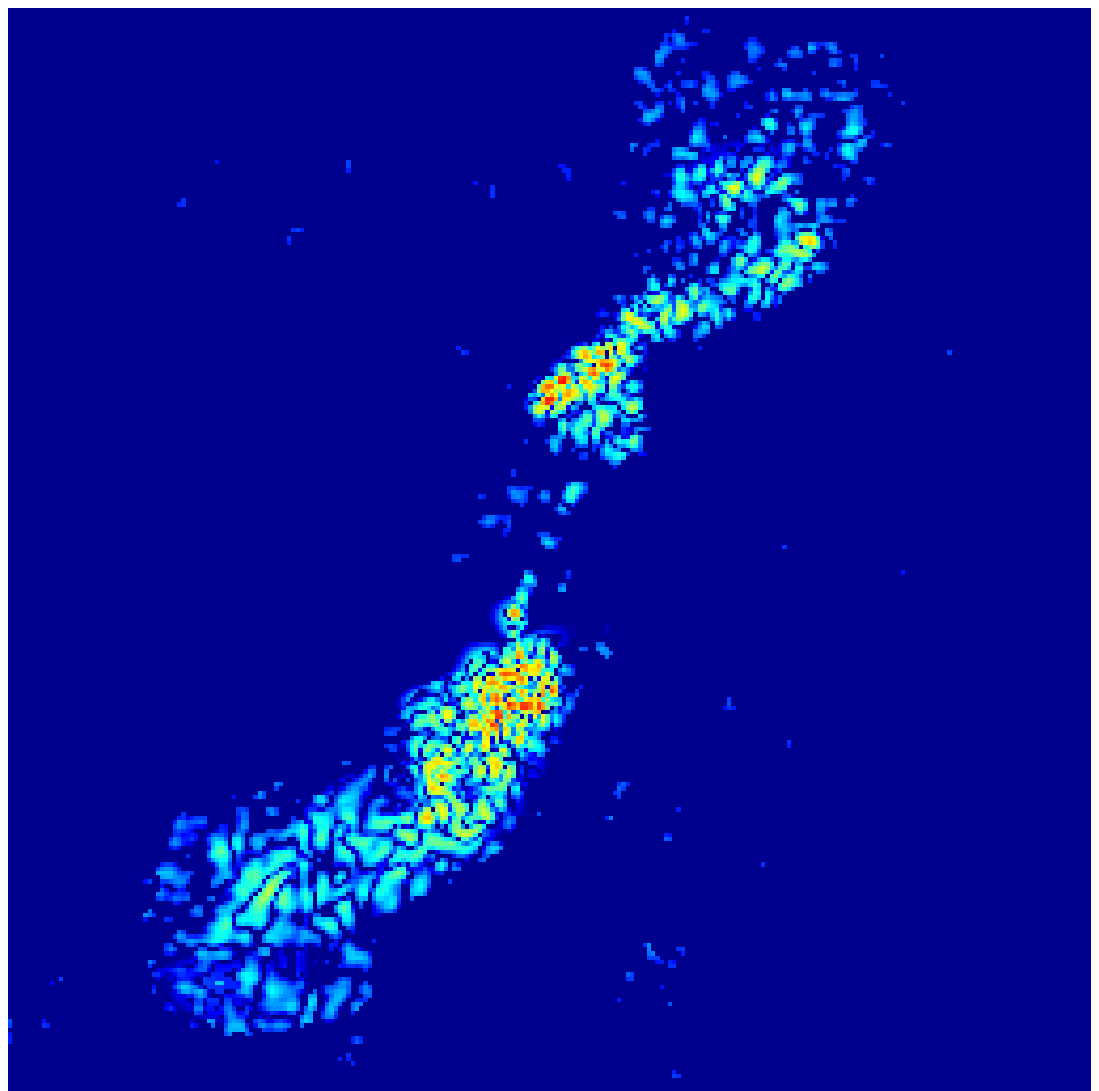} &
\hspace*{0.3cm}\includegraphics[trim ={0.2cm 0 0 0cm},clip,width=4cm,align=c]{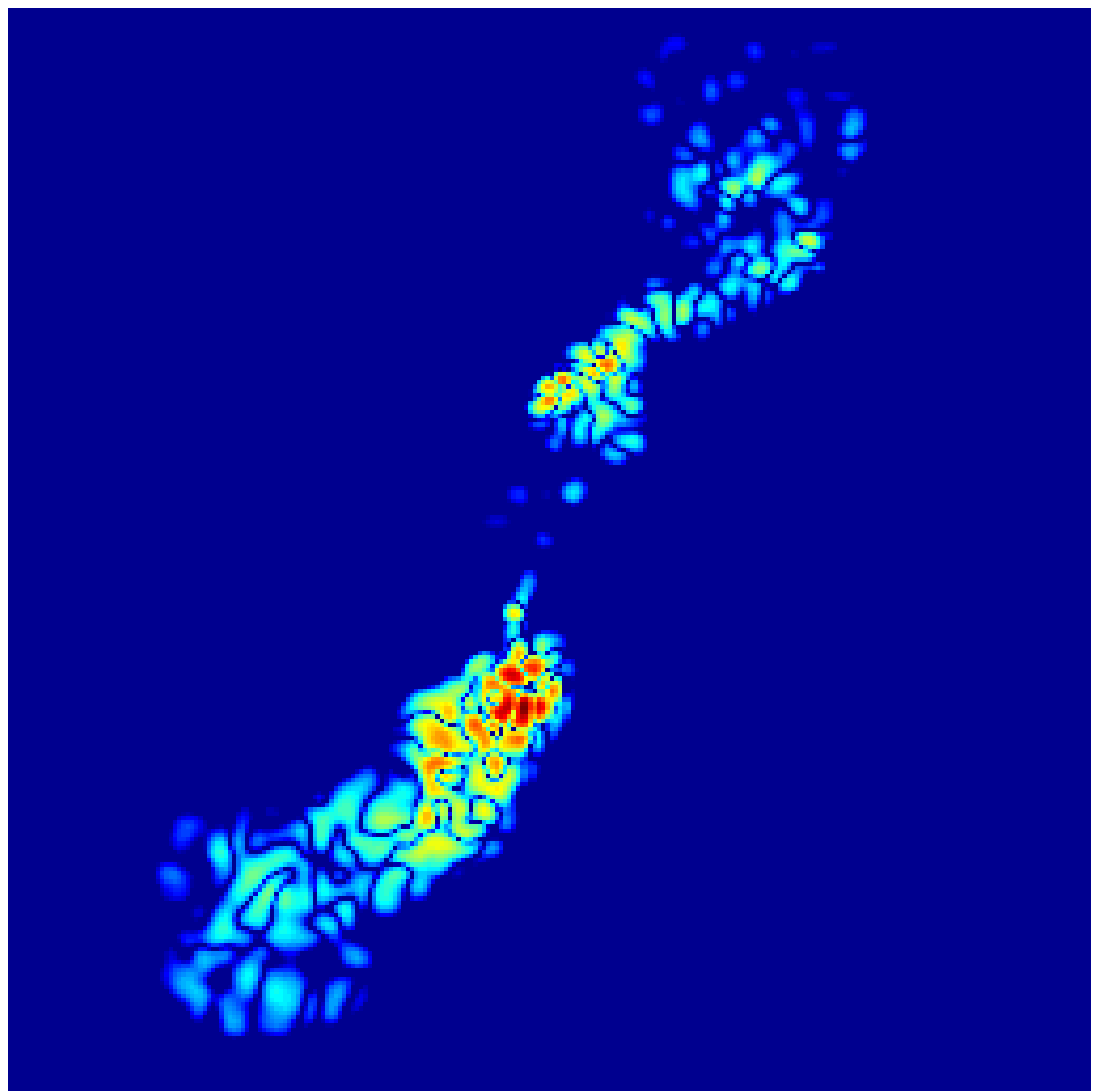} &
\hspace*{0.3cm}\includegraphics[trim ={0.2cm 0 0 0cm},clip,width=4cm,align=c]{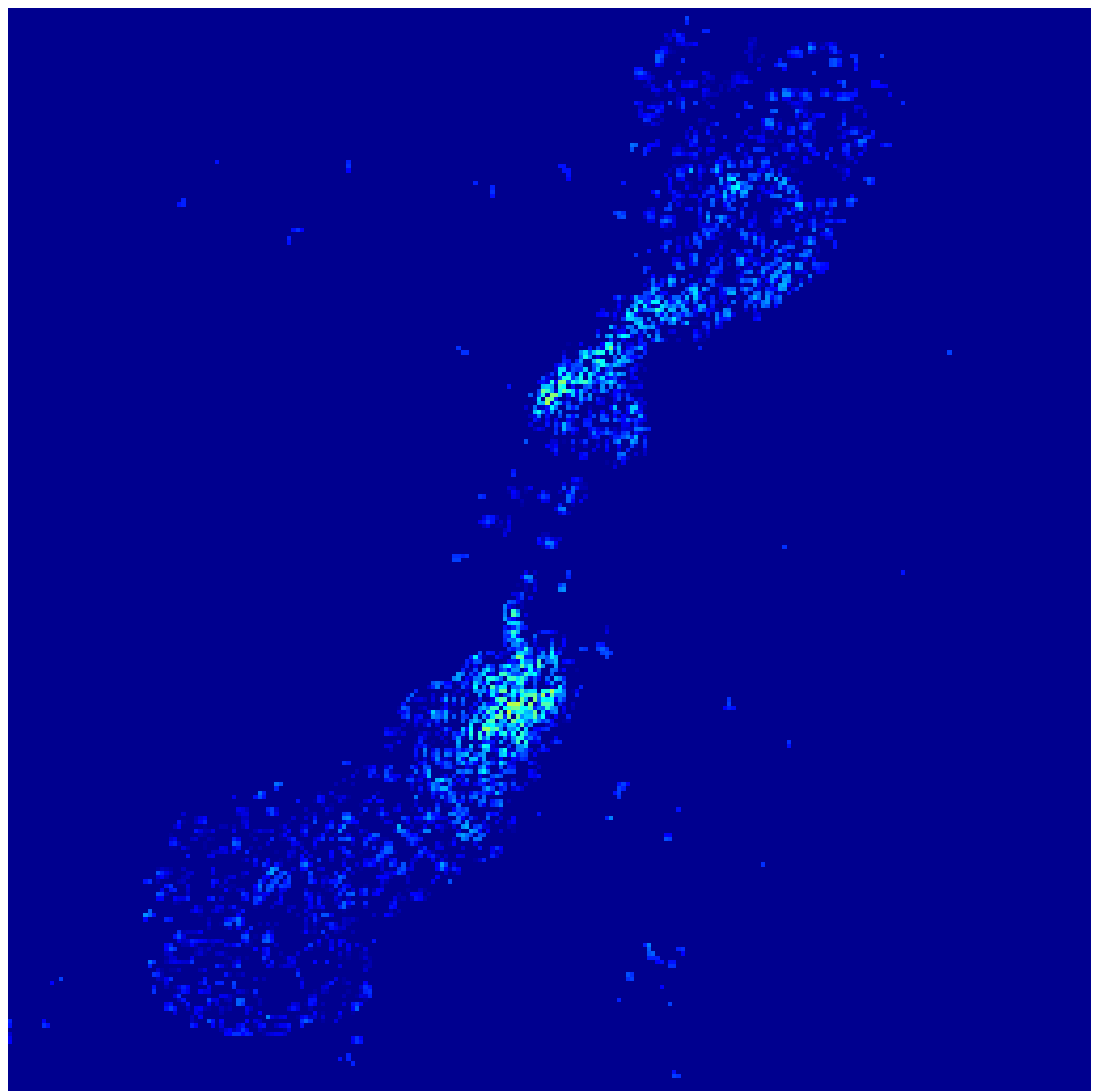} 
\end{tabular}
\caption{Hydra A Stokes $U$ true image in first row and reconstructed images (best ones over 5 performed simulations for each case) in other rows for the cases: Imaging with normalized DIEs (second row), {Joint DIE calibration and imaging (third row)}, Joint DDE calibration and imaging excluding the off-diagonal terms (fourth row), and considering full Jones matrix (fifth row). In each case, column-wise recovered images followed by their corresponding error images are displayed when imaging is performed without polarization constraint (first two columns) and with polarization constraint (last two columns). All the images are shown in log scale, with the same color range corresponding to the colorbar given in first row.}
\label{fig:hydra5_U}
\end{figure*}

The above listed observations are further supported by the visual inspection of the recovered images and the associated absolute error images, shown in Figs.~\ref{fig:cyg_a_I}-\ref{fig:hydra5_U}. While Figs.~\ref{fig:cyg_a_I},~\ref{fig:cyg_a_Q} and~\ref{fig:cyg_a_U} display the Cygnus A images for Stokes $I$, $Q$ and $U$, respectively, the corresponding images for Hydra A are shown in Figs.~\ref{fig:hydra5_I},~\ref{fig:hydra5_Q} and~\ref{fig:hydra5_U}. The error images are obtained by computing the absolute difference between the ground truth and reconstructed images. In each of these figures, the recovered images (first and third columns) followed by their absolute error images (second and fourth columns) are shown.
The shown images correspond to the reconstructions (and the associated error images) obtained when imaging is performed with: (second row) normalized DIEs, (1.a) without polarization constraint and (1.b) with polarization constraint, and  (third row) DIE calibration (2.a) without and (2.b) with the enforcement of the polarization constraint.
Similarly, the reconstructed images (and the associated error images) for the case of joint DDE calibration and imaging, excluding the off-diagonal terms and performing full Jones matrix calibration are presented respectively in third and fourth rows, both for (3.a) (and (3.c)) without polarization constraint and (3.b) (and (3.d)) with polarization constraint. It can be observed that joint DDE calibration and imaging offers remarkable advantage over considering only DIEs, mitigating the artefacts occurring because of the calibration errors. Moreover, calibration of the off-diagonal terms is crucial for high dynamic range imaging by diminishing the diffused calibration artefacts in the background. 
In addition to these remarks, the quality of reconstruction is promoted by considering the polarization constraint in the reconstruction process. While the enforcement of this constraint yields lesser residual in the error images, a careful examination of the recovered images also indicates that this prior is able to produce highly resolved images with finer details as opposed to the case when the constraint is not taken into account.

%%%%%%%  HYDRA-A: Stokes I  %%%%%%%%%%%%%%%%%

%%%%%%%%%%%%%%%%%%%%%%%%
%%%%%%%%%%%%%%%%%%%%%%%%
%%%%%%%  HYDRA-A: Stokes Q  %%%%%%%%%%%%%%%%%

%%%%%%%%%%%%%%%%%%%%%%%%

%%%%%%%%%%%%%%%%%%%%%%%%
%%%%%%%  HYDRA-A: Stokes U  %%%%%%%%%%%%%%%%%

%%%%%%%%%%%%%%%%%%%%%%%%%

%% file: conclusion.tex
\section{Conclusions}
\label{sec:conc}

We have presented a joint calibration and imaging technique taking into account the full polarization model for radio interferometry. The proposed technique, dubbed Polca SARA, unifies the estimation of the DDEs for the full Jones matrix and the Stokes images of interest within a global algorithmic structure, exploiting the same optimization framework for both calibration and imaging. In particular, it solves the underlying non-convex minimization problem employing a block-coordinate forward-backward algorithm, thereby following a forward-backward scheme for estimation of each of the variables. {The \textsc{matlab} code of the proposed method will soon be made available on GitHub (https://basp-group.github.io/Polca-SARA/).}

While our approach is shipped with convergence guarantees, it can also be adapted to incorporate suitable regularization priors for the variables under consideration. Thanks to this flexibility, for the imaging step in the global algorithm, we have employed an approach specifically developed for full Stokes imaging enforcing the physical polarization constraint \citep{Birdi2018b}. These features offered by our method are in contrast with the existing calibration and imaging algorithms in RI which (i) do not benefit from global convergence and (ii) use in fact Stokes $I$ imaging based techniques even for polarimetric imaging. The significance of the latter remark is further highlighted by the results obtained from various numerical simulations performed, achieving better SNR and higher dynamic range with the enforcement of the polarization constraint. Furthermore, in terms of the calibration, the results have shown the importance of calibrating for full Jones matrix, including DDEs and off-diagonal terms, to mitigate the artefacts appearing otherwise in the reconstructed images. Calibration of off-diagonal terms in Jones matrices, denoting the polarization leakage, has been shown to be particularly crucial to obtain high dynamic range images. 

Although in this first presentation of the proposed algorithm we have dealt with simulated data, the good performance obtained by this approach unveil it as a promising joint calibration and imaging technique in RI. Keeping this in mind, we plan to investigate its performance on real RI data sets in near future. We further note that in practice, the polarization angle deduced from reconstructed Stokes $Q$ and $U$ images can suffer from a unitary rotational ambiguity in the measurement model \citep{hamaker2000understanding, Smirnov2011}. A way to tackle this problem could be to add in the minimization problem some prior information about this rotation, that can be obtained from external calibration.

Another possible direction of work is to extend the proposed framework to account for hyperspectral data. While on the calibration part it would involve incorporation of DIEs/DDEs variation with observation frequency \citep{Dabbech2019}, the imaging part would need to be extended for wide-band polarimetric imaging, leveraging the adaptability offered by the currently developed approach.